\documentclass[aps,prb,english,twocolumn,showpacs,preprintnumbers,amsmath,amssymb,floatfix]{revtex4-1}
\usepackage[T1]{fontenc}
\usepackage[latin9]{inputenc}
\usepackage{graphicx}
\usepackage{amssymb}

\usepackage{babel}
\makeatletter

\usepackage{esint}

\numberwithin{equation}{section}

\makeatother

\usepackage{babel}
\makeatother

\begin{document}

\title{Electromagnetic normal modes and Casimir effects in layered structures}

\author{Bo E. Sernelius}

\affiliation{Division of Theory and Modeling, Department of Physics, Chemistry
and Biology, Link\"{o}ping University, SE-581 83 Link\"{o}ping, Sweden}

\email{bos@ifm.liu.se}

\begin{abstract}
We derive a general procedure for finding the electromagnetic normal modes in layered structures. We apply this procedure to planar, spherical and cylindrical structures. These normal modes are important in a variety of applications. They are the only input needed in calculations of  Casimir interactions. We present explicit expression for the condition for modes and Casimir energy for a large number of specific geometries. The layers are allowed to be two-dimensional so graphene and graphene-like sheets as well as two-dimensional electron gases can be handled within the formalism. Also forces on atoms in layered structures are obtained. One side-result is the van der Waals and Casimir-Polder interaction between two atoms.
\end{abstract}


\maketitle

\section{Introduction}

An electromagnetic normal mode is a solution to Maxwell's equations in absence of external sources; the mode feeds itself; the electromagnetic fields are self-sustained. Electromagnetic fields contain energy. In systems with macroscopic objects the modes can be divided into three groups: bulk modes, surface modes and vacuum modes. All these modes can be utilized in various applications. They can also be used to find the interaction energy in the system as caused by correlation effects. This interaction is traditionally obtained using many-body theory, often in the form of  diagrammatic perturbation theory based on Feynman diagrams.\,\cite{Mah,Fet} In many systems it is easier to obtain the same result using the normal-mode formulation instead. One can, e.g., show\,\cite{Ser} that the exchange- and correlation energy in a metal is nothing but the change in the zero-point energy of the longitudinal electromagnetic normal modes in the system when the interaction is turned on.

The bulk modes, modes confined to the interior of the objects, give rise to interaction energies, like the exchange- and correlation- energies, that are important for the stability and binding of a piece of material. 

The surface modes,\,\cite{Ser,Board,Agra} modes localized to the surface of objects or  to interfaces within or between objects, are responsible for the surface energy and surface tension, quantities that determine, e.g., wetting, the shape of flexible objects, energy of adhesion, energy of cohesion and the mechanical strength of composites. The field strengths and gradients can be very high near a surface. This may lead to catalytic effects, utilized in, e.g., catalytic converters. The surface modes are responsible for the van der Waals interaction between mesoscopic and macroscopic objects.

The vacuum modes, modes in empty space, are responsible for the Lamb shift.\,\cite{Lamb,Feyn} They are also responsible for the Casimir interaction.\,\cite{Casi} There are speculations that they might even be behind the dark energy.

Electromagnetic normal modes are very important in many scientific areas. We briefly touch upon a small selection of examples:

We have mentioned that the field strength from a normal mode at the surface of an object can be very high, leading to catalytic effects since this implies a potential possibility to reduce the energy. The field strength is extra high at edges and corners not to mention at small objects like nano-particles. The field strength can be strongly enhanced by letting the modes become populated either by raising the temperature or by using external electromagnetic fields of the right frequency.
One may stimulate the growth of noble metal colloidal particles, such as silver and gold by using a light source to populate the modes. Using monochromatic light of different laser wavelengths to irradiate an initial solution of seed crystals, the size and shape of the products can be controlled. The final size and shape is found to depend on laser wavelength and power. Nanoparticles of many different shapes may be produced. A mechanism based on a wavelength-dependent self-limiting process governed by the surface plasmon resonance controlling the photochemical reduction of particles was suggested in Ref.\,[\onlinecite{Zheng}]. A coalescence phase occurs due to strong induced optical forces.\,\cite{Hallock,Gunn} These forces show resonances at dipolar plasmon wavelengths.

For similar reasons adding nanoparticles into a chemical brew can stimulate and speed up chemical reactions.

This same effect makes the normal modes useful in sensor applications.

Due to the normal modes nano-particles can be used in medical applications. Multilayered, multifunctional nanoparticles that were assembled via a layer-by-layer technique were explored as important new systems for systemic  drug and gene delivery for tumor targeting in a study by Poon {\it et al.}\,\cite{Poon}.

A whole research field has been formed around the electromagnetic normal mode plasmon or surface plasmon viz. Plasmonics\,\cite{Barnes}
and one device developed in the field is the plasmonic solar cell. Plasmonic solar cells are a class of photovoltaic devices that convert light into electricity by using plasmons.\,\cite{Cat} Another competing device with the same purpose is the nano antenna or nantenna. A nantenna is a nanoscopic rectifying antenna, an experimental technology being developed to convert light to electric power. The idea was first proposed by Robert L. Bailey in 1972.\,\cite{Cor}

The normal modes can also cause a problem. In nano-science the van der Waals and Casimir forces are often the dominating forces and can become large. If two parts of a nano- or micro-machine come too close together they may stick and it can be very difficult or impossible to move them apart again. This is called stiction. This ends the very short list of examples showing a small fraction of all situstions where the electromagnetic normal modes are of importance.

The original formulation of the Casimir effect was for planar structures. Of interest here is a work on the van der Waals interaction in multilayer systems from 1970.\,\cite{NinPar} Progress on Casimir force calculations for other geometries has been slower in coming since these calculations are more demanding. Spherical and cylindrical geometries have naturally been objects of focus. Only in 1981 was the Casimir energy of an infinitely long perfectly conducting cylindrical shell calculated\,\cite{Naka} and the more physical but also much more involved case of a dielectric cylinder was considered only in recent years.\,\cite{Wu,Buh,Scheel,Buh2,Gorz,Ell,Drex}

The motivation for the present work is twofold: The first motivation is to give a general prescription for finding the normal modes in layered structures. This result will benefit many different scientific areas. The second is more specific, viz. to find the mode condition function which then may be inserted into the integrals giving the Casimir interaction. For small enough number of layers one may find analytical results for the mode condition function. When the number of layers increases the result quickly becomes too complicated to handle analytically. The expressions will be huge. Then the result is intended to be found numerically by using a computer.

The material is organized in the following way: Sec. \ref{CasvdW} is devoted to how the Casimir and van der Waals interactions are related to the electromagnetic normal modes of the system; Sec. \ref{basics} presents the basic formalism used to find the normal modes of layered structures; Secs. \ref{planar}, \ref{spherical}, and \ref{cylindrical} treat the three specific geometries included in this work, viz., planar, spherical and cylindrical, respectively. Each of these three sections have four subsections, A, B, C, and D. The subsections A and C contain the general non-retarded and retarded, respectively, results for the geometry at hand. The subsections B and D each contains a handful of illustrating examples. Finally, Sec. \ref{summary} is a summary and conclusion section.  The examples in subsections B and D are illustrated by figures where the schematic geometry used as input to the formalism is shown. In some figures there is a cartoon to the right showing the actual problem. In some figures there are two cartoons to the right. Then the lower cartoon shows the actual problem we address while the upper shows the layered structure we use in the derivation.

\section{\label{CasvdW}Casimir and van der Waals interactions in terms of electromagnetic normal modes}
At zero temperature the interaction energy, or Casimir energy, of a system can be expressed as the sum of the zero-point energies of all electromagnetic normal modes of the system 
\begin{equation}
E = \sum\limits_i {\frac{1}{2}\hbar {\omega _i}} .
\label{II1}
\end{equation}
(A remark is in place here: It is rather the shift of the zero-point energies when the interactions, one is concerned with, are "turned on" that should appear in the equation. See Ch.\,3 of Ref.\,[\onlinecite{Ser}].) In a simple system with a small number of well-defined modes this summation
may be performed directly.  In most cases it is more complicated. The complications can e.g.  be that the modes form continua
 or that it is difficult to find the zero-point energies explicitly.  An extension of the so-called Argument Principle\,\cite{Ser,Mahan2,vanKamp} can then be used to find the results.  

In what follows we let $z$ denote a general point in the complex frequency plane, $\omega$ a point along the real axis, and $i\xi$ a point along the imaginary axis, respectively.

\begin{figure}
\includegraphics[width=6cm]{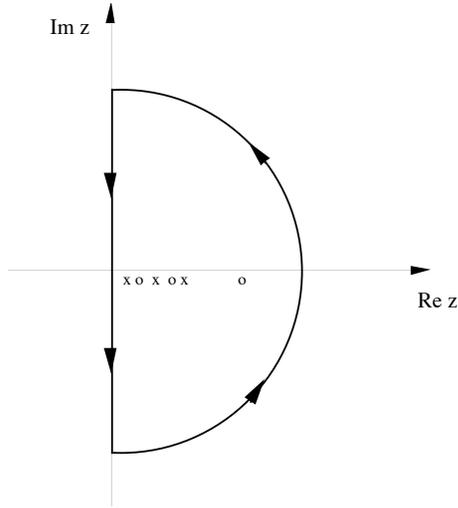}
\caption{Integration contour in the complex $z$-plane suited for zero 
temperature calculations.  Crosses and circles
are poles and zeros, respectively, of the function $f(z)$.  The radius of the
circle is let to go to infinity.}
\label{figu1}
\end{figure}

\begin{figure}
\includegraphics[width=5.5cm]{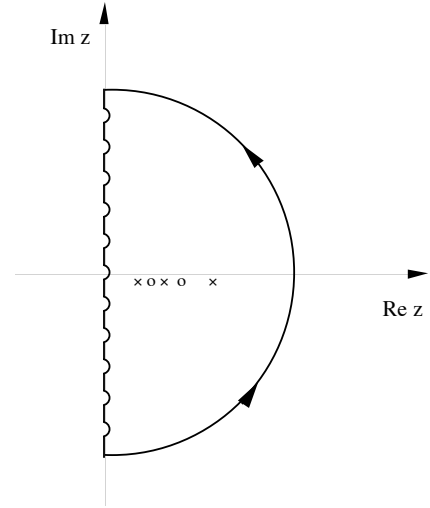}
\caption{Integration contour in the complex $z$-plane suited for finite
temperature calculations.  Crosses and circles are poles and zeros,
respectively, of the function $f(z)$.  The small semi circles are centered
at the poles of the coth function in the integrand. The radius of the
large semi circle is let to go to infinity.}
\label{figu2}
\end{figure}

Let us study a region in the complex frequency plane where
two functions are defined; one, $\varphi {\rm{(z)}}$, is
analytic in the whole region; one, $f(z)$, has poles and zeros inside the
region.  The following relation holds for an integration path around the
region:
\begin{equation}
\frac{1}{{2\pi i}}\oint {dz\varphi \left( z \right)\frac{d}{{dz}}\ln
f\left( z \right)} = \sum {\varphi \left( {z_o } \right)} - \sum {\varphi
\left( {z_\infty } \right)},
\label{II2}
\end{equation}
where $z_0 $ and $z_\infty $ are the zeros and poles, respectively, of
function $f(z)$.   If we choose the function $f(z)$ to be the function
in the defining equation for the normal modes of the system, $f\left(
{\omega _i } \right) = 0$, the function $\varphi \left( z \right)$ as
$\hbar z/2$, and let the contour enclose all the zeros 
and poles of the function $f(z)$ then Eq.\,(\ref{II2}) produces the energy
in Eq.\,(\ref{II1}).  The second term on the right hand side is just the subtraction of the zero-point energies in absence of the interactions as discussed in the remark below Eq.\,(\ref{II1}). In the original Argument Principle the function $\varphi
{\rm{(z)}}$ is replaced by unity and the right hand side then equals the
number of zeros minus the number of poles of the function $f(z)$ inside
the integration path. By using this theorem we end up with integrating along
a closed contour in the complex frequency plane.  In most cases it is
fruitful to choose the contour shown in Fig.\,\ref{figu1}.
We have the freedom to multiply the function $f(z)$ with an arbitrary
constant without changing the result on the right hand side of Eq.\,(\ref{II2}).  If we choose the constant carefully we can make the
contribution from the curved part of the contour vanish and we are only
left with an integration along the imaginary frequency axis:
\begin{equation}
 E =  \frac{\hbar }{4\pi}\int\limits_{ -
 \infty }^\infty {d\xi } \ln f\left( {i\xi } \right), 
 \label{II3}
\end{equation}
where the result was obtained from an integration by parts.  At finite
temperatures it is Helmholtz' free energy,
\begin{equation}
\begin{gathered}
\mathfrak{F} = \sum\limits_i {\tfrac{1} {2}\hbar \omega _i \left( r \right)
+ \tfrac{1} {\beta }\ln \left( {1 - e^{ - \beta \hbar \omega _i \left( r
\right)} } \right)} \hfill \\ 
\quad = \sum\limits_i
{\tfrac{1} {\beta }\ln \left( {2\sinh \tfrac{1} {2}\beta \hbar \omega _i }
\right)}, \hfill \\
\end{gathered} 
\label{II4}
\end{equation}
that is of interest.
Also here we may use the generalized Argument Principle
but now with $\ln \left[ {2\sinh \left( {\beta \hbar z/2} \right)} \right]/\beta $ instead of $\hbar z{\rm{/}}2$ for
 $\varphi {\rm{(z)}}$ in the integrand.\,\cite{NinParWei,Ser}  There is one complication. 
 This new function has poles of its own in the complex frequency plane.  We
 have to choose our contour so that it includes all poles and zeros of the
 function $f(z)$ but excludes the poles of $\varphi {\rm{(z)}}$.  The poles of
 function $\varphi {\rm{(z)}}$ all fall on the imaginary frequency axis.  The same contour as in Fig.\,\ref{figu1}, is used but now we let the straight
 part of the contour lie just to the right of, and infinitesimally close
 to, the imaginary axis.  We have
\begin{equation}
\begin{gathered}
  \mathfrak{F} = \frac{1}
{{2\pi i}}\int\limits_\infty ^{ - \infty } {d\left( {i\xi } \right)\frac{1}
{\beta }\ln \left( {2\sinh \tfrac{1}
{2}\beta \hbar i\xi } \right)\frac{d}
{{d\left( {i\xi } \right)}}\ln f\left( {i\xi } \right)}  \hfill \\
  \quad  = \frac{\hbar }
{{4\pi}}\int\limits_{ - \infty }^{ + \infty } {d\xi \coth \left( {\tfrac{1}
{2}\beta \hbar i\xi } \right)\ln f\left( {i\xi } \right)}.  \hfill \\ 
\end{gathered} 
\label{II5}
\end{equation}
The coth function has poles on the imaginary $z$-axis and they should not be
inside the contour.  The poles are at 
\begin{equation}
{z_n} = i{\xi _n} = i\frac{{2\pi n}}{{\hbar \beta }};n = 0, \pm 1, \pm 2, \ldots ,
\label{II6}
\end{equation}
and all residues are the same, equal to $2/\hbar \beta $. The
integration is performed along the imaginary axis and the path is deformed along small
semicircles around each pole.  The integration path is illustrated in Fig.\,\ref{figu2}.  The 
integration along the axis results in zero since the
integrand is odd with respect to $\xi$.  The only surviving
contributions are the ones from the small semicircles.  The result is
\begin{equation}
\begin{gathered}
\mathfrak{F} = \frac{\hbar } {{4\pi i}}\sum\limits_{\xi _n }
{\frac{{2\pi i}} {2}\frac{2} {{\hbar \beta }}} \ln f\left( {i\xi _n }
\right) \hfill \\ \quad = \frac{1} {{2\beta }}\sum\limits_{\xi _n } {\ln
f\left( {i\xi _n } \right)} ; \xi _n = \frac{{2\pi n}} {{\hbar \beta
}} ; n = 0, \pm 1, \pm 2, \ldots \hfill \\
\end{gathered} 
\label{II7}
\end{equation}
Since the summand is even in $n$ we can write this as
\begin{equation}
\mathfrak{F} = \frac{1}
{\beta }\sum\limits_{\xi _n } {^{'}\ln f\left( {i\xi _n } \right)} ;
\xi _n = \frac{{2\pi n}} {{\hbar \beta }} ; n = 0,1,2, \ldots ,
\label{II8}
\end{equation}
where the prime on the summation sign indicates that the $n = 0$ term is 
multiplied by a factor of one half.  This factor of one half is because
there is only one term with $\left| n \right| = 0$ in the original summation but two for
all other integers.  When the temperature goes to zero the spacing between
the discrete frequencies goes to zero and the summation may be replaced by
an integration:
\begin{equation}
\begin{gathered}
\mathfrak{F} = \frac{1} {\beta }\sum\limits_{\xi _n }
{^{'}\ln f\left( {i\xi _n } \right)} \xrightarrow[{T \to
0}]{}\frac{{\hbar \beta }} {{2\pi }}\frac{1} {\beta }\int\limits_0^\infty
{d\xi } \ln f\left( {i\xi} \right) \hfill \\ \quad = \hbar
\int\limits_0^\infty {\frac{{d\xi }} {{2\pi }}} \ln f\left( {i\xi
} \right) = E, \hfill \\
\end{gathered}
\label{II9}
\end{equation}
and we regain the contribution to the internal energy from the interactions,
the change in zero-point energy of the modes.

To summarize so far, at zero temperature the internal interaction energy is obtained from Eq.\,(\ref{II3}) and at finite temperature the Helmholtz free interaction energy is obtained from Eq.\,(\ref{II8}). The only input from the system is the mode condition function,  $f(z)$.
Casimir forces, pressures, surface tensions, works of adhesion and cohesion, and so on are obtained from how these energies vary when parameters of the system are changed. 

All derivations of the mode conditions are simplified if retardation effects are neglected. The forces obtained if this is done are van der Waals forces. If retardation is included in all steps the result span the whole separation region covering both Casimir and van der Waals forces. Since the non-retarded derivations are so much simpler to perform, the results so much simpler to handle and since the distances in the system often are small enough for retardation effects to be negligible we derive the results both without and with retardation effects included. The treatment in this work is limited to objects of a certain class of geometrical shape. One of the coordinates of a proper chosen coordinate system should be constant at the interface between two media. There are eleven coordinate systems in which the Helmholtz equation is separable so there are quite a few shapes where the treatment is applicable. One should note that the thickness of the layers are not constant in all geometries. They are in the three specific geometries that we apply the theory to here.

\section{\label{basics}General layered structures}
\begin{figure}
\includegraphics[width=8cm]{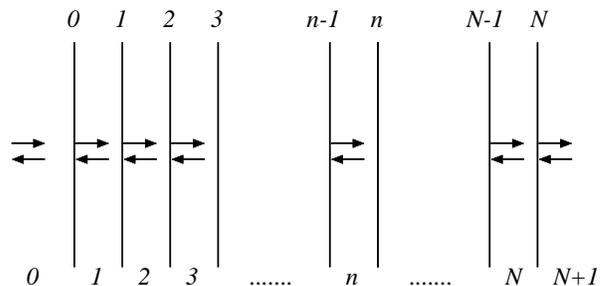}
\caption{Schematic illustration of the layered structure. The numbering of the $N+2$ media are indicated at the bottom of the figure and the numbering of the $N+1$ interfaces at the top. In the general solution of Maxwell's equations  there are one wave moving towards the right and one towards the left inside each medium. If a normal mode is excited there is no wave moving to the right in medium $0$ which is the ambient medium; if the medium number  $N+1$ is unlimited there is no wave moving to the left in that medium. See the text for more details.}
\label{figu3}
\end{figure}

Let the object we study have $N$ layers. A layer is a region bounded by two interfaces. In the system there are two more regions, each with just one boundary, a boundary in common with one of the layers. Of these two we choose the ambient to be the neighbor to layer number $1$. Thus there are $N$ layers, $N+1$ interfaces and  $N+2$ media. This is illustrated in Fig\,\ref{figu3}. The layers are numbered from $1$ to $N$, the media from $0$ to $N+1$ and the interfaces from  $0$ to $N$. This means that layer number $n$ is filled with medium number $n$ and interface number $n$ is the interface to the right of layer number $n$.  In the general solution of Maxwell's equations  there are one wave moving towards the right and one towards the left inside each medium. If a normal mode is excited there is no wave moving to the right in medium $0$ which is the ambient medium; if the medium number  $N+1$ is unlimited there is no wave moving to the left in that medium. We have here somewhat extended the concept of moving. When we say that a wave moves in a direction it either really moves or its amplitude decreases in that direction. In the retarded treatment there are Transverse Electric (TE) and Transverse Magnetic (TM) modes. In the planar and spherical geometries these are not mixed when crossing an interface. Then we may solve for these mode types separately. In other geometries like the circular cylindrical they do mix. Then we will have two modes, one TE and one TM, moving towards the right and two moving towards the left in each medium.

Let us start with the general procedure when the TE- and  TM-modes do not mix. We denote the variable that is constant on each interface by $x$. Then in a general medium $n$ we have the wave $a^n R\left( x \right) + b^n L\left( x \right)$. The boundary conditions at each interface are the standard ones that the tangential components of  $\bf E$ and $\bf H$ and the normal components of $\bf D$ and $\bf B$ are continuous across the interface. Only two are needed; the other two lead to redundant results. Making use of the boundary conditions at interface $n$ gives 
\begin{equation}
{\bf{\tilde A}}_n \left( {x_n } \right)\cdot\left( {\begin{array}{*{20}c}
   {a^n }  \\
   {b^n }  \\
\end{array}} \right) = {\bf{\tilde A}}_{n + 1} \left( {x_n } \right) \cdot \left( {\begin{array}{*{20}c}
   {a^{n + 1} }  \\
   {b^{n + 1} }  \\
\end{array}} \right), \\
\label{III1}
\end{equation}
where ${\bf{\tilde A}}_n$ is a two by two matrix which depends on the dielectric function in medium $n$. Operating from the left with the inverse of this matrix gives
\begin{equation}
\left( {\begin{array}{*{20}c}
   {a^n }  \\
   {b^n }  \\
\end{array}} \right) = {\bf{\tilde M}}_n \cdot\left( {\begin{array}{*{20}c}
   {a^{n + 1} }  \\
   {b^{n + 1} }  \\
\end{array}} \right),
\label{III2}
\end{equation}
where
\begin{equation}
{\bf{\tilde M}}_n  = {\bf{\tilde A}}_n ^{ - 1} \left( {x_n } \right)\cdot{\bf{\tilde A}}_{n + 1} \left( {x_n } \right).
\label{III3}
\end{equation}
We may now find a relation between the coefficients in the left-most and right-most media
\begin{equation}
\left( {\begin{array}{*{20}c}
   {a^0 }  \\
   {b^0 }  \\
\end{array}} \right) = {\bf{\tilde M}}\cdot\left( {\begin{array}{*{20}c}
   {a^{N + 1} }  \\
   {b^{N + 1} }  \\
\end{array}} \right),
\label{III4}
\end{equation}
where
\begin{equation}
{\bf{\tilde M}} = {\bf{\tilde M}}_0  \cdot {\bf{\tilde M}}_1  \cdots {\bf{\tilde M}}_N  = \left( {\begin{array}{*{20}c}
   {M_{11} } & {M_{12} }  \\
   {M_{21} } & {M_{22} }  \\
\end{array}} \right). 
\label{III5}
\end{equation}
Now we want to find the relation between $ {a^0 }$ and $ {b^0 }$. This relation depends on the boundary conditions at the outermost interfaces in  Fig.\,\ref{figu3}. In order to have self-sustained fields or normal modes we must not have any incoming fields from outside the object. In all cases this means that  $a^0=0$. In the planar case also the rightmost interface is the boundary to the outside which means that $b^{N + 1}=0$. In the spherical and cylindrical cases the rightmost region is the core and the boundary condition is that the waves are finite. What effect this has on the amplitudes of the waves depends on the choice of functions we make. In our non-retarded treatment it turns out that also for spherical and cylindrical objects $b^{N + 1}=0$. This leads to ${a^0} = {b^0}\left( {{M_{11}}{\rm{/}}{M_{21}}} \right)$. In our retarded treatment on the other hand $b^{N + 1}=a^{N + 1} $ follows from the condition of finite fields. This leads to ${a^0} = {b^0}\left[ {\left( {{M_{11}} + {M_{12}}} \right){\rm{/}}\left( {{M_{21}} + {M_{22}}} \right)} \right]$.  The only way we can have a non-zero $ {b^0 }$ at the same time as $ {a^0 }$ vanishes is that the factor multiplying $ {b^0 }$ vanishes. Thus the function $f\left( \omega  \right)$ in the mode condition is
\begin{equation}
\begin{array}{*{20}{l}}
{f\left( \omega  \right) = {M_{11}},\,{b^{N + 1}} = 0;}\\
{f\left( \omega  \right) = {M_{11}} + {M_{12}},\,{b^{N + 1}} = {a^{N + 1}}.}
\end{array}
\label{III6}
\end{equation}
Let us now continue with the general procedure when the TE- and one TM-modes do mix. Then in a general medium $n$ we have the wave $a_1^n R_1 \left( x \right) + b_1^n L_1 \left( x \right) + a_2^n R_2 \left( x \right) + b_2^n L_2 \left( x \right)$, where the subscript $1$ and $2$ refers to TM- and TE-waves, respectively.  Making use of the boundary conditions at interface $n$ gives 
\begin{equation}
{\bf{\tilde A}}_n \left( {x_n } \right)\cdot\left( {\begin{array}{*{20}c}
   {a_1^n }  \\
   {b_1^n }  \\
   {a_2^n }  \\
   {b_2^n }  \\
\end{array}} \right) = {\bf{\tilde A}}_{n + 1}\left( {x_n } \right)\cdot \left( {\begin{array}{*{20}c}
   {a_1^{n + 1} }  \\
   {b_1^{n + 1} }  \\
   {a_2^{n + 1} }  \\
   {b_2^{n + 1} }  \\
\end{array}} \right),
\label{III7}
\end{equation}
where ${\bf{\tilde A}}_n$ is now a four by four matrix which depends on the dielectric function in medium $n$. Operating from the left with the inverse of this matrix gives
\begin{equation}
\left( {\begin{array}{*{20}c}
   {a_1^n }  \\
   {b_1^n }  \\
   {a_2^n }  \\
   {b_2^n }  \\
\end{array}} \right) = {\bf{\tilde M}}_n \cdot \left( {\begin{array}{*{20}c}
   {a_1^{n + 1} }  \\
   {b_1^{n + 1} }  \\
   {a_2^{n + 1} }  \\
   {b_2^{n + 1} }  \\
\end{array}} \right),
\label{III8}
\end{equation}
where
\begin{equation}
{\bf{\tilde M}}_n  = {\bf{\tilde A}}_n ^{ - 1} \left( {x_n } \right)\cdot {\bf{\tilde A}}_{n + 1} \left( {x_n } \right).
\label{III9}
\end{equation}
We may now find a relation between the coefficients in left-most and right-most media
\begin{equation}
\left( {\begin{array}{*{20}c}
   {a_1^0 }  \\
   {b_1^0 }  \\
   {a_2^0 }  \\
   {b_2^0 }  \\
\end{array}} \right) = {\bf{\tilde M}}\cdot \left( {\begin{array}{*{20}c}
   {a_1^{N + 1} }  \\
   {b_1^{N + 1} }  \\
   {a_2^{N + 1} }  \\
   {b_2^{N + 1} }  \\
\end{array}} \right),
\label{III10}
\end{equation}
where
\begin{equation}
{\bf{\tilde M}} = {\bf{\tilde M}}_0  \cdot {\bf{\tilde M}}_1  \cdots {\bf{\tilde M}}_N  = \left( {\begin{array}{*{20}c}
   {M_{11} } & {M_{12} } & {M_{13} } & {M_{14} }  \\
   {M_{21} } & {M_{22} } & {M_{23} } & {M_{24} }  \\
   {M_{31} } & {M_{32} } & {M_{33} } & {M_{34} }  \\
   {M_{41} } & {M_{42} } & {M_{43} } & {M_{44} }  \\
\end{array}} \right). 
\label{III11}
\end{equation}
Now we want to find the relation between $\left( {\begin{array}{*{20}c}
   {a_1^0 }  \\
   {a_2^0 }  \\
\end{array}} \right)$ and $\left( {\begin{array}{*{20}c}
   {b_1^0 }  \\
   {b_2^0 }  \\
\end{array}} \right)$. 
This relation depends on if $b_i^{N + 1} = 0$ or not. If it is, like in the planar case and in the non-retarded spherical and cylindrical cases, then 
\begin{equation}
\left( {\begin{array}{*{20}{c}}
{a_1^0}\\
{a_2^0}
\end{array}} \right) = \left( {\begin{array}{*{20}{c}}
{{M_{11}}}&{{M_{13}}}\\
{{M_{31}}}&{{M_{33}}}
\end{array}} \right) \cdot {\left( {\begin{array}{*{20}{c}}
{{M_{21}}}&{{M_{23}}}\\
{{M_{41}}}&{{M_{43}}}
\end{array}} \right)^{ - 1}} \cdot \left( {\begin{array}{*{20}{c}}
{b_1^0}\\
{b_2^0}
\end{array}} \right).
\label{III12}
\end{equation}
If $b_i^{N + 1} = a_i^{N + 1}$, like in the retarded spherical and cylindrical cases then  
\begin{equation}
\begin{array}{*{20}{l}}
{\left( {\begin{array}{*{20}{c}}
{a_1^0}\\
{a_2^0}
\end{array}} \right) = \left( {\begin{array}{*{20}{c}}
{\left( {{M_{11}} + {M_{12}}} \right)}&{\left( {{M_{13}} + {M_{14}}} \right)}\\
{\left( {{M_{31}} + {M_{32}}} \right)}&{\left( {{M_{33}} + {M_{34}}} \right)}
\end{array}} \right)}\\
{\,\,\,\,\,\,\,\,\,\,\,\,\,\,\,\,\,\,\,\,\, \cdot {{\left( {\begin{array}{*{20}{c}}
{\left( {{M_{21}} + {M_{22}}} \right)}&{\left( {{M_{23}} + {M_{24}}} \right)}\\
{\left( {{M_{41}} + {M_{42}}} \right)}&{\left( {{M_{43}} + {M_{44}}} \right)}
\end{array}} \right)}^{ - 1}} \cdot \left( {\begin{array}{*{20}{c}}
{b_1^0}\\
{b_2^0}
\end{array}} \right).}
\end{array}
\label{III13}
\end{equation}
In order to have self-sustained fields or normal modes we must not have any incoming fields from outside the object, i.e., ${a_1^0 }$ and ${a_2^0 }$ must be zero. The only way we can have a non-zero $ {b_1^0 }$ and/or $ {b_2^0 }$ at the same time as ${a_1^0 }$ and ${a_2^0 }$  vanish is that the determinant of the matrix in front of $\left( {\begin{array}{*{20}c} {b_1^0 }  \\ {b_2^0 }  \\
\end{array}} \right)$ vanishes. Thus, the condition for modes is
\begin{equation}
\left| {\begin{array}{*{20}{c}}
{{M_{11}}}&{{M_{13}}}\\
{{M_{31}}}&{{M_{33}}}
\end{array}} \right| = 0,
\label{III14}
\end{equation}
if $b_i^{N + 1} = 0$ and
\begin{equation}
\left| {\begin{array}{*{20}{c}}
{\left( {{M_{11}} + {M_{12}}} \right)}&{\left( {{M_{13}} + {M_{14}}} \right)}\\
{\left( {{M_{31}} + {M_{32}}} \right)}&{\left( {{M_{33}} + {M_{34}}} \right)}
\end{array}} \right| = 0,
\label{III15}
\end{equation}
if $b_i^{N + 1} = a_i^{N + 1}$.

Now we will describe how the waves we have discussed are obtained. We treat metals and dielectrics on the same footing, i.e., induced current and charge densities have contributions from both bound electrons and conduction electrons. The dielectric function for a metallic system is
\begin{equation}
\tilde \varepsilon \left( \omega  \right) = \varepsilon \left( \omega  \right) + 4\pi i\sigma \left( \omega  \right){\rm{/}}\omega ,
\label{III16}
\end{equation}
where $\varepsilon \left( \omega  \right)$ would be the dielectric function if it were not for the conduction carriers. These contribute to the screening through the dynamical conductivity, ${\sigma \left( \omega  \right)}$. With this choice the Maxwell's equations (ME) read
\begin{equation}
\begin{gathered}
\begin{array}{l}
\nabla  \cdot {\bf{D}} = 4\pi {\rho _{ext}}\\
\nabla  \cdot {\bf{B}} = 0\\
\nabla  \times {\bf{E}} =  - \frac{1}{c}\frac{{\partial {\bf{B}}}}{{\partial t}}\\
\nabla  \times {\bf{H}} = \frac{{4\pi }}{c}{{\bf{J}}_{ext}} + \frac{1}{c}\frac{{\partial {\bf{D}}}}{{\partial t}}.
\end{array}
 \end{gathered}
\label{III17}
\end{equation}
The external charge and current densities are absent in our system. Furthermore since we are concerned with normal modes the time dependence of each field is given by a factor $\exp \left( { - i\omega t} \right)$ and we have
\begin{equation}
\begin{array}{l}
\nabla  \cdot {\bf{D}} = 0\\
\nabla  \cdot {\bf{B}} = 0\\
\nabla  \times {\bf{E}} = i\left( {\omega /c} \right){\bf{B}}\\
\nabla  \times {\bf{H}} =  - i\left( {\omega /c} \right){\bf{D}}.
\end{array}
\label{III18}
\end{equation}
We assume non-magnetic materials and let $\mu =1$, where $\mu$ is the magnetic permeability;  we are not interested in longitudinal bulk modes and assume that $\tilde\varepsilon \left( \omega  \right) \ne 0$. We want to keep one electric and one magnetic field. Since the $\bf E$- and $\bf H$-fields have the same boundary conditions at an interface we keep these. Thus we have
\begin{equation}
 \begin{array}{l}
\nabla  \cdot {\bf{E}} = 0\\
\nabla  \cdot {\bf{H}} = 0\\
\nabla  \times {\bf{E}} = i\left( {\omega /c} \right){\bf{H}}\\
\nabla  \times {\bf{H}} =  - i\tilde \varepsilon \left( \omega  \right)\left( {\omega /c} \right){\bf{E}}.
\end{array}
\label{III19}
\end{equation}
Neglecting retardation means letting the speed of light go to infinity. Then the MEs reduce to
\begin{equation}
\begin{gathered}
 \nabla \cdot{\bf{E}} = 0 \\ 
 \nabla \cdot{\bf{H}} = 0 \\ 
 \nabla  \times {\bf{E}} = 0 \\ 
 \nabla  \times {\bf{H}} = 0. \\ 
 \end{gathered}
\label{III20}
\end{equation}
Eqs.\,(\ref{III19}) and (\ref{III20}) are the basic equations we are starting from in all structures, Eq.\,(\ref{III19}) in the fully retarded calculations and  Eq.\,(\ref{III20}) when retardation is neglected.

In the non-retarded treatment, since $\nabla  \times {\bf{E}} = 0$ the $\bf E$- field is conservative and we may define a scalar potential, $\Phi $ such that ${\bf{E}} =  - \nabla \Phi $. Using the first line of Eq.\,(\ref{III20}) then leads to Laplace's equation,
\begin{equation}
\nabla ^2 \Phi  = 0.
\label{III21}
\end{equation}
So, when we neglect retardation effects we just solve Laplace's equation in each medium and use the proper boundary conditions at each interface to find the normal modes.

In the fully retarded treatment we take the curl of the last two lines of Eq.\,(\ref{III19}) and make use of the other relations to find
\begin{equation}
\begin{array}{l}
{\nabla ^2}{\bf{E}} + \tilde \varepsilon \left( \omega  \right){\left( {\omega /c} \right)^2}{\bf{E}} = 0\\
{\nabla ^2}{\bf{H}} + \tilde \varepsilon \left( \omega  \right){\left( {\omega /c} \right)^2}{\bf{H}} = 0.
\end{array}
\label{III22}
\end{equation}
Thus both the $\bf E$- and $\bf H$-fields obey the vector wave equation, the vector Helmholtz equation. In the planar case it is straight forward to solve these in each region but in other geometries it is not a trivial task.  One can solve the problem by introducing Hertz-Debye potentials $\pi _1 $ and $\pi _2 $. They are solutions to the scalar wave equation,
\begin{equation}
{\nabla ^2}\pi ' + {q^2}\pi ' = 0;\,\,\pi  = \pi '{e^{ - i\omega t}};\,\,{q^2} = \tilde \varepsilon \left( \omega  \right){\left( {\omega /c} \right)^2}.
\label{III23}
\end{equation}
 We let $\pi _1 $ be the potential that generates TM modes and $\pi _2 $ be the potential that generates TE modes.

Now we are done with the general formalism and turn to the three geometries we have chosen to concentrate on. We treat planar, spherical and cylindrical geometries in that order.

\section{\label{planar}Planar structures}
We assume that the spatial extension of the interfaces is very large compared to the thickness of the layers so that we may treat the interfaces as infinite in two directions. If the thickness of the object is finite the rightmost medium, $n=N+1$, in Fig.\,\ref{figu3} is the ambient as well as the leftmost, $n=0$. If not we have a multiple coated half space. In both situations the modes are solutions with the boundary conditions that there are no incoming waves in the two outer regions, i.e. there is no wave moving towards the right in medium $n=0$ and no wave moving towards the left in medium $n=N+1$. The fields are self-sustained; no fields are coming in from outside. We first treat the simplest case, the non-retarded. 

\subsection{\label{planarnonretmain}Non-retarded main results}

In the non-retarded treatment of a planar structure we let the waves represent solutions to Laplace's equation, Eq.\,(\ref{III21}), in cartesian coordinates, for the scalar potential, $\Phi$. The interfaces are parallel to the $xy$-plane and the $z$-coordinate is the coordinate that is constant on each interface. The solutions are of the form
\begin{equation}
{\Phi _{\bf{k}}}\left( {{\bf{r}},z} \right) = {e^{i{\bf{k}} \cdot {\bf{r}}}}{e^{ \pm kz}},
\label{IVA1}
\end{equation}
where ${\bf{k}}$ is the two-dimensional wave vector in the plane of the interfaces.
We let $z$ increase towards the right in  Fig.\,\ref{figu3}. We want to find the normal modes for a specific wave vector  ${\bf{k}}$. Then all waves have the common factor  $\exp \left( {i{\bf{k}} \cdot {\bf{r}}} \right)$. We suppress this factor here. Then
\begin{equation}
R\left( z \right) = {e^{ - kz}};\,L\left( z \right) = {e^{ + kz}}.
\label{IVA2}
\end{equation}
Using the boundary conditions that the potential and the normal component of the ${\bf{D}}$-field are continuous across interface $n$ gives
\begin{equation}
\begin{array}{*{20}{l}}
{{a^n}{e^{ - k{z_n}}} + {b^n}{e^{k{z_n}}} = {a^{n + 1}}{e^{ - k{z_n}}} + {b^{n + 1}}{e^{k{z_n}}}}\\
{{{a^n}{{\tilde \varepsilon }_n}{e^{ - k{z_n}}} - {b^n}{{\tilde \varepsilon }_n}{e^{k{z_n}}}}}\\
\quad \quad \quad \quad ={ {a^{n + 1}}{{\tilde \varepsilon }_{n + 1}}{e^{ - k{z_n}}} - {b^{n + 1}}{{\tilde \varepsilon }_{n + 1}}{e^{k{z_n}}}},
\end{array}
\label{IVA3}
\end{equation}
and we may identify the matrix ${{{\bf{\tilde A}}}_n}\left( {{z_n}} \right)$ as
\begin{equation}
{{{\bf{\tilde A}}}_n}\left( {{z_n}} \right) = \left( {\begin{array}{*{20}{c}}
{{e^{ - k{z_n}}}}&{{e^{k{z_n}}}}\\
{{{\tilde \varepsilon }_n}{e^{ - k{z_n}}}}&{ - {{\tilde \varepsilon }_n}{e^{k{z_n}}}}
\end{array}} \right)
\label{IVA4}
\end{equation}
and the matrix ${{{\bf{\tilde M}}}_n}$ as
\begin{equation}
{{{\bf{\tilde M}}}_n} = \frac{1}{{2{{\tilde \varepsilon }_n}}}\left( {\begin{array}{*{20}{c}}
{{{\tilde \varepsilon }_n} + {{\tilde \varepsilon }_{n + 1}}}&{{e^{2k{z_n}}}\left( {{{\tilde \varepsilon }_n} - {{\tilde \varepsilon }_{n + 1}}} \right)}\\
{{e^{ - 2k{z_n}}}\left( {{{\tilde \varepsilon }_n} - {{\tilde \varepsilon }_{n + 1}}} \right)}&{{{\tilde \varepsilon }_n} + {{\tilde \varepsilon }_{n + 1}}}
\end{array}} \right).
\label{IVA5}
\end{equation}
Now we have all we need to determine the non-retarded normal modes in a layered planar structure. We give some examples in the following section.

\subsection{\label{planarnonretspecial}Non-retarded special results}
\begin{figure}
\includegraphics[width=1.6cm]{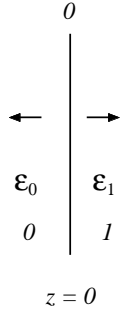}
\caption{The geometry of a single planar interface.}
\label{figu4}
\end{figure}

\subsubsection{\label{interfacen}Single planar interface (no layer)}

For a single interface, as illustrated in Fig.\,\ref{figu4}, at $z=0$ between two media with dielectric functions ${{{\tilde \varepsilon }_0}}$ and ${{{\tilde \varepsilon }_1}}$ we have
\begin{equation}
{\bf{\tilde M}} = {{{\bf{\tilde M}}}_0} = \frac{1}{{2{{\tilde \varepsilon }_0}}}\left( {\begin{array}{*{20}{c}}
{{{\tilde \varepsilon }_0} + {{\tilde \varepsilon }_1}}&{{{\tilde \varepsilon }_0} - {{\tilde \varepsilon }_1}}\\
{{{\tilde \varepsilon }_0} - {{\tilde \varepsilon }_1}}&{{{\tilde \varepsilon }_0} + {{\tilde \varepsilon }_1}}
\end{array}} \right),
\label{IVB1.1}
\end{equation}
and the mode condition is 
\begin{equation}
{\tilde \varepsilon _0}\left( \omega  \right) + {\tilde \varepsilon _1}\left( \omega  \right) = 0.
\label{IVB1.2}
\end{equation}

\subsubsection{\label{slabn}Slab or planar gap (one layer)}

\begin{figure}
\includegraphics[width=2.8cm]{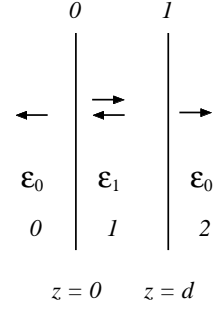}
\caption{The one planar layer geometry.}
\label{figu5}
\end{figure}

For a slab, Fig.\,\ref{figu5}, with interfaces at $z=0$ and $z=d$ made of a medium with dielectric function ${{\tilde \varepsilon }_1}$ in an ambient medium with dielectric function ${{\tilde \varepsilon }_0}$ we have 
\begin{equation}
\begin{array}{l}
{\bf{\tilde M}} = {{{\bf{\tilde M}}}_0} \cdot {{{\bf{\tilde M}}}_1} = \frac{1}{{2{{\tilde \varepsilon }_0}}}\left( {\begin{array}{*{20}{c}}
{{{\tilde \varepsilon }_0} + {{\tilde \varepsilon }_1}}&{{{\tilde \varepsilon }_0} - {{\tilde \varepsilon }_1}}\\
{{{\tilde \varepsilon }_0} - {{\tilde \varepsilon }_1}}&{{{\tilde \varepsilon }_0} + {{\tilde \varepsilon }_1}}
\end{array}} \right)\\
\quad  \times \frac{1}{{2{{\tilde \varepsilon }_1}}}\left( {\begin{array}{*{20}{c}}
{{{\tilde \varepsilon }_1} + {{\tilde \varepsilon }_0}}&{{e^{2kd}}\left( {{{\tilde \varepsilon }_1} - {{\tilde \varepsilon }_0}} \right)}\\
{{e^{ - 2kd}}\left( {{{\tilde \varepsilon }_1} - {{\tilde \varepsilon }_0}} \right)}&{{{\tilde \varepsilon }_1} + {{\tilde \varepsilon }_0}}
\end{array}} \right)
\end{array},
\label{IVB2.1}
\end{equation}
and the mode condition becomes
\begin{equation}
{\left[ {{{\tilde \varepsilon }_0}\left( \omega  \right) + {{\tilde \varepsilon }_1}\left( \omega  \right)} \right]^2} - {e^{ - 2kd}}{\left[ {{{\tilde \varepsilon }_0}\left( \omega  \right) - {{\tilde \varepsilon }_1}\left( \omega  \right)} \right]^2} = 0.
\label{IVB2.2}
\end{equation}
For a gap, of size $d$, filled with a medium with dielectric function ${{\tilde \varepsilon }_0}$ between two half spaces of material with dielectric function  ${{{\tilde \varepsilon }_1}}$ we may reuse the above result with the interchange of the two dielectric functions. We note that the result will not change. If the half spaces are made up by two different materials with ${{{\tilde \varepsilon }_1}}$ and ${{{\tilde \varepsilon }_2}}$ we find
\begin{equation}
\left[ {{{\tilde \varepsilon }_1} + {{\tilde \varepsilon }_0}} \right]\left[ {{{\tilde \varepsilon }_2} + {{\tilde \varepsilon }_0}} \right] - {e^{ - 2kd}}\left[ {{{\tilde \varepsilon }_1} - {{\tilde \varepsilon }_0}} \right]\left[ {{{\tilde \varepsilon }_2} - {{\tilde \varepsilon }_0}} \right] = 0,
\label{IVB2.3}
\end{equation}
where all dielectric function arguments, $\left( \omega  \right)$, have been suppressed.
Eq.\,(\ref{IVB2.3}) can then be used to find the zero temperature van der Waals energy per unit area as
\begin{equation}
\begin{array}{l}
 E = \frac{\hbar }{2}\int {\frac{{d^2 k}}{{\left( {2\pi } \right)^2 }}} \int\limits_{ - \infty }^\infty  {\frac{{d\xi }}{{2\pi }}} \left[ {\ln f_{\bf{k}} \left( {i\xi } \right) - \ln f_{\bf{k}}^\infty  \left( {i\xi } \right)} \right] \\ 
 \quad  = \frac{\hbar }{2}\int {\frac{{d^2 k}}{{\left( {2\pi } \right)^2 }}} \int\limits_{ - \infty }^\infty  {\frac{{d\xi }}{{2\pi }}} \ln \tilde f_{\bf{k}} \left( {i\xi } \right), \\ 
 \end{array}
\label{IVB2.4}
\end{equation}
and the finite temperature result as
\begin{equation}
\begin{array}{l}
 \mathfrak{F} = \frac{1}{\beta }\int {\frac{{d^2 k}}{{\left( {2\pi } \right)^2 }}} \sum\limits_{\xi _n }{^{'}{\left[ {\ln f_{\bf{k}} \left( {i\xi _n } \right) - \ln f_{\bf{k}}^\infty  \left( {i\xi _n } \right)} \right]} } \\ 
 \quad  = \frac{1}{\beta }\int {\frac{{d^2 k}}{{\left( {2\pi } \right)^2 }}} \sum\limits_{\xi _n } {^{'}{\ln \tilde f_{\bf{k}} \left( {i\xi _n } \right)} ;\:\xi _n  = \frac{{2\pi n}}{{\hbar \beta }}};n = 0,1,2, \ldots,  \\ 
 \end{array}
\label{IVB2.5}
\end{equation}
respectively, where ${f_{\bf{k}}^\infty  \left( {\omega} \right)}$ is the mode condition function at infinite separation and
\begin{equation}
\tilde f_{\bf{k}} \left( \omega  \right) = 1 - e^{ - 2kd} \frac{{\left[ {{\tilde \varepsilon }_1 \left( \omega  \right) - {\tilde \varepsilon }_0 \left( \omega  \right)} \right]\left[ {{\tilde \varepsilon }_2 \left( \omega  \right) - {\tilde \varepsilon }_0 \left( \omega  \right)} \right]}}{{\left[ {{\tilde \varepsilon }_1 \left( \omega  \right) + {\tilde \varepsilon }_0 \left( \omega  \right)} \right]\left[ {{\tilde \varepsilon }_2 \left( \omega  \right) + {\tilde \varepsilon }_0 \left( \omega  \right)} \right]}}
\label{IVB2.6}
\end{equation}
is the mode condition function divided by the function at infinite separation. The expressions in Eqs.\,(\ref{IVB2.4}) and (\ref{IVB2.5}) mean that we have chosen the reference system to be the system when the gap is infinitely wide.

The van der Waals force per unit area is obtained as minus the derivative of these energies with respect to the separation $d$.
\subsubsection{\label{gasfilmn}Thin planar diluted gas film (one layer)}
It is of interest to find the van der Waals force on an atom in a layered structure. We can obtain this by studying the force on a thin layer of a diluted gas with dielectric function ${\varepsilon _g}\left( \omega  \right) = 1 + 4\pi n\alpha^{at} \left( \omega  \right)$, where $\alpha^{at}$ is the polarizability of one atom and $n$ the density of atoms (we have assumed that the atom is surrounded by vacuum; if not  the $1$ should be replaced by the dielectric function of the ambient medium and the atomic polarizability should be replaced by the excess polarizability). For a diluted gas layer the atoms do not interact with each other and the force on the layer is just the sum of the forces on the individual atoms. So by dividing with the number of atoms in the film we get the force on one atom. The layer has to be thin in order to have a well defined $z$- value of the atom.  Since we will derive the force on an atom in different planar geometries it is fruitful to derive the matrix for a thin diluted gas film. This result can be directly used in the derivation of the van der Waals force on an atom in different planar geometries.

We let the film have the thickness $\delta$ and be placed in the general position $z$. We only keep terms up to linear order in $\delta$ and linear order in $n$. We find the result is
\begin{equation}
\begin{array}{l}
{{{\bf{\tilde M}}}_{{\rm{gaslayer}}}} = {{{\bf{\tilde M}}}_0} \cdot {{{\bf{\tilde M}}}_1}\\
 = \left( {\begin{array}{*{20}{c}}
1&0\\
0&1
\end{array}} \right) + \left( {\delta n} \right){\alpha ^{at}}4\pi k\left( {\begin{array}{*{20}{c}}
0&{{e^{2kz}}}\\
{ - {e^{ - 2kz}}}&0
\end{array}} \right).
\end{array}
\label{IVB3.1}
\end{equation}

Now we are done with the gas layer. We will use these results later in calculating the van der Waals force on an atom in planar layered structures.

\subsubsection{\label{2Dfilmn}2D planar film (one layer)}
In many situations one is dealing with very thin films. These may be considered 2D (two dimensional). Important examples are a graphene sheet and a 2D electron gas. In the derivation we let the film have finite thickness $\delta$ and be characterized by a 3D dielectric function ${\tilde \varepsilon ^{3D}}$. We then let the thickness go towards zero. The 3D dielectric function depends on $\delta$ as ${\tilde\varepsilon ^{3D}} \sim 1/\delta $ for small $\delta$ and $\delta {\tilde\varepsilon ^{3D}} \to 2{\tilde\alpha ^{2D}}/k$ as  $\delta$ goes towards zero.\,\cite{grap,arx} ${\tilde \alpha ^{2D}}\left( {{\bf{k}},\omega } \right)$ is the 2D polarizability of the film. We obtain
\begin{equation}
\begin{array}{l}
{{{\bf{\tilde M}}}_{{\rm{2D}}}} = {{{\bf{\tilde M}}}_0} \cdot {{{\bf{\tilde M}}}_1}\\
\quad \quad  = \left( {\begin{array}{*{20}{c}}
1&0\\
0&1
\end{array}} \right) + \frac{{k\left( {\delta {\tilde\varepsilon ^{3D}}} \right)}}{2}\left( {\begin{array}{*{20}{c}}
1&{{e^{2kz}}}\\
{ - {e^{ - 2kz}}}&{ - 1}
\end{array}} \right);\\
\quad \quad  = \left( {\begin{array}{*{20}{c}}
1&0\\
0&1
\end{array}} \right) + {\tilde\alpha ^{2D}}\left( {\begin{array}{*{20}{c}}
1&{{e^{2kz}}}\\
{ - {e^{ - 2kz}}}&{ - 1}
\end{array}} \right).
\end{array}
\label{IVB4.1}
\end{equation}

\subsubsection{\label{atomplanen}Force on an atom next to a substrate (two layers)}
\begin{figure}
\includegraphics[width=6.0cm]{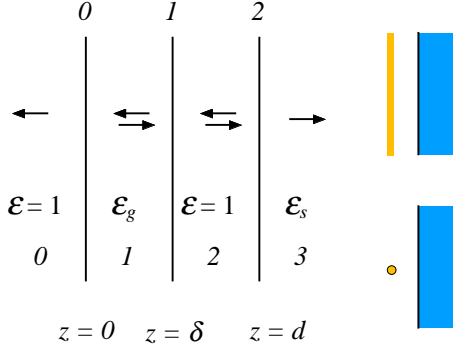}
\caption{The two planar layer geometry. This geometry is used to find the atom substrate interaction as illustrated by the cartoon in the lower right corner.  We start from the gas layer substrate geometry in the upper right corner. }
\label{figu6}
\end{figure}

The multiple layer results can be used to solve other problems like in this case the atom substrate force. We start from the two layer structure in Fig.\,\ref{figu6}. We let the ambient be vacuum. The first layer is a thin layer, of thickness $\delta$, of a diluted gas of atoms of the kind we consider. Its dielectric function is ${\varepsilon _g}\left( \omega  \right) = 1 + 4\pi n\alpha^{at} \left( \omega  \right)$, where $\alpha^{at}$ is the polarizability of one atom. The density of gas atoms, $n$, is very low. We let the first interface be at $z=0$ and hence the second at $z=\delta$. The second layer is a vacuum layer of thickness $d$. The remaining medium is the substrate which we let be infinitely thick and have the dielectric function ${{\tilde \varepsilon }_s}\left( \omega  \right)$. In what follows we only keep lowest order terms in $\delta$ and in $n$.

The matrix becomes ${\bf{\tilde M}} = {{\bf{\tilde M}}_0} \cdot {{\bf{\tilde M}}_1} \cdot {{\bf{\tilde M}}_2} = {{\bf{\tilde M}}_{{\rm{gaslayer}}}} \cdot {{\bf{\tilde M}}_2}$ where ${{\bf{\tilde M}}_{{\rm{gaslayer}}}}$ is given in Eq.\,(\ref{IVB3.1}) with $z=0$ and
\begin{equation}
{{\bf{\tilde M}}_2} = \frac{1}{2}\left( {\begin{array}{*{20}{c}}
{\left( {{{\tilde \varepsilon }_s} + 1} \right)}&{ - {e^{2kd}}\left( {{{\tilde \varepsilon }_s} - 1} \right)}\\
{ - {e^{ - 2kd}}\left( {{{\tilde \varepsilon }_s} - 1} \right)}&{\left( {{{\tilde \varepsilon }_s} + 1} \right)}
\end{array}} \right).
\label{IVB5.1}
\end{equation}
Now, the matrix element of interest is
\begin{equation}
{M_{11}} = \frac{1}{2}\left[ {\left( {{{\tilde \varepsilon }_s} + 1} \right) - 4\pi k\alpha^{at} \delta n{e^{ - 2kd}}\left( {{{\tilde \varepsilon }_s} - 1} \right)} \right],
\label{IVB5.2}
\end{equation}
and the condition for modes is
%
\begin{equation}
\left( {{{\tilde \varepsilon }_s} + 1} \right) - 4\pi k{\alpha ^{at}}\delta n{e^{ - 2kd}}\left( {{{\tilde \varepsilon }_s} - 1} \right) = 0.
\label{IVB5.3}
\end{equation}
The first part of the mode condition function is what one would have in absence of the atom. It gives the surface modes of the substrate.  We find
\begin{equation}
{{\tilde f}_{\bf{k}}}\left( \omega  \right) = 1 - 4\pi k\alpha^{at} \left( \omega  \right)\delta n{e^{ - 2kd}}\frac{{\left[ {{{\tilde \varepsilon }_s}\left( \omega  \right) - 1} \right]}}{{\left[ {{{\tilde \varepsilon }_s}\left( \omega  \right) + 1} \right]}},
\label{IVB5.4}
\end{equation}
where we have chosen the reference system as the system when the atom is at infinite distance from the substrate. The interaction energy per atom is
\begin{equation}
\begin{array}{l}
\frac{E}{{n\delta }}
 = \frac{\hbar }{{2n\delta }}\int {\frac{{{d^2}k}}{{{{\left( {2\pi } \right)}^2}}}} \int\limits_{ - \infty }^\infty  {\frac{{d\xi }}{{2\pi }}} \\
\times  \ln \left[ {1 - 4\pi k\alpha^{at} \left( {i\xi } \right)\delta n{e^{ - 2kd}}\frac{{\left[ {{{\tilde \varepsilon }_s}\left( {i\xi } \right) - 1} \right]}}{{\left[ {{{\tilde \varepsilon }_s}\left( {i\xi } \right) + 1} \right]}}} \right]\\
\quad  \approx  - \frac{\hbar }{2}\int {\frac{{{d^2}k}}{{{{\left( {2\pi } \right)}^2}}}} 4\pi k{e^{ - 2kd}}\int\limits_{ - \infty }^\infty  {\frac{{d\xi }}{{2\pi }}} \alpha^{at} \left( {i\xi } \right)\frac{{\left[ {{{\tilde \varepsilon }_s}\left( {i\xi } \right) - 1} \right]}}{{\left[ {{{\tilde \varepsilon }_s}\left( {i\xi } \right) + 1} \right]}}\\
\quad  =  - \underbrace {\int\limits_0^\infty  {dk{k^2}2{e^{ - 2kd}}} }_{\frac{1}{{2{d^3}}}}\frac{\hbar }{2}\int\limits_{ - \infty }^\infty  {\frac{{d\xi }}{{2\pi }}} \alpha^{at} \left( {i\xi } \right)\frac{{\left[ {{{\tilde \varepsilon }_s}\left( {i\xi } \right) - 1} \right]}}{{\left[ {{{\tilde \varepsilon }_s}\left( {i\xi } \right) + 1} \right]}}\\
\quad  =  - \frac{\hbar }{{4{d^3}}}\int\limits_{ - \infty }^\infty  {\frac{{d\xi }}{{2\pi }}} \alpha^{at} \left( {i\xi } \right)\frac{{\left[ {{{\tilde \varepsilon }_s}\left( {i\xi } \right) - 1} \right]}}{{\left[ {{{\tilde \varepsilon }_s}\left( {i\xi } \right) + 1} \right]}},
\end{array}
\label{IVB5.5}
\end{equation}
where we have divided the energy per unit area with the number of gas atoms per unit area resulting in the energy per atom. We have furthermore  let the number of atoms per unit area go towards zero and expanded the logarithm ($\ln \left( {1 + x} \right) \to x$). 

Thus, the force between an atom a distance $d$ from a substrate is at zero temperature
\begin{equation}
F\left( d \right) =  - \frac{{3\hbar }}{{2{d^4}}}\int\limits_0^\infty  {\frac{{d\xi }}{{2\pi }}} \alpha^{at} \left( {i\xi } \right)\frac{{\left[ {{{\tilde \varepsilon }_s}\left( {i\xi } \right) - 1} \right]}}{{\left[ {{{\tilde \varepsilon }_s}\left( {i\xi } \right) + 1} \right]}},
\label{IVB5.6}
\end{equation}
and at finite temperature it is
\begin{equation}
F\left( d \right) =  - \frac{{3}}{{2{d^4}}}\frac{1}{\beta }\sum\limits_{{\xi _n}}{^{'} {\alpha^{at} \left( {i{\xi _n}} \right)\frac{{\left[ {{{\tilde \varepsilon }_s}\left( {i{\xi _n}} \right) - 1} \right]}}{{\left[ {{{\tilde \varepsilon }_s}\left( {i{\xi _n}} \right) + 1} \right]}}}}.
\label{IVB5.7}
\end{equation}

\subsubsection{\label{atomplanesn}Force on an atom in between two planar surfaces (three layers)}
\begin{figure}
\includegraphics[width=8.0cm]{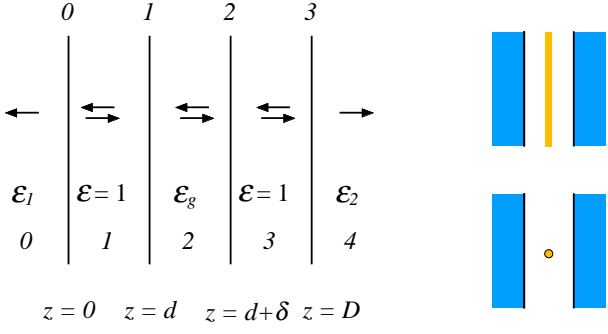}
\caption{The three planar layer geometry.}
\label{figu7}
\end{figure}
We refer to Fig.\,\ref{figu7} and let the first interface be located at $z=0$ separating one plate with dielectric function ${{\tilde \varepsilon }_1}$ from the ambient medium which we let be vacuum. Next interface, at $z=d$, is the left interface of the gas layer with dielectric function ${\varepsilon _g}$ and thickness $\delta $. Thus the third interface is at $z=d+\delta$. The forth interface is located at $z=D$ and separates vacuum from the second plate with dielectric function ${{\tilde \varepsilon }_2}$. Just as in the previous section we only keep lowest order terms in $\delta$ and in $n$. The matrix becomes ${\bf{\tilde M}} = {{\bf{\tilde M}}_0} \cdot {{\bf{\tilde M}}_1} \cdot {{\bf{\tilde M}}_2} \cdot {{\bf{\tilde M}}_3} = {{\bf{\tilde M}}_0} \cdot {{\bf{\tilde M}}_{{\rm{gaslayer}}}} \cdot {{\bf{\tilde M}}_3}$ where ${{{\bf{\tilde M}}}_{{\rm{gaslayer}}}}$ is given in Eq.\,(\ref{IVB3.1}) with $z=d$ and

\begin{equation}
\begin{array}{l}
{{{\bf{\tilde M}}}_0} = \frac{1}{{2{{\tilde \varepsilon }_1}}}\left( {\begin{array}{*{20}{c}}
{\left( {{{\tilde \varepsilon }_1} + 1} \right)}&{\left( {{{\tilde \varepsilon }_1} - 1} \right)}\\
{\left( {{{\tilde \varepsilon }_1} - 1} \right)}&{\left( {{{\tilde \varepsilon }_1} + 1} \right)}
\end{array}} \right);\\
{{{\bf{\tilde M}}}_3} = \frac{1}{2}\left( {\begin{array}{*{20}{c}}
{\left( {{{\tilde \varepsilon }_2} + 1} \right)}&{ - {e^{2kD}}\left( {{{\tilde \varepsilon }_2} - 1} \right)}\\
{ - {e^{ - 2kD}}\left( {{{\tilde \varepsilon }_2} - 1} \right)}&{\left( {{{\tilde \varepsilon }_2} + 1} \right)}
\end{array}} \right).
\end{array}
\label{IVB6.1}
\end{equation}
Now,
\begin{equation}
{M_{11}} = \left( {\begin{array}{*{20}{c}}
{M_{11}^0}&{M_{12}^0}
\end{array}} \right) \cdot {{{\bf{\tilde M}}}_{{\rm{gaslayer}}}} \cdot \left( {\begin{array}{*{20}{c}}
{M_{11}^3}\\
{M_{21}^3}
\end{array}} \right),
\label{IVB6.2}
\end{equation}
where we have moved the matrix subscript to the superscript position to make room for the element subscripts.
The matrix element of interest is
\begin{equation}
\begin{array}{*{20}{l}}
\begin{array}{l}
{M_{11}}\\
 = \frac{1}{{4{{\tilde \varepsilon }_1}}}\left\{ {\left[ {\left( {{{\tilde \varepsilon }_1} + 1} \right)\left( {{{\tilde \varepsilon }_2} + 1} \right) - {e^{ - 2kD}}\left( {{{\tilde \varepsilon }_1} - 1} \right)\left( {{{\tilde \varepsilon }_2} - 1} \right)} \right]} \right.
\end{array}\\
{\quad \quad \quad  - 4\pi k{\alpha ^{at}}\delta n\left[ {{e^{ - 2kd}}\left( {{{\tilde \varepsilon }_1} - 1} \right)\left( {{{\tilde \varepsilon }_2} + 1} \right)} \right.}\\
{\left. {\left. {\quad \quad \quad \quad  + {e^{ - 2k\left( {D - d} \right)}}\left( {{{\tilde \varepsilon }_1} + 1} \right)\left( {{{\tilde \varepsilon }_2} - 1} \right)} \right]} \right\}.}
\end{array}
\label{IVB6.3}
\end{equation}
The mode condition function after division with the function in absence of the gas layer is
 \begin{equation}
{{\tilde f}_{\bf{k}}} = 1 - 4\pi k\alpha^{at} \delta n\frac{{\left[ {{e^{ - 2kd}}\frac{{\left( {{{\tilde \varepsilon }_1} - 1} \right)}}{{\left( {{{\tilde \varepsilon }_1} + 1} \right)}} + {e^{ - 2k\left( {D - d} \right)}}\frac{{\left( {{{\tilde \varepsilon }_2} - 1} \right)}}{{\left( {{{\tilde \varepsilon }_2} + 1} \right)}}} \right]}}{{\left[ {1 - {e^{ - 2kD}}\frac{{\left( {{{\tilde \varepsilon }_1} - 1} \right)\left( {{{\tilde \varepsilon }_2} - 1} \right)}}{{\left( {{{\tilde \varepsilon }_1} + 1} \right)\left( {{{\tilde \varepsilon }_2} + 1} \right)}}} \right]}},
\label{IVB6.4}
\end{equation}
and the interaction energy per atom becomes
\begin{equation}
\begin{array}{*{20}{l}}
{\frac{E}{{n\delta }} = \frac{\hbar }{{2n\delta }}\int {\frac{{{d^2}k}}{{{{\left( {2\pi } \right)}^2}}}} \int\limits_{ - \infty }^\infty  {\frac{{d\xi }}{{2\pi }}} \ln \left[ {{{\tilde f}_{\bf{k}}}\left( {i\xi } \right)} \right]}\\
{ \approx  - \frac{\hbar }{2}\int {\frac{{{d^2}k}}{{{{\left( {2\pi } \right)}^2}}}} 4\pi k\int\limits_{ - \infty }^\infty  {\frac{{d\xi }}{{2\pi }}} {\alpha ^{at}}\frac{{\left[ {{e^{ - 2kd}}\frac{{\left( {{{\tilde \varepsilon }_1} - 1} \right)}}{{\left( {{{\tilde \varepsilon }_1} + 1} \right)}} + {e^{ - 2k\left( {D - d} \right)}}\frac{{\left( {{{\tilde \varepsilon }_2} - 1} \right)}}{{\left( {{{\tilde \varepsilon }_2} + 1} \right)}}} \right]}}{{\left[ {1 - {e^{ - 2kD}}\frac{{\left( {{{\tilde \varepsilon }_1} - 1} \right)\left( {{{\tilde \varepsilon }_2} - 1} \right)}}{{\left( {{{\tilde \varepsilon }_1} + 1} \right)\left( {{{\tilde \varepsilon }_2} + 1} \right)}}} \right]}}}\\
{ =  - \hbar \int\limits_0^\infty  {dk{k^2}} \int\limits_{ - \infty }^\infty  {\frac{{d\xi }}{{2\pi }}} {\alpha ^{at}}\frac{{\left[ {{e^{ - 2kd}}\frac{{\left( {{{\tilde \varepsilon }_1} - 1} \right)}}{{\left( {{{\tilde \varepsilon }_1} + 1} \right)}} + {e^{ - 2k\left( {D - d} \right)}}\frac{{\left( {{{\tilde \varepsilon }_2} - 1} \right)}}{{\left( {{{\tilde \varepsilon }_2} + 1} \right)}}} \right]}}{{\left[ {1 - {e^{ - 2kD}}\frac{{\left( {{{\tilde \varepsilon }_1} - 1} \right)\left( {{{\tilde \varepsilon }_2} - 1} \right)}}{{\left( {{{\tilde \varepsilon }_1} + 1} \right)\left( {{{\tilde \varepsilon }_2} + 1} \right)}}} \right]}}.}
\end{array}
\label{IVB6.5}
\end{equation}
 Thus, the force on the atom is 
%
\begin{equation}
\begin{array}{*{20}{l}}
{F\left( d \right) =  - 4\hbar \int\limits_0^\infty  {dk{k^3}} \int\limits_0^\infty  {\frac{{d\xi }}{{2\pi }}} {\alpha ^{at}}\left( {i\xi } \right)}\\
{\quad \quad \quad \quad \quad \quad  \times \frac{{\frac{{\left[ {{{\tilde \varepsilon }_1}\left( {i\xi } \right) - 1} \right]}}{{\left[ {{{\tilde \varepsilon }_1}\left( {i\xi } \right) + 1} \right]}}{e^{ - 2kd}} - \frac{{\left[ {{{\tilde \varepsilon }_2}\left( {i\xi } \right) - 1} \right]}}{{\left[ {{{\tilde \varepsilon }_2}\left( {i\xi } \right) + 1} \right]}}{e^{ - 2k\left( {D - d} \right)}}}}{{1 - \frac{{\left[ {{{\tilde \varepsilon }_1}\left( {i\xi } \right) - 1} \right]\left[ {{{\tilde \varepsilon }_2}\left( {i\xi } \right) - 1} \right]}}{{\left[ {{{\tilde \varepsilon }_1}\left( {i\xi } \right) + 1} \right]\left[ {{{\tilde \varepsilon }_2}\left( {i\xi } \right) + 1} \right]}}{e^{ - 2kD}}}},}
\end{array}
\label{IVB6.6}
\end{equation}
and at finite temperature it is
\begin{equation}
\begin{array}{l}
F\left( d \right) =  - \frac{4}{\beta }\int\limits_0^\infty  {dk{k^3}} \sum\limits_{{\xi _n}} {'{\alpha ^{at}}\left( {i{\xi _n}} \right)} \\
\quad \quad \quad \quad \quad  \times \frac{{\frac{{\left[ {{{\tilde \varepsilon }_1}\left( {i{\xi _n}} \right) - 1} \right]}}{{\left[ {{{\tilde \varepsilon }_1}\left( {i{\xi _n}} \right) + 1} \right]}}{e^{ - 2kd}} - \frac{{\left[ {{{\tilde \varepsilon }_2}\left( {i{\xi _n}} \right) - 1} \right]}}{{\left[ {{{\tilde \varepsilon }_2}\left( {i{\xi _n}} \right) + 1} \right]}}{e^{ - 2k\left( {D - d} \right)}}}}{{1 - \frac{{\left[ {{{\tilde \varepsilon }_1}\left( {i{\xi _n}} \right) - 1} \right]\left[ {{{\tilde \varepsilon }_2}\left( {i{\xi _n}} \right) - 1} \right]}}{{\left[ {{{\tilde \varepsilon }_1}\left( {i{\xi _n}} \right) + 1} \right]\left[ {{{\tilde \varepsilon }_2}\left( {i{\xi _n}} \right) + 1} \right]}}{e^{ - 2kD}}}}.
\end{array}
\label{IVB6.7}
\end{equation}

\subsubsection{\label{atomfilmnonret}Force on an atom next to a 2D planar film (three layers)}

\begin{figure}
\includegraphics[width=8cm]{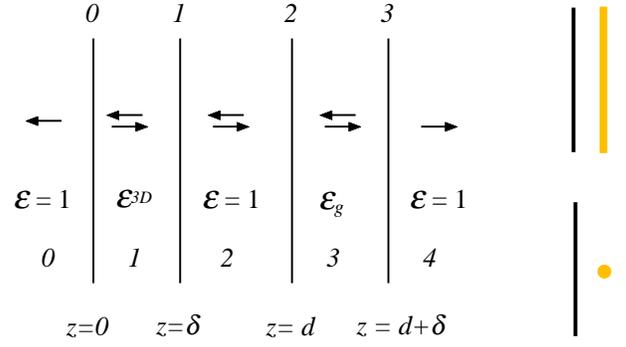}
\caption{(Color online) The geometry of a thin gas layer the distance $d$ from a thin film. This geometry is used to find the interaction between an atom and a thin film.}
\label{figu8}
\end{figure}
In this section we derive the van der Waals interaction of an atom near a very thin film. We start from the three layer structure in Fig.\,\ref{figu8}. We take the limit when the thickness of the film goes to zero.  The matrix becomes ${\bf{\tilde M}} = {{\bf{\tilde M}}_0} \cdot {{\bf{\tilde M}}_1} \cdot {{\bf{\tilde M}}_2}\cdot {{\bf{\tilde M}}_3}$, where $ {{\bf{\tilde M}}_0}\cdot {{\bf{\tilde M}}_1}$ is the matrix for the thin film,  and $ {{\bf{\tilde M}}_2}\cdot {{\bf{\tilde M}}_3}$ is the matrix for the gas film. These matrices are given in Eqs.\,(\ref{IVB4.1}) and (\ref{IVB3.1}), respectively.

The matrices are
\begin{equation}
\begin{array}{l}
{{{\bf{\tilde M}}}_0} \cdot {{{\bf{\tilde M}}}_1} = \left( {\begin{array}{*{20}{c}}
1&0\\
0&1
\end{array}} \right) + {{\tilde \alpha }^{2D}}\left( {\begin{array}{*{20}{c}}
1&1\\
{ - 1}&{ - 1}
\end{array}} \right);\\
{{{\bf{\tilde M}}}_2} \cdot {{{\bf{\tilde M}}}_3} = \left( {\begin{array}{*{20}{c}}
1&0\\
0&1
\end{array}} \right) + \left( {\delta n} \right){\alpha ^{at}}4\pi k\left( {\begin{array}{*{20}{c}}
0&{{e^{2kd}}}\\
{ - {e^{ - 2kd}}}&0
\end{array}} \right),
\end{array}
\label{IVB7.1}
\end{equation}
and the element of interest to us is
\begin{equation}
{M_{11}} = 1 + {{\tilde \alpha }^{2D}} - \left( {\delta n} \right){\alpha ^{at}}4\pi k{{\tilde \alpha }^{2D}}{e^{ - 2kd}}.
\label{IVB7.2}
\end{equation}
The first two terms produce the modes in the thin film alone and are not affected by the atom. We choose as our reference system the system when the atom is at infinite distance from the film. To get the mode condition function we divide ${M_{11}}$ with the first two terms.
The mode condition function becomes
\begin{equation}
{{\tilde f}_k} = 1 - \left( {\delta n} \right)4\pi k{\alpha ^{at}}\frac{{{{\tilde \alpha }^{2D}}}}{{1 + {{\tilde \alpha }^{2D}}}}{e^{ - 2kd}}.
\label{IVB7.3}
\end{equation}
From this we find the energy per atom is
\begin{equation}
\begin{array}{l}
\frac{E}{{n\delta }} = \frac{\hbar }{{2n\delta }}\int {\frac{{{d^2}k}}{{{{\left( {2\pi } \right)}^2}}}} \int\limits_{ - \infty }^\infty  {\frac{{d\xi }}{{2\pi }}} \ln \left[ {{{\tilde f}_{\bf{k}}}\left( {i\xi } \right)} \right]\\
\quad  \approx  - \hbar \int\limits_0^\infty  {dk{k^2}{e^{ - 2kd}}} \int\limits_{ - \infty }^\infty  {\frac{{d\xi }}{{2\pi }}} {\alpha ^{at}}\left( {i\xi } \right)\frac{{{{\tilde \alpha }^{2D}}\left( {k,i\xi } \right)}}{{1 + {{\tilde \alpha }^{2D}}\left( {k,i\xi } \right)}},
\end{array}
\label{IVB7.4}
\end{equation}
and the force on the atom is
\begin{equation}
F\left( d \right) =  - 2\hbar \int\limits_0^\infty  {dk{k^3}{e^{ - 2kd}}} \int\limits_{ - \infty }^\infty  {\frac{{d\xi }}{{2\pi }}} {\alpha ^{at}}\left( {i\xi } \right)\frac{{{{\tilde \alpha }^{2D}}\left( {k,i\xi } \right)}}{{1 + {{\tilde \alpha }^{2D}}\left( {k,i\xi } \right)}}.
\label{IVB7.5}
\end{equation}

\subsubsection{\label{twofilmsnopnret}Interaction between two 2D planar films (three layers)}
 We start from the three layer structure in Fig.\,\ref{figu8}. We take the limit when the thickness of the films goes to zero.  The matrix becomes ${\bf{\tilde M}} = {{\bf{\tilde M}}_0} \cdot {{\bf{\tilde M}}_1} \cdot {{\bf{\tilde M}}_2}\cdot {{\bf{\tilde M}}_3}$, where $ {{\bf{\tilde M}}_0}\cdot {{\bf{\tilde M}}_1}$ is the matrix for one of the two thin films,  and $ {{\bf{\tilde M}}_2}\cdot {{\bf{\tilde M}}_3}$ is the matrix for the other. These matrices are given in Eq.\,(\ref{IVB4.1}) the first for $z=0$ and the second for $z=d$.

The matrices are
\begin{equation}
\begin{array}{*{20}{l}}
{{{{\bf{\tilde M}}}_0} \cdot {{{\bf{\tilde M}}}_1} = \left( {\begin{array}{*{20}{c}}
1&0\\
0&1
\end{array}} \right) + {{\tilde \alpha }^{2D}}\left( {\begin{array}{*{20}{c}}
1&1\\
{ - 1}&{ - 1}
\end{array}} \right)};\\
{{{{\bf{\tilde M}}}_2} \cdot {{{\bf{\tilde M}}}_3} = \left( {\begin{array}{*{20}{c}}
1&0\\
0&1
\end{array}} \right) + {{\tilde \alpha }^{2D}}\left( {\begin{array}{*{20}{c}}
1&{{e^{2kd}}}\\
{ - {e^{ - 2kd}}}&{ - 1}
\end{array}} \right),}
\end{array}
\label{IVB8.1}
\end{equation}
and the element of interest to us is
\begin{equation}
\begin{array}{l}
{M_{11}} = 1 + 2{{\tilde \alpha }^{2D}} + \left( {1 - {e^{ - 2kd}}} \right){\left( {{{\tilde \alpha }^{2D}}} \right)^2}\\
\quad  = {\left( {1 + {{\tilde \alpha }^{2D}}} \right)^2} - {e^{ - 2kd}}{\left( {{{\tilde \alpha }^{2D}}} \right)^2}.
\end{array}
\label{IVB8.2}
\end{equation}
The first term produces the modes in the two thin films if they are so far apart that they are not affecting each other. We choose as our reference system the system when the two films are at infinite distance from each other. To get the mode condition function we divide ${M_{11}}$ with the first  term.
The mode condition function becomes
\begin{equation}
{{\tilde f}_k} = 1 - {e^{ - 2kd}}{\left( {\frac{{{{\tilde \alpha }^{2D}}}}{{1 + {{\tilde \alpha }^{2D}}}}} \right)^2}
\label{IVB8.3}
\end{equation}
From this we find the energy per unit area
\begin{equation}
\begin{array}{l}
E = \frac{\hbar }{2}\int {\frac{{{d^2}k}}{{{{\left( {2\pi } \right)}^2}}}} \int\limits_{ - \infty }^\infty  {\frac{{d\xi }}{{2\pi }}} \ln \left[ {{{\tilde f}_{\bf{k}}}\left( {i\xi } \right)} \right]\\
\quad  = \frac{\hbar }{2}\int {\frac{{{d^2}k}}{{{{\left( {2\pi } \right)}^2}}}} \int\limits_{ - \infty }^\infty  {\frac{{d\xi }}{{2\pi }}} \ln \left[ {1 - {e^{ - 2kd}}{{\left( {\frac{{{{\tilde \alpha }^{2D}}\left( {k,i\xi } \right)}}{{1 + {{\tilde \alpha }^{2D}}\left( {k,i\xi } \right)}}} \right)}^2}} \right].
\end{array}
\label{IVB8.4}
\end{equation}
This agrees completely with the results of Ref.\,[\onlinecite{SerBjo}].

\subsubsection{\label{atom2filmsn}Force on an atom in between two 2D planar films (five layers)}

Here we let the first 2D film be located at $z=0$, the thin diluted gas film at $z=d$, and the second 2D film at $D$. There is vacuum between the three films. Thus the matrix becomes  ${\bf{\tilde M}} = {{\bf{\tilde M}}_0} \cdot {{\bf{\tilde M}}_1} \cdot {{\bf{\tilde M}}_2}$ where
\begin{equation}
\begin{array}{*{20}{l}}
{{{{\bf{\tilde M}}}_0} = \left( {\begin{array}{*{20}{c}}
1&0\\
0&1
\end{array}} \right) + {{\tilde \alpha }^{2D}}\left( {\begin{array}{*{20}{c}}
1&1\\
{ - 1}&{ - 1}
\end{array}} \right);}\\
\begin{array}{l}
{{{\bf{\tilde M}}}_1} = \left( {\begin{array}{*{20}{c}}
1&0\\
0&1
\end{array}} \right) + \delta n{\alpha ^{at}}4\pi k\left( {\begin{array}{*{20}{c}}
0&{{e^{2kd}}}\\
{ - {e^{ - 2kd}}}&0
\end{array}} \right);\\
{{{\bf{\tilde M}}}_2} = \left( {\begin{array}{*{20}{c}}
1&0\\
0&1
\end{array}} \right) + {{\tilde \alpha }^{2D}}\left( {\begin{array}{*{20}{c}}
1&{{e^{2kD}}}\\
{ - {e^{ - 2kD}}}&{ - 1}
\end{array}} \right).
\end{array}
\end{array}
\label{IVB9.1}
\end{equation}
The matrix element of interest is
\begin{equation}
\begin{array}{l}
{M_{11}} = {\left( {1 + {{\tilde \alpha }^{2D}}} \right)^2} - {e^{ - 2kD}}{\left( {{{\tilde \alpha }^{2D}}} \right)^2}\\
\quad \quad  - \delta n{\alpha ^{at}}4\pi k{{\tilde \alpha }^{2D}}\left( {1 + {{\tilde \alpha }^{2D}}} \right)\left( {{e^{ - 2kd}} + {e^{ - 2k\left( {D - d} \right)}}} \right).
\end{array}
\label{IVB9.2}
\end{equation}
The first term is the mode condition for the two films at infinite separation in absence of the gas layer. The first two terms is the mode condition in absence of the gas layer (see Eq.\.(\ref{IVB8.2}). The mode condition function after division with the function in absence of the gas layer is
 \begin{equation}
{{\tilde f}_{\bf{k}}} = 1 - 4\pi k{\alpha ^{at}}\delta n\frac{{\frac{{{{\tilde \alpha }^{2D}}}}{{1 + {{\tilde \alpha }^{2D}}}}\left[ {{e^{ - 2kd}} + {e^{ - 2k\left( {D - d} \right)}}} \right]}}{{\left[ {1 - {e^{ - 2kD}}{{\left( {\frac{{{{\tilde \alpha }^{2D}}}}{{1 + {{\tilde \alpha }^{2D}}}}} \right)}^2}} \right]}},
\label{IVB9.3}
\end{equation}
and the interaction energy per atom becomes
\begin{equation}
\begin{array}{*{20}{l}}
{\frac{E}{{n\delta }} = \frac{\hbar }{{2n\delta }}\int {\frac{{{d^2}k}}{{{{\left( {2\pi } \right)}^2}}}} \int\limits_{ - \infty }^\infty  {\frac{{d\xi }}{{2\pi }}} \ln \left[ {{{\tilde f}_{\bf{k}}}\left( {i\xi } \right)} \right]}\\
{ \approx  - \frac{\hbar }{2}\int {\frac{{{d^2}k}}{{{{\left( {2\pi } \right)}^2}}}} 4\pi k\int\limits_{ - \infty }^\infty  {\frac{{d\xi }}{{2\pi }}} {\alpha ^{at}}\frac{{\frac{{{{\tilde \alpha }^{2D}}}}{{1 + {{\tilde \alpha }^{2D}}}}\left[ {{e^{ - 2kd}} + {e^{ - 2k\left( {D - d} \right)}}} \right]}}{{\left[ {1 - {e^{ - 2kD}}{{\left( {\frac{{{{\tilde \alpha }^{2D}}}}{{1 + {{\tilde \alpha }^{2D}}}}} \right)}^2}} \right]}}}\\
{ =  - \hbar \int\limits_0^\infty  {dk{k^2}} \int\limits_{ - \infty }^\infty  {\frac{{d\xi }}{{2\pi }}} {\alpha ^{at}}\left( {i\xi } \right)\frac{{\frac{{{{\tilde \alpha }^{2D}}\left( {k,i\xi } \right)}}{{1 + {{\tilde \alpha }^{2D}}\left( {k,i\xi } \right)}}\left[ {{e^{ - 2kd}} + {e^{ - 2k\left( {D - d} \right)}}} \right]}}{{\left[ {1 - {e^{ - 2kD}}{{\left( {\frac{{{{\tilde \alpha }^{2D}}\left( {k,i\xi } \right)}}{{1 + {{\tilde \alpha }^{2D}}\left( {k,i\xi } \right)}}} \right)}^2}} \right]}}.}
\end{array}
\label{IVB9.4}
\end{equation}
 Thus, the force on the atom is 
%
\begin{equation}
\begin{array}{*{20}{l}}
{F\left( d \right) =  - 4\hbar \int\limits_0^\infty  {dk{k^3}} \int\limits_0^\infty  {\frac{{d\xi }}{{2\pi }}} {\alpha ^{at}}\left( {i\xi } \right)}\\
{\quad \quad \quad \quad \quad \quad  \times \frac{{\frac{{{{\tilde \alpha }^{2D}}\left( {k,i\xi } \right)}}{{1 + {{\tilde \alpha }^{2D}}\left( {k,i\xi } \right)}}\left[ {{e^{ - 2kd}} - {e^{ - 2k\left( {D - d} \right)}}} \right]}}{{\left[ {1 - {e^{ - 2kD}}{{\left( {\frac{{{{\tilde \alpha }^{2D}}\left( {k,i\xi } \right)}}{{1 + {{\tilde \alpha }^{2D}}\left( {k,i\xi } \right)}}} \right)}^2}} \right]}},}
\end{array}
\label{IVB9.5}
\end{equation}
and at finite temperature it is
\begin{equation}
\begin{array}{l}
F\left( d \right) =  - \frac{4}{\beta }\int\limits_0^\infty  {dk{k^3}} \sum\limits_{{\xi _n}} {'{\alpha ^{at}}\left( {i{\xi _n}} \right)} \\
\quad \quad \quad \quad \quad  \times \frac{{\frac{{{{\tilde \alpha }^{2D}}\left( {k,i{\xi _n}} \right)}}{{1 + {{\tilde \alpha }^{2D}}\left( {k,i{\xi _n}} \right)}}\left[ {{e^{ - 2kd}} - {e^{ - 2k\left( {D - d} \right)}}} \right]}}{{\left[ {1 - {e^{ - 2kD}}{{\left( {\frac{{{{\tilde \alpha }^{2D}}\left( {k,i{\xi _n}} \right)}}{{1 + {{\tilde \alpha }^{2D}}\left( {k,i{\xi _n}} \right)}}} \right)}^2}} \right]}}.
\end{array}
\label{IVB9.6}
\end{equation}

Now we are done with the non-retarded results for planar structures and turn to the more complicated fully retarded treatment.

\subsection{\label{planarretmain}Retarded main results}
In the planar geometry nothing is gained by introducing the two Hertz-Debye potentials, $\pi_1$ and $\pi_2$. We study the fields themselves. The solution to the vector Helmholtz equation is a field with the spatial variation $\exp \left( {i{\bf{k}} \cdot {\bf{r}}} \right)\exp \left( { \pm {\gamma _i}kz} \right)$, where $\bf k$ is a two dimensional wave vector in the plane of the interfaces, $\bf r$ is the two dimensional component of the position vector in the $xy$-plane and 
\begin{equation}
\begin{array}{l}
{\gamma _i} = \sqrt {1 - {{\tilde \varepsilon }_i}{{\left( {\omega /ck} \right)}^2}} ;\\
{\gamma ^{\left( 0 \right)}} = \sqrt {1 - {{\left( {\omega /ck} \right)}^2}} .
\end{array}
\label{IVC1}
\end{equation}
%
\begin{figure}
\includegraphics[width=5cm]{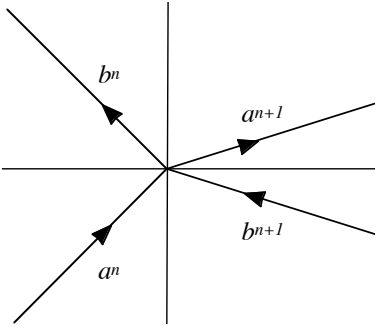}
\caption{The amplitudes of the waves at the interface number $n$.}
\label{figu9}
\end{figure}
We will make use of the Fresnel coefficients. The amplitude transmission and reflection coefficients for waves impinging on an interface between medium $i$ and $j$ from the $i$- side are
\begin{equation}
\begin{gathered}
t_{i,j}^s = \frac{{2{n_i}\cos {\theta _i}}}{{{n_i}\cos {\theta _i} + {n_j}\cos {\theta _j}}} = \frac{{2{\gamma _i}}}{{{\gamma _i} + {\gamma _j}}};\\
r_{i,j}^s = \frac{{{n_i}\cos {\theta _i} - {n_j}\cos {\theta _j}}}{{{n_i}\cos {\theta _i} + {n_j}\cos {\theta _j}}} = \frac{{{\gamma _i} - {\gamma _j}}}{{{\gamma _i} + {\gamma _j}}};\\
t_{i,j}^p = \frac{{2{n_i}\cos {\theta _i}}}{{{n_j}\cos {\theta _i} + {n_i}\cos {\theta _j}}} = \frac{{2\sqrt {{{\tilde \varepsilon }_i}{{\tilde \varepsilon }_j}} {\gamma _i}}}{{{{\tilde \varepsilon }_j}{\gamma _i} + {{\tilde \varepsilon }_i}{\gamma _j}}};\\
r_{i,j}^p = \frac{{{n_j}\cos {\theta _i} - {n_i}\cos {\theta _j}}}{{{n_j}\cos {\theta _i} + {n_i}\cos {\theta _j}}} = \frac{{{{\tilde \varepsilon }_j}{\gamma _i} - {{\tilde \varepsilon }_i}{\gamma _j}}}{{{{\tilde \varepsilon }_j}{\gamma _i} + {{\tilde \varepsilon }_i}{\gamma _j}}},
\end{gathered}
\label{IVC2}
\end{equation}
where $s$ and $p$ stands for $s$- and $p$-polarization, respectively or TE and TM, respectively. Now the wave in Fig.\,\ref{figu9} with amplitude ${a_{n + 1}}$ gets contribution from a transmitted part of the wave with amplitude ${a_n}$ and a reflected part from the wave with amplitude ${b_{n + 1}}$. Similarly the wave with amplitude ${b_n}$ gets contribution from a transmitted part of the wave with amplitude ${b_{n + 1}}$ and a reflected part of the wave with amplitude ${a_n}$. The fresnel coefficients are valid in our formalism if the interface is at $z=0$. Then we have
\begin{equation}
\begin{array}{l}
{a^{n + 1}} = {a^n}{t_{n,n + 1}} + {b^{n + 1}}{r_{n + 1,n}}\\
{b^n} = {a^{n + 1}}{r_{n,n + 1}} + {b^{n + 1}}{t_{n + 1,n}}
\end{array},
\label{IVC3}
\end{equation}
and after rearrangement and making use of the general relation ${t_{n,n + 1}}{t_{n + 1,n}} - {r_{n,n + 1}}{r_{n + 1,n}} = 1$ we find
\begin{equation}
\left( {\begin{array}{*{20}{c}}
{{a^n}}\\
{{b^n}}
\end{array}} \right) = \frac{1}{{{t_{n,n + 1}}}}\left( {\begin{array}{*{20}{c}}
1&{{r_{n,n + 1}}}\\
{{r_{n,n + 1}}}&1
\end{array}} \right)\cdot \left( {\begin{array}{*{20}{c}}
{{a^{n + 1}}}\\
{{b^{n + 1}}}
\end{array}} \right).
\label{IVC4}
\end{equation}

Now, with the position of the interface at $z=z_n$ we have
\begin{equation}
\begin{array}{l}
\left( {\begin{array}{*{20}{c}}
{{a^n}{e^{ - {\gamma _n}{kz_n}}}}\\
{{b^n}{e^{{\gamma _n}{kz_n}}}}
\end{array}} \right) =\\
 \quad \quad \quad \frac{1}{{{t_{n,n + 1}}}}\left( {\begin{array}{*{20}{c}}
1&{{r_{n,n + 1}}}\\
{{r_{n,n + 1}}}&1
\end{array}} \right) \cdot \left( {\begin{array}{*{20}{c}}
{{a^{n + 1}}{e^{ - {\gamma _{n + 1}}{kz_n}}}}\\
{{b^{n + 1}}{e^{{\gamma _{n + 1}}{kz_n}}}}
\end{array}} \right),
\end{array}
\label{IVC5}
\end{equation}
or
\begin{equation}
\begin{array}{l}
 \left( {\begin{array}{*{20}c}
   {a^n }  \\
   {b^n }  \\
\end{array}} \right) = {\bf{\tilde M}}_n \left( {\begin{array}{*{20}c}
   {a^{n + 1} }  \\
   {b^{n + 1} }  \\
\end{array}} \right); \\ 
 {\bf{\tilde M}}_n  = \\
 \frac{1}{{t_{n,n + 1} }}
 \left( {\begin{array}{*{20}c}
   {e^{ - \left( {\gamma _{n + 1}  - \gamma _n } \right)kz_n } } & {e^{\left( {\gamma _{n + 1}  + \gamma _n } \right)kz_n } r_{n,n + 1} }  \\
   {e^{ - \left( {\gamma _{n + 1}  + \gamma _n } \right)kz_n } r_{n,n + 1} } & {e^{\left( {\gamma _{n + 1}  - \gamma _n } \right)kz_n } }  \\
\end{array}} \right). \\ 
 \end{array}
\label{IVC6}
\end{equation}

Now we have all we need to determine the  fully retarded normal modes in a layered planar structure. We give some examples in the following section.

\subsection{\label{planarretspecial}Retarded special results}

\subsubsection{\label{interface}Single planar interface (no layer)}

For a single interface at $z=0$ between two media with dielectric functions ${{{\tilde \varepsilon }_0}}$ and ${{{\tilde \varepsilon }_1}}$, as illustrated in Fig.\,\ref{figu4}, we have
\begin{equation}
{\bf{\tilde M}} = {{{\bf{\tilde M}}}_0} = \frac{1}{{{t_{0,1}}}}\left( {\begin{array}{*{20}{c}}
1&{{r_{0,1}}}\\
{{r_{0,1}}}&1
\end{array}} \right).
\label{IVD1.1}
\end{equation}
For TE modes the condition for modes is 
\begin{equation}
{\gamma _0}\left( {k,\omega } \right) + {\gamma _1}\left( {k,\omega } \right) = 0.
\label{IVD1.2}
\end{equation}
This equation has no solution so there are no TE modes at a single interface.
For the TM modes  the condition for modes is
\begin{equation}
{\tilde \varepsilon _1}\left( \omega  \right){\gamma _0}\left( {k,\omega } \right) + {\tilde \varepsilon _0}\left( \omega  \right){\gamma _1}\left( {k,\omega } \right) = 0.
\label{IVD1.3}
\end{equation}
This equation has solutions, so there are  TM modes at a single interface.

\subsubsection{\label{slab}Slab or a planar gap (one layer)}

For a slab (see Fig.\,\ref{figu5}) with interfaces at $z=0$ and $z=d$ made of a medium with dielectric function ${{\tilde \varepsilon }_1}$ in an ambient medium with dielectric function ${{\tilde \varepsilon }_0}$ we have 
\begin{equation}
\begin{array}{l}
 {\bf{\tilde M}} = {\bf{\tilde M}}_0 \cdot{\bf{\tilde M}}_1  = \frac{1}{{t_{0,1} }}\left( {\begin{array}{*{20}c}
   1 & {r_{0,1} }  \\
   {r_{0,1} } & 1  \\
\end{array}} \right) \\ 
 \quad \quad  \times \frac{1}{{t_{1,0} }}\left( {\begin{array}{*{20}c}
   {e^{ - \left( {\gamma _0  - \gamma _1 } \right)kd} } & {e^{\left( {\gamma _0  + \gamma _1 } \right)kd} r_{1,0} }  \\
   {e^{ - \left( {\gamma _0  + \gamma _1 } \right)kd} r_{1,0} } & {e^{\left( {\gamma _0  - \gamma _1 } \right)kd} }  \\
\end{array}} \right), \\ 
 \end{array}
\label{IVD2.1}
\end{equation}
and the matrix elements are
\begin{equation}
\begin{array}{l}
{M_{11}} = \frac{1}{{{t_{0,1}}{t_{1,0}}}}\left[ {{e^{ - \left( {{\gamma _0} - {\gamma _1}} \right)kd}} + {e^{ - \left( {{\gamma _0} + {\gamma _1}} \right)kd}}{r_{0,1}}{r_{1,0}}} \right];\\
{M_{12}} = \frac{1}{{{t_{0,1}}{t_{1,0}}}}\left[ {{e^{\left( {{\gamma _0} + {\gamma _1}} \right)kd}}{r_{1,0}} + {e^{\left( {{\gamma _0} - {\gamma _1}} \right)kd}}{r_{0,1}}} \right];\\
{M_{21}} = \frac{1}{{{t_{0,1}}{t_{1,0}}}}\left[ {{e^{ - \left( {{\gamma _0} - {\gamma _1}} \right)kd}}{r_{0,1}} + {e^{ - \left( {{\gamma _0} + {\gamma _1}} \right)kd}}{r_{1,0}}} \right];\\
{M_{22}} = \frac{1}{{{t_{0,1}}{t_{1,0}}}}\left[ {{e^{\left( {{\gamma _0} + {\gamma _1}} \right)kd}}{r_{0,1}}{r_{1,0}} + {e^{\left( {{\gamma _0} - {\gamma _1}} \right)kd}}} \right].
\end{array}
\label{IVD2.2}
\end{equation}
For TE modes we have
%
\begin{equation}
{\left( {{\gamma _0} + {\gamma _1}} \right)^2} - {e^{ - 2{\gamma _1}kd}}{\left( {{\gamma _0} - {\gamma _1}} \right)^2} = 0,
\label{IVD2.3}
\end{equation}
and the mode condition function is
\begin{equation}
f_{\bf{k}}^{{\rm{TE}}}\left( \omega  \right) = {\left[ {{\gamma _0}\left( \omega  \right) + {\gamma _1}\left( \omega  \right)} \right]^2} - {e^{ - 2{\gamma _1}\left( \omega  \right)kd}}{\left[ {{\gamma _1}\left( \omega  \right) - {\gamma _0}\left( \omega  \right)} \right]^2}.
\label{IVD2.4}
\end{equation}
Note that we, as before, have identified the mode condition function as the numerator of the expression in the condition for modes.

For TM modes we have
\begin{equation}
{\left( {{{\tilde \varepsilon }_1}{\gamma _0} + {{\tilde \varepsilon }_0}{\gamma _1}} \right)^2} - {e^{ - 2{\gamma _1}kd}}{\left( {{{\tilde \varepsilon }_1}{\gamma _0} - {{\tilde \varepsilon }_0}{\gamma _1}} \right)^2} = 0,
\label{IVD2.5}
\end{equation}
and the mode condition function is
\begin{equation}
\begin{array}{l}
f_{\bf{k}}^{{\rm{TM}}}\left( \omega  \right) = {\left[ {{{\tilde \varepsilon }_1}\left( \omega  \right){\gamma _0}\left( \omega  \right) + {{\tilde \varepsilon }_0}\left( \omega  \right){\gamma _1}\left( \omega  \right)} \right]^2}\\
 \quad \quad \quad \quad \quad \quad- {e^{ - 2{\gamma _1}\left( \omega  \right)kd}}{\left[ {{{\tilde \varepsilon }_0}\left( \omega  \right){\gamma _1}\left( \omega  \right) - {{\tilde \varepsilon }_1}\left( \omega  \right){\gamma _0}\left( \omega  \right)} \right]^2}.
\end{array}
\label{IVD2.6}
\end{equation}

For a gap, of size $d$, filled with a medium with dielectric function ${{\tilde \varepsilon }_0}$ between two half spaces of material with dielectric function  ${{{\tilde \varepsilon }_1}}$ we may reuse the above result with the interchange of the two dielectric functions. We note that in this case, when retardation is included, the result will change. It did not in the non-retarded treatment. We have
\begin{equation}
\begin{array}{l}
f_{\bf{k}}^{{{\rm{TE}}}}\left( \omega  \right) = {\left[ {{\gamma _0}\left( \omega  \right) + {\gamma _1}\left( \omega  \right)} \right]^2} \\
\quad \quad \quad \quad \quad - {e^{ - 2{\gamma _0}\left( \omega  \right)kd}}{\left[ {{\gamma _1}\left( \omega  \right) - {\gamma _0}\left( \omega  \right)} \right]^2};\\
f_{\bf{k}}^{{\rm{TM}}}\left( \omega  \right) = {\left[ {{{\tilde \varepsilon }_1}\left( \omega  \right){\gamma _0}\left( \omega  \right) + {{\tilde \varepsilon }_0}\left( \omega  \right){\gamma _1}\left( \omega  \right)} \right]^2}\\
 \quad \quad \quad \quad \quad - {e^{ - 2{\gamma _0}\left( \omega  \right)kd}}{\left[ {{{\tilde \varepsilon }_0}\left( \omega  \right){\gamma _1}\left( \omega  \right) - {{\tilde \varepsilon }_1}\left( \omega  \right){\gamma _0}\left( \omega  \right)} \right]^2}.
\end{array}
\label{IVD2.7}
\end{equation}

If the half spaces are made up from two different materials with ${{{\tilde \varepsilon }_1}}$ and ${{{\tilde \varepsilon }_2}}$ we find
\begin{equation}
\begin{array}{l}
f_{\bf{k}}^{{{\rm{TE}}}}\left( \omega  \right) = \left[ {{\gamma _0}\left( \omega  \right) + {\gamma _1}\left( \omega  \right)} \right]\left[ {{\gamma _0}\left( \omega  \right) + {\gamma _2}\left( \omega  \right)} \right]\\
\quad \quad \quad  \quad- {e^{ - 2{\gamma _0}\left( \omega  \right)kd}}\left[ {{\gamma _1}\left( \omega  \right) - {\gamma _0}\left( \omega  \right)} \right]\left[ {{\gamma _2}\left( \omega  \right) - {\gamma _0}\left( \omega  \right)} \right];\\
f_{\bf{k}}^{{\rm{TM}}}\left( \omega  \right) = \left[ {{{\tilde \varepsilon }_1}\left( \omega  \right){\gamma _0}\left( \omega  \right) + {{\tilde \varepsilon }_0}\left( \omega  \right){\gamma _1}\left( \omega  \right)} \right]\\
 \quad \quad \quad \quad \quad \times \left[ {{{\tilde \varepsilon }_2}\left( \omega  \right){\gamma _0}\left( \omega  \right) + {{\tilde \varepsilon }_0}\left( \omega  \right){\gamma _2}\left( \omega  \right)} \right]\\
\quad \quad\quad  \quad- {e^{ - 2{\gamma _0}\left( \omega  \right)kd}}\left[ {{{\tilde \varepsilon }_0}\left( \omega  \right){\gamma _1}\left( \omega  \right) - {{\tilde \varepsilon }_1}\left( \omega  \right){\gamma _0}\left( \omega  \right)} \right]\\
\quad \quad \quad \quad \quad \times \left[ {{{\tilde \varepsilon }_2}\left( \omega  \right){\gamma _0}\left( \omega  \right) + {{\tilde \varepsilon }_0}\left( \omega  \right){\gamma _2}\left( \omega  \right)} \right].
\end{array}
\label{IVD2.8}
\end{equation}
\subsubsection{\label{gasfilmret}Thin planar diluted gas film (one layer)}
It is of interest to find the force on an atom in a layered structure. We can obtain this by studying the force on a thin layer of a diluted gas with dielectric function ${\varepsilon _g}\left( \omega  \right) = 1 + 4\pi n\alpha^{at} \left( \omega  \right)$, where $\alpha^{at}$ is the polarizability of one atom and $n$ the density of atoms (we have assumed that the atom is surrounded by vacuum; if not  the $1$ should be replaced by the dielectric function of the ambient medium and the atomic polarizability should be replaced by the excess polarizability). For a diluted gas layer the atoms do not interact with each other and the force on the layer is just the sum of the forces on the individual atoms. So by dividing with the number of atoms in the film we get the force on one atom. The layer has to be thin in order to have a well defined $z$- value of the atom.  Since we will derive the force on an atom in different planar geometries it is fruitful to derive the matrix for a thin diluted gas. This result can be directly used in the derivation of the force on an atom in different planar geometries.

We let the film have the thickness $\delta$ and be placed in the general position $z$. We only keep terms up to linear order in $\delta$ and linear order in $n$. We find the result up to linear order in $\delta$ is
\begin{equation}
\begin{array}{l}
{{{\bf{\tilde M}}}_0} \cdot {{{\bf{\tilde M}}}_1} = \left( {\begin{array}{*{20}{c}}
1&0\\
0&1
\end{array}} \right)\\
 + \delta k\left( {\begin{array}{*{20}{c}}
{ - {\gamma _0} + {\gamma _1}\left[ {\frac{{1 + {{\left( {{r_{0,1}}} \right)}^2}}}{{{t_{0,1}}{t_{1,0}}}}} \right]}&{ - \frac{{2{\gamma _1}{r_{0,1}}}}{{{t_{0,1}}{t_{1,0}}}}{e^{2{\gamma _0}kz}}}\\
{\frac{{2{\gamma _1}{r_{0,1}}}}{{{t_{0,1}}{t_{1,0}}}}{e^{ - 2{\gamma _0}kz}}}&{{\gamma _0} - {\gamma _1}\left[ {\frac{{1 + {{\left( {{r_{0,1}}} \right)}^2}}}{{{t_{0,1}}{t_{1,0}}}}} \right]}
\end{array}} \right).
\end{array}
\label{IVD3.1}
\end{equation}
To go further and find the result to lowest order in $n$ we have to specify the mode type. 

For TM modes we get

\begin{equation}
\begin{array}{*{20}{l}}
{{\bf{\tilde M}}_{{\rm{gaslayer}}}^{{\rm{TM}}} = \left( {\begin{array}{*{20}{c}}
1&0\\
0&1
\end{array}} \right) + \frac{{\left( {\delta n} \right)2\pi k{\alpha ^{at}}}}{{{\gamma ^{\left( 0 \right)}}}}}\\
{ \times \left( {\begin{array}{*{20}{c}}
{ - {{\left( {\frac{\omega }{{ck}}} \right)}^2}}&{ - \left[ {2 - {{\left( {\frac{\omega }{{ck}}} \right)}^2}} \right]{e^{2{\gamma ^{\left( 0 \right)}}kz}}}\\
{\left[ {2 - {{\left( {\frac{\omega }{{ck}}} \right)}^2}} \right]{e^{ - 2{\gamma ^{\left( 0 \right)}}kz}}}&{{{\left( {\frac{\omega }{{ck}}} \right)}^2}}
\end{array}} \right),}
\end{array}
\label{IVD3.2}
\end{equation}

and for TE modes

\begin{equation}
\begin{array}{*{20}{l}}
{{\bf{\tilde M}}_{{\rm{gaslayer}}}^{{\rm{TE}}} = \left( {\begin{array}{*{20}{c}}
1&0\\
0&1
\end{array}} \right)}\\
{\quad \quad  + \frac{{\left( {\delta n} \right)2\pi k{\alpha ^{at}}{{\left( {\omega /ck} \right)}^2}}}{{{\gamma ^{\left( 0 \right)}}}}\left( {\begin{array}{*{20}{c}}
{ - 1}&{ - {e^{2{\gamma ^{\left( 0 \right)}}kz}}}\\
{{e^{ - 2{\gamma ^{\left( 0 \right)}}kz}}}&1
\end{array}} \right).}
\end{array}
\label{IVD3.3}
\end{equation}
Now we are done with the gas layer. We will use these results later in calculating the force on an atom in planar layered structures.

\subsubsection{\label{2Dfilmret}2D planar film (one layer)}
In many situations one is dealing with very thin films. These may be considered 2D (two dimensional). Important examples are a graphene sheet and a 2D electron gas. In the derivation we let the film have finite thickness $\delta$ and be characterized by a 3D dielectric function ${\varepsilon ^{3D}}$. We then let the thickness go towards zero. The 3D dielectric function depends on $\delta$ as ${\varepsilon ^{3D}} \sim 1/\delta $ for small $\delta$ and $\delta {\varepsilon ^{3D}} \to 2{\alpha ^{2D}}/k$ as  $\delta$ goes towards zero.\,\cite{grap,arx} We may now start from Eq.(\ref{IVD3.1}) use the proper reflection and transmission coefficients for the mode type under consideration and let $\delta$ go towards zero. 

For the TM modes we obtain

\begin{equation}
\begin{array}{*{20}{l}}
{{\bf{\tilde M}}_{{\rm{2D}}}^{{\rm{TM}}} = \left( {\begin{array}{*{20}{c}}
1&0\\
0&1
\end{array}} \right) + \frac{{{\gamma ^{\left( 0 \right)}}k\left( {\delta {{\tilde \varepsilon }^{3D}}} \right)}}{2}\left( {\begin{array}{*{20}{c}}
1&{ - {e^{2{\gamma ^{\left( 0 \right)}}kz}}}\\
{{e^{ - 2{\gamma ^{\left( 0 \right)}}kz}}}&{ - 1}
\end{array}} \right)}\\
{\quad  = \left( {\begin{array}{*{20}{c}}
1&0\\
0&1
\end{array}} \right) + {{\tilde \alpha }^{2D}}{\gamma ^{\left( 0 \right)}}\left( {\begin{array}{*{20}{c}}
1&{ - {e^{2{\gamma ^{\left( 0 \right)}}kz}}}\\
{{e^{ - 2{\gamma ^{\left( 0 \right)}}kz}}}&{ - 1}
\end{array}} \right),}
\end{array}
\label{IVD4.1}
\end{equation}
and for the TE modes we find
\begin{equation}
\begin{array}{*{20}{l}}
{{\bf{\tilde M}}_{{\rm{2D}}}^{{\rm{TE}}} = \left( {\begin{array}{*{20}{c}}
1&0\\
0&1
\end{array}} \right) - \frac{{\left( {\delta {{\tilde \varepsilon }^{3D}}} \right)k{{\left( {\omega /ck} \right)}^2}}}{{2{\gamma ^{\left( 0 \right)}}}}\left( {\begin{array}{*{20}{c}}
1&{{e^{2{\gamma ^{\left( 0 \right)}}kz}}}\\
{ - {e^{ - 2{\gamma ^{\left( 0 \right)}}kz}}}&{ - 1}
\end{array}} \right)}\\
{\quad  = \left( {\begin{array}{*{20}{c}}
1&0\\
0&1
\end{array}} \right) - {{\tilde \alpha }^{2D}}\frac{{{{\left( {\omega /ck} \right)}^2}}}{{{\gamma ^{\left( 0 \right)}}}}\left( {\begin{array}{*{20}{c}}
1&{{e^{2{\gamma ^{\left( 0 \right)}}kz}}}\\
{ - {e^{ - 2{\gamma ^{\left( 0 \right)}}kz}}}&{ - 1}
\end{array}} \right).}
\end{array}
\label{IVD4.2}
\end{equation}

\subsubsection{\label{atomplane}Force on an atom next to a planar substrate (two layers)}
We start from the two layer structure in Fig.\,\ref{figu6}. We let the ambient be vacuum. The first layer is the thin gas layer treated in Sec. \ref{gasfilmret}. We place it at $z=0$. The second layer is a vacuum layer of thickness $d$. The remaining medium is the substrate which we let be infinitely thick and have the dielectric function ${{\tilde \varepsilon }_s}\left( \omega  \right)$.  The matrix becomes ${\bf{\tilde M}} = {{\bf{\tilde M}}_0} \cdot {{\bf{\tilde M}}_1} \cdot {{\bf{\tilde M}}_2} = {{\bf{\tilde M}}_{{\rm{gaslayer}}}} \cdot {{\bf{\tilde M}}_2}$ where we already know the first matrix. Now,
\begin{equation}
\begin{array}{l}
{{{\bf{\tilde M}}}_2} = \frac{1}{{{t_{2,3}}}}\left( {\begin{array}{*{20}{c}}
{{e^{ - \left( {{\gamma _3} - {\gamma _2}} \right)kd}}}&{{e^{\left( {{\gamma _3} + {\gamma _2}} \right)kd}}{r_{2,3}}}\\
{{e^{ - \left( {{\gamma _3} + {\gamma _2}} \right)kd}}{r_{2,3}}}&{{e^{\left( {{\gamma _3} - {\gamma _2}} \right)kd}}}
\end{array}} \right)\\
\quad \; = \frac{1}{{{t_{2,3}}}}\left( {\begin{array}{*{20}{c}}
{{e^{ - \left( {{\gamma _s} - {\gamma ^{(0)}}} \right)kd}}}&{{e^{\left( {{\gamma _s} + {\gamma ^{(0)}}} \right)kd}}{r_{2,3}}}\\
{{e^{ - \left( {{\gamma _s} + {\gamma ^{(0)}}} \right)kd}}{r_{2,3}}}&{{e^{\left( {{\gamma _s} - {\gamma ^{(0)}}} \right)kd}}}
\end{array}} \right).
\end{array}
\label{IVD5.1}
\end{equation}
For TM modes it becomes
\begin{equation}
\begin{array}{l}
{\bf{\tilde M}}_2^{{\rm{TM}}} = \frac{{{{\tilde \varepsilon }_s}{\gamma ^{(0)}} + {\gamma _s}}}{{2\sqrt {{{\tilde \varepsilon }_s}} {\gamma ^{(0)}}}}\\
 \times \left( {\begin{array}{*{20}{c}}
{{e^{ - \left( {{\gamma _s} - {\gamma ^{(0)}}} \right)kd}}}&{{e^{\left( {{\gamma _s} + {\gamma ^{(0)}}} \right)kd}}\frac{{{{\tilde \varepsilon }_s}{\gamma ^{(0)}} - {\gamma _s}}}{{{{\tilde \varepsilon }_s}{\gamma ^{(0)}} + {\gamma _s}}}}\\
{{e^{ - \left( {{\gamma _s} + {\gamma ^{(0)}}} \right)kd}}\frac{{{{\tilde \varepsilon }_s}{\gamma ^{(0)}} - {\gamma _s}}}{{{{\tilde \varepsilon }_s}{\gamma ^{(0)}} + {\gamma _s}}}}&{{e^{\left( {{\gamma _s} - {\gamma ^{(0)}}} \right)kd}}}
\end{array}} \right),
\end{array}
\label{IVD5.2}
\end{equation}
and for TE modes
\begin{equation}
\begin{array}{l}
{\bf{\tilde M}}_2^{{\rm{TE}}} = \frac{{{\gamma ^{(0)}} + {\gamma _s}}}{{2{\gamma ^{(0)}}}}\\
 \times \left( {\begin{array}{*{20}{c}}
{{e^{ - \left( {{\gamma _s} - {\gamma ^{(0)}}} \right)kd}}}&{{e^{\left( {{\gamma _s} + {\gamma ^{(0)}}} \right)kd}}\frac{{{\gamma ^{(0)}} - {\gamma _s}}}{{{\gamma ^{(0)}} + {\gamma _s}}}}\\
{{e^{ - \left( {{\gamma _s} + {\gamma ^{(0)}}} \right)kd}}\frac{{{\gamma ^{(0)}} - {\gamma _s}}}{{{\gamma ^{(0)}} + {\gamma _s}}}}&{{e^{\left( {{\gamma _s} - {\gamma ^{(0)}}} \right)kd}}}
\end{array}} \right).
\end{array}
\label{IVD5.3}
\end{equation}
So the condition for TM modes is
\begin{equation}
\begin{array}{*{20}{l}}
{1 - \frac{{\left( {\delta n} \right)2\pi k{\alpha ^{at}}}}{{{\gamma ^{\left( 0 \right)}}}}\left\{ {{{\left( {\frac{\omega }{{ck}}} \right)}^2}} \right.}\\
{\quad \quad \quad \left. { + {e^{ - 2{\gamma ^{(0)}}kd}}\left[ {2 - {{\left( {\frac{\omega }{{ck}}} \right)}^2}} \right]\frac{{{{\tilde \varepsilon }_s}{\gamma ^{(0)}} - {\gamma _s}}}{{{{\tilde \varepsilon }_s}{\gamma ^{(0)}} + {\gamma _s}}}} \right\} = 0,}
\end{array}
\label{IVD5.4}
\end{equation}
and for TE modes
\begin{equation}
1 - \frac{{\left( {\delta n} \right)2\pi k{\alpha ^{at}}{{\left( {\omega /ck} \right)}^2}}}{{{\gamma ^{\left( 0 \right)}}}}\left[ {1 + {e^{ - 2{\gamma ^{(0)}}kd}}\frac{{{\gamma ^{(0)}} - {\gamma _s}}}{{{\gamma ^{(0)}} + {\gamma _s}}}} \right] = 0.
\label{IVD5.5}
\end{equation}

The mode condition function for TM modes is
\begin{equation}
\begin{array}{*{20}{l}}
{f_{\bf{k}}^{{\rm{TM}}} = 1 - \frac{{\left( {\delta n} \right)2\pi k{\alpha ^{at}}}}{{{\gamma ^{\left( 0 \right)}}}}\left\{ {{{\left( {\frac{\omega }{{ck}}} \right)}^2}} \right.}\\
{\quad \quad \quad \quad \left. { + {e^{ - 2{\gamma ^{(0)}}kd}}\left[ {2 - {{\left( {\frac{\omega }{{ck}}} \right)}^2}} \right]\frac{{{{\tilde \varepsilon }_s}{\gamma ^{(0)}} - {\gamma _s}}}{{{{\tilde \varepsilon }_s}{\gamma ^{(0)}} + {\gamma _s}}}} \right\},}
\end{array}
\label{IVD5.6}
\end{equation}
and for TE modes
\begin{equation}
f_{\bf{k}}^{{\rm{TE}}} = 1 - \frac{{\left( {\delta n} \right)2\pi k{\alpha ^{at}}{{\left( {\omega /ck} \right)}^2}}}{{{\gamma ^{\left( 0 \right)}}}}\left[ {1 + {e^{ - 2{\gamma ^{(0)}}kd}}\frac{{{\gamma ^{(0)}} - {\gamma _s}}}{{{\gamma ^{(0)}} + {\gamma _s}}}} \right].
\label{IVD5.7}
\end{equation}
Note the first part in each mode condition function,
\begin{equation}
f_{\bf{k}}^{\rm{0}} = 1 - \left( {\delta n} \right){\alpha ^{at}}\frac{{2\pi k{{\left( {\omega /ck} \right)}^2}}}{{{\gamma ^{\left( 0 \right)}}}}.
\label{IVD5.8}
\end{equation}
It does not depend on the distance of the atom to the substrate. It is the effect of the contribution from the atom to the screening and the resulting change of the dispersion curves for the vacuum modes. This type of interaction was used by Feynman to derive the Lamb shift of the hydrogen atom.\,\cite{Power} We see that it contributes the same in both type of modes. We divide with this function since it leads to a constant energy, independent of the atom distance to the interface. So the relevant mode condition functions relative infinite separation are
\begin{equation}
\begin{array}{*{20}{l}}
{\tilde f_{\bf{k}}^{{\rm{TM}}} = 1 - \left( {\delta n} \right){\alpha ^{at}}\frac{{2\pi k\left[ {2 - {{\left( {\omega /ck} \right)}^2}} \right]}}{{{\gamma ^{\left( 0 \right)}}}}{e^{ - 2{\gamma ^{(0)}}kd}}\frac{{{{\tilde \varepsilon }_s}{\gamma ^{(0)}} - {\gamma _s}}}{{{{\tilde \varepsilon }_s}{\gamma ^{(0)}} + {\gamma _s}}};}\\
{\tilde f_{\bf{k}}^{{\rm{TE}}} = 1 - \left( {\delta n} \right){\alpha ^{at}}\frac{{2\pi k{{\left( {\omega /ck} \right)}^2}}}{{{\gamma ^{\left( 0 \right)}}}}{e^{ - 2{\gamma ^{(0)}}kd}}\frac{{{\gamma ^{(0)}} - {\gamma _s}}}{{{\gamma ^{(0)}} + {\gamma _s}}}.}
\end{array}
\label{IVD5.9}
\end{equation}
The interaction energy per atom becomes
\begin{equation}
\begin{array}{*{20}{l}}
{\frac{E}{{n\delta }} = \frac{\hbar }{{n\delta }}\int {\frac{{{d^2}k}}{{{{\left( {2\pi } \right)}^2}}}} \int\limits_0^\infty  {\frac{{d\xi }}{{2\pi }}} \left\{ {\ln \left[ {\tilde f_{\bf{k}}^{{\rm{TE}}}\left( {i\xi } \right)} \right] + \ln \left[ {\tilde f_{\bf{k}}^{{\rm{TM}}}\left( {i\xi } \right)} \right]} \right\}}\\
{ = 2\pi \hbar \int {\frac{{{d^2}k}}{{{{\left( {2\pi } \right)}^2}}}k} \int\limits_0^\infty  {\frac{{d\xi }}{{2\pi }}} {\alpha ^{at}}\left( {i\xi } \right){e^{ - 2{\gamma ^{(0)}}\left( {i\xi } \right)kd}}}\\
{\quad \quad  \times \left\{ {\frac{{{{\left( {\xi /ck} \right)}^2}}}{{{\gamma ^{(0)}}\left( {i\xi } \right)}}\frac{{{\gamma ^{(0)}}\left( {i\xi } \right) - {\gamma _s}\left( {i\xi } \right)}}{{{\gamma ^{(0)}}\left( {i\xi } \right) + {\gamma _s}\left( {i\xi } \right)}}} \right.}\\
{\quad \quad \quad \quad  + \left. {\frac{{\left[ {2 + {{\left( {\xi /ck} \right)}^2}} \right]}}{{{\gamma ^{(0)}}\left( {i\xi } \right)}}\frac{{{\gamma _s}\left( {i\xi } \right) - {{\tilde \varepsilon }_s}\left( {i\xi } \right){\gamma ^{(0)}}\left( {i\xi } \right)}}{{{\gamma _s}\left( {i\xi } \right) + {{\tilde \varepsilon }_s}\left( {i\xi } \right){\gamma ^{(0)}}\left( {i\xi } \right)}}} \right\},}
\end{array}
\label{IVD5.10}
\end{equation}
where we have taken the limit when $\left( {\delta n} \right)$ goes towards zero. The force on the atom is
\begin{equation}
\begin{array}{*{20}{l}}
{F\left( d \right) }\\
{ \quad= 4\pi \hbar \int {\frac{{{d^2}k}}{{{{\left( {2\pi } \right)}^2}}}} {k^2}\int\limits_0^\infty  {\frac{{d\xi }}{{2\pi }}} {\alpha ^{at}}\left( {i\xi } \right){e^{ - 2{\gamma ^{(0)}}\left( {i\xi } \right)kd}}}\\
{\quad \quad  \times \left\{ {{{\left( {\xi {\rm{/}}ck} \right)}^2}\frac{{{\gamma ^{(0)}}\left( {i\xi } \right) - {\gamma _s}\left( {i\xi } \right)}}{{{\gamma ^{(0)}}\left( {i\xi } \right) + {\gamma _s}\left( {i\xi } \right)}}} \right.}\\
{\quad \quad \quad \quad  + \left. {\left[ {2 + {{\left( {\xi /ck} \right)}^2}} \right]\frac{{{\gamma _s}\left( {i\xi } \right) - {{\tilde \varepsilon }_s}\left( {i\xi } \right){\gamma ^{(0)}}\left( {i\xi } \right)}}{{{\gamma _s}\left( {i\xi } \right) + {{\tilde \varepsilon }_s}\left( {i\xi } \right){\gamma ^{(0)}}\left( {i\xi } \right)}}} \right\}.}
\end{array}
\label{IVD5.11}
\end{equation}
At finite temperature it is
\begin{equation}
\begin{array}{l}
F\left( d \right)\\
\quad  = \frac{{4\pi }}{\beta }\int {\frac{{{d^2}k}}{{{{\left( {2\pi } \right)}^2}}}{k^2}} \sum\limits_{{\xi _n}} {'{\alpha ^{at}}\left( {i{\xi _n}} \right){e^{ - 2{\gamma ^{(0)}}\left( {i{\xi _n}} \right)d}}} \\
\quad \quad \quad  \times \left\{ {{{\left( {{\xi _n}/ck} \right)}^2}\frac{{{\gamma ^{(0)}}\left( {i{\xi _n}} \right) - {\gamma _s}\left( {i{\xi _n}} \right)}}{{{\gamma ^{(0)}}\left( {i{\xi _n}} \right) + {\gamma _s}\left( {i{\xi _n}} \right)}}} \right.\\
\quad \quad \quad \quad  + \left. {\left[ {2 + {{\left( {{\xi _n}/ck} \right)}^2}} \right]\frac{{{\gamma _s}\left( {i{\xi _n}} \right) - {{\tilde \varepsilon }_s}\left( {i{\xi _n}} \right){\gamma ^{(0)}}\left( {i{\xi _n}} \right)}}{{{\gamma _s}\left( {i{\xi _n}} \right) + {{\tilde \varepsilon }_s}\left( {i{\xi _n}} \right){\gamma ^{(0)}}\left( {i{\xi _n}} \right)}}} \right\}.
\end{array}
\label{IVD5.12}
\end{equation}
These results are in complete agreement with the results in Ref.\,[\onlinecite{Saturation}].

\subsubsection{\label{atomplanes}Force on an atom in between two planar surfaces (three layers)}

We refer to Fig.\,\ref{figu7} and let the first interface be located at $z=0$ separating one plate with dielectric function ${{\tilde \varepsilon }_1}$ from the ambient medium which we let be vacuum. Next interface, at $z=d$, is the left interface of the gas layer with dielectric function ${\varepsilon _g}$ and thickness $\delta $. Thus the third interface is at $z=d+\delta$. The forth interface is located at $z=D$ and separates vacuum from the second plate with dielectric function ${{\tilde \varepsilon }_2}$.  The matrix becomes ${\bf{\tilde M}} = {{\bf{\tilde M}}_0} \cdot {{\bf{\tilde M}}_1} \cdot {{\bf{\tilde M}}_2} \cdot {{\bf{\tilde M}}_3} = {{\bf{\tilde M}}_0} \cdot {{\bf{\tilde M}}_{{\rm{gaslayer}}}} \cdot {{\bf{\tilde M}}_3}$ where ${{\bf{\tilde M}}_{{\rm{gaslayer}}}}$ is taken from Eq.\,(\ref{IVD3.2}) or (\ref{IVD3.3}) for TM modes and TE modes, respectively, and
\begin{equation}
\begin{array}{*{20}{l}}
{{{{\bf{\tilde M}}}_0} = \frac{1}{{{t_{0,1}}}}\left( {\begin{array}{*{20}{c}}
1&{{r_{0,1}}}\\
{{r_{0,1}}}&1
\end{array}} \right);}\\
{{{{\bf{\tilde M}}}_3} = \frac{1}{{{t_{3,4}}}}\left( {\begin{array}{*{20}{c}}
{{e^{ - \left( {{\gamma _s} - {\gamma ^{\left( 0 \right)}}} \right)kD}}}&{{e^{\left( {{\gamma _s} + {\gamma ^{\left( 0 \right)}}} \right)kD}}{r_{3,4}}}\\
{{e^{ - \left( {{\gamma _s} + {\gamma ^{\left( 0 \right)}}} \right)kD}}{r_{3,4}}}&{{e^{\left( {{\gamma _s} - {\gamma ^{\left( 0 \right)}}} \right)kD}}}
\end{array}} \right).}
\end{array}
\label{IVD6.1}
\end{equation}
Now,
\begin{equation}
{M_{11}} = \left( {\begin{array}{*{20}{c}}
{M_{11}^0}&{M_{12}^0}
\end{array}} \right) \cdot {{{\bf{\tilde M}}}_{{\rm{gaslayer}}}} \cdot \left( {\begin{array}{*{20}{c}}
{M_{11}^3}\\
{M_{21}^3}
\end{array}} \right).
\label{IVD6.2}
\end{equation}
To find a general expression valid for both mode types we use the expression in Eq.\,(\ref{IVD3.1}) for ${{\bf{\tilde M}}_{{\rm{gaslayer}}}}$. It is the expression to linear order in $\delta$ but before the lowest order in $n$ is taken. Then
 \begin{equation}
\begin{array}{*{20}{l}}
{{M_{11}} = \frac{{{e^{ - \left( {{\gamma _s} - {\gamma ^{\left( 0 \right)}}} \right)kD}}}}{{{t_{0,1}}{t_{3,4}}}}\left\{ {1 + {e^{ - 2{\gamma ^{\left( 0 \right)}}kD}}{r_{0,1}}{r_{3,4}}} \right.}\\
{ + \delta k\left[ {\left( {{\gamma _g} - {\gamma ^{\left( 0 \right)}}} \right)\left( {1 - {e^{ - 2{\gamma ^{\left( 0 \right)}}kD}}{r_{0,1}}{r_{3,4}}} \right)} \right.}\\
{ - \left. {\left. {2{\gamma ^{\left( 0 \right)}}{r_{2,3}}\left( {{e^{ - 2{\gamma ^{\left( 0 \right)}}kd}}{r_{0,1}} - {e^{2{\gamma ^{\left( 0 \right)}}k\left( {d - D} \right)}}{r_{3,4}}} \right)} \right]} \right\}.}
\end{array}
\label{IVD6.3}
\end{equation}
The condition for modes is
\begin{equation}
\begin{array}{*{20}{l}}
{\begin{array}{*{20}{l}}
{1 + {e^{ - 2{\gamma ^{\left( 0 \right)}}kD}}{r_{0,1}}{r_{3,4}}}\\
{\quad  + \delta k\left[ {\left( {{\gamma _g} - {\gamma ^{\left( 0 \right)}}} \right)\left( {1 - {e^{ - 2{\gamma ^{\left( 0 \right)}}kD}}{r_{0,1}}{r_{3,4}}} \right)} \right.}
\end{array}}\\
{\left. {\quad \quad  - 2{\gamma ^{\left( 0 \right)}}{r_{2,3}}\left( {{e^{ - 2{\gamma ^{\left( 0 \right)}}kd}}{r_{0,1}} - {e^{2{\gamma ^{\left( 0 \right)}}k\left( {d - D} \right)}}{r_{3,4}}} \right)} \right] = 0.}
\end{array}
\label{IVD6.4}
\end{equation}
The mode condition function is
%
\begin{equation}
\begin{array}{*{20}{l}}
{\begin{array}{*{20}{l}}
{{{\tilde f}_{\bf{k}}} = 1 + \frac{{\delta k}}{{1 + {e^{ - 2{\gamma ^{\left( 0 \right)}}kD}}{r_{0,1}}{r_{3,4}}}}}\\
{ \times \left[ {\left( {{\gamma _g} - {\gamma ^{\left( 0 \right)}}} \right)\left( {1 - {e^{ - 2{\gamma ^{\left( 0 \right)}}kD}}{r_{0,1}}{r_{3,4}}} \right)} \right.}
\end{array}}\\
{ - \left. {2{\gamma ^{\left( 0 \right)}}{r_{2,3}}\left( {{e^{ - 2{\gamma ^{\left( 0 \right)}}kd}}{r_{0,1}} - {e^{2{\gamma ^{\left( 0 \right)}}k\left( {d - D} \right)}}{r_{3,4}}} \right)} \right]}\\
{\begin{array}{*{20}{l}}
{ = 1 + \frac{{\delta kn{\alpha ^{at}}}}{{{\gamma ^{\left( 0 \right)}}\left[ {1 + {e^{ - 2{\gamma ^{\left( 0 \right)}}D}}{r_{0,1}}{r_{3,4}}} \right]}}}\\
{ \times \left[ { - 2\pi {{\left( {\omega /ck} \right)}^2}\left( {1 - {e^{ - 2{\gamma ^{\left( 0 \right)}}kD}}{r_{0,1}}{r_{3,4}}} \right)} \right.}
\end{array}}\\
{ - \left. {2{{\left( {{\gamma ^{\left( 0 \right)}}} \right)}^2}\left( {\frac{{{r_{2,3}}}}{{n{\alpha ^{at}}}}} \right)\left( {{e^{ - 2{\gamma ^{\left( 0 \right)}}kd}}{r_{0,1}} - {e^{2{\gamma ^{\left( 0 \right)}}k\left( {d - D} \right)}}{r_{3,4}}} \right)} \right].}
\end{array}
\label{IVD6.5}
\end{equation}
Just as for the atom next to a substrate a part of this function does not depend on the position of the gas layer. The energy change is due to the screening of the vacuum caused by the polarizable atom.  It is interesting to note that this effect is modified by the presence of the two planar surfaces. We intend to investigate this further in a forthcoming publication. We divide with the function
\begin{equation}
\begin{array}{*{20}{l}}
{f_{\bf{k}}^0 = 1 - \frac{{\delta k\left( {1 - {e^{ - 2{\gamma ^{\left( 0 \right)}}kD}}{r_{0,1}}{r_{3,4}}} \right)}}{{1 + {e^{ - 2{\gamma ^{\left( 0 \right)}}kD}}{r_{0,1}}{r_{3,4}}}}\left( {{\gamma ^{(0)}} - {\gamma _g}} \right)}\\
{ = 1 - \delta kn{\alpha ^{at}}\frac{{2\pi }}{{{\gamma ^{(0)}}}}{{\left( {\frac{\omega }{{ck}}} \right)}^2}\frac{{\left( {1 - {e^{ - 2{\gamma ^{\left( 0 \right)}}kD}}{r_{0,1}}{r_{3,4}}} \right)}}{{\left( {1 + {e^{ - 2{\gamma ^{\left( 0 \right)}}kD}}{r_{0,1}}{r_{3,4}}} \right)}}.}
\end{array}
\label{IVD6.6}
\end{equation}
For the energy this means that we subtract a term that is independent of the position of the atom and will not effect the force on the atom. We find
\begin{equation}
\begin{array}{*{20}{l}}
{{{\tilde{ \tilde{ f}}}_{\bf{k}}} = 1 + \delta kn{\alpha ^{at}}}\\
{\quad  \times \frac{{\left[ { - 2\left( {{\gamma ^{\left( 0 \right)}}} \right)\left( {{r_{2,3}}/n{\alpha ^{at}}} \right)\left( {{e^{ - 2{\gamma ^{\left( 0 \right)}}kd}}{r_{0,1}} - {e^{2{\gamma ^{\left( 0 \right)}}k\left( {d - D} \right)}}{r_{3,4}}} \right)} \right]}}{{\left[ {1 + {e^{ - 2{\gamma ^{\left( 0 \right)}}kD}}{r_{0,1}}{r_{3,4}}} \right]}},}
\end{array}
\label{IVD6.7}
\end{equation}
and the interaction energy per atom is
\begin{equation}
\frac{E}{{n\delta }} = \hbar \int {\frac{{{d^2}k}}{{{{\left( {2\pi } \right)}^2}}}} \int\limits_0^\infty  {\frac{{d\xi }}{{2\pi }}} \left[ {{I^{{\rm{TM}}}}\left( {k,i\xi } \right) + {I^{{{\rm{TE}}}}}\left( {k,i\xi } \right)} \right],
\label{IVD6.8}
\end{equation}
where
\begin{equation}
\begin{array}{*{20}{l}}
{{I^{{\rm{TM}}}}\left( {k,\omega } \right)}\\
{ = {\alpha ^{at}}\frac{{2\pi k\left[ {2 - {{\left( {\omega /ck} \right)}^2}} \right]\left[ {{e^{ - 2{\gamma ^{\left( 0 \right)}}kd}}r_{0,1}^{{\rm{TM}}} - {e^{ - 2{\gamma ^{\left( 0 \right)}}k\left( {D - d} \right)}}r_{3,4}^{{\rm{TM}}}} \right]}}{{{\gamma ^{\left( 0 \right)}}\left[ {1 + {e^{ - 2{\gamma ^{\left( 0 \right)}}kD}}r_{0,1}^{{\rm{TM}}}r_{3,4}^{{\rm{TM}}}} \right]}};}\\
{{I^{{\rm{TE}}}}\left( {k,\omega } \right) = {\alpha ^{at}}\frac{{2\pi k{{\left( {\omega /ck} \right)}^2}\left[ {{e^{ - 2{\gamma ^{\left( 0 \right)}}kd}}r_{0,1}^{{\rm{TE}}} - {e^{ - 2{\gamma ^{\left( 0 \right)}}k\left( {D - d} \right)}}r_{3,4}^{{\rm{TE}}}} \right]}}{{{\gamma ^{\left( 0 \right)}}\left[ {1 + {e^{ - 2{\gamma ^{\left( 0 \right)}}kD}}r_{0,1}^{{\rm{TE}}}r_{3,4}^{{\rm{TE}}}} \right]}}.}
\end{array}
\label{IVD6.9}
\end{equation}
The explicit expressions for the entering reflection coefficients are obtained from Eq.\,(\ref{IVC2}). They are
\begin{equation}
\begin{array}{*{20}{l}}
{r_{0,1}^{{\rm{TM}}}\left( {k,\omega } \right) = \frac{{\sqrt {1 - {{\tilde \varepsilon }_1}\left( \omega  \right){{\left( {\omega /ck} \right)}^2}}  - {{\tilde \varepsilon }_1}\left( \omega  \right)\sqrt {1 - {{\left( {\omega /ck} \right)}^2}} }}{{\sqrt {1 - {{\tilde \varepsilon }_1}\left( \omega  \right){{\left( {\omega /ck} \right)}^2}}  + {{\tilde \varepsilon }_1}\left( \omega  \right)\sqrt {1 - {{\left( {\omega /ck} \right)}^2}} }};}\\
{r_{3,4}^{{\rm{TM}}}\left( {k,\omega } \right) = \frac{{{{\tilde \varepsilon }_2}\left( \omega  \right)\sqrt {1 - {{\left( {\omega /ck} \right)}^2}}  - \sqrt {1 - {{\tilde \varepsilon }_2}{{\left( {\omega /ck} \right)}^2}} }}{{{{\tilde \varepsilon }_2}\left( \omega  \right)\sqrt {1 - {{\left( {\omega /ck} \right)}^2}}  + \sqrt {1 - {{\tilde \varepsilon }_2}{{\left( {\omega /ck} \right)}^2}} }};}\\
{r_{0,1}^{{\rm{TE}}}\left( {k,\omega } \right) = \frac{{\sqrt {1 - {{\tilde \varepsilon }_1}\left( \omega  \right){{\left( {\omega /ck} \right)}^2}}  - \sqrt {1 - {{\left( {\omega /ck} \right)}^2}} }}{{\sqrt {1 - {{\tilde \varepsilon }_1}\left( \omega  \right){{\left( {\omega /ck} \right)}^2}}  + \sqrt {1 - {{\left( {\omega /ck} \right)}^2}} }};}\\
{r_{3,4}^{{\rm{TE}}}\left( {k,\omega } \right) = \frac{{\sqrt {1 - {{\left( {\omega /ck} \right)}^2}}  - \sqrt {1 - {{\tilde \varepsilon }_2}{{\left( {\omega /ck} \right)}^2}} }}{{\sqrt {1 - {{\left( {\omega /ck} \right)}^2}}  + \sqrt {1 - {{\tilde \varepsilon }_2}{{\left( {\omega /ck} \right)}^2}} }}.}
\end{array}
\label{IVD6.10}
\end{equation}
 Thus the force on the atom is 
%
\begin{equation}
\begin{array}{l}
F\left( d \right) = \hbar \int {\frac{{{d^2}k}}{{{{\left( {2\pi } \right)}^2}}}} \int\limits_0^\infty  {\frac{{d\xi }}{{2\pi }}} \left[ {{J^{{\rm{TM}}}}\left( {k,i\xi } \right) + {J^{{{\rm{TE}}}}}\left( {k,i\xi } \right)} \right]
\end{array},
\label{IVD6.11}
\end{equation}
where
\begin{equation}
\begin{array}{*{20}{l}}
{{J^{{\rm{TM}}}}\left( {k,\omega } \right)}\\
{ = {\alpha ^{at}}\frac{{4\pi {k^2}\left[ {2 - {{\left( {\omega /ck} \right)}^2}} \right]\left[ {{e^{ - 2{\gamma ^{\left( 0 \right)}}kd}}r_{0,1}^{{\rm{TM}}} + {e^{ - 2{\gamma ^{\left( 0 \right)}}k\left( {D - d} \right)}}r_{3,4}^{{\rm{TM}}}} \right]}}{{\left[ {1 + {e^{ - 2{\gamma ^{\left( 0 \right)}}kD}}r_{0,1}^{{\rm{TM}}}r_{3,4}^{{\rm{TM}}}} \right]}};}\\
{{J^{{\rm{TE}}}}\left( {k,\omega } \right) = {\alpha ^{at}}\frac{{4\pi {k^2}{{\left( {\omega /ck} \right)}^2}\left[ {{e^{ - 2{\gamma ^{\left( 0 \right)}}kd}}r_{0,1}^{{\rm{TE}}} + {e^{ - 2{\gamma ^{\left( 0 \right)}}k\left( {D - d} \right)}}r_{3,4}^{{\rm{TE}}}} \right]}}{{\left[ {1 + {e^{ - 2{\gamma ^{\left( 0 \right)}}kD}}r_{0,1}^{{\rm{TE}}}r_{3,4}^{{\rm{TE}}}} \right]}},}
\end{array}
\label{IVD6.12}
\end{equation}
and at finite temperature it is
\begin{equation}
F\left( d \right) = \frac{1}{\beta }\int {\frac{{{d^2}k}}{{{{\left( {2\pi } \right)}^2}}}} \sum\limits_{{\xi _n}} {^{'}} \left[ {{J^{{\rm{TM}}}}\left( {k,i{\xi _n}} \right) + {J^{{{\rm{TE}}}}}\left( {k,i{\xi _n}} \right)} \right].
\label{IVD6.13}
\end{equation}

\subsubsection{\label{atomfilmret}Force on an atom next to a planar 2D film (three layers)}

In this section we derive the Casimir interaction of an atom near a very thin film. We proceed along he lines of Sec.\,\ref{atomfilmnonret}. We start from the three layer structure in Fig.\,\ref{figu8}.  The matrix becomes ${\bf{\tilde M}} = {{\bf{\tilde M}}_0} \cdot {{\bf{\tilde M}}_1} \cdot {{\bf{\tilde M}}_2}\cdot {{\bf{\tilde M}}_3}$, where $ {{\bf{\tilde M}}_0}\cdot {{\bf{\tilde M}}_1}$ is the matrix for the thin film,  and $ {{\bf{\tilde M}}_2}\cdot {{\bf{\tilde M}}_3}$ is the matrix for the gas film. These matrices we have derived before. We have two mode types, TM and TE. 

We start with the TM modes. The resulting matrices for the two thin layers were given in Eqs.\,(\ref{IVD4.1}) and (\ref{IVD3.2}), respectively. They are
\begin{equation}
\begin{array}{*{20}{l}}
{{{{\bf{\tilde M}}}_0} \cdot {{{\bf{\tilde M}}}_1} = \left( {\begin{array}{*{20}{c}}
1&0\\
0&1
\end{array}} \right) + \frac{{{\gamma ^{\left( 0 \right)}}k\left( {\delta {{\tilde \varepsilon }^{3D}}} \right)}}{2}\left( {\begin{array}{*{20}{c}}
1&{ - 1}\\
1&{ - 1}
\end{array}} \right);}\\
{{{{\bf{\tilde M}}}_2} \cdot {{{\bf{\tilde M}}}_3} = \left( {\begin{array}{*{20}{c}}
1&0\\
0&1
\end{array}} \right) + \frac{{\left( {\delta n} \right)2\pi k{\alpha ^{at}}}}{{{\gamma ^{\left( 0 \right)}}}}}\\
{ \times \left( {\begin{array}{*{20}{c}}
{ - {{\left( {\frac{\omega }{{ck}}} \right)}^2}}&{ - {e^{2{\gamma ^{\left( 0 \right)}}kd}}\left[ {2 - {{\left( {\frac{\omega }{{ck}}} \right)}^2}} \right]}\\
{{e^{ - 2{\gamma ^{\left( 0 \right)}}kd}}\left[ {2 - {{\left( {\frac{\omega }{{ck}}} \right)}^2}} \right]}&{{{\left( {\frac{\omega }{{ck}}} \right)}^2}}
\end{array}} \right),}
\end{array}
\label{IVD7.1}
\end{equation}
and the resulting matrix element that interests us is
\begin{equation}
\begin{array}{*{20}{l}}
{{M_{11}} = 1 + \frac{{{\gamma ^{\left( 0 \right)}}k\left( {\delta {{\tilde \varepsilon }^{3D}}} \right)}}{2} - \frac{{\left( {\delta n} \right)2\pi k{\alpha ^{at}}}}{{{\gamma ^{\left( 0 \right)}}}}{{\left( {\frac{\omega }{{ck}}} \right)}^2}}\\
{ - \left( {\delta n} \right)\pi {k^2}{\alpha ^{at}}\left( {\delta {{\tilde \varepsilon }^{3D}}} \right)\left\{ {{{\left( {\frac{\omega }{{ck}}} \right)}^2} + {e^{ - 2{\gamma ^{\left( 0 \right)}}kd}}\left( {2{k^2} - {{\left( {\frac{\omega }{{ck}}} \right)}^2}} \right)} \right\}.}
\end{array}
\label{IVD7.2}
\end{equation}

The TM mode condition function becomes
\begin{equation}
\begin{array}{*{20}{l}}
{\begin{array}{*{20}{l}}
{\tilde f_{\bf{k}}^{{\rm{TM}}}\left( \omega  \right)}\\
{\quad  = 1 - \frac{{\left( {\delta n} \right)\pi {k^2}{\alpha ^{at}}\left( {\delta {{\tilde \varepsilon }^{3D}}} \right)\left( {2 - {{\left( {\frac{\omega }{{ck}}} \right)}^2}} \right){e^{ - 2{\gamma ^{\left( 0 \right)}}kd}}}}{{1 + \frac{{{\gamma ^{\left( 0 \right)}}k\left( {\delta {{\tilde \varepsilon }^{3D}}} \right)}}{2} - \frac{{2\pi k\left( {\delta n} \right){\alpha ^{at}}}}{{{\gamma ^{\left( 0 \right)}}}}{{\left( {\frac{\omega }{{ck}}} \right)}^2} - \left( {\delta n} \right)\pi {k^2}{\alpha ^{at}}\left( {\delta {{\tilde \varepsilon }^{3D}}} \right){{\left( {\frac{\omega }{{ck}}} \right)}^2}}}}
\end{array}}\\
{\begin{array}{*{20}{l}}
{\quad  \approx 1 - \frac{{\left( {\delta n} \right)\pi {k^2}{\alpha ^{at}}\left( {\delta {{\tilde \varepsilon }^{3D}}} \right)\left[ {2 - {{\left( {\frac{\omega }{{ck}}} \right)}^2}} \right]{e^{ - 2{\gamma ^{\left( 0 \right)}}kd}}}}{{1 + \frac{{{\gamma ^{\left( 0 \right)}}k\left( {\delta {{\tilde \varepsilon }^{3D}}} \right)}}{2}}}}\\
{ = 1 - \frac{{\left( {\delta n} \right)2\pi k{\alpha ^{at}}{{\tilde \alpha }^{2D}}\left( {{\bf{k}},\omega } \right)\left[ {2 - {{\left( {\frac{\omega }{{ck}}} \right)}^2}} \right]{e^{ - 2{\gamma ^{\left( 0 \right)}}kd}}}}{{1 + {\gamma ^{\left( 0 \right)}}{{\tilde \alpha }^{2D}}\left( {{\bf{k}},\omega } \right)}}.}
\end{array}}
\end{array}
\label{IVD7.3}
\end{equation}

For the TE modes the matrices for the two films from Eqs.\,(\ref{IVD4.2}) and (\ref{IVD3.3}), respectively  are

\begin{equation}
\begin{array}{*{20}{l}}
{{{{\bf{\tilde M}}}_0} \cdot {{{\bf{\tilde M}}}_1} = \left( {\begin{array}{*{20}{c}}
1&0\\
0&1
\end{array}} \right) - \frac{{\left( {\delta {{\tilde \varepsilon }^{3D}}} \right)k{{\left( {\omega /ck} \right)}^2}}}{{2{\gamma ^{\left( 0 \right)}}}}\left( {\begin{array}{*{20}{c}}
1&1\\
{ - 1}&{ - 1}
\end{array}} \right);}\\
{{{{\bf{\tilde M}}}_2} \cdot {{{\bf{\tilde M}}}_3} = \left( {\begin{array}{*{20}{c}}
1&0\\
0&1
\end{array}} \right)}\\
{\quad \quad  - \frac{{\left( {\delta n} \right){{\left( {\omega /ck} \right)}^2}2\pi k{\alpha ^{at}}}}{{{\gamma ^{\left( 0 \right)}}}}\left( {\begin{array}{*{20}{c}}
1&{{e^{2{\gamma ^{\left( 0 \right)}}kd}}}\\
{ - {e^{ - 2{\gamma ^{\left( 0 \right)}}kd}}}&{ - 1}
\end{array}} \right),}
\end{array}
\label{IVD7.4}
\end{equation}
and the resulting matrix element that interests us is
\begin{equation}
\begin{array}{*{20}{l}}
{{M_{11}} = 1 - \frac{{\left( {\delta {{\tilde \varepsilon }^{3D}}} \right)k{{\left( {\omega /ck} \right)}^2}}}{{2{\gamma ^{\left( 0 \right)}}}} - \frac{{{{\left( {\omega /ck} \right)}^2}2\pi k\left( {\delta n} \right){\alpha ^{at}}}}{{{\gamma ^{\left( 0 \right)}}}}}\\
{\quad \quad \quad \quad  + \frac{{{{\left( {\omega /ck} \right)}^4}\pi {k^2}\left( {\delta n} \right){\alpha ^{at}}\left( {\delta {{\tilde \varepsilon }^{3D}}} \right)}}{{{{\left[ {{\gamma ^{\left( 0 \right)}}} \right]}^2}}}\left( {1 - {e^{ - 2{\gamma ^{\left( 0 \right)}}kd}}} \right).}
\end{array}
\label{IVD7.5}
\end{equation}
The TE mode condition function is
\begin{equation}
\begin{array}{*{20}{l}}
{\begin{array}{*{20}{l}}
{\tilde f_{\bf{k}}^{{\rm{TE}}}\left( \omega  \right)}\\
{\; = 1 - \frac{{\frac{{{{\left( {\omega /ck} \right)}^4}\pi {k^2}\left( {\delta n} \right){\alpha ^{at}}\left( {\delta {{\tilde \varepsilon }^{3D}}} \right)}}{{{{\left[ {{\gamma ^{\left( 0 \right)}}} \right]}^2}}}{e^{ - 2{\gamma ^{\left( 0 \right)}}kd}}}}{{1 - \frac{{\left( {\delta {{\tilde \varepsilon }^{3D}}} \right)k{{\left( {\omega /ck} \right)}^2}}}{{2{\gamma ^{\left( 0 \right)}}}} - \frac{{\left( {\delta n} \right){{\left( {\omega /ck} \right)}^2}2\pi k{\alpha ^{at}}}}{{{\gamma ^{\left( 0 \right)}}}} + \frac{{\left( {\delta n} \right){{\left( {\omega /ck} \right)}^4}\pi {k^2}{\alpha ^{at}}\left( {\delta {{\tilde \varepsilon }^{3D}}} \right)}}{{{{\left[ {{\gamma ^{\left( 0 \right)}}} \right]}^2}}}}}}
\end{array}}\\
{\begin{array}{*{20}{l}}
{\quad \quad  \approx 1 - \frac{{\left( {\delta n} \right){{\left( {\omega /ck} \right)}^4}\pi {k^2}{\alpha ^{at}}\left( {\delta {{\tilde \varepsilon }^{3D}}} \right)}}{{{\gamma ^{\left( 0 \right)}}\left( {{\gamma ^{\left( 0 \right)}} - \frac{{\left( {\delta {{\tilde \varepsilon }^{3D}}} \right)k{{\left( {\omega /ck} \right)}^2}}}{2}} \right)}}{e^{ - 2{\gamma ^{\left( 0 \right)}}kd}}}\\
{\quad \quad  = 1 - \frac{{\left( {\delta n} \right)2\pi k{\alpha ^{at}}{{\tilde \alpha }^{2D}}\left( {{\bf{k}},\omega } \right){{\left( {\omega /ck} \right)}^4}}}{{\left( {{\gamma ^{\left( 0 \right)}}} \right)\left[ {\left( {{\gamma ^{\left( 0 \right)}}} \right) - {{\tilde \alpha }^{2D}}\left( {{\bf{k}},\omega } \right){{\left( {\omega /ck} \right)}^2}} \right]}}{e^{ - 2{\gamma ^{\left( 0 \right)}}kd}}.}
\end{array}}
\end{array}
\label{IVD7.6}
\end{equation}

The interaction energy per atom is
\begin{equation}
\begin{array}{l}
\frac{E}{{n\delta }} = \frac{\hbar }{{n\delta }}\int {\frac{{{d^2}k}}{{{{\left( {2\pi } \right)}^2}}}} \int\limits_0^\infty  {\frac{{d\xi }}{{2\pi }}} \left\{ {\ln \left[ {\tilde f_{\bf{k}}^{{\rm{TM}}}\left( {i\xi } \right)} \right] + \ln \left[ {\tilde f_{\bf{k}}^{{\rm{TE}}}\left( {i\xi } \right)} \right]} \right\}\\
 \approx \hbar \int {\frac{{{d^2}k}}{{{{\left( {2\pi } \right)}^2}}}} \int\limits_0^\infty  {\frac{{d\xi }}{{2\pi }}} \frac{{\left[ {\tilde f_{\bf{k}}^{{\rm{TM}}}\left( {i\xi } \right) - 1} \right] + \left[ {\tilde f_{\bf{k}}^{{\rm{TE}}}\left( {i\xi } \right) - 1} \right]}}{{n\delta }}\\
 =  - \frac{\hbar }{{2\pi }}\int {dk{k^2}} \int\limits_0^\infty  {d\xi } {\alpha ^{at}}\left( {i\xi } \right){{\tilde \alpha }^{2D}}\left( {k,i\xi } \right){e^{ - 2\sqrt {1 + {{\left( {\xi /ck} \right)}^2}} kd}}\\
\quad \quad \quad \quad \quad \quad \quad \quad \quad \quad \quad  \times \left[ {\frac{{\left[ {2 + {{\left( {\xi /ck} \right)}^2}} \right]}}{{1 + \sqrt {1 + {{\left( {\xi /ck} \right)}^2}} {{\tilde \alpha }^{2D}}\left( {k,i\xi } \right)}}} \right.\\
\left. {\quad \quad \quad \quad \quad  + \frac{{{{\left( {\xi /ck} \right)}^4}}}{{\sqrt {1 + {{\left( {\xi /ck} \right)}^2}} \left[ {\sqrt {1 + {{\left( {\xi /ck} \right)}^2}}  + {{\tilde \alpha }^{2D}}\left( {k,i\xi } \right){{\left( {\xi /ck} \right)}^2}} \right]}}} \right],
\end{array}
\label{IVD7.7}
\end{equation}
and the force on the atom becomes
\begin{equation}
\begin{array}{l}
F\left( d \right)\\
\quad  =  - \frac{\hbar }{\pi }\int {dk{k^3}} \int\limits_0^\infty  {d\xi } {\alpha ^{at}}\left( {i\xi } \right){{\tilde \alpha }^{2D}}\left( {k,i\xi } \right){e^{ - 2\sqrt {1 + {{\left( {\xi /ck} \right)}^2}} kd}}\\
\quad \quad \quad \quad \quad \quad \quad \quad  \times \left[ {\frac{{\left[ {2 + {{\left( {\xi /ck} \right)}^2}} \right]\sqrt {1 + {{\left( {\xi /ck} \right)}^2}} }}{{1 + \sqrt {1 + {{\left( {\xi /ck} \right)}^2}} {{\tilde \alpha }^{2D}}\left( {k,i\xi } \right)}}} \right.\\
\left. {\quad \quad \quad \quad \quad \quad  + \frac{{{{\left( {\xi /ck} \right)}^4}}}{{\left[ {\sqrt {1 + {{\left( {\xi /ck} \right)}^2}}  + {{\tilde \alpha }^{2D}}\left( {k,i\xi } \right){{\left( {\xi /ck} \right)}^2}} \right]}}} \right].
\end{array}
\label{IVD7.8}
\end{equation}

\subsubsection{\label{twofilmsret}Interaction between two 2D planar films (three layers)}
In this section we derive the Casimir interaction between two thin films. We proceed along he lines of the preceding section. We start from the three layer structure in Fig.\,\ref{figu8}. We take the limit when the thickness of the film goes to zero.  The matrix becomes ${\bf{\tilde M}} = {{\bf{\tilde M}}_0} \cdot {{\bf{\tilde M}}_1} \cdot {{\bf{\tilde M}}_2}\cdot {{\bf{\tilde M}}_3}$, where $ {{\bf{\tilde M}}_0}\cdot {{\bf{\tilde M}}_1}$ is the matrix for one of the thin films,  and $ {{\bf{\tilde M}}_2}\cdot {{\bf{\tilde M}}_3}$ is the matrix for the other. 

We start with the TM modes. The matrices from Eq.\,(\ref{IVD4.1}), one for $z=0$ and one for $z=d$ are
\begin{equation}
\begin{array}{*{20}{l}}
{{\bf{\tilde M}}_0^{{\rm{TM}}} \cdot {\bf{\tilde M}}_1^{{\rm{TM}}} = \left( {\begin{array}{*{20}{c}}
1&0\\
0&1
\end{array}} \right) + {\alpha ^{2D}}{\gamma ^{\left( 0 \right)}}\left( {\begin{array}{*{20}{c}}
1&{ - 1}\\
1&{ - 1}
\end{array}} \right)}\\
{\begin{array}{*{20}{l}}
{{\bf{\tilde M}}_2^{{\rm{TM}}} \cdot {\bf{\tilde M}}_3^{{\rm{TM}}} = \left( {\begin{array}{*{20}{c}}
1&0\\
0&1
\end{array}} \right)}\\
{\quad \quad \quad \quad \quad  + {\alpha ^{2D}}{\gamma ^{\left( 0 \right)}}\left( {\begin{array}{*{20}{c}}
1&{ - {e^{2{\gamma ^{\left( 0 \right)}}kd}}}\\
{{e^{ - 2{\gamma ^{\left( 0 \right)}}kd}}}&{ - 1}
\end{array}} \right).}
\end{array}}
\end{array}
\label{IVD8.1}
\end{equation}
The matrix element of interest to us is
\begin{equation}
{M_{11}} = {\left( {1 + {\alpha ^{2D}}{\gamma ^{\left( 0 \right)}}} \right)^2} - {\left( {{\alpha ^{2D}}{\gamma ^{\left( 0 \right)}}} \right)^2}{e^{ - 2{\gamma ^{\left( 0 \right)}}kd}},
\label{IVD8.2}
\end{equation}
and the mode condition function for TM modes is
\begin{equation}
{\tilde f^{{\rm{TM}}}} = 1 - {e^{ - 2{\gamma ^{\left( 0 \right)}}d}}\frac{{{{\left( {{\alpha ^{2D}}{\gamma ^{\left( 0 \right)}}} \right)}^2}}}{{{{\left( {1 + {\alpha ^{2D}}{\gamma ^{\left( 0 \right)}}} \right)}^2}}}.
\label{IVD8.3}
\end{equation}
Now, we continue with the TE modes. The matrices from Eq.\,(\ref{IVD4.2}), one for $z=0$ and one for $z=d$ are
\begin{equation}
\begin{array}{*{20}{l}}
{{\bf{\tilde M}}_0^{{\rm{TE}}} \cdot {\bf{\tilde M}}_1^{{\rm{TE}}} = \left( {\begin{array}{*{20}{c}}
1&0\\
0&1
\end{array}} \right) - {\alpha ^{2D}}\frac{{{{\left( {\omega /ck} \right)}^2}}}{{{\gamma ^{\left( 0 \right)}}}}\left( {\begin{array}{*{20}{c}}
1&1\\
{ - 1}&{ - 1}
\end{array}} \right);}\\
{\begin{array}{*{20}{l}}
{{\bf{\tilde M}}_2^{{\rm{TE}}} \cdot {\bf{\tilde M}}_3^{{\rm{TE}}} = \left( {\begin{array}{*{20}{c}}
1&0\\
0&1
\end{array}} \right)}\\
{\quad \quad \quad \quad  - {\alpha ^{2D}}\frac{{{{\left( {\omega /ck} \right)}^2}}}{{{\gamma ^{\left( 0 \right)}}}}\left( {\begin{array}{*{20}{c}}
1&{{e^{2{\gamma ^{\left( 0 \right)}}kd}}}\\
{ - {e^{ - 2{\gamma ^{\left( 0 \right)}}kd}}}&{ - 1}
\end{array}} \right).}
\end{array}}
\end{array}
\label{IVD8.4}
\end{equation}
The matrix element of interest to us is
\begin{equation}
{M_{11}} = {\left( {1 - {\alpha ^{2D}}\frac{{{{\left( {\omega /ck} \right)}^2}}}{{{\gamma ^{\left( 0 \right)}}}}} \right)^2} - {\left( {{\alpha ^{2D}}\frac{{{{\left( {\omega /ck} \right)}^2}}}{{{\gamma ^{\left( 0 \right)}}}}} \right)^2}{e^{ - 2{\gamma ^{\left( 0 \right)}}kd}},
\label{IVD8.5}
\end{equation}
and the mode condition function for TE modes is
\begin{equation}
{\tilde f^{{\rm{TE}}}} = 1 - {e^{ - 2{\gamma ^{\left( 0 \right)}}kd}}\frac{{{{\left( {{\alpha ^{2D}}\frac{{{{\left( {\omega /ck} \right)}^2}}}{{{\gamma ^{\left( 0 \right)}}}}} \right)}^2}}}{{{{\left( {1 - {\alpha ^{2D}}\frac{{{{\left( {\omega /ck} \right)}^2}}}{{{\gamma ^{\left( 0 \right)}}}}} \right)}^2}}}.
\label{IVD8.6}
\end{equation}
From this we find the energy per unit area
\begin{equation}
\begin{array}{*{20}{l}}
{E = \frac{\hbar }{2}\int {\frac{{{d^2}k}}{{{{\left( {2\pi } \right)}^2}}}} \int\limits_{ - \infty }^\infty  {\frac{{d\xi }}{{2\pi }}} \left\{ {\ln \left[ {\tilde f_{\bf{k}}^{{\rm{TM}}}\left( {i\xi } \right)} \right] + \ln \left[ {\tilde f_{\bf{k}}^{{\rm{TE}}}\left( {i\xi } \right)} \right]} \right\}}\\
{\begin{array}{*{20}{l}}
{ = \frac{\hbar }{2}\int {\frac{{{d^2}k}}{{{{\left( {2\pi } \right)}^2}}}} \int\limits_{ - \infty }^\infty  {\frac{{d\xi }}{{2\pi }}} \ln \left[ {1 - {e^{ - 2{\gamma ^{\left( 0 \right)}}\left( {i\xi } \right)kd}}} \right.}\\
{\left. {\quad \quad \quad \quad \quad \quad \quad \quad  \times {{\left( {\frac{{{{\tilde \alpha }^{2D}}\left( {k,i\xi } \right){\gamma ^{\left( 0 \right)}}\left( {k,i\xi } \right)}}{{1 + {{\tilde \alpha }^{2D}}\left( {k,i\xi } \right){\gamma ^{\left( 0 \right)}}\left( {k,i\xi } \right)}}} \right)}^2}} \right]}\\
{ + \frac{\hbar }{2}\int {\frac{{{d^2}k}}{{{{\left( {2\pi } \right)}^2}}}} \int\limits_{ - \infty }^\infty  {\frac{{d\xi }}{{2\pi }}} \ln \left[ {1 - {e^{ - 2{\gamma ^{\left( 0 \right)}}\left( {i\xi } \right)kd}}} \right.}\\
{\left. {\quad \quad \quad \quad \quad \quad \quad  \times {{\left( {\frac{{ - {{\tilde \alpha }^{2D}}\left( {k,i\xi } \right){{\left( {\xi /ck} \right)}^2}}}{{{\gamma ^{\left( 0 \right)}}\left( {k,i\xi } \right) + {{\tilde \alpha }^{2D}}\left( {k,i\xi } \right){{\left( {\xi /ck} \right)}^2}}}} \right)}^2}} \right].}
\end{array}}
\end{array}
\label{IVD8.7}
\end{equation}
This agrees completely with the results of Refs.\,[\onlinecite{SerBjo}] and [\onlinecite{SerRetGrap}].

\subsubsection{\label{atom2filmsret}Force on an atom in between two 2D films (five layers)}

Here we let the first 2D film be located at $z=0$, the thin diluted gas film at $z=d$, and the second 2D film at $D$. There is vacuum between the three films. Thus the matrix becomes  ${\bf{\tilde M}} = {{\bf{\tilde M}}_0} \cdot {{\bf{\tilde M}}_1} \cdot {{\bf{\tilde M}}_2}$. These matrices we have derived before. We have two mode types, TM and TE. 

We start with the TM modes. The resulting matrices for the two 2D films were given in Eq.\,(\ref{IVD4.1}) and for the gas film in (\ref{IVD3.2}), respectively. They are
\begin{equation}
\begin{array}{*{20}{l}}
{{\bf{\tilde M}}_0^{{\rm{TM}}} = \left( {\begin{array}{*{20}{c}}
1&0\\
0&1
\end{array}} \right) + {{\tilde \alpha }^{2D}}{\gamma ^{\left( 0 \right)}}\left( {\begin{array}{*{20}{c}}
1&{ - 1}\\
1&{ - 1}
\end{array}} \right);}\\
{{\bf{\tilde M}}_1^{{\rm{TM}}} = \left( {\begin{array}{*{20}{c}}
1&0\\
0&1
\end{array}} \right) + \frac{{\left( {\delta n} \right)2\pi k{\alpha ^{at}}}}{{{\gamma ^{\left( 0 \right)}}}}}\\
{ \times \left( {\begin{array}{*{20}{c}}
{ - {{\left( {\frac{\omega }{{ck}}} \right)}^2}}&{ - \left[ {2 - {{\left( {\frac{\omega }{{ck}}} \right)}^2}} \right]{e^{2{\gamma ^{\left( 0 \right)}}kd}}}\\
{\left[ {2 - {{\left( {\frac{\omega }{{ck}}} \right)}^2}} \right]{e^{ - 2{\gamma ^{\left( 0 \right)}}kd}}}&{{{\left( {\frac{\omega }{{ck}}} \right)}^2}}
\end{array}} \right);}\\
{{\bf{\tilde M}}_2^{{\rm{TM}}} = \left( {\begin{array}{*{20}{c}}
1&0\\
0&1
\end{array}} \right) + {{\tilde \alpha }^{2D}}{\gamma ^{\left( 0 \right)}}\left( {\begin{array}{*{20}{c}}
1&{ - {e^{2{\gamma ^{\left( 0 \right)}}kD}}}\\
{{e^{ - 2{\gamma ^{\left( 0 \right)}}kD}}}&{ - 1}
\end{array}} \right).}
\end{array}
\label{IVD9.1}
\end{equation}
The matrix element of interest is
\begin{equation}
\begin{array}{*{20}{l}}
{M_{11}^{{\rm{TM}}} = {{\left( {1 + {\gamma ^{\left( 0 \right)}}{{\tilde \alpha }^{2D}}} \right)}^2}\left[ {1 - \frac{{2\pi k\delta n{\alpha ^{at}}}}{{{\gamma ^{\left( 0 \right)}}}}{{\left( {\frac{\omega }{{ck}}} \right)}^2}} \right]}\\
{ - {e^{ - 2{\gamma ^{\left( 0 \right)}}kD}}{{\left( {{\gamma ^{\left( 0 \right)}}{{\tilde \alpha }^{2D}}} \right)}^2}\left[ {1 + \frac{{2\pi k\delta n{\alpha ^{at}}}}{{{\gamma ^{\left( 0 \right)}}}}{{\left( {\frac{\omega }{{ck}}} \right)}^2}} \right]}\\
{ + \frac{{2\pi k\delta n{\alpha ^{at}}}}{{{\gamma ^{\left( 0 \right)}}}} \times }\\
{\begin{array}{*{20}{l}}
{\left\{ {{e^{ - 2{\gamma ^{\left( 0 \right)}}k\left( {D - d} \right)}}\left( {{\gamma ^{\left( 0 \right)}}{{\tilde \alpha }^{2D}}} \right)} \right.}\\
{\quad \quad \quad \quad \quad \quad  \times \left[ {\left( {{\gamma ^{\left( 0 \right)}}{{\tilde \alpha }^{2D}}} \right) + 1} \right]\left[ {{{\left( {\frac{\omega }{{ck}}} \right)}^2} - 2} \right]}
\end{array}}\\
{\left. { + {e^{ - 2{\gamma ^{\left( 0 \right)}}kd}}\left( {{\gamma ^{\left( 0 \right)}}{{\tilde \alpha }^{2D}}} \right)\left[ {\left( {{\gamma ^{\left( 0 \right)}}{{\tilde \alpha }^{2D}}} \right) + 1} \right]\left[ {{{\left( {\frac{\omega }{{ck}}} \right)}^2} - 2} \right]} \right\}.}
\end{array}
\label{IVD9.2}
\end{equation}

The first term is the mode condition when all three films are at infinite distance from each other. The mode condition function after division with the part of the function that is independent of the position of the gas layer is
 \begin{equation}
\begin{array}{*{20}{l}}
{\tilde f_{\bf{k}}^{{\rm{TM}}} = 1 + \frac{{2\pi k\delta n{\alpha ^{at}}\left( \omega  \right)}}{{{\gamma ^{\left( 0 \right)}}}}}\\
{ \times \frac{{\left[ {{e^{ - 2{\gamma ^{\left( 0 \right)}}k\left( {D - d} \right)}} + {e^{ - 2{\gamma ^{\left( 0 \right)}}kd}}} \right]\left[ {{{\left( {\frac{\omega }{{ck}}} \right)}^2} - 2} \right]\frac{{{\gamma ^{\left( 0 \right)}}{{\tilde \alpha }^{2D}}}}{{1 + {\gamma ^{\left( 0 \right)}}{{\tilde \alpha }^{2D}}}}}}{{1 - {e^{ - 2{\gamma ^{\left( 0 \right)}}kD}}{{\left( {\frac{{{\gamma ^{\left( 0 \right)}}{{\tilde \alpha }^{2D}}}}{{1 + {\gamma ^{\left( 0 \right)}}{{\tilde \alpha }^{2D}}}}} \right)}^2}}},}
\end{array}
\label{IVD9.3}
\end{equation}
where the suppressed function arguments are $\left( {k,\omega } \right)$. If we instead had divided by the function in absence of the atom we would get one extra energy term that is independent of the position of the atom and does not affect the force. It gives a different contribution to the energy than the result from Eq.\,(\ref{IVD5.8}) due to the presence of the two 2D sheets. It means a modification of the atom self-energy.
The interaction energy per atom from the TM modes becomes
\begin{equation}
\begin{array}{*{20}{l}}
{\frac{{{E^{{\rm{TM}}}}}}{{n\delta }} =  - \hbar \int\limits_0^\infty  {dk{k^2}} \int\limits_{ - \infty }^\infty  {\frac{{d\xi }}{{2\pi }}} \frac{1}{{2{\gamma ^{\left( 0 \right)}}}}{\alpha ^{at}}\left( {i\xi } \right)\left[ {{{\left( {\frac{\xi }{{ck}}} \right)}^2} + 2} \right]}\\
{\quad \quad  \times \frac{{\frac{{{\gamma ^{\left( 0 \right)}}{{\tilde \alpha }^{2D}}}}{{1 + {\gamma ^{\left( 0 \right)}}{{\tilde \alpha }^{2D}}}}\left[ {{e^{ - 2{\gamma ^{\left( 0 \right)}}kd}} + {e^{ - 2{\gamma ^{\left( 0 \right)}}k\left( {D - d} \right)}}} \right]}}{{\left[ {1 - {e^{ - 2{\gamma ^{\left( 0 \right)}}kD}}{{\left( {\frac{{{\gamma ^{\left( 0 \right)}}{{\tilde \alpha }^{2D}}}}{{1 + {\gamma ^{\left( 0 \right)}}{{\tilde \alpha }^{2D}}}}} \right)}^2}} \right]}},}
\end{array}
\label{IVD9.4}
\end{equation}
where the suppressed function arguments are $\left( {k,i\xi } \right)$.
 Thus, the force on the atom from the TM modes is 
%
\begin{equation}
\begin{array}{*{20}{l}}
{{F^{{\rm{TM}}}}\left( d \right) =  - \hbar \int\limits_0^\infty  {dk{k^3}} \int\limits_{ - \infty }^\infty  {\frac{{d\xi }}{{2\pi }}} {\alpha ^{at}}\left( {i\xi } \right)\left[ {{{\left( {\frac{\xi }{{ck}}} \right)}^2} + 2} \right]}\\
{\quad \quad  \times \frac{{\frac{{{\gamma ^{\left( 0 \right)}}{{\tilde \alpha }^{2D}}}}{{1 + {\gamma ^{\left( 0 \right)}}{{\tilde \alpha }^{2D}}}}\left[ {{e^{ - 2{\gamma ^{\left( 0 \right)}}kd}} - {e^{ - 2{\gamma ^{\left( 0 \right)}}k\left( {D - d} \right)}}} \right]}}{{\left[ {1 - {e^{ - 2{\gamma ^{\left( 0 \right)}}kD}}{{\left( {\frac{{{\gamma ^{\left( 0 \right)}}{{\tilde \alpha }^{2D}}}}{{1 + {\gamma ^{\left( 0 \right)}}{{\tilde \alpha }^{2D}}}}} \right)}^2}} \right]}},}
\end{array}
\label{IVD9.5}
\end{equation}
and at finite temperature it is
\begin{equation}
\begin{array}{l}
{F^{{\rm{TM}}}}\left( d \right) =  - \frac{2}{\beta }\int\limits_0^\infty  {dkk^3} \sum\limits_{{\xi _n}} {'{\alpha ^{at}}\left( {i{\xi _n}} \right)\left[ {{{\left( {\frac{{{\xi _n}}}{ck}} \right)}^2} + 2} \right]} \\
\quad \quad  \times \frac{{\frac{{{{\gamma ^{\left( 0 \right)}}}{{\tilde \alpha }^{2D}}}}{{1 + {{\gamma ^{\left( 0 \right)}}}{{\tilde \alpha }^{2D}}}}\left[ {{e^{ - 2{\gamma ^{\left( 0 \right)}}kd}} - {e^{ - 2{\gamma ^{\left( 0 \right)}}k\left( {D - d} \right)}}} \right]}}{{\left[ {1 - {e^{ - 2{\gamma ^{\left( 0 \right)}}kD}}{{\left( {\frac{{{{\gamma ^{\left( 0 \right)}}}{{\tilde \alpha }^{2D}}}}{{1 + {{\gamma ^{\left( 0 \right)}}}{{\tilde \alpha }^{2D}}}}} \right)}^2}} \right]}},
\end{array}
\label{IVD9.6}
\end{equation}
where the suppressed function arguments are $\left( {k,i{\xi _n}} \right)$.

Now, we continue with the TE modes.  The resulting matrices for the two 2D films were given in Eq.\,(\ref{IVD4.1}) and for the gas film in (\ref{IVD3.2}), respectively. They are
\begin{equation}
\begin{array}{*{20}{l}}
{{\bf{\tilde M}}_0^{{\rm{TE}}} = \left( {\begin{array}{*{20}{c}}
1&0\\
0&1
\end{array}} \right) + {{\tilde \alpha }^{2D}}{{\left( {\frac{\omega }{{ck}}} \right)}^2}\frac{1}{{{\gamma ^{\left( 0 \right)}}}}\left( {\begin{array}{*{20}{c}}
{ - 1}&{ - 1}\\
1&1
\end{array}} \right);}\\
{{\bf{\tilde M}}_1^{{\rm{TE}}} = \left( {\begin{array}{*{20}{c}}
1&0\\
0&1
\end{array}} \right) + \frac{{\left( {\delta n} \right)2\pi k{\alpha ^{at}}}}{{{\gamma ^{\left( 0 \right)}}}}{{\left( {\frac{\omega }{{ck}}} \right)}^2}}\\
{\quad \quad \quad \quad \quad \quad \quad  \times \left( {\begin{array}{*{20}{c}}
{ - 1}&{ - {e^{2{\gamma ^{\left( 0 \right)}}kd}}}\\
{{e^{ - 2{\gamma ^{\left( 0 \right)}}kd}}}&1
\end{array}} \right);}\\
{\begin{array}{*{20}{l}}
{{\bf{\tilde M}}_2^{{\rm{TE}}} = \left( {\begin{array}{*{20}{c}}
1&0\\
0&1
\end{array}} \right)}\\
{\quad \quad  + {{\tilde \alpha }^{2D}}{{\left( {\frac{\omega }{{ck}}} \right)}^2}\frac{1}{{{\gamma ^{\left( 0 \right)}}}}\left( {\begin{array}{*{20}{c}}
{ - 1}&{ - {e^{2{\gamma ^{\left( 0 \right)}}kD}}}\\
{{e^{ - 2{\gamma ^{\left( 0 \right)}}kD}}}&1
\end{array}} \right).}
\end{array}}
\end{array}
\label{IVD9.7}
\end{equation}
The matrix element of interest is
\begin{equation}
\begin{array}{*{20}{l}}
{\begin{array}{*{20}{l}}
{M_{11}^{{\rm{TE}}} = {{\left[ {1 - {{\tilde \alpha }^{2D}}{{\left( {\frac{\omega }{{ck}}} \right)}^2}\frac{1}{{{\gamma ^{\left( 0 \right)}}}}} \right]}^2}}\\
{ - {{\left( {{{\tilde \alpha }^{2D}}{{\left( {\frac{\omega }{{ck}}} \right)}^2}\frac{1}{{{\gamma ^{\left( 0 \right)}}}}} \right)}^2}{e^{ - 2{\gamma ^{\left( 0 \right)}}kD}}}
\end{array}}\\
{ - \frac{{\left( {\delta n} \right)2\pi k{\alpha ^{at}}}}{{{\gamma ^{\left( 0 \right)}}}}{{\left( {\frac{\omega }{{ck}}} \right)}^2}{{\left[ {1 - 2{{\tilde \alpha }^{2D}}{{\left( {\frac{\omega }{{ck}}} \right)}^2}\frac{1}{{{\gamma ^{\left( 0 \right)}}}}} \right]}^2}}\\
{ - \frac{{\left( {\delta n} \right)2\pi k{\alpha ^{at}}}}{{{\gamma ^{\left( 0 \right)}}}}{{\left( {\frac{\omega }{{ck}}} \right)}^2}{{\tilde \alpha }^{2D}}{{\left( {\frac{\omega }{{ck}}} \right)}^2}\frac{1}{{{\gamma ^{\left( 0 \right)}}}}\left( {{e^{ - 2{\gamma ^{\left( 0 \right)}}kd}} + {e^{ - 2{\gamma ^{\left( 0 \right)}}k\left( {D - d} \right)}}} \right)}\\
{ + \frac{{\left( {\delta n} \right)2\pi k{\alpha ^{at}}}}{{{\gamma ^{\left( 0 \right)}}}}{{\left( {\frac{\omega }{{ck}}} \right)}^2}{{\left[ {{{\tilde \alpha }^{2D}}{{\left( {\frac{\omega }{{ck}}} \right)}^2}} \right]}^2}}\\
{\quad \quad  \times \left[ {{e^{ - 2{\gamma ^{\left( 0 \right)}}kd}} - {e^{ - 2{\gamma ^{\left( 0 \right)}}kD}} + {e^{ - 2{\gamma ^{\left( 0 \right)}}k\left( {D - d} \right)}}} \right].}
\end{array}
\label{IVD9.8}
\end{equation}

The first term is the mode condition when all three films are at infinite distance from each other. The mode condition function after division with the part of the function that is independent of the position of the gas layer is
 \begin{equation}
\begin{array}{*{20}{l}}
{\tilde f_k^{{\rm{TE}}} = 1 - \frac{{\left( {\delta n} \right)2\pi k{\alpha ^{at}}}}{{{\gamma ^{\left( 0 \right)}}}}{{\left( {\frac{\omega }{{ck}}} \right)}^2}}\\
{\quad  \times \frac{{\frac{{{{\tilde \alpha }^{2D}}{{\left( {\frac{\omega }{{ck}}} \right)}^2}}}{{\left[ {{\gamma ^{\left( 0 \right)}} - {{\tilde \alpha }^{2D}}{{\left( {\frac{\omega }{{ck}}} \right)}^2}} \right]}}\left( {{e^{ - 2{\gamma ^{\left( 0 \right)}}kd}} + {e^{ - 2{\gamma ^{\left( 0 \right)}}k\left( {D - d} \right)}}} \right)}}{{1 - {{\left[ {\frac{{{{\tilde \alpha }^{2D}}{{\left( {\frac{\omega }{{ck}}} \right)}^2}}}{{{\gamma ^{\left( 0 \right)}} - {{\tilde \alpha }^{2D}}{{\left( {\frac{\omega }{{ck}}} \right)}^2}}}} \right]}^2}{e^{ - 2{\gamma ^{\left( 0 \right)}}kD}}}}}
\end{array}
\label{IVD9.9}
\end{equation}
where the suppressed function arguments are $\left( {k,\omega } \right)$.
The interaction energy per atom from the TE modes becomes
\begin{equation}
\begin{array}{*{20}{l}}
{\frac{{{E^{{\rm{TE}}}}}}{{n\delta }} =  - \hbar \int\limits_0^\infty  {dk{k^2}} \int\limits_{ - \infty }^\infty  {\frac{{d\xi }}{{2\pi }}\frac{{{\alpha ^{at}}}}{{{\gamma ^{\left( 0 \right)}}}}{{\left( {\frac{\xi }{{ck}}} \right)}^2}} }\\
{\quad  \times \frac{{\frac{{{{\tilde \alpha }^{2D}}{{\left( {\frac{\xi }{{ck}}} \right)}^2}}}{{\left[ {{\gamma ^{\left( 0 \right)}} + {{\tilde \alpha }^{2D}}{{\left( {\frac{\xi }{{ck}}} \right)}^2}} \right]}}\left( {{e^{ - 2{\gamma ^{\left( 0 \right)}}kd}} + {e^{ - 2{\gamma ^{\left( 0 \right)}}k\left( {D - d} \right)}}} \right)}}{{1 - {{\left[ {\frac{{{{\tilde \alpha }^{2D}}{{\left( {\frac{\xi }{{ck}}} \right)}^2}}}{{{\gamma ^{\left( 0 \right)}} + {{\tilde \alpha }^{2D}}{{\left( {\frac{\xi }{{ck}}} \right)}^2}}}} \right]}^2}{e^{ - 2{\gamma ^{\left( 0 \right)}}kD}}}},}
\end{array}
\label{IVD9.10}
\end{equation}
where the suppressed function arguments are $\left( {k,i\xi } \right)$.
 Thus, the force on the atom from the TE modes is 
%
\begin{equation}
\begin{array}{*{20}{l}}
{{F^{TE}}\left( d \right) =  - 2\hbar \int\limits_0^\infty  {dk{k^3}} \int\limits_{ - \infty }^\infty  {\frac{{d\xi }}{{2\pi }}{\alpha ^{at}}{{\left( {\frac{\xi }{{ck}}} \right)}^2}} }\\
{\quad  \times \frac{{\frac{{{{\tilde \alpha }^{2D}}{{\left( {\frac{\xi }{{ck}}} \right)}^2}}}{{\left[ {{\gamma ^{\left( 0 \right)}} + {{\tilde \alpha }^{2D}}{{\left( {\frac{\xi }{{ck}}} \right)}^2}} \right]}}\left( {{e^{ - 2{\gamma ^{\left( 0 \right)}}kd}} - {e^{ - 2{\gamma ^{\left( 0 \right)}}k\left( {D - d} \right)}}} \right)}}{{1 - {{\left[ {\frac{{{{\tilde \alpha }^{2D}}{{\left( {\frac{\xi }{{ck}}} \right)}^2}}}{{{\gamma ^{\left( 0 \right)}} + {{\tilde \alpha }^{2D}}{{\left( {\frac{\xi }{{ck}}} \right)}^2}}}} \right]}^2}{e^{ - 2{\gamma ^{\left( 0 \right)}}kD}}}},}
\end{array}
\label{IVD9.11}
\end{equation}
and at finite temperature it is
\begin{equation}
\begin{array}{l}
{F^{TE}}\left( d \right) =  - \frac{1}{\beta }\int\limits_0^\infty  {dkk^3} \sum\limits_{{\xi _n}} {'4{\alpha ^{at}}{{\left( {\frac{{{\xi _n}}}{ck}} \right)}^2}} \\
{\quad  \times \frac{{\frac{{{{\tilde \alpha }^{2D}}{{\left( {\frac{\xi }{{ck}}} \right)}^2}}}{{\left[ {{\gamma ^{\left( 0 \right)}} + {{\tilde \alpha }^{2D}}{{\left( {\frac{\xi }{{ck}}} \right)}^2}} \right]}}\left( {{e^{ - 2{\gamma ^{\left( 0 \right)}}kd}} - {e^{ - 2{\gamma ^{\left( 0 \right)}}k\left( {D - d} \right)}}} \right)}}{{1 - {{\left[ {\frac{{{{\tilde \alpha }^{2D}}{{\left( {\frac{\xi }{{ck}}} \right)}^2}}}{{{\gamma ^{\left( 0 \right)}} + {{\tilde \alpha }^{2D}}{{\left( {\frac{\xi }{{ck}}} \right)}^2}}}} \right]}^2}{e^{ - 2{\gamma ^{\left( 0 \right)}}kD}}}},}
\end{array}
\label{IVD9.12}
\end{equation}
where the suppressed function arguments are $\left( {k,i{\xi _n}} \right)$.

\section{\label{spherical}Spherical structures}
For a spherical object the rightmost medium, $n=N+1$, in Fig.\,\ref{figu3} is the core. The leftmost, $n=0$, is the ambient. The boundary condition is that there are no incoming waves, i.e. there is no wave moving towards the right in medium $n=0$. The fields are self-sustained; no fields are coming in from outside.

\subsection{\label{sphericalnonretmain}Non-retarded main results}

In the non-retarded treatment of a spherical structure we let the waves represent solutions to Laplace's equation, Eq.\,(\ref{III21}), in spherical coordinates, ($r,\theta ,\varphi $), for the scalar potential, $\Phi$. The interfaces are spherical surfaces and the $r$-coordinate is the coordinate that is constant on each interface. The solutions are of the form
\begin{equation}
{\Phi _{l,m}}\left( {r,\theta ,\varphi } \right) = {r^{\begin{array}{*{20}{c}}
{ + l}\\
{ - \left( {l + 1} \right)}
\end{array}}}{Y_{l,m}}\left( {\theta ,\varphi } \right),
\label{VA1}
\end{equation}
where the functions ${Y_{l,m}}\left( {\theta ,\varphi } \right)$ are so-called Spherical Harmonics.
We let $r$ increase towards the left in  Fig.\,\ref{figu3}. We want to find the normal modes for a specific set of $l$ and $m$ values. Then all waves have the common factor  ${Y_{l,m}}\left( {\theta ,\varphi } \right)$. We suppress this factor here. Then
\begin{equation}
R\left( r \right) = {r^l};\,L\left( r \right) = {r^{ - \left( {l + 1} \right)}}.
\label{VA2}
\end{equation}
Using the boundary conditions that the potential and the normal component of the ${\bf{D}}$-field are continuous across an interface $n$ gives
\begin{equation}
\begin{array}{l}
{a^n}r_n^l + {b^n}r_n^{ - \left( {l + 1} \right)} = {a^{n + 1}}r_n^l + {b^{n + 1}}r_n^{ - \left( {l + 1} \right)}\\
{a^n}{{\tilde \varepsilon }_n}lr_n^{l - 1} - {b^n}{{\tilde \varepsilon }_n}\left( {l + 1} \right)r_n^{ - \left( {l + 2} \right)}\\
\quad \quad  = {a^{n + 1}}{{\tilde \varepsilon }_{n + 1}}lr_n^{l - 1} - {b^{n + 1}}{{\tilde \varepsilon }_{n + 1}}\left( {l + 1} \right)r_n^{ - \left( {l + 2} \right)},
\end{array}
\label{VA3}
\end{equation}
and we may identify the matrix ${{{\bf{\tilde A}}}_n}\left( {{r_n}} \right)$ as
\begin{equation}
{{{\bf{\tilde A}}}_n}\left( {{r_n}} \right) = \left( {\begin{array}{*{20}{c}}
{r_n^l}&{r_n^{ - \left( {l + 1} \right)}}\\
{{{\tilde \varepsilon }_n}lr_n^{l - 1}}&{ - {{\tilde \varepsilon }_n}\left( {l + 1} \right)r_n^{ - \left( {l + 2} \right)}}
\end{array}} \right).
\label{VA4}
\end{equation}
\begin{figure}
\includegraphics[width=5cm]{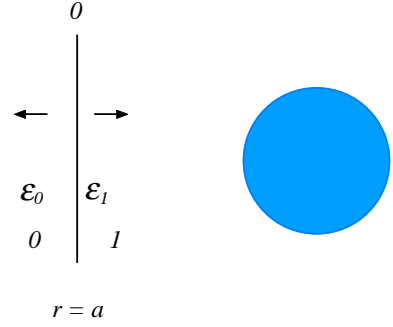}
\caption{(Color online) The geometry of a solid sphere or a solid cylinder in the non-retarded treatment.}
\label{figu10}
\end{figure}

The matrix ${{{\bf{\tilde M}}}_n}$ is
\begin{equation}
\begin{array}{l}
{{{\bf{\tilde M}}}_n} = \frac{1}{{\left( {2l + 1} \right){{\tilde \varepsilon }_n}}}\\
\times \left( {\begin{array}{*{20}{c}}
{{{\tilde \varepsilon }_n}\left( {l + 1} \right) + {{\tilde \varepsilon }_{n + 1}}l}&{\left( {l + 1} \right)\left( {{{\tilde \varepsilon }_n} - {{\tilde \varepsilon }_{n + 1}}} \right)r_n^{ - \left( {2l + 1} \right)}}\\
{l\left( {{{\tilde \varepsilon }_n} - {{\tilde \varepsilon }_{n + 1}}} \right)r_n^{2l + 1}}&{{{\tilde \varepsilon }_{n + 1}}\left( {l + 1} \right) + {{\tilde \varepsilon }_n}l}
\end{array}} \right).
\end{array}
\label{VA5}
\end{equation}
Since the function $L\left( r \right)$ in Eq.\,(\ref{VA2}) diverges at the origin it is excluded from the core region and hence we have no wave moving towards the left in that region. According to  Eq.\,(\ref{III6}) this means that
\begin{equation}
{f_{l,m}}\left( \omega  \right) = {M_{11}}.
\label{VA6}
\end{equation}
Before we end this section we introduce the ${2^l}$ pole polarizabilities ${\alpha _l^n}$ and $\alpha _l^{n\left( 2 \right)}$ for the spherical interface since these appear repeatedly in the sections that follow. The first is valid outside and the second inside. The polarizability ${\alpha _l^n}= - {b^n}/{a^n}$ under the assumption that ${b^{n+1}} = 0$. One obtains ${\alpha _l^n} =  - {M_{21}}/{M_{11}}$ and from Eq.\,(\ref{VA5}) one finds
\begin{equation}
{\alpha _l^n} =  - \frac{{{r_n}^{2l + 1}l\left( {{{\tilde \varepsilon }_n} - {{\tilde \varepsilon }_{n + 1}}} \right)}}{{{{\tilde \varepsilon }_n}\left( {l + 1} \right) + {{\tilde \varepsilon }_{n + 1}}l}}.
\label{VA7}
\end{equation}
The polarizability $\alpha _l^{n\left( 2 \right)}= - {a^{n + 1}}/{b^{n + 1}}$ under the assumption that ${a^n} = 0$. One obtains $\alpha _l^{n\left( 2 \right)} =   {M_{12}}/{M_{11}}$ and from Eq.\,(\ref{VA5}) one finds
\begin{equation}
\alpha _l^{n\left( 2 \right)} = \frac{{{r_n}^{ - \left( {2l + 1} \right)}\left( {l + 1} \right)\left( {{{\tilde \varepsilon }_n} - {{\tilde \varepsilon }_{n + 1}}} \right)}}{{{{\tilde \varepsilon }_n}\left( {l + 1} \right) + {{\tilde \varepsilon }_{n + 1}}l}}.
\label{VA8}
\end{equation}
Sometimes it is convenient to use an alternative form of the matrix ${{{{\bf{\tilde M}}}_n}}$,
\begin{equation}
\begin{array}{*{20}{l}}
{{{{\bf{\tilde M}}}_n} = \frac{{{{\tilde \varepsilon }_n}\left( {l + 1} \right) + {{\tilde \varepsilon }_{n + 1}}l}}{{\left( {2l + 1} \right){{\tilde \varepsilon }_n}}}}\\
\begin{array}{l}
 \times \left( {\begin{array}{*{20}{c}}
1&{\frac{{\left( {l + 1} \right)\left( {{{\tilde \varepsilon }_n} - {{\tilde \varepsilon }_{n + 1}}} \right)r_n^{ - \left( {2l + 1} \right)}}}{{{{\tilde \varepsilon }_n}\left( {l + 1} \right) + {{\tilde \varepsilon }_{n + 1}}l}}}\\
{\frac{{l\left( {{{\tilde \varepsilon }_n} - {{\tilde \varepsilon }_{n + 1}}} \right)r_n^{2l + 1}}}{{{{\tilde \varepsilon }_n}\left( {l + 1} \right) + {{\tilde \varepsilon }_{n + 1}}l}}}&{\frac{{{{\tilde \varepsilon }_{n + 1}}\left( {l + 1} \right) + {{\tilde \varepsilon }_n}l}}{{{{\tilde \varepsilon }_n}\left( {l + 1} \right) + {{\tilde \varepsilon }_{n + 1}}l}}}
\end{array}} \right)\\
 = M_{11}^n\left( {\begin{array}{*{20}{c}}
1&{\alpha _l^{n\left( 2 \right)}}\\
{ - \alpha _l^n}&{\frac{{{{\tilde \varepsilon }_{n + 1}}\left( {l + 1} \right) + {{\tilde \varepsilon }_n}l}}{{{{\tilde \varepsilon }_n}\left( {l + 1} \right) + {{\tilde \varepsilon }_{n + 1}}l}}}
\end{array}} \right).
\end{array}
\end{array}
\label{VA9}
\end{equation}
Now we have all we need to determine the non-retarded normal modes in a layered spherical structure. We give some examples in the following sections.

\subsection{\label{sphericalnonretpecial}Non-retarded special results}

\subsubsection{\label{solid spheren}Solid sphere (no layer)}

For a solid sphere of radius  $a$ and dielectric function ${{\tilde \varepsilon }_1}\left( \omega  \right)$ in an ambient of dielectric function ${{\tilde \varepsilon }_0}\left( \omega  \right)$, as illustrated in Fig.\,\ref{figu10}, we have
\begin{equation}
\begin{array}{l}
{\bf{\tilde M}} = {{{\bf{\tilde M}}}_0} = \frac{1}{{\left( {2l + 1} \right){{\tilde \varepsilon }_0}}}\\
\quad \quad  \times \left( {\begin{array}{*{20}{c}}
{{{\tilde \varepsilon }_0}\left( {l + 1} \right) + {{\tilde \varepsilon }_1}l}&{\left( {l + 1} \right)\left( {{{\tilde \varepsilon }_0} - {{\tilde \varepsilon }_1}} \right){a^{ - \left( {2l + 1} \right)}}}\\
{l\left( {{{\tilde \varepsilon }_0} - {{\tilde \varepsilon }_1}} \right){a^{2l + 1}}}&{{{\tilde \varepsilon }_1}\left( {l + 1} \right) + {{\tilde \varepsilon }_0}l}
\end{array}} \right),
\end{array}
\label{VB1.1}
\end{equation}
and the condition for modes is ${{\tilde \varepsilon }_1}\left(\omega \right)/{{\tilde \varepsilon }_0}\left(\omega \right)= -\left( {l + 1}\right)/l$. This result covers both solid spheres and spherical cavities. For a solid sphere of dielectric function $\tilde \varepsilon \left( \omega  \right)$ in vacuum and for a spherical cavity in a medium of dielectric function $\tilde \varepsilon \left( \omega  \right)$ the condition for modes is $\tilde \varepsilon \left( \omega  \right) =  - \left( {l + 1} \right)/l$ and $\tilde \varepsilon \left( \omega  \right) =  - l/\left( {l + 1} \right)$, respectively.

\begin{figure}
\includegraphics[width=6.7cm]{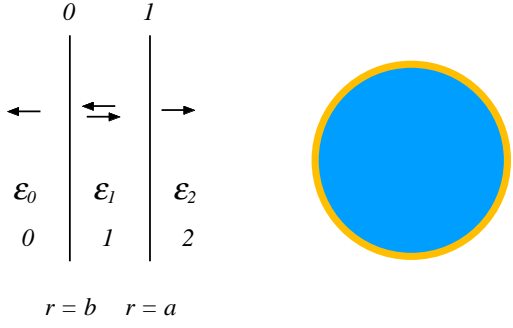}
\caption{(Color online) The geometry of a coated sphere or cylinder in the non-retarded treatment.}
\label{figu11}
\end{figure}

\subsubsection{\label{Sphericalshelln}Spherical shell or gap (one layer)}
Here we start from a more general geometry namely that of a coated sphere in a medium and get the spherical shell and gap as special limits.
For a solid sphere of dielectric function ${{\tilde \varepsilon }_2}$ with a coating of inner radius $a$ and outer radius $b$, Fig.\,\ref{figu11},  made of a medium with dielectric function ${{\tilde \varepsilon }_1}$  in an ambient medium with dielectric function ${{\tilde \varepsilon }_0}$ we have 
\begin{equation}
\begin{array}{*{20}{l}}
{{\bf{\tilde M}} = {{{\bf{\tilde M}}}_0} \cdot {{{\bf{\tilde M}}}_1} }\\
 ={\frac{1}{{\left( {2l + 1} \right){{\tilde \varepsilon }_0}}}\left( {\begin{array}{*{20}{c}}
{{{\tilde \varepsilon }_0}\left( {l + 1} \right) + {{\tilde \varepsilon }_1}l}&{\frac{{\left( {l + 1} \right)\left( {{{\tilde \varepsilon }_0} - {{\tilde \varepsilon }_1}} \right)}}{{{b^{2l + 1}}}}}\\
{l\left( {{{\tilde \varepsilon }_0} - {{\tilde \varepsilon }_1}} \right){b^{2l + 1}}}&{{{\tilde \varepsilon }_1}\left( {l + 1} \right) + {{\tilde \varepsilon }_0}l}
\end{array}} \right)}\\
{ \times \frac{1}{{\left( {2l + 1} \right){{\tilde \varepsilon }_1}}}\left( {\begin{array}{*{20}{c}}
{{{\tilde \varepsilon }_1}\left( {l + 1} \right) + {{\tilde \varepsilon }_2}l}&{\frac{{\left( {l + 1} \right)\left( {{{\tilde \varepsilon }_1} - {{\tilde \varepsilon }_2}} \right)}}{{{a^{2l + 1}}}}}\\
{l\left( {{{\tilde \varepsilon }_1} - {{\tilde \varepsilon }_2}} \right){a^{2l + 1}}}&{{{\tilde \varepsilon }_2}\left( {l + 1} \right) + {{\tilde \varepsilon }_1}l}
\end{array}} \right)},
\end{array}
\label{VB2.1}
\end{equation}
and from direct derivation of the $M_{11}$ element  the condition for modes  becomes
\begin{equation}
\begin{array}{l}
{\left( {\frac{b}{a}} \right)^{2l + 1}}\left( {\frac{{{{\tilde \varepsilon }_1}}}{{{{\tilde \varepsilon }_0}}} + \frac{{\left( {l + 1} \right)}}{l}} \right)\left( {\frac{{{{\tilde \varepsilon }_2}}}{{{{\tilde \varepsilon }_1}}} + \frac{{\left( {l + 1} \right)}}{l}} \right)\\
 \quad \quad  \quad \quad  \quad \quad = - \frac{{\left( {l + 1} \right)}}{l}\left( {\frac{{{{\tilde \varepsilon }_1}}}{{{{\tilde \varepsilon }_0}}} - 1} \right)\left( {\frac{{{{\tilde \varepsilon }_2}}}{{{{\tilde \varepsilon }_1}}} - 1} \right).
\end{array}
\label{VB2.2}
\end{equation}
Alternatively we may elaborate using the matrix version in Eq.\,(\ref{VA9})
\begin{equation}
\begin{array}{l}
{M_{11}} = M_{11}^0M_{11}^1\left( {1 - \alpha _l^{0\left( 2 \right)}\alpha _l^1} \right)\\
 = M_{11}^0M_{11}^1\left[ {1 - \frac{{{b^{ - \left( {2l + 1} \right)}}\left( {l + 1} \right)\left( {{{\tilde \varepsilon }_0} - {{\tilde \varepsilon }_1}} \right)}}{{{{\tilde \varepsilon }_0}\left( {l + 1} \right) + {{\tilde \varepsilon }_1}l}}\frac{{{a^{2l + 1}}l\left( {{{\tilde \varepsilon }_2} - {{\tilde \varepsilon }_1}} \right)}}{{{{\tilde \varepsilon }_1}\left( {l + 1} \right) + {{\tilde \varepsilon }_2}l}}} \right] = 0.
\end{array}
\label{VB2.3}
\end{equation}

Let us now study a spherical shell of inner radius $a$, outer radius $b$ and of a medium with dielectric function $\tilde \varepsilon \left( \omega  \right)$ in a medium of dielectric function ${{\tilde \varepsilon }_0}\left( \omega  \right)$. The condition for modes we get from Eq.\,(\ref{VB2.2}) by the replacements ${{\tilde \varepsilon }_2}\left( \omega  \right) \to {{\tilde \varepsilon }_0}\left( \omega  \right)$ and $\;{{\tilde \varepsilon }_1}\left( \omega  \right) \to \tilde \varepsilon \left( \omega  \right)$. 
For a spherical gap of dielectric function ${{\tilde \varepsilon }_0}\left( \omega  \right)$ in a medium of dielectric function $\tilde \varepsilon \left( \omega  \right)$ we instead make the replacements ${{\tilde \varepsilon }_0}\left( \omega  \right),\;{{\tilde \varepsilon }_2}\left( \omega  \right) \to \tilde \varepsilon \left( \omega  \right)$ and $\;{{\tilde \varepsilon }_1}\left( \omega  \right) \to {{\tilde \varepsilon }_0}\left( \omega  \right)$. For both these geometries we find the same condition for modes, viz.
\begin{equation}
\begin{array}{*{20}{l}}
{{{\left( {\frac{b}{a}} \right)}^{2l + 1}}\left[ {\tilde \varepsilon \left( \omega  \right)l + {{\tilde \varepsilon }_0}\left( \omega  \right)\left( {l + 1} \right)} \right]\left[ {{{\tilde \varepsilon }_0}\left( \omega  \right)l + \tilde \varepsilon \left( \omega  \right)\left( {l + 1} \right)} \right]}\\
{\quad \quad \quad \quad \quad \quad  = l\left( {l + 1} \right){{\left[ {\tilde \varepsilon \left( \omega  \right) - {{\tilde \varepsilon }_0}\left( \omega  \right)} \right]}^2}}.
\end{array}
\label{VB2.4}
\end{equation}

\subsubsection{\label{Sphericalgasfilmn}Thin spherical diluted gas film (one layer)}
It is of interest to find the van der Waals force on an atom in a layered structure. We can obtain this by studying the force on a thin layer of a diluted gas with dielectric function ${\varepsilon _g}\left( \omega  \right) = 1 + 4\pi n\alpha^{at} \left( \omega  \right)$, where $\alpha^{at}$ is the polarizability of one atom and $n$ the density of atoms (we have assumed that the atom is surrounded by vacuum; if not  the $1$ should be replaced by the dielectric function of the ambient medium and the atomic polarizability should be replaced by the excess polarizability). For a diluted gas layer the atoms do not interact with each other and the force on the layer is just the sum of the forces on the individual atoms. So by dividing with the number of atoms in the film we get the force on one atom. The layer has to be thin in order to have a well defined $r$- value of the atom.  Since we will derive the force on an atom in different spherical geometries it is fruitful to derive the matrix for a thin diluted gas shell. This result can be directly used in the derivation of the van der Waals force on an atom in different spherical geometries.

We let the film have the thickness $\delta$ and be of a general radius $r$. We only keep terms up to linear order in $\delta$ and linear order in $n$. We find the result is
\begin{equation}
\begin{array}{*{20}{l}}
{{{{\bf{\tilde M}}}_{{\rm{gaslayer}}}} = {{{\bf{\tilde M}}}_0} \cdot {{{\bf{\tilde M}}}_1}}\\
{ = \left( {\begin{array}{*{20}{c}}
1&0\\
0&1
\end{array}} \right) + \left( {\delta n} \right)4\pi {\alpha ^{at}}\left( {\begin{array}{*{20}{c}}
0&{\left( {l + 1} \right){r^{ - \left( {2l + 2} \right)}}}\\
{ - l{r^{\left( {2l} \right)}}}&0
\end{array}} \right)}.
\end{array}
\label{VB3.1}
\end{equation}

Now we are done with the gas layer. We will use these results later in calculating the van der Waals force on an atom in spherical layered structures.

\subsubsection{\label{Spherical2Dfilmn}2D spherical film (one layer)}
In many situations one is dealing with very thin films. These may be considered 2D (two dimensional). Important examples are a graphene sheet and a 2D electron gas. In the derivation we let the film have finite thickness $\delta$ and be characterized by a 3D dielectric function ${\tilde \varepsilon ^{3D}}$. We then let the thickness go towards zero. The 3D dielectric function depends on $\delta$ as ${\tilde\varepsilon ^{3D}} \sim 1/\delta $ for small $\delta$. In the planar structure we could in the limit when $\delta$ goes towards zero obtain a momentum dependent 2D dielectric function. Here we only keep the long wave length limit of the 2D dielectric function.\,\cite{grap,arx} We obtain
\begin{equation}
\begin{array}{l}
{{{\bf{\tilde M}}}_{{\rm{2D}}}} = {{{\bf{\tilde M}}}_0} \cdot {{{\bf{\tilde M}}}_1}\\
 = \left( {\begin{array}{*{20}{c}}
1&0\\
0&1
\end{array}} \right) + \frac{{\left( {\delta {{\tilde \varepsilon }^{3D}}} \right)l\left( {l + 1} \right)}}{{\left( {2l + 1} \right)r}}\left( {\begin{array}{*{20}{c}}
1&{{r^{ - \left( {2l + 1} \right)}}}\\
{ - {r^{\left( {2l + 1} \right)}}}&{ - 1}
\end{array}} \right).
\end{array}
\label{VB4.1}
\end{equation}

We will also need the ${2^l}$ pole polarizability of the thin spherical shell in vacuum.  It can be obtained from Eq.\,(\ref{VB4.1}). The polarizability is $ - {b^0}/{a^0}$ under the assumption that ${b^1} = 0$. One obtains ${\alpha _l^{2D}} =  - {M_{21}}/{M_{11}}$. We find
\begin{equation}
{\alpha _l^{2D}}\left( {a;\omega } \right) = \frac{{\delta {{\tilde \varepsilon }^{3D}}l\left( {l + 1} \right){a^{2l + 1}}}}{{\left( {2l + 1} \right)a + \delta {{\tilde \varepsilon }^{3D}}l\left( {l + 1} \right)}},
\label{VB4.2}
\end{equation}
where we have reserved the first argument before the semicolon for the radius of the spherical film.
Note that for a perfectly reflecting thin spherical shell  the ${2^l}$ pole polarizability, ${\alpha _l^{2D}} = {a^{2l + 1}}$, coincides with that for a perfectly reflecting sphere of the same radius (compare with Eq.\,(\ref{VA7})) and the interaction is the same. This is what one would expect. It is further convenient to define the ${2^l}$ pole susceptibility\,\cite{Lang} as the polarizability stripped by the factor ${a^{2l + 1}}$,
\begin{equation}
{\chi_l^{2D}}\left( {a;\omega } \right) = \frac{{\delta {{\tilde \varepsilon }^{3D}}l\left( {l + 1} \right)}}{{\left( {2l + 1} \right)a + \delta {{\tilde \varepsilon }^{3D}}l\left( {l + 1} \right)}},
\label{VB4.3}
\end{equation}

The ${2^l}$ pole polarizability "seen from inside the shell" we get  from Eq.\,(\ref{VB4.1}). The polarizability is $ - {a^1}/{b^1}$ under the assumption that ${a^0} = 0$. One obtains $\alpha _l^{2D\left( 2 \right)} =   {M_{12}}/{M_{11}}$ and
\begin{equation}
\alpha _l^{2D\left( 2 \right)}\left( {a;\omega } \right) = \frac{{\left( {\delta {{\tilde \varepsilon }^{3D}}} \right)l\left( {l + 1} \right){a^{ - \left( {2l + 1} \right)}}}}{{\left( {2l + 1} \right)a + \left( {\delta {{\tilde \varepsilon }^{3D}}} \right)l\left( {l + 1} \right)}}.
\label{VB4.4}
\end{equation}
Note that it is the same as the ordinary ${2^l}$ pole polarizability Eq.\,(\ref{VB4.2}), for a thin spherical shell except that now the radius of the sphere has been inverted in the numerator. Thus ${\alpha _l^{2D}}\left( {a;\omega } \right) ={a^{2l + 1}}{\chi_l^{2D}}\left( {a;\omega } \right)$ and $\alpha _l^{2D\left( 2 \right)}\left( {a;\omega } \right)  ={{{a^{ - \left( {2l + 1} \right)}}}}{\chi_l^{2D}}\left( {a;\omega } \right)$.

Sometimes it is convenient to use an alternative form of the matrix ${{{\bf{\tilde M}}}_{{\rm{2D}}}}$,
\begin{equation}
\begin{array}{l}
{{{\bf{\tilde M}}}_{{\rm{2D}}}} = \frac{{\left( {2l + 1} \right)r + \left( {\delta {{\tilde \varepsilon }^{3D}}} \right)l\left( {l + 1} \right)}}{{\left( {2l + 1} \right)r}}\\
 \times \left( {\begin{array}{*{20}{c}}
1&{\frac{{\left( {\delta {{\tilde \varepsilon }^{3D}}} \right)l\left( {l + 1} \right){r^{ - \left( {2l + 1} \right)}}}}{{\left( {2l + 1} \right)r + \left( {\delta {{\tilde \varepsilon }^{3D}}} \right)l\left( {l + 1} \right)}}}\\
{ - \frac{{\left( {\delta {{\tilde \varepsilon }^{3D}}} \right)l\left( {l + 1} \right){r^{\left( {2l + 1} \right)}}}}{{\left( {2l + 1} \right)r + \left( {\delta {{\tilde \varepsilon }^{3D}}} \right)l\left( {l + 1} \right)}}}&{\frac{{\left( {2l + 1} \right)r - \left( {\delta {{\tilde \varepsilon }^{3D}}} \right)l\left( {l + 1} \right)}}{{\left( {2l + 1} \right)r + \left( {\delta {{\tilde \varepsilon }^{3D}}} \right)l\left( {l + 1} \right)}}}
\end{array}} \right)\\
 = M_{11}^{2D}\left( {\begin{array}{*{20}{c}}
1&{\alpha _l^{2D\left( 2 \right)}}\\
{ - \alpha _l^{2D}}&{\frac{{\left( {2l + 1} \right)r - \left( {\delta {{\tilde \varepsilon }^{3D}}} \right)l\left( {l + 1} \right)}}{{\left( {2l + 1} \right)r + \left( {\delta {{\tilde \varepsilon }^{3D}}} \right)l\left( {l + 1} \right)}}}
\end{array}} \right).
\end{array}
\label{VB4.5}
\end{equation}
\subsubsection{\label{atomspheren}Force on an atom outside a sphere (two layers)}
\begin{figure}
\includegraphics[width=8cm]{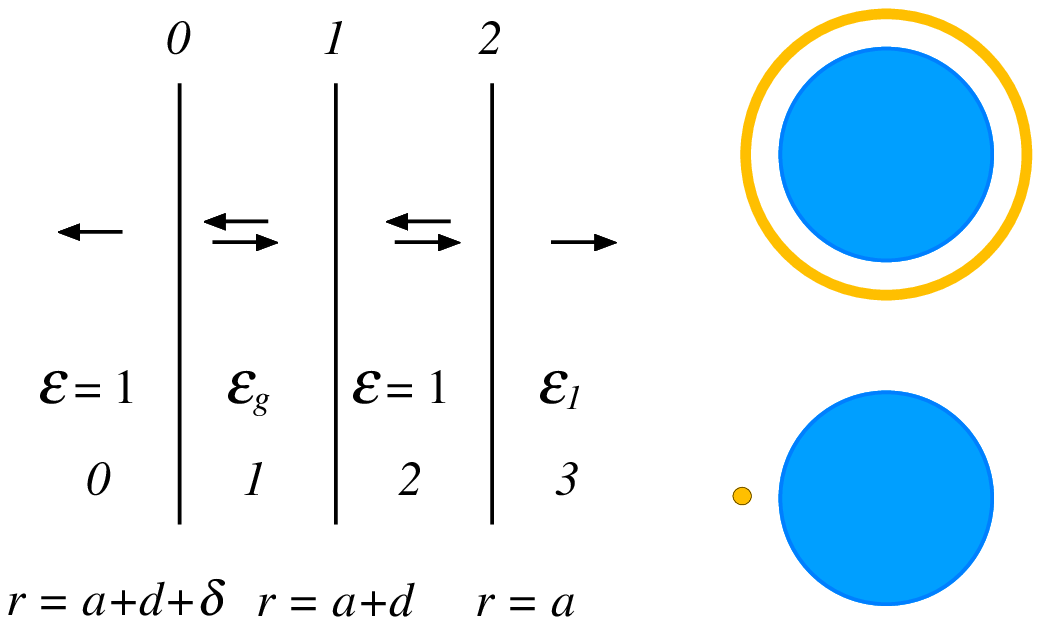}
\caption{(Color online) The geometry of a thin gas layer the distance $d$ from a sphere or  cylinder of radius $a$ in the non-retarded treatment.}
\label{figu12}
\end{figure}

We let the atom be at the distance $d$ from the sphere of radius $a$ and at the distance $b$ from the center of the sphere.
We start from the two layer structure in Fig.\,\ref{figu12}. We let the ambient be vacuum. The first layer is a thin layer, of thickness $\delta$, of a diluted gas of atoms of the kind we consider. Its dielectric function is ${\varepsilon _g}\left( \omega  \right) = 1 + 4\pi n\alpha^{at} \left( \omega  \right)$, where $\alpha^{at}$ is the polarizability of one atom. The density of gas atoms, $n$, is very low. We let the first interface be at $r={b + \delta }$ and hence the second at $r={b  }$, where $b=a+d$. The second layer is a vacuum layer of thickness $d$. The remaining medium is the sphere of radius $a$ with the dielectric function ${{\tilde \varepsilon }_1}\left( \omega  \right)$. In what follows we only keep lowest order terms in $\delta$ and in $n$.

The matrix becomes ${\bf{\tilde M}} = {{\bf{\tilde M}}_0} \cdot {{\bf{\tilde M}}_1} \cdot {{\bf{\tilde M}}_2} = {{\bf{\tilde M}}_{{\rm{gaslayer}}}} \cdot {{\bf{\tilde M}}_2}$ where ${{\bf{\tilde M}}_{{\rm{gaslayer}}}}$ is the matrix in Eq.\,(\ref{VB3.1}) now for the $r$ value $b$ and 
\begin{equation}
\begin{array}{l}
{{\bf{M}}_2} = \frac{1}{{\left( {2l + 1} \right)}}\\
\quad  \times \left( {\begin{array}{*{20}{c}}
{\left( {l + 1} \right) + {{\tilde \varepsilon }_1}l}&{\left( {l + 1} \right)\left( {1 - {{\tilde \varepsilon }_1}} \right){a^{ - \left( {2l + 1} \right)}}}\\
{l\left( {1 - {{\tilde \varepsilon }_1}} \right){a^{2l + 1}}}&{{{\tilde \varepsilon }_1}\left( {l + 1} \right) + l}
\end{array}} \right),
\end{array}
\label{VB5.1}
\end{equation}
the matrix in Eq.\,(\ref{VB1.1}) with the replacement ${{\tilde \varepsilon }_0} \to 1$.
The matrix element of interest is
\begin{equation}
\begin{array}{*{20}{l}}
{{M_{11}} \approx \frac{1}{{\left( {2l + 1} \right)}}\left\{ {\left( {l + 1} \right) + {{\tilde \varepsilon }_1}l} \right.}\\
{\quad \quad  - \left. {\left( {\delta n} \right)4\pi {\alpha ^{at}}l\left( {l + 1} \right){b^{ - \left( {2l + 2} \right)}}{a^{2l + 1}}\left( {{{\tilde \varepsilon }_1} - 1} \right)} \right\}.}
\end{array}
\label{VB5.2}
\end{equation}
The mode condition function when the reference system is that when the atom is at infinite distance from the sphere then becomes
\begin{equation}
{{\tilde f}_{l,m}} = 1 - \left( {\delta n} \right)4\pi {\alpha ^{at}}\left[ {\frac{{l\left( {l + 1} \right){b^{ - \left( {2l + 2} \right)}}{a^{2l + 1}}\left( {{{\tilde \varepsilon }_1} - 1} \right)}}{{\left( {l + 1} \right) + {{\tilde \varepsilon }_1}l}}} \right].
\label{VB5.3}
\end{equation}
The interaction energy per atom we get by dividing the energy with the number of atoms in the gas shell. It is
\begin{equation}
\begin{array}{*{20}{l}}
{\frac{E}{{4\pi {b^2}\delta n}}}\\
{ = \frac{\hbar }{{4\pi {b^2}\delta n}}\int\limits_0^\infty  {\frac{{d\xi }}{{2\pi }}} \sum\limits_{l = 0}^\infty  {\sum\limits_{m =  - l}^l {\ln \left[ {{{\tilde f}_{l,m}}\left( {i\xi } \right)} \right]} } }\\
{ \approx  - \frac{\hbar }{{4\pi {b^2}\delta n}}\int\limits_0^\infty  {\frac{{d\xi }}{{2\pi }}} \sum\limits_{l = 0}^\infty  {\sum\limits_{m =  - l}^l {4\pi n{\alpha ^{at}}\delta } } }\\
{\quad \quad \quad \quad  \times \left[ {\frac{{l\left( {l + 1} \right){b^{ - \left( {2l + 2} \right)}}{a^{2l + 1}}\left( {{{\tilde \varepsilon }_1} - 1} \right)}}{{\left( {l + 1} \right) + {{\tilde \varepsilon }_1}l}}} \right]}\\
{ =  - \hbar \int\limits_0^\infty  {\frac{{d\xi }}{{2\pi }}} \sum\limits_{l = 0}^\infty  {{\alpha ^{at}}\left( {i\xi } \right)\frac{{\left( {2l + 1} \right)\left( {l + 1} \right)}}{{{b^{2\left( {l + 2} \right)}}}}\frac{{{a^{2l + 1}}l\left[ {{{\tilde \varepsilon }_1}\left( {i\xi } \right) - 1} \right]}}{{\left( {l + 1} \right) + {{\tilde \varepsilon }_1}\left( {i\xi } \right)l}}} }\\
{\begin{array}{*{20}{l}}
{ =  - \hbar \int\limits_0^\infty  {\frac{{d\xi }}{{2\pi }}} \sum\limits_{l = 0}^\infty  {\frac{{\left( {2l + 1} \right)\left( {l + 1} \right)}}{{{b^{2\left( {l + 2} \right)}}}}{\alpha ^{at}}\left( {i\xi } \right){\alpha _l}\left( {a;i\xi } \right)} }\\
{ =  - \hbar \int\limits_0^\infty  {\frac{{d\xi }}{{2\pi }}} \sum\limits_{l = 0}^\infty  {\frac{{\left[ {2l + 2} \right]!}}{{\left[ {2l} \right]!\left[ 2 \right]!}}\frac{{{\alpha ^{at}}\left( {i\xi } \right){\alpha _l}\left( {a;i\xi } \right)}}{{{b^{2\left( {l + 2} \right)}}}}} },
\end{array}}
\end{array}
\label{VB5.4}
\end{equation}
where $b=a+d$, and
\begin{equation}
{\alpha _l}\left( {a;i\xi } \right) = \frac{{{a^{2l + 1}}l\left[ {{{\tilde \varepsilon }_1}\left( {i\xi } \right) - 1} \right]}}{{\left( {l + 1} \right) + {{\tilde \varepsilon }_1}\left( {i\xi } \right)l}}
\label{VB5.5}
\end{equation}
is the ${2^l}$ pole polarizability,   introduced in Eq.\,(\ref{VA7}), of the sphere in vacuum (Ref.\,[\onlinecite{Ser}], Eq.\,(5.68)). 
Note that the $l=0$ term does not contribute to the interaction.

The force on the atom is obtained as minus the derivative of the result in Eq.\,(\ref{VB5.4}) with respect to $d$, i.e.
\begin{equation}
F\left( b \right) =  - \hbar \int\limits_0^\infty  {\frac{{d\xi }}{{2\pi }}} \sum\limits_{l = 0}^\infty  {\frac{{\left[ {2l + 2} \right]!}}{{\left[ {2l} \right]!\left[ 2 \right]!}}2\left( {l + 2} \right)\frac{{{\alpha ^{at}}\left( {i\xi } \right){\alpha _l}\left( {a;i\xi } \right)}}{{{b^{2l + 5}}}}}.
\label{VB5.6}
\end{equation}

\subsubsection{\label{atomsphericalcavityn}Force on an atom in a spherical cavity (two layers)}
\begin{figure}
\includegraphics[width=8cm]{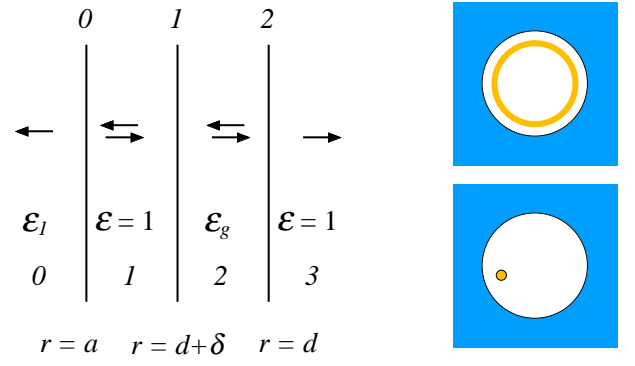}
\caption{(Color online) The geometry of a thin gas layer at radius $d$ inside a spherical or cylindrical cavity of radius $a$ in the non-retarded treatment.}
\label{figu13}
\end{figure}
We let the atom be at the distance $d$ from the center of the spherical cavity, of radius $a$.
We start from the two layer structure in Fig.\,\ref{figu13}. We let the medium surrounding the cavity have dielectric function ${{\tilde \varepsilon }_1}\left( \omega  \right)$. The first layer is a vacuum layer. The second is a thin layer, of thickness $\delta$, of a diluted gas of atoms of the kind we consider. Its dielectric function is ${\varepsilon _g}\left( \omega  \right) = 1 + 4\pi n\alpha^{at} \left( \omega  \right)$, where $\alpha^{at}$ is the polarizability of one atom. The density of gas atoms, $n$, is very low. We let the first interface be at $r=a$ and hence the second at $r={{d + \delta }  }$ and the third at  $r=d$. In what follows we only keep lowest order terms in $\delta$ and in $n$.

The matrix becomes ${\bf{\tilde M}} = {{\bf{\tilde M}}_0} \cdot {{\bf{\tilde M}}_1} \cdot {{\bf{\tilde M}}_2} = {{\bf{\tilde M}}_0} \cdot {{\bf{\tilde M}}_{{\rm{gaslayer}}}}$ where ${{\bf{\tilde M}}_0}$ is obtained from Eq.\,(\ref{VB1.1}) with the replacements ${{\tilde \varepsilon }_1} \to 1$ and ${{\tilde \varepsilon }_0} \to {{\tilde \varepsilon }_1}$. ${{{\bf{\tilde M}}}_{{\rm{gaslayer}}}}$ is obtained from Eq.\,(\ref{VB3.1}) with $r=d$. We find
\begin{equation}
\begin{array}{*{20}{l}}
{{M_{11}} \approx \frac{1}{{\left( {2l + 1} \right){{\tilde \varepsilon }_1}}}\left[ {\left( {{{\tilde \varepsilon }_1}\left( {l + 1} \right) + l} \right)} \right.}\\
{\quad \quad  - \left. {\delta n\frac{{4\pi {\alpha ^{at}}}}{d}\left( {l + 1} \right)l\left( {{{\tilde \varepsilon }_1} - 1} \right){{\left( {d/a} \right)}^{2l + 1}}} \right].}
\end{array}
\label{VB6.1}
\end{equation}
The mode condition function when the reference system is that in absence of the atom is 
\begin{equation}
\begin{array}{*{20}{l}}
{{{\tilde f}_{l,m}} \approx 1 - \left( {\delta n} \right)4\pi {\alpha ^{at}}\frac{{\left( {l + 1} \right)l\left( {{{\tilde \varepsilon }_1} - 1} \right){{\left( {d/a} \right)}^{2l + 1}}}}{{d\left[ {{{\tilde \varepsilon }_1}\left( {l + 1} \right) + l} \right]}}}\\
{\quad \quad  = 1 - \left( {\delta n} \right)4\pi {\alpha ^{at}}\alpha _l^{\left( 2 \right)}\left( {a;\omega } \right)l{d^{2l}},}
\end{array}
\label{VB6.2}
\end{equation}
where we have identified the new ${2^l}$ pole polarizability that was in introduced in Eq.\,(\ref{VA8})
\begin{equation}
\alpha _l^{\left( 2 \right)}\left( {a;\omega } \right) =  - \frac{{\left( {l + 1} \right)\left( {1 - {{\tilde \varepsilon }_1}} \right){{\left( {1/a} \right)}^{2l + 1}}}}{{l + {{\tilde \varepsilon }_1}\left( {l + 1} \right)}},
\label{VB6.3}
\end{equation}
for the spherical cavity of radius $a$ "as seen from the inside."

From this we find the interaction energy per atom becomes
\begin{equation}
\begin{array}{l}
\frac{E}{{4\pi {d^2}\delta n}} = \frac{\hbar }{{4\pi {d^2}\delta n}}\int\limits_0^\infty  {\frac{{d\xi }}{{2\pi }}} \sum\limits_{l = 0}^\infty  {\sum\limits_{m =  - l}^l {\ln \left[ {{{\tilde f}_{l,m}}\left( {i\xi } \right)} \right]} } \\
 \approx  - \frac{\hbar }{{4\pi {d^2}\delta n}}\int\limits_0^\infty  {\frac{{d\xi }}{{2\pi }}} \sum\limits_{l = 0}^\infty  {\sum\limits_{m =  - l}^l {4\pi \left( {\delta n} \right){\alpha ^{at}}\left( {i\xi } \right)\alpha _l^{\left( 2 \right)}\left( {a;i\xi } \right)l{d^{2l}}} } \\
 =  - \hbar \int\limits_0^\infty  {\frac{{d\xi }}{{2\pi }}} \sum\limits_{l = 0}^\infty  {{\alpha ^{at}}\left( {i\xi } \right)\alpha _l^{\left( 2 \right)}\left( {a;i\xi } \right)\left( {2l + 1} \right)l{d^{2\left( {l - 1} \right)}}} \\
 =  - \hbar \int\limits_0^\infty  {\frac{{d\xi }}{{2\pi }}} \sum\limits_{l = 1}^\infty  {\frac{{\left[ {2l + 1} \right]!}}{{\left[ {2l - 1} \right]!\left[ 2 \right]!}}{\alpha ^{at}}\left( {i\xi } \right)\alpha _l^{\left( 2 \right)}\left( {a;i\xi } \right){d^{2\left( {l - 1} \right)}}}, 
\end{array}
\label{VB6.4}
\end{equation}
and the force on the atom is
\begin{equation}
\begin{array}{l}
F =  - \hbar \int\limits_0^\infty  {\frac{{d\xi }}{{2\pi }}} \sum\limits_{l = 1}^\infty  {\frac{{\left[ {2l + 1} \right]!}}{{\left[ {2l - 1} \right]!\left[ 2 \right]!}}2\left( {l - 1} \right){d^{2l - 3}}} \\
\quad \quad \quad \quad \quad \quad  \times {\alpha ^{at}}\left( {i\xi } \right)\alpha _l^{\left( 2 \right)}\left( {a;i\xi } \right).
\end{array}
\label{VB6.5}
\end{equation}
Note that the $l=0$ and $l=1$ terms do not contribute to the interaction.
\subsubsection{\label{vdW}Van der Waals interaction between two atoms (two layers)}

Here we may use the result from the previous section to derive the van der Waals interaction between two atoms. We let the atom outside the sphere be of type 1.  We let the spherical core of the structure be made up of a diluted gas of atoms of type 2. Then we let the density of the gas go towards zero and at the same time let the radius of the sphere go to zero ($b$ goes towards $d$). We furthermore only keep the $l=1$ term in the expansion; we are only interested in dipole dipole interactions. So we divide the energy in Eq.\,(\ref{VB5.4}) further with the number of atoms of the other species contained in the sphere and take the limits
%
\begin{equation}
\begin{array}{*{20}{l}}
{\frac{E}{{4\pi {b^2}\delta {n_1}\left( {{n_2}4\pi {a^3}/3} \right)}}}\\
{ \approx  - \frac{\hbar }{{{b^2}\left( {{n_2}4\pi {a^3}/3} \right)}}\int\limits_0^\infty  {\frac{{d\xi }}{{2\pi }}} \alpha _1^{at}\left( {i\xi } \right)\frac{{3 \cdot 2}}{{{b^4}}}\frac{{{a^3}\left[ {4\pi {n_2}\alpha _2^{at}\left( {i\xi } \right)} \right]}}{3}}\\
{ \approx  - \frac{{6\hbar }}{{{d^6}}}\int\limits_0^\infty  {\frac{{d\xi }}{{2\pi }}} \alpha _1^{at}\left( {i\xi } \right)\alpha _2^{at}\left( {i\xi } \right),}
\end{array}
\label{VB7.1}
\end{equation}
which is the van der Waals result (Ref.\,[\onlinecite{Ser}], Eq.\,(6.39)).

\subsubsection{\label{spherespheren}Force between two spherical objects}

In Sec.\,\ref{atomspheren} we obtained the force between an atom and a spherical object. We kept contributions from dipolar fluctuations in the atom, only. The dipolar and all higher order fluctuations of the sphere were included. This means that the results are valid for separations large compared to the size of the atom. Eq.\,(\ref{VB5.4}) is the first term of the more general expression,
\begin{equation}
E =  - 2\hbar \int\limits_0^\infty  {\frac{{d\xi }}{{2\pi }}} \sum\limits_{l' = 0}^\infty  {\sum\limits_{l = 0}^\infty  {\frac{{\left[ {2l + 2l'} \right]!}}{{\left[ {2l} \right]!\left[ {2l'} \right]!}}\frac{{\alpha _l^1\left( {i\xi } \right)\alpha _{l'}^2\left( {i\xi } \right)}}{{{d^{2\left( {l + l' + 1} \right)}}}}} },
\label{VB8.1}
\end{equation}
which is valid for all spherical objects. Here $d$ denotes the distance between the centers of the spheres.

\subsubsection{\label{atomsphericalgapn}Force on an atom in a spherical gap (three layers)}

Here we study an atom in a spherical vacuum gap with the outer and inner radii $b$ and $a$, respectively. The medium outside the gap has dielectric function ${{\tilde \varepsilon }_1}\left( \omega  \right)$ and the medium inside the dielectric function  ${{\tilde \varepsilon }_2}\left( \omega  \right)$.  The atom is at the distance $r$ from the center. The matrix for this geometry is ${\bf{\tilde M}} = {{{\bf{\tilde M}}}_0} \cdot {{{\bf{\tilde M}}}_1} \cdot {{{\bf{\tilde M}}}_2}$, where

\begin{equation}
\begin{array}{l}
{{{\bf{\tilde M}}}_0} = \frac{{{{\tilde \varepsilon }_1}\left( {l + 1} \right) + l}}{{\left( {2l + 1} \right){{\tilde \varepsilon }_1}}}\left( {\begin{array}{*{20}{c}}
1&{\alpha _l^{0\left( 2 \right)}}\\
{ - \alpha _l^0}&{\frac{{\left( {l + 1} \right) + {{\tilde \varepsilon }_1}l}}{{{{\tilde \varepsilon }_1}\left( {l + 1} \right) + l}}}
\end{array}} \right);\\
{{{\bf{\tilde M}}}_1} = \\
\left( {\begin{array}{*{20}{c}}
1&0\\
0&1
\end{array}} \right) + \left( {\delta n} \right)4\pi {\alpha ^{at}}\left( {\begin{array}{*{20}{c}}
0&{\left( {l + 1} \right){r^{ - \left( {2l + 2} \right)}}}\\
{ - l{r^{\left( {2l} \right)}}}&0
\end{array}} \right);\\
{{{\bf{\tilde M}}}_2} = \frac{{\left( {l + 1} \right) + {{\tilde \varepsilon }_2}l}}{{\left( {2l + 1} \right)}}\left( {\begin{array}{*{20}{c}}
1&{\alpha _l^{2\left( 2 \right)}}\\
{ - \alpha _l^2}&{\frac{{{{\tilde \varepsilon }_2}\left( {l + 1} \right) + l}}{{\left( {l + 1} \right) + {{\tilde \varepsilon }_2}l}}}
\end{array}} \right).
\end{array}
\label{VB10.1}
\end{equation}

The matrix element of interest is
\begin{equation}
\begin{array}{l}
{M_{11}} = M_{11}^0M_{11}^2\\
 \times \left\{ {1 - \alpha _l^2\alpha _l^{0\left( 2 \right)}} \right.\\
\left. { - \left( {\delta n} \right)4\pi {\alpha ^{at}}\left[ {\left( {l + 1} \right){r^{ - \left( {2l + 2} \right)}}\alpha _l^2 + l{r^{\left( {2l} \right)}}\alpha _l^{0\left( 2 \right)}} \right]} \right\}.
\end{array}
\label{VB10.2}
\end{equation}
This leads to the following proper mode condition function
\begin{equation}
{{\tilde f}_{l,m}} = 1 - \left( {\delta n} \right)4\pi {\alpha ^{at}}\frac{{\left[ {\left( {l + 1} \right){r^{ - \left( {2l + 2} \right)}}\alpha _l^2 + l{r^{\left( {2l} \right)}}\alpha _l^{0\left( 2 \right)}} \right]}}{{1 - \alpha _l^2\alpha _l^{0\left( 2 \right)}}},
\label{VB10.3}
\end{equation}
where the reference system is the spherical gap in absence of the atom. The two ${2^l}$ pole polarizabilities ${\alpha _l^2}$ and ${\alpha _l^{0\left( 2 \right)}}$ defined in Eqs.\,(\ref{VA7}) and (\ref{VA8}) are
\begin{equation}
\begin{array}{*{20}{l}}
{\alpha _l^2 = \frac{{l\left( {{{\tilde \varepsilon }_2} - 1} \right){a^{2l + 1}}}}{{{{\tilde \varepsilon }_2}l + \left( {l + 1} \right)}};}\\
{\alpha _l^{0\left( 2 \right)} =  - \frac{{\left( {l + 1} \right)\left( {1 - {{\tilde \varepsilon }_1}} \right){{\left( {1/b} \right)}^{2l + 1}}}}{{l + {{\tilde \varepsilon }_1}\left( {l + 1} \right)}}.}
\end{array}
\label{VB10.4}
\end{equation}
The energy per atom is
\begin{equation}
\begin{array}{l}
\frac{E}{{4\pi {r^2}\delta n}} = \frac{\hbar }{{4\pi {r^2}\delta n}}\int\limits_0^\infty  {\frac{{d\xi }}{{2\pi }}} \sum\limits_{l = 0}^\infty  {\sum\limits_{m =  - l}^l {\ln \left[ {{{\tilde f}_{l,m}}\left( {i\xi } \right)} \right]} } \\
 \approx  - \frac{\hbar }{{4\pi {r^2}\delta n}}\int\limits_0^\infty  {\frac{{d\xi }}{{2\pi }}} \sum\limits_{l = 0}^\infty  {\sum\limits_{m =  - l}^l {\left( {\delta n} \right)4\pi {\alpha ^{at}}} } \\
\quad  \times \frac{{\left( {\left( {l + 1} \right)\alpha _l^2{r^{ - \left( {2l + 2} \right)}} + l\alpha _l^{\left( 2 \right)}{r^{2l}}} \right)}}{{1 - \alpha _l^2\alpha _l^{0\left( 2 \right)}}}\\
 =  - \hbar \int\limits_0^\infty  {\frac{{d\xi }}{{2\pi }}} \sum\limits_{l = 0}^\infty  {{\alpha ^{at}}\frac{{\left[ {2l + 2} \right]!}}{{\left[ {2l} \right]!\left[ 2 \right]!}}\frac{1}{{{r^3}}}\frac{{\left( {\alpha _l^2{r^{ - \left( {2l + 1} \right)}} + \frac{l}{{l + 1}}\alpha _l^{0\left( 2 \right)}{r^{2l + 1}}} \right)}}{{1 - \alpha _l^2\alpha _l^{0\left( 2 \right)}}}},
\end{array}
\label{VB10.5}
\end{equation}
and the force on the atom is
\begin{equation}
\begin{array}{l}
F\left( r \right) =  - \hbar \int\limits_0^\infty  {\frac{{d\xi }}{{2\pi }}} \sum\limits_{l = 0}^\infty  {{\alpha ^{at}}\frac{{\left[ {2l + 2} \right]!}}{{\left[ {2l} \right]!\left[ 2 \right]!}}} \\
\quad \quad  \times \frac{{\left( {\left( {2l + 4} \right)\alpha _l^2{r^{ - \left( {2l + 5} \right)}} - \left( {2l - 2} \right)\frac{l}{{l + 1}}\alpha _l^{0\left( 2 \right)}{r^{2l - 3}}} \right)}}{{1 - \alpha _l^2\alpha _l^{0\left( 2 \right)}}}.
\end{array}
\label{VB10.6}
\end{equation}

\begin{figure}
\includegraphics[width=8cm]{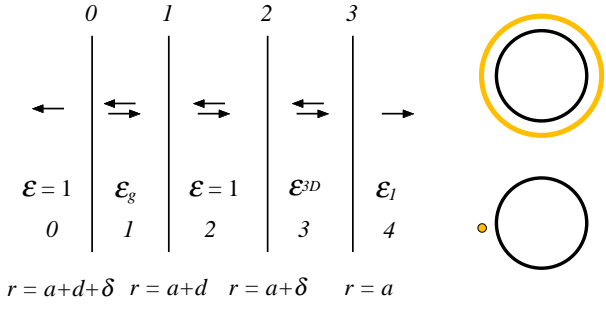}
\caption{(Color online) The geometry of a thin gas layer the distance $d$ from thin spherical or  cylindrical shell of radius $a$ in the non-retarded treatment.}
\label{figu14}
\end{figure}
\subsubsection{\label{atomsphericalshelln}Force on an atom outside a 2D spherical shell (three layers)}

In this section we derive the van der Waals interaction on an atom outside a very thin spherical shell. We start from the three layer structure in Fig.\,\ref{figu14}. We take the limit when the thickness goes to zero. The 3D dielectric function of the shell material then goes to infinity. We follow the procedure in Sec.\,\ref{atomspheren} but now there is one extra matrix. The matrix becomes ${\bf{\tilde M}} = {{\bf{\tilde M}}_0} \cdot {{\bf{\tilde M}}_1} \cdot {{\bf{\tilde M}}_2}\cdot {{\bf{\tilde M}}_3}$, where ${{{\bf{\tilde M}}}_0} \cdot {{{\bf{\tilde M}}}_1}$ is the matrix for a gas layer in Eq.\,(\ref{VB3.1}) with $r=b=a+d$ and ${{\bf{\tilde M}}_2}\cdot {{\bf{\tilde M}}_3}$ is the matrix for a thin film in Eq.\,(\ref{VB4.5}) with $r=a$. The matrix element of interest for us is
\begin{equation}
\begin{array}{l}
{M_{11}} = \frac{{\left( {2l + 1} \right)a + \left( {\delta {{\tilde \varepsilon }^{3D}}} \right)l\left( {l + 1} \right)}}{{\left( {2l + 1} \right)a}}\\
 \times \left[ {1 - \left( {\delta n} \right)4\pi {\alpha ^{at}}\left( \omega  \right)\alpha _l^{2D}\left( {a;\omega } \right)\left( {l + 1} \right){b^{ - \left( {2l + 2} \right)}}} \right].
\end{array}\
\label{VB11.1}
\end{equation}

The mode condition function when the reference system is that in absence of the atom is 

\begin{equation}
{{\tilde f}_{l,m}}\left( {i\xi } \right) = 1 - \left( {\delta n} \right)4\pi {\alpha ^{at}}\left( {i\xi } \right)\alpha _l^{2D}\left( {a;i\xi } \right)\left( {l + 1} \right){b^{ - \left( {2l + 2} \right)}}.
\label{VB11.2}
\end{equation}
We may identify the ${2^l}$ pole polarizability of the thin spherical shell in vacuum given in Eq.\,(\ref{VB4.2}).

Two examples where these results can be applied are a sphere made of a graphene like film and a thin metal film, respectively. Then\,\cite{grap,arx} 
\begin{equation}
\delta {{\tilde \varepsilon }^{3D}}\left( {i\xi } \right) \approx \delta {\alpha ^{3D}}\left( {i\xi } \right) \approx \left\{ {\begin{array}{*{20}{l}}
{\frac{{\pi {e^2}}}{{\hbar \left| \xi  \right|}},\;{\rm{graphene}}\;{\rm{like}}\;{\rm{film}}}\\
{\frac{{4\pi {n^{2D}}{e^2}}}{{{m^*}{m_e}{\xi ^2}}},\;{\rm{metal}}\;{\rm{film.}}}
\end{array}} \right.
\label{VB11.3}
\end{equation}
The final result is independent of $\delta$ and is the 2D limit. 

The energy per atom is
\begin{equation}
\begin{array}{l}
\frac{E}{{4\pi {b^2}\delta n}} = \frac{\hbar }{{4\pi {b^2}\delta n}}\int\limits_0^\infty  {\frac{{d\xi }}{{2\pi }}} \sum\limits_{l = 0}^\infty  {\sum\limits_{m =  - l}^l {\ln \left[ {{{\tilde f}_{l,m}}\left( {i\xi } \right)} \right]} } \\
 \approx  - \frac{\hbar }{{4\pi {b^2}\delta n}}\\
 \times \int\limits_0^\infty  {\frac{{d\xi }}{{2\pi }}} \sum\limits_{l = 0}^\infty  {\sum\limits_{m =  - l}^l {\left( {\delta n} \right)4\pi {\alpha ^{at}}\left( {i\xi } \right)\alpha _l^{2D}\left( {a;i\xi } \right)} } \\
\quad \quad \quad \quad \quad \quad \quad \quad \quad \quad \quad \quad  \times \left( {l + 1} \right){b^{ - \left( {2l + 2} \right)}}\\
 =  - \hbar \int\limits_0^\infty  {\frac{{d\xi }}{{2\pi }}} \sum\limits_{l = 0}^\infty  {{\alpha ^{at}}\left( {i\xi } \right)\alpha _l^{2D}\left( {a;i\xi } \right)\frac{{\left( {2l + 1} \right)\left( {l + 1} \right)}}{{{b^{2\left( {l + 2} \right)}}}}} \\
 =  - \hbar \int\limits_0^\infty  {\frac{{d\xi }}{{2\pi }}} \sum\limits_{l = 0}^\infty  {\frac{{\left[ {2l + 2} \right]!}}{{\left[ {2l} \right]!\left[ 2 \right]!}}\frac{1}{{{b^{2\left( {l + 2} \right)}}}}{\alpha ^{at}}\left( {i\xi } \right)\alpha _l^{2D}\left( {a;i\xi } \right)} .
\end{array}
\label{VB11.4}
\end{equation}
and the force on the atom is
\begin{equation}
F\left( b \right) =  - \hbar \int\limits_0^\infty  {\frac{{d\xi }}{{2\pi }}} \sum\limits_{l = 0}^\infty  {\frac{{\left[ {2l + 2} \right]!}}{{\left[ {2l} \right]!\left[ 2 \right]!}}\frac{{2\left( {l + 2} \right)}}{{{b^{2l + 5}}}}{\alpha ^{at}}\left( {i\xi } \right)\alpha _l^{2D}\left( {a;i\xi } \right)}.
\label{VB115}
\end{equation}
\begin{figure}
\includegraphics[width=8cm]{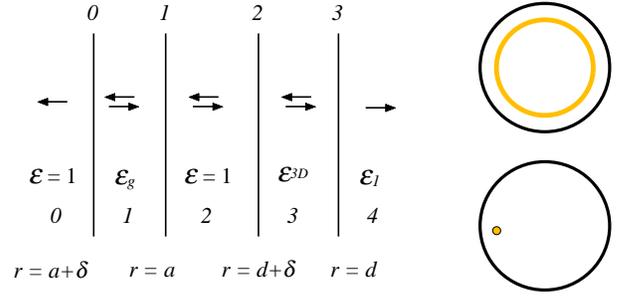}
\caption{(Color online) The geometry of a thin gas layer at radius $d$ inside a thin spherical or cylindrical shell of radius $a$ in the non-retarded treatment.}
\label{figu15}
\end{figure}
\subsubsection{\label{atomsphericalshellcavityn}Force on an atom inside a 2D spherical shell (three layers)}

In this section we derive the van der Waals interaction on an atom inside a very thin spherical shell. We start from the three layer structure in Fig.\,\ref{figu15}. We take the limit when the thickness goes to zero. The derivation is analogous to the one in Sec.\,\ref{atomsphericalshelln}. The matrix becomes ${\bf{\tilde M}} = {{\bf{\tilde M}}_0} \cdot {{\bf{\tilde M}}_1} \cdot {{\bf{\tilde M}}_2}\cdot {{\bf{\tilde M}}_3}$, where $ {{\bf{\tilde M}}_0}\cdot {{\bf{\tilde M}}_1}$ is the matrix for the thin film, given in Eq.\,(\ref{VB4.5}) for $r=a$, and $ {{\bf{\tilde M}}_2}\cdot {{\bf{\tilde M}}_3}$ is the matrix for the gas film, given in Eq.\,(\ref{VB3.1}) for $r=d$. The matrix element of interest for us is
\begin{equation}
\begin{array}{l}
{M_{11}} = \frac{{\left( {2l + 1} \right)a + \left( {\delta {{\tilde \varepsilon }^{3D}}} \right)l\left( {l + 1} \right)}}{{\left( {2l + 1} \right)a}}\\
 \times \left[ {1 - \left( {\delta n} \right)4\pi {\alpha ^{at}}\left( \omega  \right)\alpha _l^{2D\left( 2 \right)}\left( {a;\omega } \right)l{d^{2l}}} \right]
\end{array}
\label{VB12.1}
\end{equation}

The mode condition function when the reference system is that in absence of the atom is 
\begin{equation}
{{\tilde f}_{l,m}}\left( {i\xi } \right) = 1 - \left( {\delta n} \right)4\pi {\alpha ^{at}}\left( {i\xi } \right)\alpha _l^{2D\left( 2 \right)}\left( {a;i\xi } \right)l{d^{2l}},
\label{VB12.2}
\end{equation}
where we have introduced the ${2^l}$ pole polarizability for a thin spherical shell as "seen from the inside," given in Eq.\,(\ref{VB4.3}).

The energy per atom is
\begin{equation}
\begin{array}{*{20}{l}}
{\frac{E}{{4\pi {d^2}\delta n}} = \frac{\hbar }{{4\pi {d^2}\delta n}}\int\limits_0^\infty  {\frac{{d\xi }}{{2\pi }}} \sum\limits_{l = 0}^\infty  {\sum\limits_{m =  - l}^l {\ln \left[ {{{\tilde f}_{l,m}}\left( {i\xi } \right)} \right]} } }\\
{ \approx  - \frac{\hbar }{{4\pi {d^2}\delta n}}\int\limits_0^\infty  {\frac{{d\xi }}{{2\pi }}} \sum\limits_{l = 0}^\infty  {\sum\limits_{m =  - l}^l {4\pi \delta n{\alpha ^{at}}} } }\\
{\quad \quad \quad \quad  \times {\alpha ^{at}}\left( {i\xi } \right)\alpha _l^{2D\left( 2 \right)}\left( {a;i\xi } \right)l{d^{2l}}}\\
{ =  - \hbar \int\limits_0^\infty  {\frac{{d\xi }}{{2\pi }}} \sum\limits_{l = 0}^\infty  {\sum\limits_{m =  - l}^l {{\alpha ^{at}}\left( {i\xi } \right)\alpha _l^{2D\left( 2 \right)}\left( {a;i\xi } \right)l{d^{2\left( {l - 1} \right)}}} } }\\
{ =  - \hbar \int\limits_0^\infty  {\frac{{d\xi }}{{2\pi }}} \sum\limits_{l = 1}^\infty  {\frac{{\left[ {2l + 1} \right]!}}{{\left[ {2l - 1} \right]!2!}}{d^{2\left( {l - 1} \right)}}{\alpha ^{at}}\left( {i\xi } \right)\alpha _l^{2D\left( 2 \right)}\left( {a;i\xi } \right)} ,}
\end{array}
\label{VB12.3}
\end{equation}
and the force on the atom is
\begin{equation}
\begin{array}{l}
F =  - \hbar \int\limits_0^\infty  {\frac{{d\xi }}{{2\pi }}} \sum\limits_{l = 1}^\infty  {\frac{{\left[ {2l + 1} \right]!}}{{\left[ {2l - 1} \right]!\left[ 2 \right]!}}2\left( {l - 1} \right){d^{2l - 3}}} \\
\quad \quad \quad \quad \quad \quad  \times {\alpha ^{at}}\left( {i\xi } \right)\alpha _l^{2D\left( 2 \right)}\left( {a;i\xi } \right).
\end{array}
\label{VB12.4}
\end{equation}
Note that the $l=0$ and $l=1$ terms do not contribute to the interaction.

\subsubsection{\label{twosphericalfilmsn}Interaction between two 2D concentric spherical shells (three layers)}

We consider two concentric thin spherical shells. The outer shell has radius $b$ and the inner radius $a$. Here the matrix is ${\bf{\tilde M}} = {{{\bf{\tilde M}}}_0} \cdot {{{\bf{\tilde M}}}_1}$ where
\begin{equation}
\begin{array}{l}
{{{\bf{\tilde M}}}_0} = \left( {\begin{array}{*{20}{c}}
1&0\\
0&1
\end{array}} \right) + \frac{{\left( {\delta {{\tilde \varepsilon }^{3D}}} \right)l\left( {l + 1} \right)}}{{\left( {2l + 1} \right)b}}\left( {\begin{array}{*{20}{c}}
1&{{b^{ - \left( {2l + 1} \right)}}}\\
{ - {b^{\left( {2l + 1} \right)}}}&{ - 1}
\end{array}} \right);\\
{{{\bf{\tilde M}}}_1} = \left( {\begin{array}{*{20}{c}}
1&0\\
0&1
\end{array}} \right) + \frac{{\left( {\delta {{\tilde \varepsilon }^{3D}}} \right)l\left( {l + 1} \right)}}{{\left( {2l + 1} \right)a}}\left( {\begin{array}{*{20}{c}}
1&{{a^{ - \left( {2l + 1} \right)}}}\\
{ - {a^{\left( {2l + 1} \right)}}}&{ - 1}
\end{array}} \right),
\end{array}
\label{VB13.1}
\end{equation}
and the element of interest is
\begin{equation}
\begin{array}{l}
{M_{11}} = 1 + \frac{{\left( {\delta {{\tilde \varepsilon }^{3D}}} \right)l\left( {l + 1} \right)}}{{\left( {2l + 1} \right)}}\left( {\frac{1}{a} + \frac{1}{b}} \right)\\
\quad\quad \quad \quad  + {\left[ {\frac{{\left( {\delta {{\tilde \varepsilon }^{3D}}} \right)l\left( {l + 1} \right)}}{{\left( {2l + 1} \right)}}} \right]^2}\frac{1}{{ab}}\left[ {1 - {{\left( {\frac{a}{b}} \right)}^{2l + 1}}} \right].
\end{array}
\label{VB13.2}
\end{equation}
The proper mode condition function becomes
\begin{equation}
\begin{array}{*{20}{l}}
{{{\tilde f}_{l,m}}\left( {i\xi } \right)}\\
\begin{array}{l}
\quad  = 1 - \left[ {\frac{{\delta {{\tilde \varepsilon }^{3D}}l\left( {l + 1} \right)}}{{a\left( {2l + 1} \right) + \delta {{\tilde \varepsilon }^{3D}}l\left( {l + 1} \right)}}} \right]\\
\quad \quad \quad  \times \left[ {\frac{{\delta {{\tilde \varepsilon }^{3D}}l\left( {l + 1} \right)}}{{b\left( {2l + 1} \right) + \delta {{\tilde \varepsilon }^{3D}}l\left( {l + 1} \right)}}} \right]{\left( {\frac{a}{b}} \right)^{2l + 1}}
\end{array}\\
{\quad  = 1 - {\alpha _l}\left( {a;i\xi } \right)\alpha _l^{\left( 2 \right)}\left( {b;i\xi } \right),}
\end{array}
\label{VB13.3}
\end{equation}
where $\alpha _l$ is the ${2^l}$ pole polarizability of a thin spherical shell of radius $a$ in vacuum according to Eq.\,(\ref{VB10.4}) and ${\alpha _l^{\left( 2 \right)}}$ is the ${2^l}$ pole polarizability of a thin spherical shell of radius $b$ in vacuum  as seen from inside according to Eq.\,(\ref{VB10.4}). We have chosen as reference system a system where the two shells are separated from each other and at infinite distance from each other. The energy obtained by using this mode condition function is the energy change when bringing the two shells at infinite separation together and putting the inner shell inside the outer shell.

\subsubsection{\label{atomspherical2filmsn}Force on an atom in between two 2D spherical films (five layers)}
Here we study an atom in between two spherical films in vacuum. The outer and inner films are of radii $b$ and $a$, respectively. The atom is at the distance $r$ from the center. The matrix for this geometry is ${\bf{\tilde M}} = {{{\bf{\tilde M}}}_0} \cdot {{{\bf{\tilde M}}}_1} \cdot {{{\bf{\tilde M}}}_2}$, where
\begin{equation}
\begin{array}{*{20}{l}}
\begin{array}{l}
{{{\bf{\tilde M}}}_0} = \frac{{\left( {2l + 1} \right)b + \left( {\delta {{\tilde \varepsilon }^{3D}}} \right)l\left( {l + 1} \right)}}{{\left( {2l + 1} \right)b}}\\
\quad \quad  \times \left( {\begin{array}{*{20}{c}}
1&{\alpha _l^{2D\left( 2 \right)}\left( {b;\omega } \right)}\\
{ - \alpha _l^{2D}\left( {b;\omega } \right)}&{\frac{{\left( {2l + 1} \right)b - \left( {\delta {{\tilde \varepsilon }^{3D}}} \right)l\left( {l + 1} \right)}}{{\left( {2l + 1} \right)b + \left( {\delta {{\tilde \varepsilon }^{3D}}} \right)l\left( {l + 1} \right)}}}
\end{array}} \right);
\end{array}\\
{{{{\bf{\tilde M}}}_1} = \left( {\begin{array}{*{20}{c}}
1&0\\
0&1
\end{array}} \right) + }\\
{\quad \quad \quad \quad \left( {\delta n} \right)4\pi {\alpha ^{at}}\left( {\begin{array}{*{20}{c}}
0&{\left( {l + 1} \right){r^{ - \left( {2l + 2} \right)}}}\\
{ - l{r^{\left( {2l} \right)}}}&0
\end{array}} \right);}\\
\begin{array}{l}
{{{\bf{\tilde M}}}_2} = \frac{{\left( {2l + 1} \right)a + \left( {\delta {{\tilde \varepsilon }^{3D}}} \right)l\left( {l + 1} \right)}}{{\left( {2l + 1} \right)a}}\\
\quad \quad \quad  \times \left( {\begin{array}{*{20}{c}}
1&{\alpha _l^{2D\left( 2 \right)}\left( {a;\omega } \right)}\\
{ - \alpha _l^{2D}\left( {a;\omega } \right)}&{\frac{{\left( {2l + 1} \right)a - \left( {\delta {{\tilde \varepsilon }^{3D}}} \right)l\left( {l + 1} \right)}}{{\left( {2l + 1} \right)a + \left( {\delta {{\tilde \varepsilon }^{3D}}} \right)l\left( {l + 1} \right)}}}
\end{array}} \right)
\end{array}.
\end{array}
\label{VB14.1}
\end{equation}
The matrix element of interest to us is
\begin{equation}
\begin{array}{*{20}{l}}
\begin{array}{l}
{M_{11}}\\
 = \frac{{\left( {2l + 1} \right)b + \left( {\delta {{\tilde \varepsilon }^{3D}}} \right)l\left( {l + 1} \right)}}{{\left( {2l + 1} \right)b}}\frac{{\left( {2l + 1} \right)a + \left( {\delta {{\tilde \varepsilon }^{3D}}} \right)l\left( {l + 1} \right)}}{{\left( {2l + 1} \right)a}}
\end{array}\\
{\quad  \times \left\{ {1 - \alpha _l^{2D\left( 2 \right)}\left( {b;\omega } \right)\alpha _l^{2D}\left( {a;\omega } \right)} \right. - \left( {\delta n} \right)4\pi {\alpha ^{at}}}\\
{\quad \left. { \times \left[ {\alpha _l^{2D}\left( {a;\omega } \right)\left( {l + 1} \right){r^{ - \left( {2l + 2} \right)}} + \alpha _l^{2D\left( 2 \right)}\left( {b;\omega } \right)l{r^{\left( {2l} \right)}}} \right]} \right\},}
\end{array}
\label{VB14.2}
\end{equation}
which results in the following, proper mode condition function:
\begin{equation}
\begin{array}{l}
{{\tilde f}_{l,m}}\left( {i\xi } \right) = 1 - \left( {\delta n} \right)4\pi {\alpha ^{at}}\left( {i\xi } \right)\\
\quad \quad \quad \quad  \times \frac{{\alpha _l^{2D}\left( {a;i\xi } \right)\left( {l + 1} \right){r^{ - \left( {2l + 2} \right)}} + \alpha _l^{2D\left( 2 \right)}\left( {b;i\xi } \right)l{r^{2l}}}}{{1 - \alpha _l^{2D\left( 2 \right)}\left( {b;i\xi } \right)\alpha _l^{2D}\left( {a;i\xi } \right)}}.
\end{array}
\label{VB14.3}
\end{equation}
From this we obtain the interaction energy of the atom. It is
\begin{equation}
\begin{array}{*{20}{l}}
{\frac{E}{{4\pi {r^2}\delta n}} = \frac{\hbar }{{4\pi {r^2}\delta n}}\int\limits_0^\infty  {\frac{{d\xi }}{{2\pi }}} \sum\limits_{l = 0}^\infty  {\sum\limits_{m =  - l}^l {\ln \left[ {{{\tilde f}_{l,m}}\left( {i\xi } \right)} \right]} } }\\
{ \approx  - \frac{\hbar }{{4\pi {r^2}\delta n}}\int\limits_0^\infty  {\frac{{d\xi }}{{2\pi }}} \sum\limits_{l = 0}^\infty  {\sum\limits_{m =  - l}^l {4\pi \delta n{\alpha ^{at}}} } }\\
{\quad \quad  \times \frac{{\alpha _l^{2D}\left( {a;i\xi } \right)\left( {l + 1} \right){r^{ - \left( {2l + 2} \right)}} + \alpha _l^{2D\left( 2 \right)}\left( {b;i\xi } \right)l{r^{2l}}}}{{1 - \alpha _l^{2D\left( 2 \right)}\left( {b;i\xi } \right)\alpha _l^{2D}\left( {a;i\xi } \right)}}}\\
\begin{array}{l}
 =  - \hbar \int\limits_0^\infty  {\frac{{d\xi }}{{2\pi }}} \sum\limits_{l = 0}^\infty  {\sum\limits_{m =  - l}^l {\left[ {\frac{{\alpha _l^{2D}\left( {a;i\xi } \right)\left( {l + 1} \right){r^{ - 2\left( {l + 2} \right)}}}}{{1 - \alpha _l^{2D\left( 2 \right)}\left( {b;i\xi } \right)\alpha _l^{2D}\left( {a;i\xi } \right)}}} \right.} } \\
\left. {\quad \quad \quad \quad \quad \quad  + \frac{{\alpha _l^{2D\left( 2 \right)}\left( {b;i\xi } \right)l{r^{2\left( {l - 1} \right)}}}}{{1 - \alpha _l^{2D\left( 2 \right)}\left( {b;i\xi } \right)\alpha _l^{2D}\left( {a;i\xi } \right)}}} \right]
\end{array}\\
\begin{array}{l}
 =  - \hbar \int\limits_0^\infty  {\frac{{d\xi }}{{2\pi }}} \sum\limits_{l = 1}^\infty  {\left[ {\frac{{\frac{{\left[ {2l + 2} \right]!}}{{\left[ {2l} \right]!\left[ 2 \right]!}}\alpha _l^{2D}\left( {a;i\xi } \right){r^{ - 2\left( {l + 2} \right)}}}}{{1 - \alpha _l^{2D\left( 2 \right)}\left( {b;i\xi } \right)\alpha _l^{2D}\left( {a;i\xi } \right)}}} \right.} \\
\left. {\quad \quad \quad \quad \quad \quad  + \frac{{\frac{{\left[ {2l + 1} \right]!}}{{\left[ {2l - 1} \right]!\left[ 2 \right]!}}\alpha _l^{2D\left( 2 \right)}\left( {b;i\xi } \right){r^{2\left( {l - 1} \right)}}}}{{1 - \alpha _l^{2D\left( 2 \right)}\left( {b;i\xi } \right)\alpha _l^{2D}\left( {a;i\xi } \right)}}} \right].
\end{array}
\end{array}
\label{VB14.4}
\end{equation}
Note that this expression agrees  formally with the results for an atom in a spherical gap, Eq.\,(\ref{VB10.5}) but the $2^l$ pole polarizabilities are of course different. 
The force on the atom becomes
\begin{equation}
\begin{array}{*{20}{l}}
{F\left( r \right) =  - \hbar \int\limits_0^\infty  {\frac{{d\xi }}{{2\pi }}} \sum\limits_{l = 1}^\infty  {\left[ {\frac{{\frac{{\left[ {2l + 2} \right]!}}{{\left[ {2l} \right]!\left[ 2 \right]!}}2\left( {l + 2} \right)\alpha _l^{2D}\left( {a;i\xi } \right){r^{ - 2l - 5}}}}{{1 - \alpha _l^{2D\left( 2 \right)}\left( {b;i\xi } \right)\alpha _l^{2D}\left( {a;i\xi } \right)}}} \right.} }\\
{\left. {\quad \quad \quad \quad \quad \quad  - \frac{{\frac{{\left[ {2l + 1} \right]!}}{{\left[ {2l - 1} \right]!\left[ 2 \right]!}}2\left( {l - 1} \right)\alpha _l^{2D\left( 2 \right)}\left( {b;i\xi } \right){r^{2l - 3}}}}{{1 - \alpha _l^{2D\left( 2 \right)}\left( {b;i\xi } \right)\alpha _l^{2D}\left( {a;i\xi } \right)}}} \right].}
\end{array}
\label{VB14.5}
\end{equation}
Now we are done with the non-retarded treatment of spherical systems and turn to the more complicated retarded treatment.
\subsection{\label{sphericalretmain}Retarded main results}
To find the normal modes for a layered sphere including retardation effects we need to solve the wave equation for the electric and magnetic fields in all layers and use the proper boundary conditions at the interfaces. To solve the vector wave equation the vector Helmholtz equation, Eq.\,(\ref{III22}), is not a trivial task.  Instead one may solve the problem by introducing Hertz-Debye potentials $\pi _1 $ and $\pi _2 $. They are solutions to the scalar wave equation, Eq.\,(\ref{III23}).  We let $\pi _1 $ be the potential that generates TM modes and $\pi _2 $ be the potential that generates TE modes.
Separation of variables, $\pi  = R\left( r \right)\Theta \left( \theta  \right)\Phi \left( \phi  \right)$, leads to one differential equation for each of the variables, 
\begin{equation}
\begin{array}{l}
\frac{{{d^2}\left[ {rR\left( r \right)} \right]}}{{d{r^2}}} + \left[ {{q^2} - \frac{{i\left( {i + 1} \right)}}{{{r^2}}}} \right]\left[ {rR\left( r \right)} \right] = 0;\\
\frac{1}{{\sin \theta }}\frac{d}{{d\theta }}\left[ {\sin \theta \frac{{d\Theta \left( \theta  \right)}}{{d\theta }}} \right] + \left[ {i\left( {i + 1} \right) - \frac{{{m^2}}}{{{{\sin }^2}\theta }}} \right]\Theta \left( \theta  \right) = 0;\\
\frac{{{d^2}\Phi \left( \phi  \right)}}{{d{\phi ^2}}} + {m^2}\Phi \left( \phi  \right) = 0.
\end{array}
\label{VC1}
\end{equation}

The angular equations lead to spherical harmonics and for the radial part $rR(r)$ is a solution to the Ricatti-Bessel equation,
\begin{equation}
{z^2}\frac{{{d^2}\omega }}{{d{z^2}}} + \left[ {{z^2} - i\left( {i + 1} \right)} \right]\omega  = 0.
\label{VC2}
\end{equation}
The Ricatti-Bessel equation has many different solutions:
\begin{itemize}
\item Ricatti-Bessel functions of the first kind: 
\begin{equation}
{S_i}\left( z \right) = z{j_i}\left( z \right) = \sqrt {\pi z/2} {J_{i + 1/2}}\left( z \right) = {\psi _i}\left( z \right);
\label{VC3}
\end{equation}

\item Ricatti-Bessel functions of second kind: 
\begin{equation}
{C_i}\left( z \right) =  - z{y_i}\left( z \right) = \sqrt {\pi z/2} {Y_{i + 1/2}}\left( z \right) = {\chi _i}\left( z \right);
\label{VC4}
\end{equation}

\item Ricatti-Bessel functions of the third kind: 
\begin{equation}
\begin{array}{*{20}{l}}
{{\xi _i}\left( z \right) = z{h_i}^{\left( 1 \right)}\left( z \right) = \sqrt {\pi z/2} H_{i + 1/2}^{\left( 1 \right)}\left( z \right)}\\
{\quad  = {S_i}\left( z \right) - i{C_i}\left( z \right) = z{j_i}\left( z \right) + iz{y_i}\left( z \right);}\\
{{\zeta _i}\left( z \right) = z{h_i}^{\left( 2 \right)}\left( z \right) = \sqrt {\pi z/2} H_{i + 1/2}^{\left( 2 \right)}\left( z \right)}\\
{\quad  = {S_i}\left( z \right) + i{C_i}\left( z \right) = z{j_i}\left( z \right) - iz{y_i}\left( z \right).}
\end{array}
\label{VC5}
\end{equation}
\end{itemize}
Let us study a layered sphere of radius $R$ consisting of $N$ layers and an inner spherical core. We have $N+2$ media and $N+1$ interfaces. Let the numbering be as follows. Medium $0$ is the medium surrounding the sphere, medium $1$ is the outermost layer and medium $N+1$ the innermost layer and $N+2$ the innermost spherical core region. Let ${r_n}$ be the inner radius of layer $n$. This is completely in line with the system represented by Fig.\,\ref{figu3}.
 
We will use the two Hankel versions in Eq.\,(\ref{VC5})  since they represent waves that go in either the positive or negative $r$-directions. We assume a time dependence of the form ${e^{ - i\omega t}}$. With this choice the first Ricatti-Hankel function, ${\xi _n}\left( {qr} \right){e^{ - i\omega t}} \propto {e^{i\left( {qr - \omega t} \right)}}$, represents a wave moving in the positive radial direction (towards the left in Fig.\,\ref{figu3}) while the second, ${\zeta _n}\left( {qr} \right){e^{ - i\omega t}} \propto {e^{ - i\left( {qr + \omega t} \right)}}$, represents a wave moving in the negative radial direction (towards the right in Fig.\,\ref{figu3}.)
Thus the general solution for the potentials is
\begin{equation}
r\pi  = \sum\limits_{l = 0}^\infty  {\sum\limits_{m =  - l}^l {\left[ {{a_l}{\zeta _l}\left( {qr} \right) + {b_l}{\xi _l}\left( {qr} \right)} \right]{Y_{l,m}}\left( {\theta ,\phi } \right)} } {e^{ - i\omega t}} .
\label{VC6}
\end{equation}
From the potentials we get the fields\,\cite{Kerker,Debye,Born}
\begin{equation}
\begin{array}{l}
{E_r} = {E_{1r}} + {E_{2r}} = \frac{{{\partial ^2}\left( {r{\pi _1}} \right)}}{{\partial {r^2}}} + {q^2}r{\pi _1} + 0;\\
{E_\theta } = {E_{1\theta }} + {E_{2\theta }} = \frac{1}{r}\frac{{{\partial ^2}\left( {r{\pi _1}} \right)}}{{\partial r\partial \theta }} - \frac{{i\omega }}{c}\frac{1}{{r\sin \theta }}\frac{{\partial \left( {r{\pi _2}} \right)}}{{\partial \phi }};\\
{E_\phi } = {E_{1\phi }} + {E_{2\phi }} = \frac{1}{{r\sin \theta }}\frac{{{\partial ^2}\left( {r{\pi _1}} \right)}}{{\partial r\partial \phi }} + \frac{{i\omega }}{c}\frac{1}{r}\frac{{\partial \left( {r{\pi _2}} \right)}}{{\partial \theta }};\\
{H_r} = {H_{1r}} + {H_{2r}} = 0 + \frac{{{\partial ^2}\left( {r{\pi _2}} \right)}}{{\partial {r^2}}} + {q^2}r{\pi _2};\\
{H_\theta } = {H_{1\theta }} + {H_{2\theta }} = \frac{{i\omega  {\tilde \varepsilon }}}{c}\frac{1}{{r\sin \theta }}\frac{{\partial \left( {r{\pi _1}} \right)}}{{\partial \phi }} + \frac{1}{r}\frac{{{\partial ^2}\left( {r{\pi _2}} \right)}}{{\partial r\partial \theta }};\\
{H_\phi } = {H_{1\phi }} + {H_{2\phi }} =  - \frac{{i\omega  {\tilde \varepsilon }}}{c}\frac{1}{r}\frac{{\partial \left( {r{\pi _1}} \right)}}{{\partial \theta }} + \frac{1}{{r\sin \theta }}\frac{{{\partial ^2}\left( {r{\pi _2}} \right)}}{{\partial r\partial \phi }}.
\end{array}
\label{VC7}
\end{equation}
Let us now use the boundary conditions that the tangential components of ${\bf{E}}$ and ${\bf{H}}$ are continuous at the interface between layer $n$ and $n+1$. We get
\begin{equation}
\begin{array}{l}
\left( \partial /\partial r \right){\left[ {r\pi _{_1}^n} \right]_{r = {r_n}}} = \left( \partial /\partial r \right){\left[ {r\pi _{_1}^{n + 1}} \right]_{r = {r_n}}};\\
 \left( \partial /\partial r  \right){\left[ {r\pi _{_2}^n} \right]_{r = {r_n}}} = \left( \partial /\partial r \right){\left[ {r\pi _{_2}^{n + 1}} \right]_{r = {r_n}}};\\
\left( {{{ {\tilde \varepsilon }}^n}i\omega /c} \right){\left[ {r\pi _1^n} \right]_{r = {r_n}}} = \left( {{{ {\tilde \varepsilon }}^{n + 1}}i\omega /c} \right){\left[ {r\pi _1^{n + 1}} \right]_{r = {r_n}}};\\
{ \left( \omega /c \right) }{\left[ {r\pi _2^n} \right]_{r = {r_n}}} = { \left( \omega /c \right) }{\left[ {r\pi _{_2}^{n + 1}} \right]_{r = {r_n}}}.
\end{array}
\label{VC8}
\end{equation}
This gives
\begin{equation}
\begin{array}{l}
{q_n}\left[ {a_{1,l}^n{\zeta _l}'\left( {{q_n}{r_n}} \right) + b_{1,l}^n{\xi _l}'\left( {{q_n}{r_n}} \right)} \right]\\
\quad  = {q_{n + 1}}\left[ {a_{1,l}^{n + 1}{\zeta _l}'\left( {{q_{n + 1}}{r_n}} \right) + b_{1,l}^{n + 1}{\xi _l}'\left( {{q_{n + 1}}{r_n}} \right)} \right];\\
{q_n}\left[ {a_{2,l}^n{\zeta _l}'\left( {{q_n}{r_n}} \right) + b_{2,l}^n{\xi _l}'\left( {{q_n}{r_n}} \right)} \right]\\
\quad  = {q_{n + 1}}\left[ {a_{2,l}^{n + 1}{\zeta _l}'\left( {{q_{n + 1}}{r_n}} \right) + b_{2,l}^{n + 1}{\xi _l}'\left( {{q_{n + 1}}{r_n}} \right)} \right];\\
q_n^2\left[ {a_{1,l}^n{\zeta _l}\left( {{q_n}{r_n}} \right) + b_{1,l}^n{\xi _l}\left( {{q_n}{r_n}} \right)} \right]\\
\quad  = q_{n + 1}^2\left[ {a_{1,l}^{n + 1}{\zeta _l}\left( {{q_{n + 1}}{r_n}} \right) + b_{1,l}^{n + 1}{\xi _l}\left( {{q_{n + 1}}{r_n}} \right)} \right];\\
\left[ {a_{2,l}^n{\zeta _l}\left( {{q_n}{r_n}} \right) + b_{2,l}^n{\xi _l}\left( {{q_n}{r_n}} \right)} \right]\\
\quad  = \left[ {a_{2,l}^{n + 1}{\zeta _l}\left( {{q_{n + 1}}{r_n}} \right) + b_{2,l}^{n + 1}{\xi _l}\left( {{q_{n + 1}}{r_n}} \right)} \right],
\end{array}
\label{VC9}
\end{equation}
where a prime on a function means the derivative with respect to its argument.

Let us first assume pure TM-modes. That means keeping ${\pi _1}$ only. Then we have
\begin{equation}
\begin{array}{l}
a_{1,l}^n{q_n}{\zeta _l}'\left( {{q_n}{r_n}} \right) + b_{1,l}^n{q_n}{\xi _l}'\left( {{q_n}{r_n}} \right)\\
\quad  = a_{1,l}^{n + 1}{q_{n + 1}}{\zeta _l}'\left( {{q_{n + 1}}{r_n}} \right) + b_{1,l}^{n + 1}{q_{n + 1}}{\xi _l}'\left( {{q_{n + 1}}{r_n}} \right);\\
a_{1,l}^nq_n^2{\zeta _l}\left( {{q_n}{r_n}} \right) + b_{1,l}^nq_n^2{\xi _l}\left( {{q_n}{r_n}} \right)\\
\quad  = a_{1,l}^{n + 1}q_{n + 1}^2{\zeta _l}\left( {{q_{n + 1}}{r_n}} \right) + b_{1,l}^{n + 1}q_{n + 1}^2{\xi _l}\left( {{q_{n + 1}}{r_n}} \right),
\end{array}
\label{VC10}
\end{equation}
and we may identify the matrix ${{{\bf{\tilde A}}}_n}\left( {{r_n}} \right)$ as
\begin{equation}
{\bf{\tilde A}}_n^{{\rm{TM}}}\left( {{r_n}} \right) = \left( {\begin{array}{*{20}{c}}
{{q_n}{\zeta _l}'\left( {{q_n}{r_n}} \right)}&{{q_n}{\xi _l}'\left( {{q_n}{r_n}} \right)}\\
{q_n^2{\zeta _l}\left( {{q_n}{r_n}} \right)}&{q_n^2{\xi _l}\left( {{q_n}{r_n}} \right)}
\end{array}} \right),
\label{VC11}
\end{equation}
and the matrix ${{{\bf{\tilde M}}}_n}$ as
\begin{equation}
\begin{array}{l}
{\bf{\tilde M}}_n^{{\rm{TM}}} = \frac{{{-q_{n + 1}}}}{{q_n^2\left( {  2i} \right)}}\\
 \times \left( {\begin{array}{*{20}{c}}
{{q_n}{\xi _l}{\zeta _l}{'^ + } - {q_{n + 1}}{\xi _l}'{\zeta _l}^ + }&{{q_n}{\xi _l}{\xi _l}{'^ + } - {q_{n + 1}}{\xi _l}'{\xi _l}^ + }\\
{ - {q_n}{\zeta _l}{\zeta _l}{'^ + } + {q_{n + 1}}{\zeta _l}'{\zeta _l}^ + }&{ - {q_n}{\zeta _l}{\xi _l}{'^ + } + {q_{n + 1}}{\zeta _l}'{\xi _l}^ + }
\end{array}} \right),
\end{array}
\label{VC12}
\end{equation}
where we to save space have omitted the function arguments. All functions with a + added as a superscript have the argument ${{q_{n + 1}}{r_n}}$ and the ones without the superscript have the argument ${{q_n}{r_n}}$. We have also made use of the Wronskian of the two Ricatti-Bessel functions: $W\left[ {{\zeta _l}\left( x \right),{\xi _l}\left( x \right)} \right] = {\xi _l}'\left( x \right){\zeta _l}\left( x \right) - {\xi _l}\left( x \right){\zeta _l}'\left( x \right) =   2i$.

Now we repeat the derivation for TE- modes. That means keeping ${\pi _2}$ only. Then we have
\begin{equation}
\begin{array}{l}
a_{2,l}^n{q_n}{\zeta _l}'\left( {{q_n}{r_n}} \right) + b_{2,l}^n{q_n}{\xi _l}'\left( {{q_n}{r_n}} \right)\\
\quad  = a_{2,l}^{n + 1}{q_{n + 1}}{\zeta _l}'\left( {{q_{n + 1}}{r_n}} \right) + b_{2,l}^{n + 1}{q_{n + 1}}{\xi _l}'\left( {{q_{n + 1}}{r_n}} \right);\\
a_{2,l}^n{\zeta _l}\left( {{q_n}{r_n}} \right) + b_{2,l}^n{\xi _l}\left( {{q_n}{r_n}} \right)\\
\quad  = a_{2,l}^{n + 1}{\zeta _l}\left( {{q_{n + 1}}{r_n}} \right) + b_{2,l}^{n + 1}{\xi _l}\left( {{q_{n + 1}}{r_n}} \right),
\end{array}
\label{VC13}
\end{equation}
and we may identify the matrix ${{{\bf{\tilde A}}}_n}\left( {{r_n}} \right)$ as
\begin{equation}
{\bf{\tilde A}}_n^{{{\rm{TE}}}}\left( {{r_n}} \right) = \left( {\begin{array}{*{20}{c}}
{{q_n}{\zeta _l}'\left( {{q_n}{r_n}} \right)}&{{q_n}{\xi _l}'\left( {{q_n}{r_n}} \right)}\\
{{\zeta _l}\left( {{q_n}{r_n}} \right)}&{{\xi _l}\left( {{q_n}{r_n}} \right)}
\end{array}} \right),
\label{VC14}
\end{equation}
and the matrix ${{{\bf{\tilde M}}}_n}$ as
\begin{equation}
\begin{array}{l}
{\bf{\tilde M}}_n^{{{\rm{TE}}}} = \frac{-1}{{{q_n}\left( {  2i} \right)}}\\
 \times \left( {\begin{array}{*{20}{c}}
{ - {q_n}{\xi _l}'{\zeta _l}^ +  + {q_{n + 1}}{\xi _l}{\zeta _l}{'^ + }}&{ - {q_n}{\xi _l}'{\xi _l}^ +  + {q_{n + 1}}{\xi _l}{\xi _l}{'^ + }}\\
{{q_n}{\zeta _l}'{\zeta _l}^ +  - {q_{n + 1}}{\zeta _l}{\zeta _l}{'^ + }}&{{q_n}{\zeta _l}'{\xi _l}^ +  - {q_{n + 1}}{\zeta _l}{\xi _l}{'^ + }}
\end{array}} \right).
\end{array}
\label{VC15}
\end{equation}
Of the solutions to the Ricatti-Bessel equation in Eq.\,(\ref{VC2}) the Ricatti-Bessel function of the first kind is the function that is regular at the origin. Thus this is the function we should use in the rightmost region of Fig.\,\ref{figu3}. Now, Since the function ${\psi _l}\left( z \right) = \left[ {{\xi _l}\left( x \right) + {\zeta _l}\left( x \right)} \right]/2$ we have that ${b_{N + 1}} = {a_{N + 1}}$. According to  Eq.\,(\ref{III6}) this means that
\begin{equation}
{f_{l,m}}\left( \omega  \right) = {M_{11}+M_{12}}.
\label{VC16}
\end{equation}
Before we end this section we introduce the ${2^l}$ pole polarizabilities ${\alpha _l^n}$ and $\alpha _l^{n\left( 2 \right)}$ for the spherical interface since these appear repeatedly in the sections that follow. The first is valid outside and the second inside. The polarizability ${\alpha _l^n} = - {b^n}/{a^n}$ under the assumption that ${b^{n+1}} = {a^{n+1}}$. One obtains $\alpha _l^n =  - \left( {M_{21}^n + M_{22}^n} \right)/\left( {M_{11}^n + M_{12}^n} \right)$ and from Eq.\,(\ref{VC12}) one finds that for TM modes
\begin{equation}
\begin{array}{l}
\alpha _l^{{\rm{TM,}}n}\\
 = \frac{{{q_n}{\zeta _l}\left( {{q_n}{r_n}} \right){\psi _l}'\left( {{q_{n + 1}}{r_n}} \right) - {q_{n + 1}}{\zeta _l}^\prime \left( {{q_n}{r_n}} \right){\psi _l}\left( {{q_{n + 1}}{r_n}} \right)}}{{{q_n}{\xi _l}\left( {{q_n}{r_n}} \right){\psi _l}'\left( {{q_{n + 1}}{r_n}} \right) - {q_{n + 1}}{\xi _l}^\prime \left( {{q_n}{r_n}} \right){\psi _l}\left( {{q_{n + 1}}{r_n}} \right)}}.
\end{array}
\label{VC17}
\end{equation}
In the same way one finds from Eq.\,(\ref{VC15}) that for TE modes
\begin{equation}
\begin{array}{l}
\alpha _l^{{\rm{TE,}}n}\\
 = \frac{{{q_n}{\zeta _l}'\left( {{q_n}{r_n}} \right){\psi _l}\left( {{q_{n + 1}}{r_n}} \right) - {q_{n + 1}}{\zeta _l}\left( {{q_n}{r_n}} \right){\psi _l}'\left( {{q_{n + 1}}{r_n}} \right)}}{{{q_n}{\xi _l}'\left( {{q_n}{r_n}} \right){\psi _l}\left( {{q_{n + 1}}{r_n}} \right) - {q_{n + 1}}{\xi _l}\left( {{q_n}{r_n}} \right){\psi _l}'\left( {{q_{n + 1}}{r_n}} \right)}}.
\end{array}
\label{VC18}
\end{equation}
The polarizability $\alpha _l^{n\left( 2 \right)} = - {a^{n + 1}}/{b^{n + 1}}$ under the assumption that ${a^n} = 0$. One obtains $\alpha _l^{n\left( 2 \right)} =   {M_{12}}/{M_{11}}$ and from Eq.\,(\ref{VC12}) one finds that for the TM modes
\begin{equation}
\begin{array}{l}
\alpha _l^{{\rm{TM,}}n\left( 2 \right)}\\
 = \frac{{{q_n}{\xi _l}\left( {{q_n}{r_n}} \right){\xi _l}'\left( {{q_{n + 1}}{r_n}} \right) - {q_{n + 1}}{\xi _l}'\left( {{q_n}{r_n}} \right){\xi _l}\left( {{q_{n + 1}}{r_n}} \right)}}{{{q_n}{\xi _l}\left( {{q_n}{r_n}} \right){\zeta _l}'\left( {{q_{n + 1}}{r_n}} \right) - {q_{n + 1}}{\xi _l}'\left( {{q_n}{r_n}} \right){\zeta _l}\left( {{q_{n + 1}}{r_n}} \right)}}.
\end{array}
\label{VC19}
\end{equation}
From Eq.\,(\ref{VC15}) one finds that for TE modes
\begin{equation}
\begin{array}{l}
\alpha _l^{{\rm{TE}},n\left( 2 \right)}\\
 = \frac{{{q_n}{\xi _l}'\left( {{q_n}{r_n}} \right){\xi _l}'\left( {{q_{n + 1}}{r_n}} \right) - {q_{n + 1}}{\xi _l}\left( {{q_n}{r_n}} \right){\xi _l}'\left( {{q_{n + 1}}{r_n}} \right)}}{{{q_n}{\xi _l}'\left( {{q_n}{r_n}} \right){\zeta _l}\left( {{q_{n + 1}}{r_n}} \right) - {q_{n + 1}}{\xi _l}\left( {{q_n}{r_n}} \right){\zeta _l}'\left( {{q_{n + 1}}{r_n}} \right)}}.
\end{array}
\label{VC20}
\end{equation}

When we calculate the energy by an integral along the imaginary frequency axis the arguments of the Ricatti-Bessel functions become imaginary. It may be favorable to have real-valued arguments, $\xi a/c$, and $\sqrt {{\tilde \varepsilon }\left( {i\xi } \right)} \xi a/c$ instead of  $i\xi a/c$  and $\sqrt {{\tilde \varepsilon }\left( {i\xi } \right)} i\xi a/c$, respectively. To achieve real valued arguments we transform the functions.  The transformation rules are:\,\cite{Steg}

\begin{equation}
\begin{array}{l}
{\xi _l}\left( {ix} \right) = \frac{1}{{{i^{l + 1}}}}\frac{1}{\pi }\sqrt {2\pi x} {K_{l + 1/2}}\left( x \right);\\
{\zeta _l}\left( {ix} \right) = {i^{l + 1}}\sqrt {2\pi x} \\
\quad \quad  \times \left[ {{I_{l + 1/2}}\left( x \right) + \frac{1}{\pi }{{\left( { - 1} \right)}^{l}}{K_{l + 1/2}}\left( x \right)} \right];\\
{\psi _l}\left( {ix} \right) = {i^{l + 1}}\frac{1}{2}\sqrt {2\pi x} {I_{l + 1/2}}\left( x \right);\\
{\xi _l}'\left( {ix} \right) = \frac{1}{{{i^{l + 2}}}}\frac{1}{\pi }\sqrt {2\pi x} \\
\quad \quad  \times \left[ {\frac{1}{{2x}}{K_{l + 1/2}}\left( x \right) + K{'_{l + 1/2}}\left( x \right)} \right];\\
\zeta {'_l}\left( {ix} \right) = {i^l}\sqrt {2\pi x} \\
\quad \quad  \times \left\{ {\frac{1}{{2x}}\left[ {{I_{l + 1/2}}\left( x \right) + \frac{1}{\pi }{{\left( { - 1} \right)}^{l}}{K_{l + 1/2}}\left( x \right)} \right]} \right.\\
\quad \quad \left. { + \left[ {I{'_{l + 1/2}}\left( x \right) + \frac{1}{\pi }{{\left( { - 1} \right)}^{l}}K{'_{l + 1/2}}\left( x \right)} \right]} \right\};\\
\psi {'_l}\left( {ix} \right) = {i^l}\frac{1}{2}\sqrt {2\pi x} \\
\quad \quad  \times \left[ {\frac{1}{{2x}}{I_{l + 1/2}}\left( x \right) + I{'_{l + 1/2}}\left( x \right)} \right].
\end{array}
\label{VC21}
\end{equation}

Now we have all we need to determine the fully retarded normal modes in a layered spherical structure. We give some examples in the following sections.
\subsection{\label{sphericalretpecial}Retarded special results}
\subsubsection{\label{solid sphere}Solid sphere (no layer)}
\begin{figure}
\includegraphics[width=5cm]{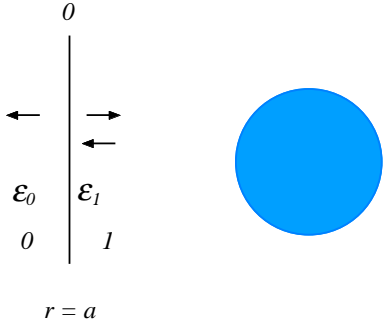}
\caption{(Color online) The geometry of a solid sphere or cylinder of radius $a$ in the fully retarded treatment.}
\label{figu16}
\end{figure}
For a solid sphere of radius  $a$ and dielectric function ${{\tilde \varepsilon }_1}\left( \omega  \right)$ in an ambient of dielectric function ${{\tilde \varepsilon }_0}\left( \omega  \right)$, as illustrated in Fig.\,\ref{figu16}, we have ${\bf{\tilde M}} = {{{\bf{\tilde M}}}_0}$, and for the TM modes we find
\begin{equation}
\begin{array}{*{20}{l}}
{{M_{11}} + {M_{12}} = \frac{{i{q_1}}}{{2q_0^2}}\left\{ {{q_0}{\xi _l}\left( {{q_0}a} \right)\left[ {{\zeta _l}^\prime \left( {{q_1}a} \right) + {\xi _l}^\prime \left( {{q_1}a} \right)} \right]} \right.}\\
{\quad \quad \quad \quad  - \left. {{q_1}{\xi _l}^\prime \left( {{q_0}a} \right)\left[ {{\zeta _l}\left( {{q_1}a} \right) + {\xi _l}\left( {{q_1}a} \right)} \right]} \right\}}\\
\begin{array}{l}
\quad \quad \quad  = \frac{{-i{q_1}}}{{2q_0^2}}\left\{ {{q_0}{\xi _l}\left( {{q_0}a} \right)2{\psi _l}^\prime \left( {{q_1}a} \right)} \right.\\
\quad \quad \quad \quad  - \left. {{q_1}{\xi _l}^\prime \left( {{q_0}a} \right)2{\psi _l}\left( {{q_1}a} \right)} \right\}
\end{array}\\
{\quad \quad \quad  = \frac{{-i{q_1}}}{{q_0^2}}\left\{ {\left[ {{{\tilde \varepsilon }_0}{{\left( {\omega /c} \right)}^2}ah_l^{\left( 1 \right)}\left( {{q_0}a} \right)} \right]{{\left[ {{q_1}a{j_l}\left( {{q_1}a} \right)} \right]}^\prime }} \right.}\\
{\quad \quad \quad \quad  - \left. {{{\left[ {{q_0}ah_l^{\left( 1 \right)}\left( {{q_0}a} \right)} \right]}^\prime }\left[ {{{\tilde \varepsilon }_1}{{\left( {\omega /c} \right)}^2}a{j_l}\left( {{q_1}a} \right)} \right]} \right\}},
\end{array}
\label{VD1.1}
\end{equation}
where we have used the relations between the different solutions to the Ricatti-Bessel equation given in Eqs.\,(\ref{VC3}) and (\ref{VC5}). We have furthermore used the relations $q_0^2 = {{\tilde \varepsilon }_0}{\left( {\omega /c} \right)^2}$ and $q_1^2 = { {\tilde \varepsilon }_1}{\left( {\omega /c} \right)^2}$.
 
The mode condition function for TM modes is 
\begin{equation}
\begin{array}{*{20}{l}}
{f_{l,m}^{{\rm{TM}}} = \left[ {{{\tilde \varepsilon }_0}\left( \omega  \right)h_l^{\left( 1 \right)}\left( {{q_0}a} \right)} \right]{{\left[ {\left( {{q_1}a} \right){j_l}\left( {{q_1}a} \right)} \right]}^\prime }}\\
{\quad \quad  - {{\left[ {\left( {{q_0}a} \right)h_l^{\left( 1 \right)}\left( {{q_0}a} \right)} \right]}^\prime }\left[ {{{\tilde \varepsilon }_1}\left( \omega  \right){j_l}\left( {{q_1}a} \right)} \right]}.
\end{array}
\label{VD1.2}
\end{equation}
This result agrees with the result of Ruppin in Eq.\,(43) on page 353, in Ref.\,[\onlinecite{Board}].
For the TE modes we find
\begin{equation}
\begin{array}{*{20}{l}}
{{M_{11}} + {M_{12}} = \frac{i}{{2{q_0}}}\left\{ { - {q_0}{\xi_l}^\prime \left( {{q_0}a} \right)\left[ {{\zeta _l}\left( {{q_1}a} \right) + {\xi _l}\left( {{q_1}a} \right)} \right]} \right.}\\
{\quad \quad \quad \quad  + \left. {{q_1}{\xi _l}\left( {{q_0}a} \right)\left[ {{\zeta _l}^\prime \left( {{q_1}a} \right) + {\xi _l}^\prime \left( {{q_1}a} \right)} \right]} \right\}}\\
\begin{array}{l}
\quad \quad \quad  = \frac{-i}{{2{q_0}}}\left\{ { - {q_0}{\xi _l}^\prime \left( {{q_0}a} \right)2{\psi _l}\left( {{q_1}a} \right)} \right.\\
\quad \quad \quad \quad  + \left. {{q_1}{\xi _l}\left( {{q_0}a} \right)2{\psi _l}^\prime \left( {{q_1}a} \right)} \right\}
\end{array}\\
{\quad \quad \quad  = -i{q_1}a\left\{ { - {{\left[ {{q_0}ah_l^{\left( 1 \right)}\left( {{q_0}a} \right)} \right]}^\prime }\left[ {{j_l}\left( {{q_1}a} \right)} \right]} \right.}\\
{\quad \quad \quad \quad  + \left. {\left[ {h_l^{\left( 1 \right)}\left( {{q_0}a} \right)} \right]{{\left[ {{q_1}a{j_l}\left( {{q_1}a} \right)} \right]}^\prime }} \right\},}
\end{array}
\label{VD1.3}
\end{equation}
and the mode condition function for TE modes is
\begin{equation}
\begin{array}{*{20}{l}}
{f_{l,m}^{{\rm{TE}}} = \left[ {h_l^{\left( 1 \right)}\left( {{q_0}a} \right)} \right]{{\left[ {\left( {{q_1}a} \right){j_l}\left( {{q_1}a} \right)} \right]}^\prime }}\\
{\quad \quad  - {{\left[ {\left( {{q_0}a} \right)h_l^{\left( 1 \right)}\left( {{q_0}a} \right)} \right]}^\prime }\left[ {{j_l}\left( {{q_1}a} \right)} \right].}
\end{array}
\label{VD1.4}
\end{equation}
This result agrees with the result of Ruppin in Eq.\,(34) on page 351, in Ref.\,[\onlinecite{Board}]. The results of Eqs.\,(\ref{VD1.2}) and (\ref{VD1.4}) can also be used for a spherical cavity in a medium if the two dielectric functions are interchanged.

\subsubsection{\label{Sphericalshellret}Spherical shell or gap (one layer)}
\begin{figure}
\includegraphics[width=6cm]{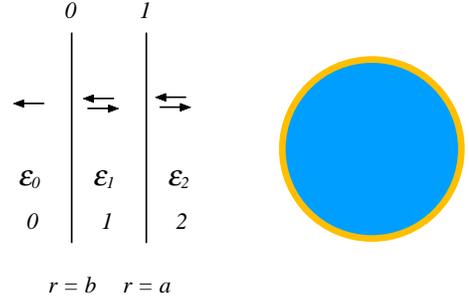}
\caption{(Color online) The geometry of a coated sphere or cylinder of radius $a$ in the fully retarded treatment.}
\label{figu17}
\end{figure}
For a spherical shell of inner radius $a$ and outer radius $b$, Fig.\,\ref{figu17},  made of a medium with dielectric function ${{\tilde \varepsilon }_1}$ in an ambient medium with dielectric function ${{\tilde \varepsilon }_0}$ we have ${\bf{\tilde M}} = {{{\bf{\tilde M}}}_0} \cdot {{{\bf{\tilde M}}}_1}$. This geometry covers the problem of a vacuum gap in the shape of a spherical shell inside an infinite medium, as treated in Ref.\,[\onlinecite{BreAar}]. 

We do not need all elements of the two matrices. We have

\begin{equation}
\begin{array}{l}
{M_{11}} + {M_{12}}\\
 = \left( {M_{11}^0,M_{12}^0} \right)\cdot \left( {\begin{array}{*{20}{c}}
{M_{11}^1 + M_{12}^1}\\
{M_{21}^1 + M_{22}^1}
\end{array}} \right)\\
 = M_{11}^0\left( {M_{11}^1 + M_{12}^1} \right)\\
 \times \left( {1,\alpha _l^{0\left( 2 \right)}} \right)\cdot \left( {\begin{array}{*{20}{c}}
1\\
{ - \alpha _l^1}
\end{array}} \right)\\
 = M_{11}^0\left( {M_{11}^1 + M_{12}^1} \right)\left( {1 - \alpha _l^{0\left( 2 \right)}\alpha _l^1} \right).
\end{array}
\label{VD2.1}
\end{equation}
We want to end up with expressions for the mode condition functions that are suitable to use on the imaginary frequency axis. This demands some manipulations. First we note that $1 - \alpha _l^{0\left( 2 \right)}\alpha _l^1 = 0 \to 1/\alpha _l^{0\left( 2 \right)} - \alpha _l^1 = 0$
and for TM modes we have
\begin{equation}
\begin{array}{l}
\frac{{{q_0}{\xi _l}\left( {{q_0}b} \right){\zeta _l}'\left( {{q_1}b} \right) - {q_1}{\xi _l}'\left( {{q_0}b} \right){\zeta _l}\left( {{q_1}b} \right)}}{{{q_0}{\xi _l}\left( {{q_0}b} \right){\xi _l}'\left( {{q_1}b} \right) - {q_1}{\xi _l}'\left( {{q_0}b} \right){\xi _l}\left( {{q_1}b} \right)}}\\
\quad\quad\quad\quad - \frac{{{q_1}{\zeta _l}\left( {{q_1}a} \right){\psi _l}'\left( {{q_0}a} \right) - {q_0}{\zeta _l}^\prime \left( {{q_1}a} \right){\psi _l}\left( {{q_0}a} \right)}}{{{q_1}{\xi _l}\left( {{q_1}a} \right){\psi _l}'\left( {{q_0}a} \right) - {q_0}{\xi _l}^\prime \left( {{q_1}a} \right){\psi _l}\left( {{q_0}a} \right)}} = 0.
\end{array}
\label{VD2.2}
\end{equation}
We may use the relation $2{\psi _l} = {\zeta _l} + {\xi _l}$ to find
\begin{equation}
\begin{array}{l}
2\frac{{{q_0}{\xi _l}\left( {{q_0}b} \right){\psi _l}'\left( {{q_1}b} \right) - {q_1}{\xi _l}'\left( {{q_0}b} \right){\psi _l}\left( {{q_1}b} \right)}}{{{q_0}{\xi _l}\left( {{q_0}b} \right){\xi _l}'\left( {{q_1}b} \right) - {q_1}{\xi _l}'\left( {{q_0}b} \right){\xi _l}\left( {{q_1}b} \right)}} - 1\\
 \quad\quad- 2\frac{{{q_1}{\psi _l}\left( {{q_1}a} \right){\psi _l}'\left( {{q_0}a} \right) - {q_0}{\psi _l}^\prime \left( {{q_1}a} \right){\psi _l}\left( {{q_0}a} \right)}}{{{q_1}{\xi _l}\left( {{q_1}a} \right){\psi _l}'\left( {{q_0}a} \right) - {q_0}{\xi _l}^\prime \left( {{q_1}a} \right){\psi _l}\left( {{q_0}a} \right)}} + 1 = 0.
\end{array}
\label{VD2.3}
\end{equation}
We find the following mode condition function for TM modes 
\begin{equation}
\begin{array}{*{20}{l}}
{\tilde f_{l,m}^{{\rm{TM}}}\left( \omega  \right) = 1 - \frac{{{q_0}{\xi _l}\left( {{q_0}b} \right){\psi _l}^\prime \left( {{q_1}b} \right) - {q_1}{\xi _l}^\prime \left( {{q_0}b} \right){\psi _l}\left( {{q_1}b} \right)}}{{{q_0}{\xi _l}\left( {{q_0}b} \right){\xi _l}^\prime \left( {{q_1}b} \right) - {q_1}{\xi _l}^\prime \left( {{q_0}b} \right){\xi _l}\left( {{q_1}b} \right)}}}\\
{\quad \quad \quad \quad\quad\quad \quad\quad \times \frac{{{q_1}{\xi _l}\left( {{q_1}a} \right){\psi _l}^\prime \left( {{q_0}a} \right) - {q_0}{\xi _l}^\prime \left( {{q_1}a} \right){\psi _l}\left( {{q_0}a} \right)}}{{{q_1}{\psi _l}\left( {{q_1}a} \right){\psi _l}^\prime \left( {{q_0}a} \right) - {q_0}{\psi _l}^\prime \left( {{q_1}a} \right){\psi _l}\left( {{q_0}a} \right)}},}
\end{array}
\label{VD2.4}
\end{equation}
and analogous manipulations for the TE modes give
\begin{equation}
\begin{array}{*{20}{l}}
{\tilde f_{l,m}^{{\rm{TE}}}\left( \omega  \right) = 1 - \frac{{{q_0}{\xi _l}^\prime \left( {{q_0}b} \right){\psi _l}\left( {{q_1}b} \right) - {q_1}{\xi _l}\left( {{q_0}b} \right){\psi _l}^\prime \left( {{q_1}b} \right)}}{{{q_0}{\xi _l}^\prime \left( {{q_0}b} \right){\xi _l}\left( {{q_1}b} \right) - {q_1}{\xi _l}\left( {{q_0}b} \right){\xi _l}^\prime \left( {{q_1}b} \right)}}}\\
{\quad \quad \quad \quad\quad\quad\quad\quad  \times \frac{{{q_1}{\xi _l}^\prime \left( {{q_1}a} \right){\psi _l}\left( {{q_0}a} \right) - {q_0}{\xi _l}\left( {{q_1}a} \right){\psi _l}^\prime \left( {{q_0}a} \right)}}{{{q_1}{\psi _l}^\prime \left( {{q_1}a} \right){\psi _l}\left( {{q_0}a} \right) - {q_0}{\psi _l}\left( {{q_1}a} \right){\psi _l}^\prime \left( {{q_0}a} \right)}}.}
\end{array}
\label{VD2.5}
\end{equation}
In these equations, ${q_0}$ is $ \sqrt {{{\tilde \varepsilon }_0}\left( \omega  \right)} \omega /c$ and ${q_1} $ is $ \sqrt {{{\tilde \varepsilon }_1}\left( \omega  \right)} \omega /c$, respectively. We have expressed the mode condition functions in terms of ${\xi _l}$ and ${\psi _l}$ since these are easier to transform from functions of imaginary arguments into functions of real arguments by following Eq.\,(\ref{VC17}). For the vacuum gap treated in Ref.\,[\onlinecite{BreAar}] one should put ${{\tilde \varepsilon }_0} = {\tilde \varepsilon }$ and
${{\tilde \varepsilon }_1} = 1$. Our results agree with those in  Ref.\,[\onlinecite{BreAar}].

\subsubsection{\label{coated sphereret}Coated sphere in a medium (one layer)}
The result for a coated sphere in a medium is obtained trivially from the previous section. One just replaces ${q_0}$ in the last factor in Eqs.\,(\ref{VD2.4}) and (\ref{VD2.5}) with ${q_2} = \sqrt {{\varepsilon _2}\left( \omega  \right)} \omega /c$ where ${{\varepsilon _2}\left( \omega  \right)}$ is the dielectric function of the sphere medium.

\subsubsection{\label{Sphericalgasfilmret}Thin spherical diluted gas film (one layer)}
It is of interest to find the Casimir force on an atom in a layered structure. We can obtain this by studying the force on a thin layer of a diluted gas with dielectric function ${\varepsilon _g}\left( \omega  \right) = 1 + 4\pi n\alpha^{at} \left( \omega  \right)$, where $\alpha^{at}$ is the polarizability of one atom and $n$ the density of atoms (we have assumed that the atom is surrounded by vacuum; if not  the $1$ should be replaced by the dielectric function of the ambient medium and the atomic polarizability should be replaced by the excess polarizability). For a diluted gas layer the atoms do not interact with each other and the force on the layer is just the sum of the forces on the individual atoms. So by dividing with the number of atoms in the film we get the force on one atom. The layer has to be thin in order to have a well defined $r$- value of the atom.  Since we will derive the force on an atom in different spherical geometries it is fruitful to derive the matrix for a thin diluted gas shell. This result can be directly used in the derivation of the Casimir force on an atom in different spherical geometries.

We let the film have the thickness $\delta$ and be of general radius $r$. We only keep terms up to linear order in $\delta$ and linear order in $n$. We find the result for TM modes is

\begin{equation}
\begin{array}{*{20}{l}}
{{\bf{\tilde M}}_{{\rm{gaslayer}}}^{{\rm{TM}}} = \left( {\begin{array}{*{20}{c}}
1&0\\
0&1
\end{array}} \right) - \left( {\delta n} \right)2\pi {\alpha ^{at}}{q_0}i}\\
{\quad \quad \quad  \times \left( {\begin{array}{*{20}{c}}
{{\xi _l}'{\zeta _l}' + {\xi _l}{\zeta _l}\frac{{l\left( {l + 1} \right)}}{{{{\left( {{q_0}r} \right)}^2}}}}&{{{\left[ {{\xi _l}'} \right]}^2} + {{\left[ {{\xi _l}} \right]}^2}\frac{{l\left( {l + 1} \right)}}{{{{\left( {{q_0}r} \right)}^2}}}}\\
{ - {{\left[ {{\zeta _l}'} \right]}^2} - {{\left[ {{\zeta _l}} \right]}^2}\frac{{l\left( {l + 1} \right)}}{{{{\left( {{q_0}r} \right)}^2}}}}&{ - {\xi _l}'{\zeta _l}' - {\xi _l}{\zeta _l}\frac{{l\left( {l + 1} \right)}}{{{{\left( {{q_0}r} \right)}^2}}}}
\end{array}} \right),}
\end{array}
\label{VD4.1}
\end{equation}
where we have suppressed the argument $\left( {{q_0}r} \right)$ in all Ricatti Bessel functions. For TE modes we find
\begin{equation}
\begin{array}{*{20}{l}}
{{\bf{\tilde M}}_{{\rm{gaslayer}}}^{{\rm{TE}}} = \left( {\begin{array}{*{20}{c}}
1&0\\
0&1
\end{array}} \right) - \left( {\delta n} \right)2\pi {\alpha ^{at}}{q_0}i}\\
{\quad \quad \quad  \times \left( {\begin{array}{*{20}{c}}
{{\xi _l}\left( {{q_0}r} \right){\zeta _l}\left( {{q_0}r} \right)}&{{{\left[ {{\xi _l}\left( {{q_0}r} \right)} \right]}^2}}\\
{ - {{\left[ {{\zeta _l}\left( {{q_0}r} \right)} \right]}^2}}&{ - {\xi _l}\left( {{q_0}r} \right){\zeta _l}\left( {{q_0}r} \right)}
\end{array}} \right).}
\end{array}
\label{VD4.2}
\end{equation}
Now we are done with the gas layer. We will use these results later in calculating the Casimir force on an atom in planar layered structures.
\subsubsection{\label{Spherical2Dfilmret}2D spherical film (one layer)}
In many situations one is dealing with very thin films. These may be considered 2D (two dimensional). Important examples are a graphene sheet and a 2D electron gas. In the derivation we let the film have finite thickness $\delta$ and be characterized by a 3D dielectric function ${\tilde \varepsilon ^{3D}}$. We then let the thickness go towards zero. The 3D dielectric function depends on $\delta$ as ${\tilde\varepsilon ^{3D}} \sim 1/\delta $ for small $\delta$. In the planar structure we could in the limit when $\delta$ goes towards zero obtain a momentum dependent 2D dielectric function. Here we only obtain the long wave length limit of the 2D dielectric function.\,\cite{grap,arx} We obtain for TM modes
\begin{equation}
\begin{array}{l}
{\bf{\tilde M}}_{{\rm{2D}}}^{{\rm{TM}}} = \left( {\begin{array}{*{20}{c}}
1&0\\
0&1
\end{array}} \right)\\
 - \frac{{\delta {{\tilde \varepsilon }^{{\rm{3D}}}}{q_0}i}}{2}\left( {\begin{array}{*{20}{c}}
{{\xi _l}'\left( {{q_0}r} \right){\zeta _l}'\left( {{q_0}r} \right)}&{{{\left[ {{\xi _l}'\left( {{q_0}r} \right)} \right]}^2}}\\
{ - {{\left[ {{\zeta _l}'\left( {{q_0}r} \right)} \right]}^2}}&{ - {\xi _l}'\left( {{q_0}r} \right){\zeta _l}'\left( {{q_0}r} \right)}
\end{array}} \right),
\end{array}
\label{VD5.1}
\end{equation}
and for TE modes
\begin{equation}
\begin{array}{l}
{\bf{\tilde M}}_{{\rm{2D}}}^{{\rm{TE}}} = \left( {\begin{array}{*{20}{c}}
1&0\\
0&1
\end{array}} \right)\\
 - \frac{{\delta {{\tilde \varepsilon }^{{\rm{3D}}}}{q_0}i}}{2}\left( {\begin{array}{*{20}{c}}
{{\xi _l}\left( {{q_0}r} \right){\zeta _l}\left( {{q_0}r} \right)}&{{{\left[ {{\xi _l}\left( {{q_0}r} \right)} \right]}^2}}\\
{ - {{\left[ {{\zeta _l}\left( {{q_0}r} \right)} \right]}^2}}&{ - {\xi _l}\left( {{q_0}r} \right){\zeta _l}\left( {{q_0}r} \right)}
\end{array}} \right).
\end{array}
\label{VD5.2}
\end{equation}
\begin{figure}
\includegraphics[width=8cm]{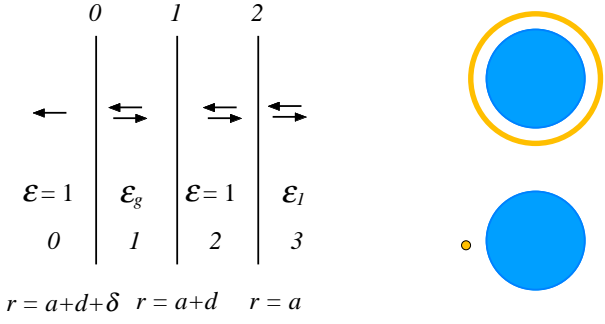}
\caption{(Color online) The geometry of a thin gas layer the distance $d$ from a sphere or  cylinder of radius $a$ in the  fully retarded treatment.}
\label{figu18}
\end{figure}

We will also need the ${2^l}$ pole polarizabilities ${\alpha _l^{\rm2D}}$ and $\alpha _l^{\rm2D\left( 2 \right)}$ for the thin spherical film since these appear repeatedly in the sections that follow. The first is valid outside and the second inside. The polarizability ${\alpha _l^{\rm2D}} = - {b^0}/{a^0}$ under the assumption that ${b^{1}} = {a^{1}}$. One obtains ${\alpha _l^{\rm2D}} =  - \left( {M_{21} + M_{22}} \right)/\left( {M_{11} + M_{12}} \right)$ and from Eq.\,(\ref{VD5.1}) one finds that for TM modes
\begin{equation}
\begin{array}{l}
\alpha _l^{{\rm{2D,TM}}}=  - \frac{{2 + \delta {{\tilde \varepsilon }^{{\rm{3D}}}}{q_0}i\left[ {{\zeta _l}^\prime {{\left( {{q_0}r} \right)}^2} + {\xi _l}^\prime \left( {{q_0}r} \right){\zeta _l}^\prime \left( {{q_0}r} \right)} \right]}}{{2 - \delta {{\tilde \varepsilon }^{{\rm{3D}}}}{q_0}i\left[ {{\xi _l}^\prime {{\left( {{q_0}r} \right)}^2} + {\xi _l}^\prime \left( {{q_0}r} \right){\zeta _l}^\prime \left( {{q_0}r} \right)} \right]}}.
\end{array}
\label{VD5.3}
\end{equation}
In the same way one finds from Eq.\,(\ref{VD5.2}) that for TE modes
\begin{equation}
\begin{array}{*{20}{l}}
{\alpha _l^{{\rm{2D,TE}}}}{ =  - \frac{{2 + \delta {{\tilde \varepsilon }^{{\rm{3D}}}}{q_0}i\left[ {{\zeta _l}{{\left( {{q_0}r} \right)}^2} + {\xi _l}\left( {{q_0}r} \right){\zeta _l}\left( {{q_0}r} \right)} \right]}}{{2 - \delta {{\tilde \varepsilon }^{{\rm{3D}}}}{q_0}i\left[ {{\xi _l}{{\left( {{q_0}r} \right)}^2} + {\xi _l}\left( {{q_0}r} \right){\zeta _l}\left( {{q_0}r} \right)} \right]}}.}
\end{array}
\label{VD5.4}
\end{equation}
The polarizability $\alpha _l^{\rm2D\left( 2 \right)} = - {a^{ 1}}/{b^{ 1}}$ under the assumption that ${a^0} = 0$. One obtains $\alpha _l^{{\rm2D\left( 2 \right)}} =   {M_{12}}/{M_{11}}$ and from Eq.\,(\ref{VD5.1}) one finds that for the TM modes
\begin{equation}
\alpha _l^{{\rm{2D}}\left( 2 \right){\rm{,TM}}} = \frac{{ - \delta {{\tilde \varepsilon }^{{\rm{3D}}}}{q_0}i{\xi _l}^\prime {{\left( {{q_0}r} \right)}^2}}}{{2 - \delta {{\tilde \varepsilon }^{{\rm{3D}}}}{q_0}i{\xi _l}^\prime \left( {{q_0}r} \right){\zeta _l}^\prime \left( {{q_0}r} \right)}}.
\label{VD5.5}
\end{equation}
From Eq.\,(\ref{VD5.2}) one finds that for TE modes
\begin{equation}
\alpha _l^{{\rm{2D}}\left( 2 \right){\rm{,TE}}} = \frac{{ - \delta {{\tilde \varepsilon }^{{\rm{3D}}}}{q_0}i{\xi _l}{{\left( {{q_0}r} \right)}^2}}}{{2 - \delta {{\tilde \varepsilon }^{{\rm{3D}}}}{q_0}i{\xi _l}\left( {{q_0}r} \right){\zeta _l}\left( {{q_0}r} \right)}}.
\label{VD5.6}
\end{equation}
\subsubsection{\label{atomsphere}Force on an atom outside a sphere (two layers)}
We let the atom be at the distance $d$ from the sphere of radius $a$ and at the distance $b$ from the center of the sphere.
For this problem we start with the geometry given in Fig.\,\ref{figu18}, where we let the shell be a very thin gas layer. We have two layers and three interfaces. The matrix ${\bf{\tilde M}} = {{{\bf{\tilde M}}}_0} \cdot {{{\bf{\tilde M}}}_1} \cdot {{{\bf{\tilde M}}}_2}$. Here we could instead of the first two matrices have used the matrix for a thin diluted gas shell as given in Eqs.\,(\ref{VD4.1}) and (\ref{VD4.2}). To vary the derivations to some extent we refrain from doing that.  
The left hand side of the condition for modes is
\begin{equation}
\begin{array}{l}
{M_{11}} + {M_{12}}\\
\quad \quad  = \left( {\begin{array}{*{20}{c}}
{M_{11}^0}&{M_{12}^0}
\end{array}} \right) \cdot {{{\bf{\tilde M}}}_1} \cdot \left( {\begin{array}{*{20}{c}}
{M_{11}^2 + M_{12}^2}\\
{M_{21}^2 + M_{22}^2}
\end{array}} \right),
\end{array}
\label{VD6.1}
\end{equation}
where we have moved the matrix subscripts to superscripts to make room for the element indices. We now list all elements needed in the above equation. We begin with the matrices for TM modes. The elements of the first matrix is
\begin{equation}
\begin{array}{*{20}{l}}
{M_{11}^0 = \frac{{i{n_g}}}{2}\left\{ {{\xi _l}\left[ {{q_0}\left( {b + \delta } \right)} \right]{\zeta _l}^\prime \left[ {{q_g}\left( {b + \delta } \right)} \right]} \right.}\\
{\quad \quad \left. { - {n_g}{\xi _l}^\prime \left[ {{q_0}\left( {b + \delta } \right)} \right]{\zeta _l}\left[ {{q_g}\left( {b + \delta } \right)} \right]} \right\};}\\
{M_{12}^0 = \frac{{i{n_g}}}{2}\left\{ {{\xi _l}\left[ {{q_0}\left( {b + \delta } \right)} \right]{\xi _l}^\prime \left[ {{q_g}\left( {b + \delta } \right)} \right]} \right.}\\
{\quad \quad \left. { - {n_g}{\xi _l}^\prime \left[ {{q_0}\left( {b + \delta } \right)} \right]{\xi _l}\left[ {{q_g}\left( {b + \delta } \right)} \right]} \right\},}
\end{array}
\label{VD6.2}
\end{equation}

and of the second
\begin{equation}
\begin{array}{*{20}{l}}
{M_{11}^1 = \frac{i}{{2n_g^2}}\left[ {{n_g}{\xi _l}\left( {{q_g}b} \right){\zeta _l}^\prime \left( {{q_0}b} \right) - {\xi _l}^\prime \left( {{q_g}b} \right){\zeta _l}\left( {{q_0}b} \right)} \right];}\\
{M_{12}^1 = \frac{i}{{2n_g^2}}\left[ {{n_g}{\xi _l}\left( {{q_g}b} \right){\xi _l}^\prime \left( {{q_0}b} \right) - {\xi _l}^\prime \left( {{q_g}b} \right){\xi _l}\left( {{q_0}b} \right)} \right];}\\
{M_{21}^1 = \frac{i}{{2n_g^2}}\left[ {{\zeta _l}^\prime \left( {{q_g}b} \right){\zeta _l}\left( {{q_0}b} \right) - {n_g}{\zeta _l}\left( {{q_g}b} \right){\zeta _l}^\prime \left( {{q_0}b} \right)} \right];}\\
{M_{22}^1 = \frac{i}{{2n_g^2}}\left[ {{\zeta _l}^\prime \left( {{q_g}b} \right){\xi _l}\left( {{q_0}b} \right) - {n_g}{\zeta _l}\left( {{q_g}b} \right){\xi _l}^\prime \left( {{q_0}b} \right)} \right],}
\end{array}
\label{VD6.3}
\end{equation}
and of the third
\begin{equation}
\begin{array}{*{20}{l}}
{M_{11}^2 + M_{12}^2}\\
{ = \frac{{i{n_1}}}{2}\left[ {{\xi _l}\left( {{q_0}a} \right){\zeta _l}^\prime \left( {{q_1}a} \right) - {n_1}{\xi _l}^\prime \left( {{q_0}a} \right){\zeta _l}\left( {{q_1}a} \right)} \right.}\\
{\quad \quad \left. { + {\xi _l}\left( {{q_0}a} \right){\xi _l}^\prime \left( {{q_1}a} \right) - {n_1}{\xi _l}^\prime \left( {{q_0}a} \right){\xi _l}\left( {{q_1}a} \right)} \right]}\\
{ = i{n_1}\left[ {{\xi _l}\left( {{q_0}a} \right){\psi _l}^\prime \left( {{q_1}a} \right) - {n_1}{\xi _l}^\prime \left( {{q_0}a} \right){\psi _l}\left( {{q_1}a} \right)} \right];}\\
{M_{21}^2 + M_{22}^2}\\
{ = \frac{{i{n_1}}}{2}\left[ {{n_1}{\zeta _l}^\prime \left( {{q_0}a} \right){\zeta _l}\left( {{q_1}a} \right) - {\zeta _l}\left( {{q_0}a} \right){\zeta _l}^\prime \left( {{q_1}a} \right)} \right.}\\
{\quad \quad \left. { + {n_1}{\zeta _l}^\prime \left( {{q_0}a} \right){\xi _l}\left( {{q_1}a} \right) - {\zeta _l}\left( {{q_0}a} \right){\xi _l}^\prime \left( {{q_1}a} \right)} \right]}\\
{ = i{n_1}\left[ {{n_1}{\zeta _l}^\prime \left( {{q_0}a} \right){\psi _l}\left( {{q_1}a} \right) - {\zeta _l}\left( {{q_0}a} \right){\psi _l}^\prime \left( {{q_1}a} \right)} \right],}
\end{array}
\label{VD6.4}
\end{equation}
where ${n_g}$ and ${n_1}$ are the refractive indices of the gas layer and the sphere, respectively.

We now make a series expansion of the first matrix  up to linear order in $\delta$. The other matrices do not depend on $\delta$. The zeroth order term multiplied with the second matrix produces the matrix $\left( {\begin{array}{*{20}{c}}1&0\end{array}} \right)$, so it contributes with ${M_{11}^2 + M_{12}^2}$ to the condition for modes. We then expand in ${\alpha _g}$, the polarizability of the gas. There is no zeroth order term in the term linear in $\delta$. The lowest order term is linear in ${\alpha _g}$. This means that we do not need to expand ${{{\bf{\tilde M}}}_1}$ in ${\alpha _g}$. The zeroth order term is just the unit matrix. Thus if we denote the term of the matrix ${{{\bf{\tilde M}}}_0}$ that is linear in both $\delta$ and  ${\alpha _g}$ with $\delta {{{\bf{\tilde M}}}_0}$ the condition for modes can be written as
\begin{equation}
\begin{array}{l}
\left( {M_{11}^2 + M_{12}^2} \right) + \delta M_{11}^0\left( {M_{11}^2 + M_{12}^2} \right)\\
\quad \quad  + \delta M_{12}^0\left( {M_{21}^2 + M_{22}^2} \right) = 0,
\end{array}
\label{VD6.5}
\end{equation}
and the mode condition function is
\begin{equation}
\begin{array}{*{20}{l}}
{\tilde f_{_l}^{{\rm{TM}}}\left( \omega  \right) = 1 + \delta M_{12}^0\frac{{\left( {M_{21}^2 + M_{22}^2} \right) - \left( {M_{11}^2 + M_{12}^2} \right)}}{{\left( {M_{11}^2 + M_{12}^2} \right)}}}\\
{ = 1 - \delta M_{12}^0{2}\frac{{{n_1}{\psi _l}^\prime \left( {{q_0}a} \right){\psi _l}\left( {{q_1}a} \right) - {\psi _l}\left( {{q_0}a} \right){\psi _l}^\prime \left( {{q_1}a} \right)}}{{{n_1}{\xi _l}^\prime \left( {{q_0}a} \right){\psi _l}\left( {{q_1}a} \right) - {\xi _l}\left( {{q_0}a} \right){\psi _l}^\prime \left( {{q_1}a} \right)}},}
\end{array}
\label{VD6.6}
\end{equation}
where
\begin{equation}
\delta M_{12}^0 =  - \delta {\alpha _g}{q_0}\frac{i}{2}\left[ {{{\left[ {{\xi _l}^\prime \left( {{q_0}b} \right)} \right]}^2} + \frac{{l\left( {l + 1} \right){{\left[ {{\xi _l}\left( {{q_0}b} \right)} \right]}^2}}}{{{{\left( {{q_0}b} \right)}^2}}}} \right].
\label{VD6.7}
\end{equation}
To obtain the mode condition function in Eq.\,(\ref{VD6.6}) we have divided the function (the left hand side of Eq.\,(\ref{VD6.5})) both with the corresponding function for the sphere alone, $M_{11}^2 + M_{12}^2$, and for the spherical shell alone, $1 + \delta M_{11}^0 + \delta M_{12}^0$. Note that the final expression does not contain any elements of matrix ${{{\bf{\tilde M}}}_1}$ and just one of $\delta {{{\bf{\tilde M}}}_0}$.

Now, we proceed with the TE modes. The elements of the first matrix is
\begin{equation}
\begin{array}{*{20}{l}}
{M_{11}^0 = \frac{i}{2}\left\{ { - {\xi _l}'\left[ {{q_0}\left( {b + \delta } \right)} \right]{\zeta _l}\left[ {{q_g}\left( {b + \delta } \right)} \right]} \right.}\\
{\quad \quad \left. { + {n_g}{\xi _l}\left[ {{q_0}\left( {b + \delta } \right)} \right]{\zeta _l}'\left[ {{q_g}\left( {b + \delta } \right)} \right]} \right\};}\\
{M_{12}^0 = \frac{i}{2}\left\{ { - {\xi _l}'\left[ {{q_0}\left( {b + \delta } \right)} \right]{\xi _l}\left[ {{q_g}\left( {b + \delta } \right)} \right]} \right.}\\
{\quad \quad \left. { +{n_g}{\xi _l}\left[ {{q_0}\left( {b + \delta } \right)} \right]{\xi _l}'\left[ {{q_g}\left( {b + \delta } \right)} \right]} \right\},}
\end{array}
\label{VD6.8}
\end{equation}
and of the second
\begin{equation}
\begin{array}{*{20}{l}}
{M_{11}^1 = \frac{i}{{2{n_g}}}\left[ { - {n_g}{\xi _l}'\left( {{q_g}b} \right){\zeta _l}\left( {{q_0}b} \right) + {\xi _l}\left( {{q_g}b} \right){\zeta _l}'\left( {{q_0}b} \right)} \right];}\\
{M_{12}^1 = \frac{i}{{2{n_g}}}\left[ { - {n_g}{\xi _l}'\left( {{q_g}b} \right){\xi _l}\left( {{q_0}b} \right) + {\xi _l}\left( {{q_g}b} \right){\xi _l}'\left( {{q_0}b} \right)} \right];}\\
{M_{21}^1 = \frac{i}{{2{n_g}}}\left[ {{n_g}{\zeta _l}'\left( {{q_g}b} \right){\zeta _l}\left( {{q_0}b} \right) - {\zeta _l}\left( {{q_g}b} \right){\zeta _l}'\left( {{q_0}b} \right)} \right];}\\
{M_{22}^1 = \frac{i}{{2{n_g}}}\left[ {{n_g}{\zeta _l}'\left( {{q_g}b} \right){\xi _l}\left( {{q_0}b} \right) - {\zeta _l}\left( {{q_g}b} \right){\xi _l}'\left( {{q_0}b} \right)} \right],}
\end{array}
\label{VD6.9}
\end{equation}
and of the third
\begin{equation}
\begin{array}{*{20}{l}}
{M_{11}^2 + M_{12}^2}\\
{ = \frac{i}{2}\left[ { - {\xi _l}^\prime \left( {{q_0}a} \right){\zeta _l}\left( {{q_1}a} \right) + {n_1}{\xi _l}\left( {{q_0}a} \right){\zeta _l}^\prime \left( {{q_1}a} \right)} \right.}\\
{\quad \quad \left. { - {\xi _l}^\prime \left( {{q_0}a} \right){\xi _l}\left( {{q_1}a} \right) + {n_1}{\xi _l}\left( {{q_0}a} \right){\xi _l}^\prime \left( {{q_1}a} \right)} \right]}\\
{ = i\left[ { - {\xi _l}^\prime \left( {{q_0}a} \right){\psi _l}\left( {{q_1}a} \right) + {n_1}{\xi _l}\left( {{q_0}a} \right){\psi _l}^\prime \left( {{q_1}a} \right)} \right];}\\
{M_{21}^2 + M_{22}^2}\\
{ = \frac{i}{2}\left[ {{\zeta _l}^\prime \left( {{q_0}a} \right){\zeta _l}\left( {{q_1}a} \right) - {n_1}{\zeta _l}\left( {{q_0}a} \right){\zeta _l}^\prime \left( {{q_1}a} \right)} \right.}\\
{\quad \quad \left. { + {\zeta _l}^\prime \left( {{q_0}a} \right){\xi _l}\left( {{q_1}a} \right) - {n_1}{\zeta _l}\left( {{q_0}a} \right){\xi _l}^\prime \left( {{q_1}a} \right)} \right]}\\
{ = i\left[ {{\zeta _l}^\prime \left( {{q_0}a} \right){\psi _l}\left( {{q_1}a} \right) - {n_1}{\zeta _l}\left( {{q_0}a} \right){\psi _l}^\prime \left( {{q_1}a} \right)} \right],}
\end{array}
\label{VD6.10}
\end{equation}
where ${n_g}$ and ${n_1}$ are the refractive indices of the gas layer and the sphere, respectively.

We now make a series expansion of the first matrix  up to linear order in $\delta$. The other matrices do not depend on $\delta$. The zeroth order term multiplied with the second matrix produces the matrix $\left( {\begin{array}{*{20}{c}}1&0\end{array}} \right)$, so it contributes with ${M_{11}^2 + M_{12}^2}$ to the condition for modes. We then expand in ${\alpha _g}$, the polarizability of the gas. There is no zeroth order term in the term linear in $\delta$. The lowest order term is linear in ${\alpha _g}$. This means that we do not need to expand ${{{\bf{\tilde M}}}_1}$ in ${\alpha _g}$. The zeroth order term is just the unit matrix. Thus if we denote the term of the matrix ${{{\bf{\tilde M}}}_0}$ that is linear in both $\delta$ and  ${\alpha _g}$ with $\delta {{{\bf{\tilde M}}}_0}$ the condition for modes can be written as
\begin{equation}
\begin{array}{l}
\left( {M_{11}^2 + M_{12}^2} \right) + \delta M_{11}^0\left( {M_{11}^2 + M_{12}^2} \right)\\
\quad \quad  + \delta M_{12}^0\left( {M_{21}^2 + M_{22}^2} \right) = 0,
\end{array}
\label{VD6.11}
\end{equation}
and the mode condition function is
\begin{equation}
\begin{array}{*{20}{l}}
{\tilde f_{_l}^{{\rm{TE}}}\left( \omega  \right) = 1 + \delta M_{12}^0\frac{{\left( {M_{21}^2 + M_{22}^2} \right) - \left( {M_{11}^2 + M_{12}^2} \right)}}{{\left( {M_{11}^2 + M_{12}^2} \right)}}}\\
{ = 1 - \delta M_{12}^0{2}\frac{{{\psi _l}^\prime \left( {{q_0}a} \right){\psi _l}\left( {{q_1}a} \right) - {n_1}{\psi _l}\left( {{q_0}a} \right){\psi _l}^\prime \left( {{q_1}a} \right)}}{{ - {\xi _l}^\prime \left( {{q_0}a} \right){\psi _l}\left( {{q_1}a} \right) + {n_1}{\xi _l}\left( {{q_0}a} \right){\psi _l}^\prime \left( {{q_1}a} \right)}},}
\end{array}
\label{VD6.12}
\end{equation}
where
\begin{equation}
\delta M_{12}^0 =  - \delta {\alpha _g}{q_0}\frac{i}{2}\left[ {{\xi _l}\left( {{q_0}b} \right){\xi _l}\left( {{q_0}b} \right)} \right].
\label{VD6.13}
\end{equation}
To obtain the mode condition function in Eq.\,(\ref{VD6.12}) we have divided the function (the left hand side of Eq.\,(\ref{VD6.11})) both with the corresponding function for the sphere alone, $M_{11}^2 + M_{12}^2$, and for the spherical shell alone, $1 + \delta M_{11}^0 + \delta M_{12}^0$.%

The interaction energy per atom is
\begin{equation}
\begin{array}{l}
\frac{E(b)}{{4\pi {b^2}\delta {n_g}}}\\
= \frac{\hbar }{{4\pi {b^2}\delta {n_g}}}\int\limits_0^\infty  {\frac{{d\xi }}{{2\pi }}} \sum\limits_{l = 0}^\infty  {\left( {2l + 1} \right)\ln \left[ {\tilde f_l^{{\rm{TM}}}\left( {i\xi } \right)\tilde f_l^{{\rm{TE}}}\left( {i\xi } \right)} \right]} \\
 \approx \frac{\hbar }{{4\pi {b^2}\delta {n_g}}}\int\limits_0^\infty  {\frac{{d\xi }}{{2\pi }}} \sum\limits_{l = 0}^\infty  {\left( {2l + 1} \right)\left\{ {\left[ {\tilde f_l^{{\rm{TM}}}\left( {i\xi } \right) - 1} \right]} \right.} \\
\left. {\quad \quad \quad \quad \quad \quad \quad \quad \quad \quad \quad  + \left[ {\tilde f_l^{{\rm{TE}}}\left( {i\xi } \right) - 1} \right]}  \right\}\\
 = \frac{\hbar }{{4\pi {b^2}\delta {n_g}}}\int\limits_0^\infty  {\frac{{d\xi }}{{2\pi }}} \sum\limits_{l = 0}^\infty  {\left( {2l + 1} \right)4\pi {n_g}{\alpha ^{at}}\left( {i\xi } \right)\delta \frac{{i\left( {i\xi b/c} \right)}}{{2b}}} \\
 \times \left\{ { - \left[ {{{\left[ {{\xi _l}^\prime \left( {i\xi b/c} \right)} \right]}^2} + \frac{{l\left( {l + 1} \right){{\left[ {{\xi _l}\left( {i\xi b/c} \right)} \right]}^2}}}{{{{\left( {i\xi b/c} \right)}^2}}}} \right]} \right.\\
 \times 2\frac{{{n_1}{\psi _l}^\prime \left( {i\xi a/c} \right){\psi _l}\left( {i\xi {n_1}a/c} \right) - {\psi _l}\left( {i\xi a/c} \right){\psi _l}^\prime \left( {i\xi {n_1}a/c} \right)}}{{{n_1}{\xi _l}^\prime \left( {i\xi a/c} \right){\psi _l}\left( {i\xi {n_1}a/c} \right) - {\xi _l}\left( {i\xi a/c} \right){\psi _l}^\prime \left( {i\xi {n_1}a/c} \right)}}\\
 + 2{\left[ {{\xi _l}\left( {i\xi b/c} \right)} \right]^2}\\
\left. { \times \frac{{{\psi _l}^\prime \left( {i\xi a/c} \right){\psi _l}\left( {i\xi {n_1}a/c} \right) - {n_1}{\psi _l}\left( {i\xi a/c} \right){\psi _l}^\prime \left( {i\xi {n_1}a/c} \right)}}{{ - {\xi _l}^\prime \left( {i\xi a/c} \right){\psi _l}\left( {i\xi {n_1}a/c} \right) + {n_1}{\xi _l}\left( {i\xi a/c} \right){\psi _l}^\prime \left( {i\xi {n_1}a/c} \right)}}} \right\}\\
 = \frac{\hbar }{{{b^2}}}\int\limits_0^\infty  {\frac{{d\xi }}{{2\pi }}} \sum\limits_{l = 0}^\infty  {\left( {2l + 1} \right){\alpha ^{at}}\left( {i\xi } \right)\frac{{i\left( {i\xi b/c} \right)}}{b}} \\
 \times \left\{ { - \left[ {{{\left[ {{\xi _l}^\prime \left( {i\xi b/c} \right)} \right]}^2} + \frac{{l\left( {l + 1} \right){{\left[ {{\xi _l}\left( {i\xi b/c} \right)} \right]}^2}}}{{{{\left( {i\xi b/c} \right)}^2}}}} \right]} \right.\\
 \times \frac{{{n_1}{\psi _l}^\prime \left( {i\xi a/c} \right){\psi _l}\left( {i\xi {n_1}a/c} \right) - {\psi _l}\left( {i\xi a/c} \right){\psi _l}^\prime \left( {i\xi {n_1}a/c} \right)}}{{{n_1}{\xi _l}^\prime \left( {i\xi a/c} \right){\psi _l}\left( {i\xi {n_1}a/c} \right) - {\xi _l}\left( {i\xi a/c} \right){\psi _l}^\prime \left( {i\xi {n_1}a/c} \right)}}\\
 + {\left[ {{\xi _l}\left( {i\xi b/c} \right)} \right]^2}\\
\left. { \times \frac{{{\psi _l}^\prime \left( {i\xi a/c} \right){\psi _l}\left( {i\xi {n_1}a/c} \right) - {n_1}{\psi _l}\left( {i\xi a/c} \right){\psi _l}^\prime \left( {i\xi {n_1}a/c} \right)}}{{ - {\xi _l}^\prime \left( {i\xi a/c} \right){\psi _l}\left( {i\xi {n_1}a/c} \right) + {n_1}{\xi _l}\left( {i\xi a/c} \right){\psi _l}^\prime \left( {i\xi {n_1}a/c} \right)}}} \right\},
\end{array}
\label{VD6.14}
\end{equation}
where now $n_g$ is the density of gas atoms in the gas shell.
The force on the atom is ${\bf{F}}\left( b \right) =  - {\bf{\hat r}}{\rm d}E(b)/{\rm d}b$.
\begin{figure}
\includegraphics[width=8cm]{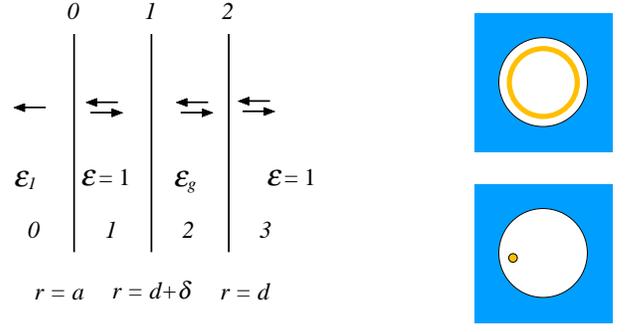}
\caption{(Color online)The geometry of a thin gas layer at radius $d$ inside a spherical or cylindrical cavity of radius $a$ in the fully retarded treatment.}
\label{figu19}
\end{figure}

\subsubsection{\label{atomsphericalcavity}Force on an atom in a spherical cavity (two layers)}
We let the atom be at the distance $d$ from the center of the spherical cavity, of radius $a$.
We start from the two layer structure in Fig.\,\ref{figu19}. We let the medium surrounding the cavity have dielectric function ${{\tilde \varepsilon }_1}\left( \omega  \right)$. The first layer is a vacuum layer. The second is a thin layer, of thickness $\delta$, of a diluted gas of atoms of the kind we consider. Its dielectric function is ${\varepsilon _g}\left( \omega  \right) = 1 + 4\pi N{\alpha ^{at}}\left( \omega  \right)$, where $\alpha ^{at}$ is the polarizability of one atom. The density of gas atoms, $N$, is very low. We let the first interface be at $r=a$ and hence the second at $r={{d + \delta }  }$ and the third at  $r=d$. In what follows we only keep lowest order terms in $\delta$ and in $N$.

Just as in Sec.\ref{atomsphere} the matrix becomes ${\bf{\tilde M}} = {{\bf{\tilde M}}_0} \cdot {{\bf{\tilde M}}_1} \cdot {{\bf{\tilde M}}_2}$ and the left hand side of the condition for modes is given by Eq.\,(\ref{VD6.1}). In this section  ${q_0} =  \omega /c$, ${q_1} = \sqrt {{{\tilde \varepsilon }_1}\left( \omega  \right)} \omega /c$ and ${q_g} = \sqrt {{\varepsilon _g}\left( \omega  \right)} \omega /c$. %


We now list all elements needed in Eq.\,(\ref{VD6.1}). We begin with the matrices for TM modes. The elements of the first matrix is
\begin{equation}
\begin{array}{*{20}{l}}
{M_{11}^0 = \frac{i}{{2{{\tilde \varepsilon }_1}}}\left\{ {{n_1}{\xi _l}\left( {{q_1}a} \right){\zeta _l}^\prime \left( {{q_0}a} \right)} \right.}\\
{\quad \quad \left. { - {\xi _l}^\prime \left( {{q_1}a} \right){\zeta _l}\left( {{q_0}a} \right)} \right\};}\\
{M_{12}^0 = \frac{i}{{2{{\tilde \varepsilon }_1}}}\left\{ {{n_1}{\xi _l}\left( {{q_1}a} \right){\xi _l}^\prime \left( {{q_0}a} \right)} \right.}\\
{\quad \quad \left. {\left. { - {\xi _l}^\prime \left( {{q_1}a} \right){\xi _l}\left( {{q_0}a} \right)} \right\}} \right\},}
\end{array}
\label{VD7.1}
\end{equation}
and of the second

\begin{equation}
\begin{array}{*{20}{l}}
\begin{array}{l}
M_{11}^1 = \frac{{i{n_g}}}{2}\left\{ {{\xi _l}\left[ {{q_0}\left( {d + \delta } \right)} \right]{\zeta _l}^\prime \left[ {{q_g}\left( {d + \delta } \right)} \right]} \right.\\
\left. {\quad \quad  - {n_g}{\xi _l}^\prime \left[ {{q_0}\left( {d + \delta } \right)} \right]{\zeta _l}\left[ {{q_g}\left( {d + \delta } \right)} \right]} \right\};
\end{array}\\
\begin{array}{l}
M_{12}^1 = \frac{{i{n_g}}}{2}\left\{ {{\xi _l}\left[ {{q_0}\left( {d + \delta } \right)} \right]{\xi _l}^\prime \left[ {{q_g}\left( {d + \delta } \right)} \right]} \right.\\
\quad \quad \left. { - {n_g}{\xi _l}^\prime \left[ {{q_0}\left( {d + \delta } \right)} \right]{\xi _l}\left[ {{q_g}\left( {d + \delta } \right)} \right]} \right\};
\end{array}\\
\begin{array}{l}
M_{21}^1 = \frac{{i{n_g}}}{2}\left\{ {{n_g}{\zeta _l}^\prime \left[ {{q_0}\left( {d + \delta } \right)} \right]{\zeta _l}\left[ {{q_g}\left( {d + \delta } \right)} \right]} \right.\\
\quad \quad \left. { - {\zeta _l}\left[ {{q_0}\left( {d + \delta } \right)} \right]{\zeta _l}^\prime \left[ {{q_g}\left( {d + \delta } \right)} \right]} \right\};
\end{array}\\
\begin{array}{l}
M_{22}^1 = \frac{{i{n_g}}}{2}\left\{ {{n_g}{\zeta _l}^\prime \left[ {{q_0}\left( {d + \delta } \right)} \right]{\xi _l}\left[ {{q_g}\left( {d + \delta } \right)} \right]} \right.\\
\quad \quad \left. { - {\zeta _l}\left[ {{q_0}\left( {d + \delta } \right)} \right]{\xi _l}^\prime \left[ {{q_g}\left( {d + \delta } \right)} \right]} \right\},
\end{array}
\end{array}
\label{VD7.2}
\end{equation}
and of the third
\begin{equation}
\begin{array}{*{20}{l}}
{M_{11}^2 + M_{12}^2}\\
{ = \frac{i}{{2{\varepsilon _g}}}\left[ {{n_g}{\xi _l}\left( {{q_g}d} \right){\zeta _l}^\prime \left( {{q_0}d} \right) - {\xi _l}^\prime \left( {{q_g}d} \right){\zeta _l}\left( {{q_0}d} \right)} \right.}\\
{\quad \quad \left. { + {n_g}{\xi _l}\left( {{q_g}d} \right){\xi _l}^\prime \left( {{q_0}d} \right) - {\xi _l}^\prime \left( {{q_g}d} \right){\xi _l}\left( {{q_0}d} \right)} \right]}\\
{ = \frac{i}{{{\varepsilon _g}}}\left[ {{n_g}{\xi _l}\left( {{q_g}d} \right){\psi _l}^\prime \left( {{q_0}d} \right) - {\xi _l}^\prime \left( {{q_g}d} \right){\psi _l}\left( {{q_0}d} \right)} \right];}\\
{M_{21}^2 + M_{22}^2}\\
{ = \frac{i}{{2{\varepsilon _g}}}\left[ {{\zeta _l}^\prime \left( {{q_g}d} \right){\zeta _l}\left( {{q_0}d} \right) - {n_g}{\zeta _l}\left( {{q_g}d} \right){\zeta _l}^\prime \left( {{q_0}d} \right)} \right.}\\
{\quad \quad \left. { + {\zeta _l}^\prime \left( {{q_g}d} \right){\xi _l}\left( {{q_0}d} \right) - {n_g}{\zeta _l}\left( {{q_g}d} \right){\xi _l}^\prime \left( {{q_0}d} \right)} \right]}\\
{ = \frac{i}{{{\varepsilon _g}}}\left[ {{\zeta _l}^\prime \left( {{q_g}d} \right){\psi _l}\left( {{q_0}d} \right) - {n_g}{\zeta _l}\left( {{q_g}d} \right){\psi _l}^\prime \left( {{q_0}d} \right)} \right],}
\end{array}
\label{VD7.3}
\end{equation}
where ${n_g}$ and ${n_1}$ are the refractive indices of the gas layer and the surrounding medium, respectively.

We now make a series expansion of the second matrix,  Eq.\,(\ref{VD7.2}), up to linear order in $\delta$. The other matrices do not depend on $\delta$. The zeroth order term multiplied with the third matrix produces the matrix $\left( {\begin{array}{*{20}{c}}
1\\
1
\end{array}} \right)$, so it contributes with ${M_{11}^0 + M_{12}^0}$ to the condition for modes. We then expand in ${\alpha _g}$, the polarizability of the gas. There is no zeroth order term in the term linear in $\delta$. The lowest order term is linear in ${\alpha _g}$. This means that we do not need to expand the third matrix in ${\alpha _g}$. Thus if we denote the term of the matrix ${{{\bf{\tilde M}}}_1}$ that is linear in both $\delta$ and  ${\alpha _g}$ with $\delta {{{\bf{\tilde M}}}_1}$ the condition for modes can be written as
\begin{equation}
\begin{array}{*{20}{l}}
{\left( {M_{11}^0 + M_{12}^0} \right) + M_{11}^0\left( {\delta M_{11}^1 + \delta M_{12}^1} \right)}\\
{\quad \quad  + M_{12}^0\left( {\delta M_{21}^1 + \delta M_{22}^1} \right) = 0}.
\end{array}
\label{VD7.4}
\end{equation}

To get mode condition function we first rewrite this as

\begin{equation}
\begin{array}{*{20}{l}}
{\left( {M_{11}^0 + M_{12}^0} \right) + \left( {M_{11}^0 + M_{12}^0} \right)\left( {\delta M_{11}^1 + \delta M_{12}^1} \right)}\\
{ + M_{12}^0\left[ {\left( {\delta M_{21}^1 + \delta M_{22}^1} \right) - \left( {\delta M_{11}^1 + \delta M_{12}^1} \right)} \right] = 0,}
\end{array}
\label{VD7.5}
\end{equation}
and the proper mode condition function becomes
\begin{equation}
\begin{array}{l}
\tilde f_l^{{\rm{TM}}}\left( \omega  \right) = 1 + M_{12}^0\frac{{\left( {\delta M_{21}^1 + \delta M_{22}^1} \right) - \left( {\delta M_{11}^1 + \delta M_{12}^1} \right)}}{{\left( {M_{11}^0 + M_{12}^0} \right)}}\\
 = 1 + 4\pi N{\alpha ^{at}}i\delta {q_0}\left\{ {{{\left[ {{\psi _l}'\left( {{q_0}d} \right)} \right]}^2} + \frac{{l\left( {l + 1} \right)}}{{{{\left( {{q_0}d} \right)}^2}}}{{\left[ {{\psi _l}\left( {{q_0}d} \right)} \right]}^2}} \right\}\\
\quad \quad  \times \frac{{{n_1}{\xi _l}\left( {{q_1}a} \right){\xi _l}^\prime \left( {{q_0}a} \right) - {\xi _l}^\prime \left( {{q_1}a} \right){\xi _l}\left( {{q_0}a} \right)}}{{{n_1}{\xi _l}\left( {{q_1}a} \right){\psi _l}^\prime \left( {{q_0}a} \right) - {\xi _l}^\prime \left( {{q_1}a} \right){\psi _l}\left( {{q_0}a} \right)}}.
\end{array}
\label{VD7.6}
\end{equation}

To obtain the mode condition function in Eq.\,(\ref{VD7.6}) we have divided the function (the left hand side of Eq.\,(\ref{VD7.5})) both with the corresponding function for the cavity alone, $M_{11}^0 + M_{12}^0$, and for the spherical shell alone, $1 + \delta M_{11}^1 + \delta M_{12}^1$.

Now, we proceed with the TE modes. The elements of the first matrix is
\begin{equation}
\begin{array}{*{20}{l}}
{M_{11}^0 = \frac{i}{{2{n_1}}}\left\{ { - {n_1}{\xi _l}'\left( {{q_1}a} \right){\zeta _l}\left( {{q_0}a} \right)} \right.}\\
{\quad \quad \left. { + {\xi _l}\left( {{q_1}a} \right){\zeta _l}'\left( {{q_0}a} \right)} \right\};}\\
{M_{12}^0 = \frac{i}{{2{n_1}}}\left\{ { - {n_1}{\xi _l}'\left( {{q_1}a} \right){\xi _l}\left( {{q_0}a} \right)} \right.}\\
{\quad \quad \left. {\left. { + {\xi _l}\left( {{q_1}a} \right){\xi _l}'\left( {{q_0}a} \right)} \right\}} \right\},}
\end{array}
\label{VD7.7}
\end{equation}
and of the second
\begin{equation}
\begin{array}{*{20}{l}}
{\begin{array}{*{20}{l}}
{M_{11}^1 = \frac{i}{2}\left\{ { - {\xi _l}'\left[ {{q_0}\left( {d + \delta } \right)} \right]{\zeta _l}\left[ {{q_g}\left( {d + \delta } \right)} \right]} \right.}\\
{\left. {\quad \quad  + {n_g}{\xi _l}\left[ {{q_0}\left( {d + \delta } \right)} \right]{\zeta _l}'\left[ {{q_g}\left( {d + \delta } \right)} \right]} \right\};}
\end{array}}\\
{\begin{array}{*{20}{l}}
{M_{12}^1 = \frac{i}{2}\left\{ { - {\xi _l}'\left[ {{q_0}\left( {d + \delta } \right)} \right]{\xi _l}\left[ {{q_g}\left( {d + \delta } \right)} \right]} \right.}\\
{\quad \quad \left. { + {n_g}{\xi _l}\left[ {{q_0}\left( {d + \delta } \right)} \right]{\xi _l}'\left[ {{q_g}\left( {d + \delta } \right)} \right]} \right\};}
\end{array}}\\
{\begin{array}{*{20}{l}}
{M_{21}^1 = \frac{i}{2}\left\{ { - {n_g}{\zeta _l}\left[ {{q_0}\left( {d + \delta } \right)} \right]{\zeta _l}'\left[ {{q_g}\left( {d + \delta } \right)} \right]} \right.}\\
{\quad \quad \left. { + {\zeta _l}'\left[ {{q_0}\left( {d + \delta } \right)} \right]{\zeta _l}\left[ {{q_g}\left( {d + \delta } \right)} \right]} \right\};}
\end{array}}\\
{\begin{array}{*{20}{l}}
{M_{22}^1 = \frac{i}{2}\left\{ { - {n_g}{\zeta _l}\left[ {{q_0}\left( {d + \delta } \right)} \right]{\xi _l}'\left[ {{q_g}\left( {d + \delta } \right)} \right]} \right.}\\
{\quad \quad \left. { + {\zeta _l}'\left[ {{q_0}\left( {d + \delta } \right)} \right]{\xi _l}\left[ {{q_g}\left( {d + \delta } \right)} \right]} \right\},}
\end{array}}
\end{array}
\label{VD7.8}
\end{equation}
and of the third
\begin{equation}
\begin{array}{*{20}{l}}
{M_{11}^2 + M_{12}^2}\\
{ = \frac{i}{{2{n_g}}}\left[ { - {n_g}{\xi _l}'\left( {{q_g}d} \right){\zeta _l}\left( {{q_0}d} \right) + {\xi _l}\left( {{q_g}d} \right){\zeta _l}'\left( {{q_0}d} \right)} \right.}\\
{\quad \quad \left. { - {n_g}{\xi _l}'\left( {{q_g}d} \right){\xi _l}\left( {{q_0}d} \right) + {\xi _l}\left( {{q_g}d} \right){\xi _l}'\left( {{q_0}d} \right)} \right]}\\
{ = \frac{i}{{{n_g}}}\left[ { - {n_g}{\xi _l}'\left( {{q_g}d} \right){\psi _l}\left( {{q_0}d} \right) + {\xi _l}\left( {{q_g}d} \right){\psi _l}'\left( {{q_0}d} \right)} \right];}\\
{M_{21}^2 + M_{22}^2}\\
{ = \frac{i}{{2{n_g}}}\left[ { - {\zeta _l}\left( {{q_g}d} \right){\zeta _l}'\left( {{q_0}d} \right) + {n_g}{\zeta _l}'\left( {{q_g}d} \right){\zeta _l}\left( {{q_0}d} \right)} \right.}\\
{\quad \quad \left. { - {\zeta _l}\left( {{q_g}d} \right){\xi _l}'\left( {{q_0}d} \right) + {n_g}{\zeta _l}'\left( {{q_g}d} \right){\xi _l}\left( {{q_0}d} \right)} \right]}\\
{ = \frac{i}{{{n_g}}}\left[ { - {\zeta _l}\left( {{q_g}d} \right){\psi _l}'\left( {{q_0}d} \right) + {n_g}{\zeta _l}'\left( {{q_g}d} \right){\psi _l}\left( {{q_0}d} \right)} \right],}
\end{array}
\label{VD7.9}
\end{equation}
where ${n_g}$ and ${n_1}$ are the refractive indices of the gas layer and the surrounding medium, respectively.

We now make a series expansion of the second matrix,  Eq.\,(\ref{VD7.8}), up to linear order in $\delta$. The other matrices do not depend on $\delta$. The zeroth order term multiplied with the third matrix produces the matrix $\left( {\begin{array}{*{20}{c}}
1\\
1
\end{array}} \right)$, so it contributes with ${M_{11}^0 + M_{12}^0}$ to the condition for modes. We then expand in ${\alpha _g}$, the polarizability of the gas. There is no zeroth order term in the term linear in $\delta$. The lowest order term is linear in ${\alpha _g}$. This means that we do not need to expand the third matrix in ${\alpha _g}$. Thus if we denote the term of the matrix ${{{\bf{\tilde M}}}_1}$ that is linear in both $\delta$ and  ${\alpha _g}$ with $\delta {{{\bf{\tilde M}}}_1}$ the condition for modes can be written as
\begin{equation}
\begin{array}{*{20}{l}}
{\left( {M_{11}^0 + M_{12}^0} \right) + M_{11}^0\left( {\delta M_{11}^1 + \delta M_{12}^1} \right)}\\
{\quad \quad  + M_{12}^0\left( {\delta M_{21}^1 + \delta M_{22}^1} \right) = 0}.
\end{array}
\label{VD7.10}
\end{equation}

To get mode condition function we first rewrite this as
\begin{equation}
\begin{array}{*{20}{l}}
{\left( {M_{11}^0 + M_{12}^0} \right) + \left( {M_{11}^0 + M_{12}^0} \right)\left( {\delta M_{11}^1 + \delta M_{12}^1} \right)}\\
{ + M_{12}^0\left[ {\left( {\delta M_{21}^1 + \delta M_{22}^1} \right) - \left( {\delta M_{11}^1 + \delta M_{12}^1} \right)} \right] = 0,}
\end{array}
\label{VD7.11}
\end{equation}
and the proper mode condition function becomes
\begin{equation}
\begin{array}{*{20}{l}}
{\tilde f_l^{{\rm{TE}}}\left( \omega  \right) = 1 + M_{12}^0\frac{{\left( {\delta M_{21}^1 + \delta M_{22}^1} \right) - \left( {\delta M_{11}^1 + \delta M_{12}^1} \right)}}{{\left( {M_{11}^0 + M_{12}^0} \right)}}}\\
{ = 1 + 4\pi N{\alpha ^{at}}i\delta {q_0}{{\left[ {{\psi _l}\left[ {{q_0}d} \right]} \right]}^2}}\\
{\quad \quad  \times \frac{{\left[ { - {n_1}{\xi _l}'\left( {{q_1}a} \right){\xi _l}\left( {{q_0}a} \right) + {\xi _l}\left( {{q_1}a} \right){\xi _l}'\left( {{q_0}a} \right)} \right]}}{{\left[ { - {n_1}{\xi _l}'\left( {{q_1}a} \right){\psi _l}\left( {{q_0}a} \right) + {\xi _l}\left( {{q_1}a} \right){\psi _l}'\left( {{q_0}a} \right)} \right]}}.}
\end{array}
\label{VD7.12}
\end{equation}

To obtain the mode condition function in Eq.\,(\ref{VD7.12}) we have divided the function (the left hand side of Eq.\,(\ref{VD7.11})) both with the corresponding function for the cavity alone, $M_{11}^0 + M_{12}^0$, and for the spherical shell alone, $1 + \delta M_{11}^1 + \delta M_{12}^1$.

The interaction energy per atom is
\begin{equation}
\begin{array}{l}
\frac{E(d)}{{4\pi N{d^2}\delta }}\\
 = \frac{\hbar }{{4\pi N{d^2}\delta }}\int\limits_0^\infty  {\frac{{d\xi }}{{2\pi }}} \sum\limits_{l = 0}^\infty  {\left( {2l + 1} \right)\ln \left[ {\tilde f_l^{{\rm{TM}}}\left( {i\xi } \right)\tilde f_l^{{\rm{TE}}}\left( {i\xi } \right)} \right]} \\
 =  - \frac{\hbar }{{{d^2}}}\int\limits_0^\infty  {\frac{{d\xi }}{{2\pi }}} \sum\limits_{l = 0}^\infty  {\left( {2l + 1} \right){\alpha ^{at}}} \frac{\xi }{c}\\
 \times \left[ {\left\{ {{{\left[ {{\psi _l}'\left( {\frac{{i\xi d}}{c}} \right)} \right]}^2} + \frac{{l\left( {l + 1} \right)}}{{{{\left( {\frac{{i\xi d}}{c}} \right)}^2}}}{{\left[ {{\psi _l}\left( {\frac{{i\xi d}}{c}} \right)} \right]}^2}} \right\}} \right.\\
 \times \frac{{{n_1}{\xi _l}\left( {\frac{{i{n_1}\xi a}}{c}} \right){\xi _l}^\prime \left( {\frac{{i\xi a}}{c}} \right) - {\xi _l}^\prime \left( {\frac{{i{n_1}\xi a}}{c}} \right){\xi _l}\left( {\frac{{i\xi a}}{c}} \right)}}{{{n_1}{\xi _l}\left( {\frac{{i{n_1}\xi a}}{c}} \right){\psi _l}^\prime \left( {\frac{{i\xi a}}{c}} \right) - {\xi _l}^\prime \left( {\frac{{i{n_1}\xi a}}{c}} \right){\psi _l}\left( {\frac{{i\xi a}}{c}} \right)}}\\
\left. { + {{\left[ {{\psi _l}\left( {\frac{{i\xi d}}{c}} \right)} \right]}^2}\frac{{{n_1}{\xi _l}'\left( {\frac{{i{n_1}\xi a}}{c}} \right){\xi _l}\left( {\frac{{i\xi a}}{c}} \right) - {\xi _l}\left( {\frac{{i{n_1}\xi a}}{c}} \right){\xi _l}'\left( {\frac{{i\xi a}}{c}} \right)}}{{{n_1}{\xi _l}'\left( {\frac{{i{n_1}\xi a}}{c}} \right){\psi _l}\left( {\frac{{i\xi a}}{c}} \right) - {\xi _l}\left( {\frac{{i{n_1}\xi a}}{c}} \right){\psi _l}'\left( {\frac{{i\xi a}}{c}} \right)}}} \right],
\end{array}
\label{VD7.13}
\end{equation}
where we have let $\delta$, the thickness of the gas layer, go to zero when passing from the second to the third line.
The force on the atom is ${\bf{F}}\left( d \right) =  - {\bf{\hat r}}{\rm d}E(d)/{\rm d}d$.

\subsubsection{\label{Casimir}Casimir Polder interaction between two atoms (two layers)}

Here, we start with the geometry in Fig.\,\ref{figu18}. We let the thin shell consist of a diluted gas of atoms of type 2 with density $N_2$ and the sphere consist of a diluted gas of atoms of type 1 with density $N_1$. We use upper case $N$ for the density here to distinguish the densities from the refractive indices that we denote by lower case $n$. The thickness of the shell and the radius of the sphere we let go toward zero at the end. This means that the interaction energy becomes the sum of the interaction energy between all pairs of atoms of type 1 and 2, all with the separation $d$. To get energy for one atom pair we divide the result by the number of atoms of type 1 and by the number of atoms of type 2. Since we let the thickness of the layer, $\delta$, go toward zero we may expand the logarithm in the integrand and keep the lowest order term, $\ln \left( {1 + x} \right) \approx x$. We are furthermore only interested in the dipole-dipole interactions which means that only the $l = 1$ term is kept in the integrand.  Both the TE and TM contributions have the same structure,

\begin{equation}
E = \hbar \int\limits_0^\infty  {\frac{{d\xi }}{{2\pi }}} \left( {2l + 1} \right)A\left( b \right){N_1}{\left. {\frac{{\partial B\left( a \right)}}{{\partial {N_1}}}} \right|_{{N_1} = 0}},
\label{VD8.1}
\end{equation}
where 
\begin{equation}
A\left( b \right) = \delta M_{12}^0\left( b \right),
\label{VD8.2}
\end{equation}
and
\begin{equation}
B\left( a \right) = \frac{{\left( {M_{21}^2 + M_{22}^2} \right) - \left( {M_{11}^2 + M_{12}^2} \right)}}{{\left( {M_{11}^2 + M_{12}^2} \right)}},
\label{VD8.3}
\end{equation}
respectively.
Now,
\begin{equation}
{N_1}{\left. {\frac{{\partial B\left( a \right)}}{{\partial {N_1}}}} \right|_{{N_1} = 0}} = {N_1}\frac{1}{2}4\pi \alpha _1^{at}{\left. {\frac{{\partial B\left( a \right)}}{{\partial {n_1}}}} \right|_{{n_1} = 1}}.
\label{VD8.4}
\end{equation}
In the contribution for TE modes we have
\begin{equation}
\begin{array}{*{20}{l}}
{{A^{{\rm{TE}}}}\left( b \right) =  - \delta {\alpha _g}\left( {i\xi } \right)\frac{{i\xi }}{c}\frac{i}{2}{{\left[ {{\xi _1}\left( {\frac{{i\xi b}}{c}} \right)} \right]}^2}}\\
{ = \delta {\alpha _g}\left( {i\xi } \right)\frac{\xi }{c}\frac{1}{2}{e^{ - 2\xi b/c}}{{\left[ {\frac{{1 + \xi b/c}}{{\xi b/c}}} \right]}^2},}
\end{array}
\label{VD8.5}
\end{equation}
and
\begin{equation}
\begin{array}{*{20}{l}}
{{B^{{\rm{TE}}}}\left( a \right)}\\
{ = 2\frac{{{\psi _l}^\prime \left( {i\xi a/c} \right){\psi _l}\left( {i{n_1}\xi a/c} \right) - {n_1}{\psi _l}\left( {i\xi a/c} \right){\psi _l}^\prime \left( {i{n_1}\xi a/c} \right)}}{{{\xi _l}^\prime \left( {i\xi a/c} \right){\psi _l}\left( {i{n_1}\xi a/c} \right) - {n_1}{\xi _l}\left( {i\xi a/c} \right){\psi _l}^\prime \left( {i{n_1}\xi a/c} \right)}}.}
\end{array}
\label{VD8.6}
\end{equation}
Now,
\begin{equation}
{\left. {\frac{{\partial {B^{{\rm{TE}}}}\left( a \right)}}{{\partial {n_1}}}} \right|_{{n_1} = 1}} = 0,
\label{VD8.7}
\end{equation}
so there is no TE contribution to the dipole dipole interaction between two polarizable atoms.
In the contribution for TM modes we have
\begin{equation}
\begin{array}{*{20}{l}}
{{A^{{\rm{TM}}}}\left( b \right) =  - \delta 4\pi {N_2}\alpha _2^{at}\left( {i\xi } \right)\frac{{i\xi }}{c}\frac{i}{2}}\\
{ \times \left[ {{{\left[ {{\xi _l}^\prime \left( {i\xi b/c} \right)} \right]}^2} + \frac{{l\left( {l + 1} \right){{\left[ {{\xi _l}\left( {i\xi b/c} \right)} \right]}^2}}}{{{{\left( {i\xi b/c} \right)}^2}}}} \right]}\\
{ =  - \delta 4\pi {N_2}\alpha _2^{at}\left( {i\xi } \right)\frac{\xi }{c}\frac{1}{2}\frac{1}{{{{\left( {\xi b/c} \right)}^4}}}{e^{ - 2\xi b/c}}}\\
{ \times \left[ {3 + 6\left( {\xi b/c} \right) + 5{{\left( {\xi b/c} \right)}^2} + 2{{\left( {\xi b/c} \right)}^3} + {{\left( {\xi b/c} \right)}^4}} \right]},
\end{array}
\label{VD8.8}
\end{equation}
and
\begin{equation}
\begin{array}{*{20}{l}}
{{B^{{\rm{TM}}}}\left( a \right) = }\\
{ - 2\frac{{{n_1}{\psi _l}^\prime \left( {i\xi a/c} \right){\psi _l}\left( {i{n_1}\xi a/c} \right) - {\psi _l}\left( {i\xi a/c} \right){\psi _l}^\prime \left( {i{n_1}\xi a/c} \right)}}{{{n_1}{\xi _l}^\prime \left( {i\xi a/c} \right){\psi _l}\left( {i{n_1}\xi a/c} \right) - {\xi _l}\left( {i\xi a/c} \right){\psi _l}^\prime \left( {i{n_1}\xi a/c} \right)}}.}
\end{array}
\label{VD8.9}
\end{equation}
Now,
\begin{equation}
{\left. {\frac{{\partial {B^{{\rm{TM}}}}\left( a \right)}}{{\partial {n_1}}}} \right|_{{n_1} = 1}} = \frac{8}{9}{\left( {\xi a/c} \right)^3},
\label{VD8.10}
\end{equation}
so the energy per atom pair is
\begin{equation}
\begin{array}{*{20}{l}}
{\frac{E}{{\left( {{N_1}4\pi {a^3}/3} \right)\left( {{N_2}4\pi {d^2}\delta } \right)}}}\\
{ = \frac{\hbar }{{\left( {{N_1}4\pi {a^3}/3} \right)\left( {{N_2}4\pi {d^2}\delta } \right)}}\int\limits_0^\infty  {\frac{{d\xi }}{{2\pi }}} 3{A^{{\rm{TM}}}}\left( b \right){N_1}{{\left. {\frac{{\partial {B^{{\rm{TM}}}}\left( a \right)}}{{\partial {N_1}}}} \right|}_{{N_1} = 0}}}\\
{ =  - \frac{{9\hbar }}{{\left( {{N_1}4\pi {a^3}} \right)\left( {{N_2}4\pi {d^2}\delta } \right)2\pi }}}\\
\begin{array}{l}
 \times \int\limits_0^\infty  {d\xi } \delta 4\pi {N_2}\alpha _2^{at}\left( {i\xi } \right)\frac{\xi }{c}\frac{1}{2}\frac{1}{{{{\left( {\xi d/c} \right)}^4}}}\\
 \times {N_1}\frac{1}{2}4\pi \alpha _1^{at}\left( {i\xi } \right)\frac{8}{9}{\left( {\xi a/c} \right)^3}{e^{ - 2\xi d/c}}
\end{array}\\
{ \times \left[ {3 + 6\left( {\xi d/c} \right) + 5{{\left( {\xi d/c} \right)}^2} + 2{{\left( {\xi d/c} \right)}^3} + {{\left( {\xi d/c} \right)}^4}} \right]}\\
{ =  - \frac{\hbar }{{{d^6}\pi }}\int\limits_0^\infty  {d\xi } \alpha _2^{at}\left( {i\xi } \right)\alpha _1^{at}\left( {i\xi } \right){e^{ - 2\xi d/c}}}\\
{ \times \left[ {3 + 6\left( {\xi d/c} \right) + 5{{\left( {\xi d/c} \right)}^2} + 2{{\left( {\xi d/c} \right)}^3} + {{\left( {\xi d/c} \right)}^4}} \right].}
\end{array}
\label{VD8.11}
\end{equation}
It is interesting to note that we reproduce the Casimir-Polder interaction between two polarizable atoms.\,\cite{Ser6,CasPol}  Thus we have three quite different methods to derive the Casimir-Polder interaction that produce identical results. To be noted is that only the TM modes contribute.

\subsubsection{\label{atomsphericalgapret}Force on an atom in a spherical gap (three layers)}
Let the outer radius be $b$, the inner radius $a$ and the radial position of the atom be $r$. The medium surrounding the vacuum gap has the dielectric function $\varepsilon \left( \omega  \right)$. This geometry involves four interfaces and in a straightforward approach the final matrix would be the product of four matrices. The matrix elements in this retarded treatment are rather bulky and difficult to put in print. We will use three matrices, where the middle one is that for the thin diluted gas shell, and take advantage of Eq.\,(\ref {VD6.1}). To make the expressions even more compact we make use of the two types of ${2^l}$ pole polarizabilities, introduced in Sec.\,\ref{sphericalretmain}. Thus, the matrix is ${\bf{\tilde M}} = {{\bf{\tilde M}}_0} \cdot {{\bf{\tilde M}}_1} \cdot {{\bf{\tilde M}}_2}$ and
\begin{equation}
\begin{array}{l}
{M_{11}} + {M_{12}} = M_{11}^0\left( {M_{11}^2 + M_{12}^2} \right)\\
 \quad \quad  \quad \quad \quad\quad\times \left( {\begin{array}{*{20}{c}}
1&{\alpha _l^{0\left( 2 \right)}}
\end{array}} \right) \cdot {{{\bf{\tilde M}}}_1} \cdot \left( {\begin{array}{*{20}{c}}
1\\
{ - \alpha _l^2}
\end{array}} \right).
\end{array}
\label{VD9.1}
\end{equation}
Now, let us introduce $\delta {{{\bf{\tilde M}}}_1}$ so that
\begin{equation}
{{{\bf{\tilde M}}}_1} = \left( {\begin{array}{*{20}{c}}
1&0\\
0&1
\end{array}} \right) + \delta {{{\bf{\tilde M}}}_1}.
\label{VD9.2}
\end{equation}
Then 
\begin{equation}
\begin{array}{*{20}{l}}
{{{\tilde f}_{l,m}} = \frac{{M_{11}^1 + \alpha _l^{0\left( 2 \right)}M_{21}^1 - \alpha _l^2\left( {M_{12}^1 + \alpha _l^{0\left( 2 \right)}M_{22}^1} \right)}}{{\left( {1 - \alpha _l^{0\left( 2 \right)}\alpha _l^2} \right)\left( {M_{11}^1 + M_{12}^1} \right)}}}\\
\begin{array}{l}
 \approx 1\\
\quad  - \frac{{\delta M_{12}^1\left[ {1 + \alpha _l^2\left( {1 - \alpha _l^{0\left( 2 \right)}} \right)} \right] - \alpha _l^{0\left( 2 \right)}\delta M_{21}^1 + \alpha _l^{0\left( 2 \right)}\alpha _l^2\left( {\delta M_{22}^1 - \delta M_{11}^1} \right)}}{{\left( {1 - \alpha _l^{0\left( 2 \right)}\alpha _l^2} \right)}},
\end{array}
\end{array}
\label{VD9.3}
\end{equation}
where we have kept terms up to linear order in the atom density. We have chosen as reference system a system with the spherical gap and the gas shell well separated from each other. Thus we have divided our mode condition function both with that for a free gas film and that for the spherical gap.

For TM modes we have
\begin{equation}
\begin{array}{l}
\delta {\bf{\tilde M}}_{\rm{1}}^{{\rm{TM}}} =  - \left( {\delta N} \right)2\pi {\alpha ^{at}}{q_0}i\\
\quad \quad  \times \left( {\begin{array}{*{20}{c}}
{{\xi _l}^\prime {\zeta _l}^\prime  + {\xi _l}{\zeta _l}\frac{{l\left( {l + 1} \right)}}{{{{\left( {{q_0}r} \right)}^2}}}}&{{{\left[ {{\xi _l}^\prime } \right]}^2} + {{\left[ {{\xi _l}} \right]}^2}\frac{{l\left( {l + 1} \right)}}{{{{\left( {{q_0}r} \right)}^2}}}}\\
{ - {{\left[ {{\zeta _l}^\prime } \right]}^2} - {{\left[ {{\zeta _l}} \right]}^2}\frac{{l\left( {l + 1} \right)}}{{{{\left( {{q_0}r} \right)}^2}}}}&{ - {\xi _l}^\prime {\zeta _l}^\prime  - {\xi _l}{\zeta _l}\frac{{l\left( {l + 1} \right)}}{{{{\left( {{q_0}r} \right)}^2}}}}
\end{array}} \right);\\
\alpha _l^{2,{\rm{TM}}} = \frac{{{\zeta _l}\left( {{q_0}a} \right){\psi _l}^\prime \left( {{q_m}a} \right) - n{\zeta _l}^\prime \left( {{q_0}a} \right){\psi _l}\left( {{q_m}a} \right)}}{{{\xi _l}\left( {{q_0}a} \right){\psi _l}^\prime \left( {{q_m}a} \right) - n{\xi _l}^\prime \left( {{q_0}a} \right){\psi _l}\left( {{q_m}a} \right)}};\\
\alpha _l^{0\left( 2 \right),{\rm{TM}}} = \frac{{n{\xi _l}\left( {{q_m}b} \right){\xi _l}^\prime \left( {{q_0}b} \right) - {\xi _l}^\prime \left( {{q_m}b} \right){\xi _l}\left( {{q_0}b} \right)}}{{n{\xi _l}\left( {{q_m}b} \right){\zeta _l}^\prime \left( {{q_0}b} \right) - {\xi _l}^\prime \left( {{q_m}b} \right){\zeta _l}\left( {{q_0}b} \right)}},
\end{array}
\label{VD9.4}
\end{equation}
where ${q_0} = \omega /c$, ${q_m} = n\omega /c$, and $n = \sqrt \varepsilon $. All Ricatti-Bessel functions in the matrix have the argument ${\left( {{q_0}r} \right)}$.
For TE modes the functions are
\begin{equation}
\begin{array}{l}
\delta {\bf{\tilde M}}_{\rm{1}}^{{\rm{TE}}} =  - \left( {\delta N} \right)2\pi {\alpha ^{at}}{q_0}i\\
\quad \quad  \times \left( {\begin{array}{*{20}{c}}
{{\xi _l}\left( {{q_0}r} \right){\zeta _l}\left( {{q_0}r} \right)}&{{{\left[ {{\xi _l}\left( {{q_0}r} \right)} \right]}^2}}\\
{ - {{\left[ {{\zeta _l}\left( {{q_0}r} \right)} \right]}^2}}&{ - {\xi _l}\left( {{q_0}r} \right){\zeta _l}\left( {{q_0}r} \right)}
\end{array}} \right);\\
\alpha _l^{2,{\rm{TE}}} = \frac{{{\zeta _l}^\prime \left( {{q_0}a} \right){\psi _l}\left( {{q_m}a} \right) - n{\zeta _l}\left( {{q_0}a} \right){\psi _l}^\prime \left( {{q_m}a} \right)}}{{{\xi _l}^\prime \left( {{q_0}a} \right){\psi _l}\left( {{q_m}a} \right) - n{\xi _l}\left( {{q_0}a} \right){\psi _l}^\prime \left( {{q_m}a} \right)}};\\
\alpha _l^{0\left( 2 \right),{\rm{TE}}} = \frac{{n{\xi _l}^\prime \left( {{q_m}b} \right){\xi _l}\left( {{q_0}b} \right) - {\xi _l}\left( {{q_m}b} \right){\xi _l}^\prime \left( {{q_0}b} \right)}}{{n{\xi _l}^\prime \left( {{q_m}b} \right){\zeta _l}\left( {{q_0}b} \right) - {\xi _l}\left( {{q_m}b} \right){\zeta _l}^\prime \left( {{q_0}b} \right)}}.
\end{array}
\label{VD9.5}
\end{equation}
Here one may take the opportunity to check the results. If we let ${\alpha _l^{0\left( 2 \right)}}=0$ we regain the results for an atom outside a solid sphere, in Sec.\,\ref{atomsphere}. If we instead let ${\alpha _l^2}=-1$ we regain the results for an atom in a spherical cavity, in Sec.\,\ref{atomsphericalcavity}.
\begin{figure}
\includegraphics[width=8cm]{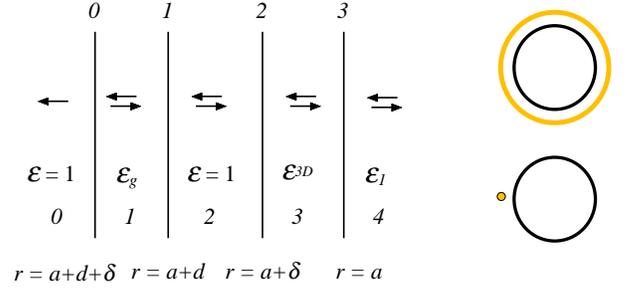}
\caption{(Color online)The geometry of a thin gas layer at distance $d$ from a thin spherical or cylindrical shell of radius $a$ in the fully retarded treatment.}
\label{figu20}
\end{figure}
%
%
%


\subsubsection{\label{atomsphericalshell}Force on an atom outside a 2D spherical shell (three layers)}
We start from the geometry in Fig.\,\ref{figu20}. We use already from the outset the matrices for a gas layer at $r=b=a+d$ and a spherical 2D film at $r=a$. These were given in Eqs.\,(\ref{VD4.1}), (\ref{VD4.2}), (\ref{VD5.1}), and (\ref{VD5.2}). We find
\begin{equation}
\begin{array}{*{20}{l}}
\begin{array}{l}
\tilde f_l^{{\rm{TM}}}\\
 = 1 - \left( {\delta n} \right)4\pi {\alpha ^{at}}{q_0}i\left[ {{{\left[ {{\xi _l}^\prime \left( {{q_0}b} \right)} \right]}^2} + {{\left[ {{\xi _l}\left( {{q_0}b} \right)} \right]}^2}\frac{{l\left( {l + 1} \right)}}{{{{\left( {{q_0}b} \right)}^2}}}} \right]
\end{array}\\
{\quad \quad \quad \quad  \times \frac{{\delta {{\tilde \varepsilon }^{{\rm{3D}}}}{q_0}i{{\left[ {{\psi _l}^\prime \left( {{q_0}a} \right)} \right]}^2}}}{{\left[ {1 - \delta {{\tilde \varepsilon }^{{\rm{3D}}}}{q_0}i{\xi _l}^\prime \left( {{q_0}a} \right){\psi _l}^\prime \left( {{q_0}a} \right)} \right]}}},
\end{array}
\label{VD10.1}
\end{equation}
and
\begin{equation}
\begin{array}{*{20}{l}}
{\tilde f_l^{{\rm{TE}}} = 1 - \left( {\delta n} \right)4\pi {\alpha ^{at}}i{q_0}{{\left[ {{\xi _l}\left( {{q_0}b} \right)} \right]}^2}}\\
{\quad \quad \quad \quad  \times \frac{{\delta {{\tilde \varepsilon }^{{\rm{3D}}}}i{q_0}{{\left[ {{\psi _l}\left( {{q_0}a} \right)} \right]}^2}}}{{1 - \delta {{\tilde \varepsilon }^{{\rm{3D}}}}i{q_0}\left[ {{\xi _l}\left( {{q_0}a} \right){\psi _l}\left( {{q_0}a} \right)} \right]}}.}
\end{array}
\label{VD10.2}
\end{equation}

We have also derived these results in the alternative way followed in the preceding sections.
The interaction energy per atom is
\begin{equation}
\begin{array}{*{20}{l}}
{\frac{E(b)}{{4\pi n{b^2}\delta }}}\\
{ = \frac{\hbar }{{4\pi n{b^2}\delta }}\int\limits_0^\infty  {\frac{{d\xi }}{{2\pi }}} \sum\limits_{l = 0}^\infty  {\left( {2l + 1} \right)\ln \left[ {\tilde f_l^{{\rm{TM}}}\left( {i\xi } \right)\tilde f_l^{{\rm{TE}}}\left( {i\xi } \right)} \right]} }\\
{ =  - \frac{\hbar }{{{b^2}}}\int\limits_0^\infty  {\frac{{d\xi }}{{2\pi }}} \sum\limits_{l = 0}^\infty  {\left( {2l + 1} \right){\alpha ^{at}}\left( {i\xi } \right)} \delta {{\tilde \varepsilon }^{{\rm{3D}}}}\left( {i\xi } \right){{\left( {\frac{\xi }{c}} \right)}^2}}\\
{ \times \left\{ {\left[ {{{\left[ {{\xi _l}^\prime \left( {i\frac{{\xi b}}{c}} \right)} \right]}^2} - {{\left[ {{\xi _l}\left( {i\frac{{\xi b}}{c}} \right)} \right]}^2}\frac{{l\left( {l + 1} \right)}}{{{{\left( {\frac{{\xi b}}{c}} \right)}^2}}}} \right]} \right.}\\
{ \times \frac{{{{\left[ {{\psi _l}^\prime \left( {i\frac{{\xi a}}{c}} \right)} \right]}^2}}}{{\left[ {1 + \delta {{\tilde \varepsilon }^{{\rm{3D}}}}\left( {i\xi } \right)\left( {\frac{\xi }{c}} \right){\xi _l}^\prime \left( {i\frac{{\xi a}}{c}} \right){\psi _l}^\prime \left( {i\frac{{\xi a}}{c}} \right)} \right]}}}\\
{\left. { + \frac{{{{\left[ {{\xi _l}\left( {\frac{{i\xi b}}{c}} \right)} \right]}^2}{{\left[ {{\psi _l}\left( {\frac{{i\xi a}}{c}} \right)} \right]}^2}}}{{1 + \delta {{\tilde \varepsilon }^{{\rm{3D}}}}\left( {i\xi } \right)\left( {\frac{\xi }{c}} \right)\left[ {{\xi _l}\left( {\frac{{i\xi a}}{c}} \right){\psi _l}\left( {\frac{{i\xi a}}{c}} \right)} \right]}}} \right\},}
\end{array}
\label{VD10.3}
\end{equation}
where we have let $\delta$, the thickness of the gas layer, go to zero when passing from the second to the third line.
The force on the atom is ${\bf{F}}\left( b \right) =  - {\bf{\hat r}}{\rm d}E(b)/{\rm d}b$.

\begin{figure}
\includegraphics[width=8cm]{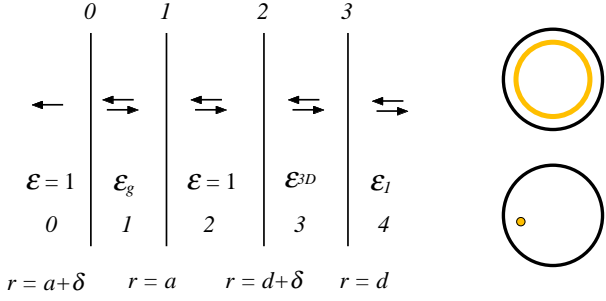}
\caption{(Color online) The geometry of a thin gas layer at radius $d$ inside a thin spherical or cylindrical shell of radius $a$ in the  fully retarded treatment.}
\label{figu21}
\end{figure}
\subsubsection{\label{atomsphericalshellcavity}Force on an atom inside a 2D spherical shell (three layers)}

We start from the geometry in Fig.\,\ref{figu21}. We use already from the outset the matrices for a spherical 2D film at $r=a$ and a gas layer at $r=d$. These were given in Eqs.\,(\ref{VD4.1}), (\ref{VD4.2}), (\ref{VD5.1}), and (\ref{VD5.2}). We find
\begin{equation}
\begin{array}{*{20}{l}}
{\tilde f_l^{{\rm{TM}}} = 1 + M_{12}^0\frac{{\left( {\delta M_{21}^1 + \delta M_{22}^1} \right) - \left( {\delta M_{11}^1 + \delta M_{12}^1} \right)}}{{M_{11}^0 + M_{12}^0}}}\\
{{\rm{ = }}1 - i{q_0}\delta {{\tilde \varepsilon }^{{\rm{3D}}}}\frac{1}{2}{{\left[ {\xi '\left( {{q_0}a} \right)} \right]}^2}}\\
{\quad  \times \frac{{\left( {\delta M_{21}^1 + \delta M_{22}^1} \right) - \left( {\delta M_{11}^1 + \delta M_{12}^1} \right)}}{{1 - i{q_0}\delta {{\tilde \varepsilon }^{{\rm{3D}}}}\psi '\left( {{q_0}a} \right){\xi _l}^\prime \left( {{q_0}a} \right)}}}\\
{ = 1 - \frac{{i{q_0}\delta {{\tilde \varepsilon }^{{\rm{3D}}}}{{\left[ {{\xi _l}^\prime \left( {{q_0}a} \right)} \right]}^2}4\pi N{\alpha ^{at}}\delta i{q_0}}}{{1 - i{q_0}\delta {{\tilde \varepsilon }^{{\rm{3D}}}}\psi '\left( {{q_0}a} \right){\xi _l}^\prime \left( {{q_0}a} \right)}}}\\
{\quad  \times \left\{ {{{\left[ {{\psi _l}^\prime \left( {{q_0}d} \right)} \right]}^2} + \frac{{l\left( {l + 1} \right)}}{{{{\left( {{q_0}d} \right)}^2}}}{{\left[ {{\psi _l}\left( {{q_0}d} \right)} \right]}^2}} \right\}},
\end{array}
\label{VD11.1}
\end{equation}
and
\begin{equation}
\begin{array}{*{20}{l}}
{\tilde f_l^{{\rm{TE}}} = 1 + M_{12}^0\frac{{\left( {\delta M_{21}^1 + \delta M_{22}^1} \right) - \left( {\delta M_{11}^1 + \delta M_{12}^1} \right)}}{{M_{11}^0 + M_{12}^0}}}\\
{ = 1 - i{q_0}\delta {{\tilde \varepsilon }^{{\rm{3D}}}}\frac{1}{2}{{\left[ {{\xi _l}\left( {{q_0}a} \right)} \right]}^2}}\\
{\quad  \times \frac{{\left( {\delta M_{21}^1 + \delta M_{22}^1} \right) - \left( {\delta M_{11}^1 + \delta M_{12}^1} \right)}}{{1 - i{q_0}\delta {{\tilde \varepsilon }^{{\rm{3D}}}}\psi \left( {{q_0}a} \right){\xi _l}\left( {{q_0}a} \right)}}}\\
{ = 1 - \frac{{i{q_0}\delta {{\tilde \varepsilon }^{{\rm{3D}}}}{{\left[ {{\xi _l}\left( {{q_0}a} \right)} \right]}^2}4\pi N{\alpha ^{at}}\delta i{q_0}{{\left[ {{\psi _l}\left( {{q_0}d} \right)} \right]}^2}}}{{1 - i{q_0}\delta {{\tilde \varepsilon }^{{\rm{3D}}}}\psi \left( {{q_0}a} \right){\xi _l}\left( {{q_0}a} \right)}}.}
\end{array}
\label{VD11.2}
\end{equation}

The interaction energy per atom is
\begin{equation}
\begin{array}{*{20}{l}}
{\frac{E(d)}{{4\pi N{d^2}\delta }}}\\
{ = \frac{\hbar }{{4\pi n{d^2}\delta }}\int\limits_0^\infty  {\frac{{d\xi }}{{2\pi }}} \sum\limits_{l = 0}^\infty  {\left( {2l + 1} \right)\ln \left[ {\tilde f_l^{{\rm{TM}}}\left( {i\xi } \right)\tilde f_l^{{\rm{TE}}}\left( {i\xi } \right)} \right]} }\\
{ =  - \frac{\hbar }{{{d^2}}}\int\limits_0^\infty  {\frac{{d\xi }}{{2\pi }}} \sum\limits_{l = 0}^\infty  {\left( {2l + 1} \right){\alpha ^{at}}\left( {i\xi } \right)} \delta {{\tilde \varepsilon }^{{\rm{3D}}}}\left( {i\xi } \right){{\left( {\frac{\xi }{c}} \right)}^2}}\\
{ \times \left\{ {\left[ {{{\left[ {{\psi _l}^\prime \left( {i\frac{{\xi d}}{c}} \right)} \right]}^2} - {{\left[ {{\psi _l}\left( {i\frac{{\xi d}}{c}} \right)} \right]}^2}\frac{{l\left( {l + 1} \right)}}{{{{\left( {\frac{{\xi d}}{c}} \right)}^2}}}} \right]} \right.}\\
{ \times \frac{{{{\left[ {{\xi _l}^\prime \left( {i\frac{{\xi a}}{c}} \right)} \right]}^2}}}{{\left[ {1 + \delta {{\tilde \varepsilon }^{{\rm{3D}}}}\left( {i\xi } \right)\left( {\frac{\xi }{c}} \right){\xi _l}^\prime \left( {i\frac{{\xi a}}{c}} \right){\psi _l}^\prime \left( {i\frac{{\xi a}}{c}} \right)} \right]}}}\\
{\left. { + \frac{{{{\left[ {{\xi _l}\left( {\frac{{i\xi a}}{c}} \right)} \right]}^2}{{\left[ {{\psi _l}\left( {\frac{{i\xi d}}{c}} \right)} \right]}^2}}}{{1 + \delta {{\tilde \varepsilon }^{{\rm{3D}}}}\left( {i\xi } \right)\left( {\frac{\xi }{c}} \right)\left[ {{\xi _l}\left( {\frac{{i\xi a}}{c}} \right){\psi _l}\left( {\frac{{i\xi a}}{c}} \right)} \right]}}} \right\},}
\end{array}
\label{VD11.3}
\end{equation}
where we have let $\delta$, the thickness of the gas layer, go to zero when passing from the second to the third line.
The force on the atom is ${\bf{F}}\left( d \right) =  - {\bf{\hat r}}{\rm d}E(d)/{\rm d}d$.

\subsubsection{\label{twosphericalfilmsret}Interaction between two 2D spherical shells (three layers)}
We consider two concentric thin spherical shells in vacuum. Since the films are in vacuum  ${q_0} = \omega /c$. The outer shell has radius $b$ and the inner radius $a$. Here the matrix is ${\bf{\tilde M}} = {{{\bf{\tilde M}}}_0} \cdot {{{\bf{\tilde M}}}_1}$ where for TM modes
\begin{equation}
\begin{array}{l}
{\bf{\tilde M}}_{\rm{0}}^{{\rm{TM}}} = \left( {\begin{array}{*{20}{c}}
1&0\\
0&1
\end{array}} \right)\\
 - \frac{{\delta {{\tilde \varepsilon }^{{\rm{3D}}}}{q_0}i}}{2}\left( {\begin{array}{*{20}{c}}
{{\xi _l}^\prime \left( {{q_0}b} \right){\zeta _l}^\prime \left( {{q_0}b} \right)}&{{{\left[ {{\xi _l}^\prime \left( {{q_0}b} \right)} \right]}^2}}\\
{ - {{\left[ {{\zeta _l}^\prime \left( {{q_0}b} \right)} \right]}^2}}&{ - {\xi _l}^\prime \left( {{q_0}b} \right){\zeta _l}^\prime \left( {{q_0}b} \right)}
\end{array}} \right);\\
{\bf{\tilde M}}_{\rm{1}}^{{\rm{TM}}} = \left( {\begin{array}{*{20}{c}}
1&0\\
0&1
\end{array}} \right)\\
 - \frac{{\delta {{\tilde \varepsilon }^{{\rm{3D}}}}{q_0}i}}{2}\left( {\begin{array}{*{20}{c}}
{{\xi _l}^\prime \left( {{q_0}a} \right){\zeta _l}^\prime \left( {{q_0}a} \right)}&{{{\left[ {{\xi _l}^\prime \left( {{q_0}a} \right)} \right]}^2}}\\
{ - {{\left[ {{\zeta _l}^\prime \left( {{q_0}a} \right)} \right]}^2}}&{ - {\xi _l}^\prime \left( {{q_0}a} \right){\zeta _l}^\prime \left( {{q_0}a} \right)}
\end{array}} \right),
\end{array}
\label{VD12.1}
\end{equation}
and the condition for modes is
\begin{equation}
\begin{array}{l}
M_{11}^{{\rm{TM}}} + M_{12}^{{\rm{TM}}} = 1 - \frac{{\delta {{\tilde \varepsilon }^{{\rm{3D}}}}{q_0}i}}{2}\left[ {{\xi _l}^\prime \left( {{q_0}b} \right){\zeta _l}^\prime \left( {{q_0}b} \right)} \right.\\
\left. {\quad \quad  + {{\left[ {{\xi _l}^\prime \left( {{q_0}b} \right)} \right]}^2} + {\xi _l}^\prime \left( {{q_0}a} \right){\zeta _l}^\prime \left( {{q_0}a} \right) + {{\left[ {{\xi _l}^\prime \left( {{q_0}a} \right)} \right]}^2}} \right]\\
 + {\left( {\frac{{\delta {{\tilde \varepsilon }^{{\rm{3D}}}}{q_0}i}}{2}} \right)^2}\left[ {{\xi _l}^\prime \left( {{q_0}b} \right){\zeta _l}^\prime \left( {{q_0}b} \right){\xi _l}^\prime \left( {{q_0}a} \right){\zeta _l}^\prime \left( {{q_0}a} \right)} \right.\\
\left. {\quad \quad \quad \quad \quad \quad \quad \quad  - {{\left[ {{\xi _l}^\prime \left( {{q_0}b} \right)} \right]}^2}{{\left[ {{\zeta _l}^\prime \left( {{q_0}a} \right)} \right]}^2}} \right]=0.
\end{array}
\label{VD12.2}
\end{equation}

For TE modes the two matrices are
\begin{equation}
\begin{array}{*{20}{l}}
{{\bf{\tilde M}}_{\rm{0}}^{{\rm{TE}}} = \left( {\begin{array}{*{20}{c}}
1&0\\
0&1
\end{array}} \right)}\\
{ - \frac{{\delta {{\tilde \varepsilon }^{{\rm{3D}}}}{q_0}i}}{2}\left( {\begin{array}{*{20}{c}}
{{\xi _l}\left( {{q_0}b} \right){\zeta _l}\left( {{q_0}b} \right)}&{{{\left[ {{\xi _l}\left( {{q_0}b} \right)} \right]}^2}}\\
{ - {{\left[ {{\zeta _l}\left( {{q_0}b} \right)} \right]}^2}}&{ - {\xi _l}\left( {{q_0}b} \right){\zeta _l}\left( {{q_0}b} \right)}
\end{array}} \right);}\\
{{\bf{\tilde M}}_{\rm{1}}^{{\rm{TE}}} = \left( {\begin{array}{*{20}{c}}
1&0\\
0&1
\end{array}} \right)}\\
{ - \frac{{\delta {{\tilde \varepsilon }^{{\rm{3D}}}}{q_0}i}}{2}\left( {\begin{array}{*{20}{c}}
{{\xi _l}\left( {{q_0}a} \right){\zeta _l}\left( {{q_0}a} \right)}&{{{\left[ {{\xi _l}\left( {{q_0}a} \right)} \right]}^2}}\\
{ - {{\left[ {{\zeta _l}\left( {{q_0}a} \right)} \right]}^2}}&{ - {\xi _l}\left( {{q_0}a} \right){\zeta _l}\left( {{q_0}a} \right)}
\end{array}} \right),}
\end{array}
\label{VD12.3}
\end{equation}
and the condition for modes becomes
\begin{equation}
\begin{array}{l}
M_{11}^{{\rm{TE}}} + M_{12}^{{\rm{TE}}} = 1 - \frac{{\delta {{\tilde \varepsilon }^{{\rm{3D}}}}{q_0}i}}{2}\left[ {{\xi _l}\left( {{q_0}b} \right){\zeta _l}\left( {{q_0}b} \right)} \right.\\
\left. {\quad \quad  + {{\left[ {{\xi _l}\left( {{q_0}b} \right)} \right]}^2} + {\xi _l}\left( {{q_0}a} \right){\zeta _l}\left( {{q_0}a} \right) + {{\left[ {{\xi _l}\left( {{q_0}a} \right)} \right]}^2}} \right]\\
 + {\left( {\frac{{\delta {{\tilde \varepsilon }^{{\rm{3D}}}}{q_0}i}}{2}} \right)^2}\left[ {{\xi _l}\left( {{q_0}b} \right){\zeta _l}\left( {{q_0}b} \right){\xi _l}\left( {{q_0}a} \right){\zeta _l}\left( {{q_0}a} \right)} \right.\\
\left. {\quad \quad \quad \quad \quad \quad \quad \quad  - {{\left[ {{\xi _l}\left( {{q_0}b} \right)} \right]}^2}{{\left[ {{\zeta _l}\left( {{q_0}a} \right)} \right]}^2}} \right]=0.
\end{array}
\label{VD12.4}
\end{equation}
The mode condition function for TM modes is
\begin{equation}
\begin{array}{*{20}{l}}
{\tilde f_{l.m}^{{\rm{TM}}}\left( {i\xi } \right) = 1 - {{\left( {\frac{{\delta {{\tilde \varepsilon }^{{\rm{3D}}}}\left( {i\xi } \right)\xi }}{{2c}}} \right)}^2}{{\left[ {{\xi _l}^\prime \left( {\frac{{i\xi b}}{c}} \right)} \right]}^2}{{\left[ {{\xi _l}^\prime \left( {\frac{{i\xi a}}{c}} \right)} \right]}^2} \times }\\
{\frac{{\left[ {2\frac{{{\psi _l}^\prime \left( {\frac{{i\xi a}}{c}} \right)}}{{{\xi _l}^\prime \left( {\frac{{i\xi a}}{c}} \right)}} - 1} \right]\left[ {4\frac{{{\psi _l}^\prime \left( {\frac{{i\xi a}}{c}} \right)}}{{{\xi _l}^\prime \left( {\frac{{i\xi a}}{c}} \right)}} - 1} \right] + \left[ {2\frac{{{\psi _l}^\prime \left( {\frac{{i\xi b}}{c}} \right)}}{{{\xi _l}^\prime \left( {\frac{{i\xi b}}{c}} \right)}} - 1} \right]}}{{\left[ {1 + \delta {{\tilde \varepsilon }^{{\rm{3D}}}}\left( {i\xi } \right)\left( {\frac{\xi }{c}} \right){\xi _l}^\prime \left( {\frac{{i\xi b}}{c}} \right){\psi _l}^\prime \left( {\frac{{i\xi b}}{c}} \right)} \right]\left[ {1 + \delta {{\tilde \varepsilon }^{{\rm{3D}}}}\left( {i\xi } \right)\left( {\frac{\xi }{c}} \right){\xi _l}^\prime \left( {\frac{{i\xi a}}{c}} \right){\psi _l}^\prime \left( {\frac{{i\xi a}}{c}} \right)} \right]}}},
\end{array}
\label{VD12.5}
\end{equation}
and for TE modes
\begin{equation}
\begin{array}{l}
\tilde f_{l.m}^{{\rm{TE}}}\left( {i\xi } \right) = 1 - {\left( {\frac{{\delta {{\tilde \varepsilon }^{{\rm{3D}}}}\left( {i\xi } \right)\xi }}{{2c}}} \right)^2}{\left[ {{\xi _l}\left( {\frac{{i\xi b}}{c}} \right)} \right]^2}{\left[ {{\xi _l}\left( {\frac{{i\xi a}}{c}} \right)} \right]^2}\\
 \times \frac{{\left[ {2\frac{{{\psi _l}\left( {\frac{{i\xi a}}{c}} \right)}}{{{\xi _l}\left( {\frac{{i\xi a}}{c}} \right)}} - 1} \right]\left[ {4\frac{{{\psi _l}\left( {\frac{{i\xi a}}{c}} \right)}}{{{\xi _l}\left( {\frac{{i\xi a}}{c}} \right)}} - 1} \right] + \left[ {2\frac{{{\psi _l}\left( {\frac{{i\xi b}}{c}} \right)}}{{{\xi _l}\left( {\frac{{i\xi b}}{c}} \right)}} - 1} \right]}}{{\left[ {1 + \delta {{\tilde \varepsilon }^{{\rm{3D}}}}\left( {i\xi } \right)\left( {\frac{\xi }{c}} \right){\xi _l}\left( {\frac{{i\xi b}}{c}} \right){\psi _l}\left( {\frac{{i\xi b}}{c}} \right)} \right]\left[ {1 + \delta {{\tilde \varepsilon }^{{\rm{3D}}}}\left( {i\xi } \right)\left( {\frac{\xi }{c}} \right){\xi _l}\left( {\frac{{i\xi a}}{c}} \right){\psi _l}\left( {\frac{{i\xi a}}{c}} \right)} \right]}}.
\end{array}
\label{VD12.6}
\end{equation}

Here, our notation may unfortunately cause some confusion. Note that $\xi $ is the variable along the imaginary frequency axis and ${{\xi _l}}$ is a Ricatti-Bessel function. We have chosen to express the mode condition functions in terms of the functions ${\xi _l}$ and ${\psi _l}$ to simplify the transformation into real valued functions of real valued arguments according to Eq.\,({\ref{VC17}).

We have chosen as reference system a system where the two shells are separated from each other and at infinite distance from each other. The energy obtained by using this mode condition function is the energy change when bringing the two shells together from at infinite separation  and putting the inner shell inside the outer shell.

\subsubsection{\label{atomspherical2filmsret}Force on an atom in between two 2D spherical films (five layers)}
Here we may reuse the results from Sec.\,\ref{atomsphericalgapret}. The only difference is the expressions for the ${2^l}$ pole polarizabilities.

The mode condition function is
\begin{equation}
\begin{array}{l}
{{\tilde f}_{l,m}} = \frac{{M_{11}^1 + \alpha _l^{{\rm{2D0}}\left( 2 \right)}M_{21}^1 - \alpha _l^{{\rm{2D}}2}\left( {M_{12}^1 + \alpha _l^{{\rm{2D0}}\left( 2 \right)}M_{22}^1} \right)}}{{\left( {1 - \alpha _l^{{\rm{2D0}}\left( 2 \right)}\alpha _l^{{\rm{2D}}2}} \right)\left( {M_{11}^1 + M_{12}^1} \right)}}\\
 \approx 1 - \frac{{\delta M_{12}^1\left[ {1 + \alpha _l^{{\rm{2D}}2}\left( {1 - \alpha _l^{{\rm{2D0}}\left( 2 \right)}} \right)} \right] - }}{{\left( {1 - \alpha _l^{{\rm{2D0}}\left( 2 \right)}\alpha _l^{{\rm{2D}}2}} \right)}}\\
\quad \quad  \cdots \frac{{ - \alpha _l^{{\rm{2D0}}\left( 2 \right)}\delta M_{21}^1 + \alpha _l^{{\rm{2D0}}\left( 2 \right)}\alpha _l^{{\rm{2D}}2}\left( {\delta M_{22}^1 - \delta M_{11}^1} \right)}}{{}},
\end{array}
\label{VD13.1}
\end{equation}
where we have kept terms up to linear order in the atom density. We have chosen as reference system a system with the spherical films and the gas shell well separated from each other. Thus we have divided our mode condition function both with that for a free gas film and that for the thin spherical films.

For TM modes we have
\begin{equation}
\begin{array}{*{20}{l}}
{\delta {\bf{\tilde M}}_{\rm{1}}^{{\rm{TM}}} =  - \left( {\delta N} \right)2\pi {\alpha ^{at}}{q_0}i}\\
{\quad \quad  \times \left( {\begin{array}{*{20}{c}}
{{\xi _l}^\prime {\zeta _l}^\prime  + {\xi _l}{\zeta _l}\frac{{l\left( {l + 1} \right)}}{{{{\left( {{q_0}r} \right)}^2}}}}&{{{\left[ {{\xi _l}^\prime } \right]}^2} + {{\left[ {{\xi _l}} \right]}^2}\frac{{l\left( {l + 1} \right)}}{{{{\left( {{q_0}r} \right)}^2}}}}\\
{ - {{\left[ {{\zeta _l}^\prime } \right]}^2} - {{\left[ {{\zeta _l}} \right]}^2}\frac{{l\left( {l + 1} \right)}}{{{{\left( {{q_0}r} \right)}^2}}}}&{ - {\xi _l}^\prime {\zeta _l}^\prime  - {\xi _l}{\zeta _l}\frac{{l\left( {l + 1} \right)}}{{{{\left( {{q_0}r} \right)}^2}}}}
\end{array}} \right)};\\
{\alpha _l^{{\rm{2D2,TM}}} =  - \frac{{2 + \delta {{\tilde \varepsilon }^{{\rm{3D}}}}{q_0}i\left[ {{\zeta _l}^\prime {{\left( {{q_0}a} \right)}^2} + {\xi _l}^\prime \left( {{q_0}a} \right){\zeta _l}^\prime \left( {{q_0}a} \right)} \right]}}{{2 - \delta {{\tilde \varepsilon }^{{\rm{3D}}}}{q_0}i\left[ {{\xi _l}^\prime {{\left( {{q_0}a} \right)}^2} + {\xi _l}^\prime \left( {{q_0}a} \right){\zeta _l}^\prime \left( {{q_0}a} \right)} \right]}}};\\
{\alpha _l^{{\rm{2D}}0\left( 2 \right),{\rm{TM}}} = \frac{{ - \delta {{\tilde \varepsilon }^{{\rm{3D}}}}{q_0}i{\xi _l}^\prime {{\left( {{q_0}b} \right)}^2}}}{{2 - \delta {{\tilde \varepsilon }^{{\rm{3D}}}}{q_0}i{\xi _l}^\prime \left( {{q_0}b} \right){\zeta _l}^\prime \left( {{q_0}b} \right)}},}
\end{array}
\label{VD13.2}
\end{equation}
where ${q_0} = \omega /c$. All Ricatti-Bessel functions in the matrix have the argument ${\left( {{q_0}r} \right)}$.
For TE modes the functions are
\begin{equation}
\begin{array}{*{20}{l}}
{\delta {\bf{\tilde M}}_{\rm{1}}^{{\rm{TE}}} =  - \left( {\delta N} \right)2\pi {\alpha ^{at}}{q_0}i}\\
{\quad \quad  \times \left( {\begin{array}{*{20}{c}}
{{\xi _l}\left( {{q_0}r} \right){\zeta _l}\left( {{q_0}r} \right)}&{{{\left[ {{\xi _l}\left( {{q_0}r} \right)} \right]}^2}}\\
{ - {{\left[ {{\zeta _l}\left( {{q_0}r} \right)} \right]}^2}}&{ - {\xi _l}\left( {{q_0}r} \right){\zeta _l}\left( {{q_0}r} \right)}
\end{array}} \right)};\\
{\alpha _l^{{\rm{2D2,TE}}} =  - \frac{{2 + \delta {{\tilde \varepsilon }^{{\rm{3D}}}}{q_0}i\left[ {{\zeta _l}{{\left( {{q_0}a} \right)}^2} + {\xi _l}\left( {{q_0}a} \right){\zeta _l}\left( {{q_0}a} \right)} \right]}}{{2 - \delta {{\tilde \varepsilon }^{{\rm{3D}}}}{q_0}i\left[ {{\xi _l}{{\left( {{q_0}a} \right)}^2} + {\xi _l}\left( {{q_0}a} \right){\zeta _l}\left( {{q_0}a} \right)} \right]}}};\\
{\alpha _l^{{\rm{2D}}0\left( 2 \right),{\rm{TE}}} = \frac{{ - \delta {{\tilde \varepsilon }^{{\rm{3D}}}}{q_0}i{\xi _l}{{\left( {{q_0}b} \right)}^2}}}{{2 - \delta {{\tilde \varepsilon }^{{\rm{3D}}}}{q_0}i{\xi _l}\left( {{q_0}b} \right){\zeta _l}\left( {{q_0}b} \right)}}.}
\end{array}
\label{VD13.32}
\end{equation}

\section{\label{cylindrical}Cylindrical structures}
The system we consider here is a layered cylinder consisting of $N$ layers and an inner cylindrical core. We have $N+2$ media and $N+1$ interfaces. Let the numbering be as follows. Medium $0$ is the medium surrounding the cylinder, medium $1$ is the outermost layer, medium $N$ the innermost layer and $N+1$ the innermost cylindrical region (the core). Let ${{r_n}}$ be the inner radius of layer $n$. The boundary condition is that there are no incoming waves, i.e. there is no wave moving towards the right in medium $n=0$ in Fig.\,\ref{figu3}. The fields are self-sustained; no fields are coming in from outside.

\subsection{\label{cylindricalnonretmain}Non-retarded main results}
In the non-retarded treatment of a cylindrical structure we let the waves represent solutions to Laplace's equation, Eq.\,(\ref{III21}), in cylindrical coordinates, ($r,\theta ,z$), for the scalar potential, $\Phi$. The interfaces are cylindrical surfaces and the $r$-coordinate is the coordinate that is constant on each interface. The solutions are of the form
\begin{equation}
{\Phi _{k,m}}\left( {r,\theta ,z} \right) = {I_m}\left( {kr} \right){e^{im\theta }}{e^{ikz}}{\rm\,{ and }}\,{K_m}\left( {kr} \right){e^{im\theta }}{e^{ikz}},
\label{VIA1}
\end{equation}
where the functions ${I_m}\left( {z} \right)$ and ${K_m}\left( {z} \right)$ are so-called modified Bessel functions. The first is bounded for small $z$ values and the second for large. They are solutions to the modified Bessel equation,
\begin{equation}
{z^2}{\partial ^2}\omega /\partial {z^2} + z\partial \omega /\partial z - \left( {{m^2} + {z^2}} \right)\omega  = 0.
\label{VIA2}
\end{equation}
Note that the variable $z$ here denotes a general complex variable and should not be mistaken for the spatial $z$-variable in Eq.\,(\ref{VIA1}). 
We let $r$ increase towards the left in  Fig.\,\ref{figu3}. We want to find the normal modes for a specific set of $k$ and $m$ values. Then all waves have the common factor  ${e^{im\theta }}{e^{ikz}}$. We suppress this factor here. Then
\begin{equation}
R\left( r \right) = {I_m}\left( {kr} \right);\,L\left( r \right) = {K_m}\left( {kr} \right).
\label{VIA3}
\end{equation}
Using the boundary conditions that the potential and the normal component of the ${\bf{D}}$-field are continuous across an interface $n$ gives
\begin{equation}
\begin{array}{*{20}{l}}
\begin{array}{l}
{a^n}{I_m}\left( {k{r_n}} \right) + {b^n}{K_m}\left( {k{r_n}} \right)\\
\quad \quad  = {a^n}{I_m}\left( {k{r_n}} \right) + {b^n}{K_m}\left( {k{r_n}} \right)
\end{array}\\
{{a^n}{{\tilde \varepsilon }_n}k{I_m}'\left( {k{r_n}} \right) + {b^n}{{\tilde \varepsilon }_n}k{K_m}'\left( {k{r_n}} \right)}\\
{\quad \quad  = {a^{n + 1}}{{\tilde \varepsilon }_{n + 1}}k{I_m}'\left( {k{r_n}} \right) + {b^{n + 1}}{{\tilde \varepsilon }_{n + 1}}k{K_m}'\left( {k{r_n}} \right),}
\end{array}
\label{VIA4}
\end{equation}
and we may identify the matrix ${{{\bf{\tilde A}}}_n}\left( {{r_n}} \right)$ as
\begin{equation}
{{{\bf{\tilde A}}}_n}\left( {{r_n}} \right) = \left( {\begin{array}{*{20}{c}}
{{I_m}\left( {k{r_n}} \right)}&{{K_m}\left( {k{r_n}} \right)}\\
{{{\tilde \varepsilon }_n}{I_m}'\left( {k{r_n}} \right)}&{{{\tilde \varepsilon }_n}{K_m}'\left( {k{r_n}} \right)}
\end{array}} \right).
\label{VIA5}
\end{equation}

The matrix ${{{\bf{\tilde M}}}_n}$ is
\begin{equation}
\begin{array}{*{20}{l}}
{{{{\bf{\tilde M}}}_n} = {\bf{\tilde A}}_n^{ - 1} \cdot {{{\bf{\tilde A}}}_{n + 1}} = \frac{1}{{W{{\tilde \varepsilon }_n}}}}\\
{ \times \left( {\begin{array}{*{20}{c}}
{{{\tilde \varepsilon }_{n + 1}}I_m^\prime {K_m} - {{\tilde \varepsilon }_n}{I_m}K_m^\prime }&{\left( {{{\tilde \varepsilon }_{n + 1}} - {{\tilde \varepsilon }_n}} \right){K_m}K_m^\prime }\\
{\left( {{{\tilde \varepsilon }_n} - {{\tilde \varepsilon }_{n + 1}}} \right){I_m}I_m^\prime }&{{{\tilde \varepsilon }_n}I_m^\prime {K_m} - {{\tilde \varepsilon }_{n + 1}}{I_m}K_m^\prime }
\end{array}} \right),}
\end{array}
\label{VIA6}
\end{equation}
where we have suppressed the argument $\left( {k{r_n}} \right)$ of all modified Bessel functions and their derivatives. As before the derivative is with respect to the argument. We have made use of the Wronskian of the two modified Bessel functions: $W\left[ {{K_m}\left( x \right),{I_m}\left( x \right)} \right] = {K_m}\left( x \right){I_m}^\prime \left( x \right) - {K_m}^\prime \left( x \right){I_m}\left( x \right) = 1/x$.

Since the function $L\left( z \right)$ in Eq.\,(\ref{VIA3}) diverges at the origin it is excluded from the core region and hence we have no wave moving towards the left in that region. According to  Eq.\,(\ref{III6}) this means that
\begin{equation}
{f_{k,m}}\left( \omega  \right) = {M_{11}}.
\label{VIA7}
\end{equation}
Before we end this section we introduce the multipole polarizabilities ${\alpha _{k,m}^n}$ and $\alpha _{k,m}^{n\left( 2 \right)}$ for the cylindrical interface since these appear repeatedly in the sections that follow. The first is valid outside and the second inside. The polarizability ${\alpha _{k,m}^n}= - {b^n}/{a^n}$ under the assumption that ${b^{n+1}} = 0$. One obtains ${\alpha _{k,m}^n} =  - {M_{21}}/{M_{11}}$ and from Eq.\,(\ref{VIA6}) one finds
\begin{equation}
\alpha _{k,m}^n\left( {{r_n};\omega } \right) = \frac{{\left( {{{\tilde \varepsilon }_{n + 1}} - {{\tilde \varepsilon }_n}} \right){I_m}I_m^\prime }}{{{{\tilde \varepsilon }_{n + 1}}I_m^\prime {K_m} - {{\tilde \varepsilon }_n}{I_m}K_m^\prime }}.
\label{VIA8}
\end{equation}
The polarizability $\alpha _{k,m}^{n\left( 2 \right)}= - {a^{n + 1}}/{b^{n + 1}}$ under the assumption that ${a^n} = 0$. One obtains $\alpha _{k,m}^{n\left( 2 \right)} =   {M_{12}}/{M_{11}}$ and from Eq.\,(\ref{VA5}) one finds
\begin{equation}
\alpha _{k,m}^{n\left( 2 \right)}\left( {{r_n};\omega } \right) = \frac{{\left( {{{\tilde \varepsilon }_{n + 1}} - {{\tilde \varepsilon }_n}} \right){K_m}K_m^\prime }}{{{{\tilde \varepsilon }_{n + 1}}I_m^\prime {K_m} - {{\tilde \varepsilon }_n}{I_m}K_m^\prime }}.
\label{VIA9}
\end{equation}
The suppressed argument of the modified Bessel functions in the  multipole polarizabilities above is $\left( {k{r_n}} \right)$.
Sometimes it is convenient to use an alternative form of the matrix ${{{{\bf{\tilde M}}}_n}}$,
\begin{equation}
{{{\bf{\tilde M}}}_n} = M_{11}^n\left( {\begin{array}{*{20}{c}}
1&{\alpha _{k,m}^{n\left( 2 \right)}}\\
{ - \alpha _{k,m}^n}&{\frac{{{{\tilde \varepsilon }_n}I_m^\prime {K_m} - {{\tilde \varepsilon }_{n + 1}}{I_m}K_m^\prime }}{{{{\tilde \varepsilon }_{n + 1}}I_m^\prime {K_m} - {{\tilde \varepsilon }_n}{I_m}K_m^\prime }}}
\end{array}} \right).
\label{VIA10}
\end{equation}

Now we have all we need to determine the non-retarded normal modes in a layered cylindrical structure. We give some examples in the following sections.

\subsection{\label{cylindricalnonretspecial}Non-retarded special results}

\subsubsection{\label{cylindern}Solid cylinder (no layer)}
For a solid cylinder of radius  $a$ and dielectric function ${{\tilde \varepsilon }_1}\left( \omega  \right)$ in an ambient of dielectric function ${{\tilde \varepsilon }_0}\left( \omega  \right)$, as illustrated in Fig.\,\ref{figu10}, we have
\begin{equation}
\begin{array}{*{20}{l}}
{{\bf{\tilde M}} = {{{\bf{\tilde M}}}_0} = \frac{{ka}}{{{{\tilde \varepsilon }_0}}}}\\
{ \times \left( {\begin{array}{*{20}{c}}
{{{\tilde \varepsilon }_1}{I_m}^\prime {K_m} - {{\tilde \varepsilon }_0}{I_m}{K_m}^\prime }&{\left( {{{\tilde \varepsilon }_1} - {{\tilde \varepsilon }_0}} \right){K_m}{K_m}^\prime }\\
{\left( {{{\tilde \varepsilon }_0} - {{\tilde \varepsilon }_1}} \right){I_m}{I_m}^\prime }&{{{\tilde \varepsilon }_0}{I_m}^\prime {K_m} - {{\tilde \varepsilon }_1}{I_m}{K_m}^\prime }
\end{array}} \right),}
\end{array}
\label{VIB1.1}
\end{equation}
where the suppressed arguments are $\left( {ka} \right)$. The condition for modes is 
\begin{equation}
\frac{{{{\tilde \varepsilon }_1}\left( \omega  \right)}}{{{{\tilde \varepsilon }_0}\left( \omega  \right)}} = \frac{{{I_m}\left( {ka} \right){K_m}'\left( {ka} \right)}}{{{I_m}'\left( {ka} \right){K_m}\left( {ka} \right)}}.
\label{VIB1.2}
\end{equation}

\subsubsection{\label{CylindricalShelln}Cylindrical shell or gap (one layer)}
Here we start from a more general geometry namely that of a coated cylinder in a medium and get the cylindrical shell and gap as special limits.
For a solid cylinder of dielectric function ${{\tilde \varepsilon }_2}$ with a coating of inner radius $a$ and outer radius $b$, Fig.\,\ref{figu11},  made of a medium with dielectric function ${{\tilde \varepsilon }_1}$  in an ambient medium with dielectric function ${{\tilde \varepsilon }_0}$ we have 
\begin{equation}
\begin{array}{*{20}{l}}
{{\bf{\tilde M}} = {{{\bf{\tilde M}}}_0} \cdot {{{\bf{\tilde M}}}_1}}\\
\begin{array}{l}
 = \frac{{kb}}{{{{\tilde \varepsilon }_0}}}\left( {{{\tilde \varepsilon }_1}I_m^\prime \left( {kb} \right){K_m}\left( {kb} \right) - {{\tilde \varepsilon }_0}{I_{m\left( {kb} \right)}}K_m^\prime \left( {kb} \right)} \right)\\
 \times \left( {\begin{array}{*{20}{c}}
1&{\alpha _{k,m}^{0\left( 2 \right)}}\\
{ - \alpha _{k,m}^0}&{\frac{{{{\tilde \varepsilon }_0}I_m^\prime \left( {kb} \right){K_m}\left( {kb} \right) - {{\tilde \varepsilon }_1}{I_m}\left( {kb} \right)K_m^\prime \left( {kb} \right)}}{{{{\tilde \varepsilon }_1}I_m^\prime \left( {kb} \right){K_m}\left( {kb} \right) - {{\tilde \varepsilon }_0}{I_m}\left( {kb} \right)K_m^\prime \left( {kb} \right)}}}
\end{array}} \right)
\end{array}\\
\begin{array}{l}
 \times \frac{{ka}}{{{{\tilde \varepsilon }_1}}}\left( {{{\tilde \varepsilon }_2}I_m^\prime \left( {ka} \right){K_m}\left( {ka} \right) - {{\tilde \varepsilon }_1}{I_m}\left( {ka} \right)K_m^\prime \left( {ka} \right)} \right)\\
 \times \left( {\begin{array}{*{20}{c}}
1&{\alpha _{k,m}^{1\left( 2 \right)}}\\
{ - \alpha _{k,m}^1}&{\frac{{{{\tilde \varepsilon }_1}I_m^\prime \left( {ka} \right){K_m}\left( {ka} \right) - {{\tilde \varepsilon }_2}{I_m}\left( {ka} \right)K_m^\prime \left( {ka} \right)}}{{{{\tilde \varepsilon }_2}I_m^\prime \left( {ka} \right){K_m}\left( {ka} \right) - {{\tilde \varepsilon }_0}{I_m}\left( {ka} \right)K_m^\prime \left( {ka} \right)}}}
\end{array}} \right),
\end{array}
\end{array}
\label{VIB2.1}
\end{equation}
and from direct derivation of the $M_{11}$ element  the condition for modes  becomes
\begin{equation}
\begin{array}{l}
0 = \left( {1 - \alpha _{k,m}^{0\left( 2 \right)}\alpha _{k,m}^1} \right)\\
\quad  = 1 - \frac{{\left( {{{\tilde \varepsilon }_1} - {{\tilde \varepsilon }_0}} \right){K_m}\left( {kb} \right)K_m^\prime \left( {kb} \right)}}{{{{\tilde \varepsilon }_1}I_m^\prime \left( {kb} \right){K_m}\left( {kb} \right) - {{\tilde \varepsilon }_0}{I_m}\left( {kb} \right)K_m^\prime \left( {kb} \right)}}\\
\quad \quad \quad \quad  \times \frac{{\left( {{{\tilde \varepsilon }_2} - {{\tilde \varepsilon }_1}} \right){I_m}\left( {ka} \right)I_m^\prime \left( {ka} \right)}}{{{{\tilde \varepsilon }_2}I_m^\prime \left( {ka} \right){K_m}\left( {ka} \right) - {{\tilde \varepsilon }_1}{I_m}\left( {ka} \right)K_m^\prime \left( {ka} \right)}}.
\end{array}
\label{VIB2.2}
\end{equation}

Let us now study a cylindrical shell of inner radius $a$, outer radius $b$ and of a medium with dielectric function $\tilde \varepsilon \left( \omega  \right)$ in a medium of dielectric function ${{\tilde \varepsilon }_0}\left( \omega  \right)$. The condition for modes we get from Eq.\,(\ref{VIB2.2}) by the replacements ${{\tilde \varepsilon }_2}\left( \omega  \right) \to {{\tilde \varepsilon }_0}\left( \omega  \right)$ and $\;{{\tilde \varepsilon }_1}\left( \omega  \right) \to \tilde \varepsilon \left( \omega  \right)$.  The result is

\begin{equation}
\begin{array}{l}
\left[ {\tilde \varepsilon \frac{{K_m^\prime \left( {ka} \right)}}{{I_m^\prime \left( {ka} \right)}} - {{\tilde \varepsilon }_0}\frac{{{K_m}\left( {ka} \right)}}{{{I_m}\left( {ka} \right)}}} \right]\left[ {\tilde \varepsilon \frac{{I_m^\prime \left( {kb} \right)}}{{K_m^\prime \left( {kb} \right)}} - {{\tilde \varepsilon }_0}\frac{{{I_m}\left( {kb} \right)}}{{{K_m}\left( {kb} \right)}}} \right]\\
 = {\left( {\tilde \varepsilon  - {{\tilde \varepsilon }_0}} \right)^2}.
\end{array}
\label{VIB2.3}
\end{equation}

For a cylindrical gap of dielectric function ${{\tilde \varepsilon }_0}\left( \omega  \right)$ in a medium of dielectric function $\tilde \varepsilon \left( \omega  \right)$ we instead make the replacements ${{\tilde \varepsilon }_0}\left( \omega  \right),\;{{\tilde \varepsilon }_2}\left( \omega  \right) \to \tilde \varepsilon \left( \omega  \right)$ and $\;{{\tilde \varepsilon }_1}\left( \omega  \right) \to {{\tilde \varepsilon }_0}\left( \omega  \right)$. The condition for modes is
\begin{equation}
\begin{array}{l}
\left[ {\tilde \varepsilon \frac{{{K_m}\left( {ka} \right)}}{{{I_m}\left( {ka} \right)}} - {{\tilde \varepsilon }_0}\frac{{K_m^\prime \left( {ka} \right)}}{{I_m^\prime \left( {ka} \right)}}} \right]\left[ {\tilde \varepsilon \frac{{{I_m}\left( {kb} \right)}}{{{K_m}\left( {kb} \right)}} - {{\tilde \varepsilon }_0}\frac{{I_m^\prime \left( {kb} \right)}}{{K_m^\prime \left( {kb} \right)}}} \right]\\
 = {\left( {\tilde \varepsilon  - {{\tilde \varepsilon }_0}} \right)^2}.
\end{array}
\label{VIB2.4}
\end{equation}

\subsubsection{\label{Cylindricalgasfilmn}Thin cylindrical diluted gas film (one layer)}
It is of interest to find the van der Waals force on an atom in a layered structure. We can obtain this by studying the force on a thin layer of a diluted gas with dielectric function ${\varepsilon _g}\left( \omega  \right) = 1 + 4\pi n\alpha^{at} \left( \omega  \right)$, where $\alpha^{at}$ is the polarizability of one atom and $n$ the density of atoms (we have assumed that the atom is surrounded by vacuum; if not  the $1$ should be replaced by the dielectric function of the ambient medium and the atomic polarizability should be replaced by the excess polarizability). For a diluted gas layer the atoms do not interact with each other and the force on the layer is just the sum of the forces on the individual atoms. So by dividing with the number of atoms in the film we get the force on one atom. The layer has to be thin in order to have a well defined $r$-value of the atom.  Since we will derive the force on an atom in different cylindrical geometries it is fruitful to derive the matrix for a thin diluted gas shell. This result can be directly used in the derivation of the van der Waals force on an atom in different cylindrical geometries.

We let the film have the thickness $\delta$ and be of a general radius $r$. We only keep terms up to linear order in $\delta$ and linear order in $n$. The matrix for the gas film is ${{{\bf{\tilde M}}}_0} \cdot {{{\bf{\tilde M}}}_1}$ where 
\begin{equation}
\begin{array}{l}
{{{\bf{\tilde M}}}_0} = \left( {\begin{array}{*{20}{c}}
1&0\\
0&1
\end{array}} \right)\\
 + 4\pi n{\alpha ^{at}}kr\left( {\begin{array}{*{20}{c}}
{I_m^\prime \left( {kr} \right){K_m}\left( {kr} \right)}&{{K_m}\left( {kr} \right)K_m^\prime \left( {kr} \right)}\\
{ - {I_m}\left( {kr} \right)I_m^\prime \left( {kr} \right)}&{ - {I_m}\left( {kr} \right)K_m^\prime \left( {kr} \right)}
\end{array}} \right).
\end{array}
\label{VIB3.1}
\end{equation}
Now, 
\begin{equation}
\begin{array}{l}
{{{\bf{\tilde M}}}_1} = \left( {\begin{array}{*{20}{c}}
1&0\\
0&1
\end{array}} \right)\\
 \quad\quad- 4\pi n{\alpha ^{at}}k\left( {r - \delta } \right)\left( {\begin{array}{*{20}{c}}
{I_m^\prime {K_m}}&{{K_m}K_m^\prime }\\
{ - {I_m}I_m^\prime }&{ - {I_m}K_m^\prime }
\end{array}} \right),
\end{array}
\label{VIB3.2}
\end{equation}
where the suppressed arguments of all modified Bessel functions are $k\left( {r - \delta } \right)$. We find
\begin{equation}
\begin{array}{*{20}{l}}
{{{{\bf{\tilde M}}}_{{\rm{gaslayer}}}} = {{{\bf{\tilde M}}}_0} \cdot {{{\bf{\tilde M}}}_1}}\\
{\quad \quad \quad \quad  = \left( {\begin{array}{*{20}{c}}
1&0\\
0&1
\end{array}} \right) + 4\pi \left( {\delta n} \right){\alpha ^{at}}k}\\
{\quad \quad \quad \quad \quad \quad \quad  \times \left( {\begin{array}{*{20}{c}}
{\frac{{d\left[ {\left( {kr} \right)I_m^\prime {K_m}} \right]}}{{d\left( {kr} \right)}}}&{\frac{{d\left[ {\left( {kr} \right)K_m^\prime {K_m}} \right]}}{{d\left( {kr} \right)}}}\\
{ - \frac{{d\left[ {\left( {kr} \right)I_m^\prime {I_m}} \right]}}{{d\left( {kr} \right)}}}&{ - \frac{{d\left[ {\left( {kr} \right){I_m}K_m^\prime } \right]}}{{d\left( {kr} \right)}}}
\end{array}} \right),}
\end{array}
\label{VIB3.3}
\end{equation}
where now the arguments of all modified Bessel functions are ${\left( {kr} \right)}$. Performing the derivatives and using the modified Bessel equation, Eq.\,(\ref{VIA2}), we find
\begin{equation}
\begin{array}{l}
{{{\bf{\tilde M}}}_{{\rm{gaslayer}}}} = \left( {\begin{array}{*{20}{c}}
1&0\\
0&1
\end{array}} \right)\\
\quad \quad  + \left( {\delta n} \right)\frac{{4\pi {\alpha ^{at}}\left[ {{m^2} + {{\left( {kr} \right)}^2}} \right]}}{r}\left( {\begin{array}{*{20}{c}}
{{I_m}{K_m}}&{{K_m}{K_m}}\\
{ - {I_m}{I_m}}&{ - {I_m}{K_m}}
\end{array}} \right)\\
\quad \quad  + \left( {\delta n} \right)\frac{{4\pi {\alpha ^{at}}{{\left( {kr} \right)}^2}}}{r}\left( {\begin{array}{*{20}{c}}
{I_m^\prime K_m^\prime }&{K_m^\prime K_m^\prime }\\
{ - I_m^\prime I_m^\prime }&{ - I_m^\prime K_m^\prime }
\end{array}} \right).
\end{array}
\label{VIB3.4}
\end{equation}

Now we are done with the gas layer. We will use these results later in calculating the van der Waals force on an atom in cylindrical layered structures.

\subsubsection{\label{Cylindrical2Dfilmn}2D cylindrical film (one layer)}
In many situations one is dealing with very thin films. These may be considered 2D. Important examples are a graphene sheet and a 2D electron gas. In the derivation we let the film have finite thickness $\delta$ and be characterized by a 3D dielectric function ${\tilde \varepsilon ^{3D}}$. We then let the thickness go towards zero. The 3D dielectric function depends on $\delta$ as ${\tilde\varepsilon ^{3D}} \sim 1/\delta $ for small $\delta$. In the planar structure we could in the limit when $\delta$ goes towards zero obtain a momentum dependent 2D dielectric function. Here we only keep the long wave length limit of the 2D dielectric function.\,\cite{grap,arx} The matrix is ${{{{\bf{\tilde M}}}_{{\rm{2D}}}} = {{{\bf{\tilde M}}}_0} \cdot {{{\bf{\tilde M}}}_1}}$. Before we derive these matrices it it convenient to introduce two  auxiliary matrices,
\begin{equation}
\begin{array}{*{20}{l}}
{{\bf{\tilde B}}\left( x \right) = kx\left( {\begin{array}{*{20}{c}}
{I_m^\prime \left( {kx} \right){K_m}\left( {kx} \right)}&{{K_m}\left( {kx} \right)K_m^\prime \left( {kx} \right)}\\
{ - {I_m}\left( {kx} \right)I_m^\prime \left( {kx} \right)}&{ - {I_m}\left( {kx} \right)K_m^\prime \left( {kx} \right)}
\end{array}} \right);}\\
\begin{array}{l}
{\bf{\tilde C}}\left( x \right)\\
\quad  = kx\left( {\begin{array}{*{20}{c}}
{ - {I_m}\left( {kx} \right)K_m^\prime \left( {kx} \right)}&{ - {K_m}\left( {kx} \right)K_m^\prime \left( {kx} \right)}\\
{{I_m}\left( {kx} \right)I_m^\prime \left( {kx} \right)}&{I_m^\prime \left( {kx} \right){K_m}\left( {kx} \right)}
\end{array}} \right).
\end{array}
\end{array}
\label{VIB4.1}
\end{equation}
Using the Wronskian for the modified Bessel function we find these matrices have the following properties:
\begin{equation}
\begin{array}{l}
{\bf{\tilde B}}\left( x \right) + {\bf{\tilde C}}\left( x \right) = \tilde 1;\\
{\bf{\tilde B}}\left( x \right) \cdot {\bf{\tilde B}}\left( x \right) = {\bf{\tilde B}}\left( x \right);\\
{\bf{\tilde C}}\left( x \right) \cdot {\bf{\tilde C}}\left( x \right) = {\bf{\tilde C}}\left( x \right);\\
{\bf{\tilde B}}\left( x \right) \cdot {\bf{\tilde C}}\left( x \right) = \tilde 0.
\end{array}
\label{VIB4.2}
\end{equation}
Now, we have
\begin{equation}
\begin{array}{l}
{{{\bf{\tilde M}}}_0} = {{\tilde \varepsilon }^{{\rm{3D}}}}{\bf{\tilde B}}\left( r \right) + {\bf{\tilde C}}\left( r \right);\\
{{{\bf{\tilde M}}}_1} = \frac{1}{{{{\tilde \varepsilon }^{{\rm{3D}}}}}}{\bf{\tilde B}}\left( {r - \delta } \right) + {\bf{\tilde C}}\left( {r - \delta } \right),
\end{array}
\label{VIB4.3}
\end{equation}
and using Eq.\,(\ref{VIB4.2}) and the modified Bessel equation,  Eq.\,(\ref{VIA2}), we arrive at
\begin{equation}
\begin{array}{l}
{{{\bf{\tilde M}}}_{{\rm{2D}}}} = {{{\bf{\tilde M}}}_0} \cdot {{{\bf{\tilde M}}}_1}\\
 = \tilde 1 - \delta {{\tilde \varepsilon }^{{\rm{3D}}}}\frac{{{m^2} + {{\left( {kr} \right)}^2}}}{r}\left( {\begin{array}{*{20}{c}}
{ - {I_m}{K_m}}&{ - {K_m}{K_m}}\\
{{I_m}{I_m}}&{{I_m}{K_m}}
\end{array}} \right),
\end{array}
\label{VIB4.4}
\end{equation}
where the suppressed arguments of the modified Bessel functions are ${\left( {kr} \right)}$.

We will also need the multipole polarizability of the thin cylindrical shell in vacuum.  It can be obtained from Eq.\,(\ref{VIB4.4}). The polarizability is $ - {b^0}/{a^0}$ under the assumption that ${b^1} = 0$. One obtains ${\alpha _l^{2D}} =  - {M_{21}}/{M_{11}}$. We find
\begin{equation}
\alpha _{k,m}^{2D}\left( {r;\omega } \right) = \frac{{\delta {{\tilde \varepsilon }^{{\rm{3D}}}}\left[ {{m^2} + {{\left( {kr} \right)}^2}} \right]{I_m}\left( {kr} \right){I_m}\left( {kr} \right)}}{{r + \delta {{\tilde \varepsilon }^{{\rm{3D}}}}\left[ {{m^2} + {{\left( {kr} \right)}^2}} \right]{I_m}\left( {kr} \right){K_m}\left( {kr} \right)}},
\label{VIB4.5}
\end{equation}
where we have reserved the first argument before the semicolon for the radius of the cylindrical film.

The multipole polarizability "seen from inside the shell" we get  from Eq.\,(\ref{VIB4.4}). The polarizability is $ - {a^1}/{b^1}$ under the assumption that ${a^0} = 0$. One obtains $\alpha _l^{2D\left( 2 \right)} =   {M_{12}}/{M_{11}}$, and
\begin{equation}
\alpha _{k,m}^{2D\left( 2 \right)}\left( {r;\omega } \right) = \frac{{\delta {{\tilde \varepsilon }^{{\rm{3D}}}}\left[ {{m^2} + {{\left( {kr} \right)}^2}} \right]{K_m}\left( {kr} \right){K_m}\left( {kr} \right)}}{{r + \delta {{\tilde \varepsilon }^{{\rm{3D}}}}\left[ {{m^2} + {{\left( {kr} \right)}^2}} \right]{I_m}\left( {kr} \right){K_m}\left( {kr} \right)}}.
\label{VIB4.6}
\end{equation}

Sometimes it is convenient to use an alternative form of the matrix ${{{\bf{\tilde M}}}_{{\rm{2D}}}}$,
\begin{equation}
{{{\bf{\tilde M}}}_{{\rm{2D}}}} = M_{11}^{2D}\left( {\begin{array}{*{20}{c}}
1&{\alpha _{k,m}^{2D\left( 2 \right)}}\\
{ - \alpha _{k,m}^{2D}}&{\frac{{r - \delta {{\tilde \varepsilon }^{{\rm{3D}}}}\left[ {{m^2} + {{\left( {kr} \right)}^2}} \right]{I_m}\left( {kr} \right){K_m}\left( {kr} \right)}}{{r + \delta {{\tilde \varepsilon }^{{\rm{3D}}}}\left[ {{m^2} + {{\left( {kr} \right)}^2}} \right]{I_m}\left( {kr} \right){K_m}\left( {kr} \right)}}}
\end{array}} \right).
\label{VIB4.7}
\end{equation}
\subsubsection{\label{atom-cylindern}Force on an atom outside a cylinder (two layers)}
In this section we derive the van der Waals interaction between a polarizable atom and an infinitely long solid cylinder of dielectric function ${{\tilde \varepsilon }_1}\left( \omega  \right)$. We assume that the atom and cylinder are in vacuum for simplicity. The geometry of the problem is shown in Fig.\,\ref{figu12}. The radius of the cylinder is $a$ and the atom is at the distance $b=a+d$ from the cylinder axis. To obtain the results we proceed as follows. We introduce a thin shell defined by the radii $b$ and $b+\delta$. We let the medium of the shell have the dielectric function $\varepsilon_g  = 1 + \alpha  = 1 + 4\pi n{\alpha ^{at}} = 1 + 4\pi {\alpha ^{at}}/\left( {2\pi b\delta L} \right) = 1 + 2{\alpha ^{at}}/\left( {b\delta L} \right)$ where ${\alpha ^{at}}$ is the polarizability of the atom and $L$ is the length of the cylinder which we let go to infinity at the end. We assume that the medium of the shell is very diluted. We let $\alpha $ go towards zero and keep only terms up to linear order before we let $\delta $ go towards zero. 

The matrix of the problem is just the matrix of the thin shell, Eq.\,(\ref{VIB3.4}), multiplying that for the cylindrical core, given in Eq.\,(\ref{VIB1.1}), ${\bf{\tilde M}} = {{\bf{\tilde M}}_{{\rm{shell}}}} \cdot {{\bf{\tilde M}}_{{\rm{core}}}}$. The element of interest is
\begin{equation}
{M_{11}} = M_{11}^{{\rm{shell}}}M_{11}^{{\rm{core}}} + M_{12}^{{\rm{shell}}}M_{21}^{{\rm{core}}}.
\label{VIB5.1}
\end{equation}
The mode condition function becomes
\begin{equation}
\begin{array}{l}
{{\tilde f}_{k,m}} = 1 + \frac{{M_{12}^{{\rm{shell}}}}}{{M_{11}^{{\rm{shell}}}}}\frac{{M_{21}^{{\rm{core}}}}}{{M_{11}^{{\rm{core}}}}}\\
\quad\quad = 1 - M_{12}^{{\rm{shell}}}\alpha _{k,m}^{{\rm{core}}}\\
\quad \quad = 1 - \left( {\delta n} \right)4\pi {\alpha ^{at}}ka\left( {\tilde \varepsilon  - 1} \right){I_m}\left( {ka} \right){I_m}^\prime \left( {ka} \right)\\
\quad\quad\quad \times \frac{{\left[ {{m^2} + {{\left( {kb} \right)}^2}} \right]{{\left[ {{K_m}\left( {kb} \right)} \right]}^2} + {{\left( {kb} \right)}^2}{{\left[ {{K_m}^\prime \left( {kb} \right)} \right]}^2}}}{{b\left[ {1 + ka\left( {\tilde \varepsilon  - 1} \right){I_m}^\prime \left( {ka} \right){K_m}\left( {ka} \right)} \right]}},
\end{array}
\label{VIB5.2}
\end{equation}
where we have taken as reference system a system where the gas shell and the core are well separated from each other.

Now, the non-retarded (van der Waals) interaction energy between an atom and a cylinder is given by
\begin{equation}
\begin{array}{*{20}{l}}
{E = \hbar \int\limits_0^\infty  {\frac{{d\xi }}{{2\pi }}} \sum\limits_{m =  - \infty }^\infty  {L\int\limits_{ - \infty }^\infty  {\frac{{dk}}{{2\pi }}} \ln \left[ {{{\tilde f}_{k,m}}\left( {i\xi } \right)} \right]} }\\
\begin{array}{l}
 = \hbar L\\
 \times \int\limits_0^\infty  {\frac{{d\xi }}{{2\pi }}} \sum\limits_{m =  - \infty }^\infty  {\int\limits_{ - \infty }^\infty  {\frac{{dk}}{{2\pi }}} \ln \left[ {1 - M_{12}^{shell}\left( {k,i\xi } \right){\alpha _{k,m}}\left( {a;i\xi } \right)} \right]} 
\end{array}\\
{ \approx  - 2\hbar L\int\limits_0^\infty  {\frac{{d\xi }}{{2\pi }}} \sum\limits_{m =  - \infty }^\infty  {\int\limits_0^\infty  {\frac{{dk}}{{2\pi }}} M_{12}^{shell}\left( {k,i\xi } \right){\alpha _{k,m}}\left( {a;i\xi } \right)} }\\
{ =  - 2\hbar L\int\limits_0^\infty  {\frac{{d\xi }}{{2\pi }}} \sum\limits_{m =  - \infty }^\infty  {\int\limits_0^\infty  {\frac{{dk}}{{2\pi }}} } \left( {\delta n} \right)4\pi {\alpha ^{at}}\left( {i\xi } \right){\alpha _{k,m}}\left( {a;i\xi } \right)}\\
{\quad  \times \left\{ {\left[ {{m^2} + {{\left( {kb} \right)}^2}} \right]{{\left[ {{K_m}\left( {kb} \right)} \right]}^2} + {{\left( {kb} \right)}^2}{{\left[ {{K_m}^\prime \left( {kb} \right)} \right]}^2}} \right\}}\\
{ =  - \frac{{4\hbar }}{{{b^2}}}\int\limits_0^\infty  {\frac{{d\xi }}{{2\pi }}} \sum\limits_{m =  - \infty }^\infty  {\int\limits_0^\infty  {\frac{{dk}}{{2\pi }}} } {\alpha ^{at}}\left( {i\xi } \right){\alpha _{k,m}}\left( {a;i\xi } \right)}\\
{\quad  \times \left\{ {\left[ {{m^2} + {{\left( {kb} \right)}^2}} \right]{{\left[ {{K_m}\left( {kb} \right)} \right]}^2} + {{\left( {kb} \right)}^2}{{\left[ {{K_m}^\prime \left( {kb} \right)} \right]}^2}} \right\},}
\end{array}
\label{VIB5.3}
\end{equation}
where
\begin{equation}
{\alpha _{k,m}}\left( {a;i\xi } \right) = \frac{{\left[ {{{\tilde \varepsilon }_1}\left( {i\xi } \right) - 1} \right]\left( {ka} \right){I_m}\left( {ka} \right){I_m}^\prime \left( {ka} \right)}}{{1 + \left[ {{{\tilde \varepsilon }_1}\left( {i\xi } \right) - 1} \right]\left( {ka} \right){K_m}\left( {ka} \right){I_m}^\prime \left( {ka} \right)}}
\label{VIB5.4}
\end{equation}
is the multipole polarizability for a cylinder in vacuum, Eq.\,(\ref{VIA8}). See Ref.\,[\onlinecite{Ser}] Eq.\,(5.77). Note that in this section we defined the density of the gas shell so that the shell contained a single atom. Then we did not have to divide the energy with the number of atoms. The force on the atom is ${\bf{F}}\left( b \right) =  - {\bf{\hat r}}{\rm d}E(b)/{\rm d}b$.
\subsubsection{\label{atom-cylinderCavityn}Force on an atom inside a cylindrical cavity (two layers)}
In this section we derive the van der Waals interaction between a polarizable atom inside an infinitely long cylindrical vacuum cavity in a medium of dielectric function ${{\tilde \varepsilon }_1}\left( \omega  \right)$. The geometry of the problem is shown in Fig.\,\ref{figu13}. The radius of the cavity is $a$ and the atom is at the distance $d$ from the cylinder axis. To obtain the results we proceed as follows. We introduce a thin shell defined by the radii $d$ and $d+\delta$. We let the medium of the shell have the dielectric function $\varepsilon_g  = 1 + \alpha  = 1 + 4\pi n{\alpha ^{at}} = 1 + 4\pi {\alpha ^{at}}/\left( {2\pi d\delta L} \right) = 1 + 2{\alpha ^{at}}/\left( {d\delta L} \right)$ where ${\alpha ^{at}}$ is the polarizability of the atom and $L$ is the length of the cylinder which we let go to infinity at the end. We assume that the medium of the shell is very diluted. We let $\alpha $ go towards zero and keep only terms up to linear order before we let $\delta $ go towards zero. 

The matrix of the problem is just the matrix of the cavity, Eq.\,(\ref{VIB1.1}), multiplying that for the diluted gas shell, given in Eq.\,(\ref{VIB3.4}), ${\bf{\tilde M}} = {{\bf{\tilde M}}_{{\rm{cavity}}}} \cdot {{\bf{\tilde M}}_{{\rm{shell}}}}$. The element of interest is
\begin{equation}
{M_{11}} = M_{11}^{{\rm{cavity}}}M_{11}^{{\rm{shell}}} + M_{12}^{{\rm{cavity}}}M_{21}^{{\rm{shell}}}.
\label{VIB6.1}
\end{equation}
The mode condition function becomes
\begin{equation}
\begin{array}{*{20}{l}}
{{{\tilde f}_{k,m}} = 1 + \frac{{M_{12}^{{\rm{cavity}}}}}{{M_{11}^{{\rm{cavity}}}}}\frac{{M_{21}^{{\rm{shell}}}}}{{M_{11}^{{\rm{shell}}}}}}\\
{\quad \quad  = 1 + \alpha _{k,m}^{{\rm{cavity}}\left( 2 \right)}M_{21}^{{\rm{shell}}}}\\
{ = 1 - \frac{{\left( {1 - {{\tilde \varepsilon }_1}} \right){K_m}\left( {ka} \right){K_m}^\prime \left( {ka} \right)}}{{{I_m}^\prime \left( {ka} \right){K_m}\left( {ka} \right) - {{\tilde \varepsilon }_1}{I_m}\left( {ka} \right){K_m}^\prime \left( {ka} \right)}}}\\
{ \times \left( {\delta n} \right)4\pi {\alpha ^{at}}\frac{{\left[ {{m^2} + {{\left( {kd} \right)}^2}} \right]{{\left[ {{K_m}\left( {kd} \right)} \right]}^2} + {{\left( {kd} \right)}^2}{{\left[ {{K_m}^\prime \left( {kd} \right)} \right]}^2}}}{d},}
\end{array}
\label{VIB6.2}
\end{equation}
where we have taken as reference system a system where the gas shell and the cavity are well separated from each other.

Now, the non-retarded (van der Waals) interaction energy for an atom inside a cylindrical cavity is given by
\begin{equation}
\begin{array}{*{20}{l}}
{E = \hbar \int\limits_0^\infty  {\frac{{d\xi }}{{2\pi }}} \sum\limits_{m =  - \infty }^\infty  {L\int\limits_{ - \infty }^\infty  {\frac{{dk}}{{2\pi }}} \ln \left[ {{{\tilde f}_{k,m}}\left( {k,i\xi } \right)} \right]} }\\
\begin{array}{l}
 = \hbar L\\
 \times \int\limits_0^\infty  {\frac{{d\xi }}{{2\pi }}} \sum\limits_{m =  - \infty }^\infty  {\int\limits_{ - \infty }^\infty  {\frac{{dk}}{{2\pi }}} \ln \left[ {1 + M_{21}^{{\rm{gasl}}.}\left( {k,i\xi } \right)\alpha _{k,m}^{\left( 2 \right)}\left( {a;i\xi } \right)} \right]} 
\end{array}\\
{ \approx \frac{{2\hbar L}}{d}\int\limits_0^\infty  {\frac{{d\xi }}{{2\pi }}} \sum\limits_{m =  - \infty }^\infty  {\int\limits_0^\infty  {\frac{{dk}}{{2\pi }}} M_{21}^{{\rm{gasl}}.}\left( {k,i\xi } \right)\alpha _{k,m}^{\left( 2 \right)}\left( {a;i\xi } \right)} }\\
{ = \frac{{ - 2\hbar L}}{d}\int\limits_0^\infty  {\frac{{d\xi }}{{2\pi }}} \sum\limits_{m =  - \infty }^\infty  {\int\limits_0^\infty  {\frac{{dk}}{{2\pi }}} } \left( {\delta n} \right)4\pi {\alpha ^{at}}\left( {i\xi } \right)\alpha _{k,m}^{\left( 2 \right)}\left( {a;i\xi } \right)}\\
{\quad  \times \left\{ {\left[ {{m^2} + {{\left( {kd} \right)}^2}} \right]{{\left[ {{K_m}\left( {kd} \right)} \right]}^2} + {{\left( {kd} \right)}^2}{{\left[ {{K_m}^\prime \left( {kd} \right)} \right]}^2}} \right\}}\\
{ =  - \frac{{4\hbar }}{{{d^2}}}\int\limits_0^\infty  {\frac{{d\xi }}{{2\pi }}} \sum\limits_{m =  - \infty }^\infty  {\int\limits_0^\infty  {\frac{{dk}}{{2\pi }}} } {\alpha ^{at}}\left( {i\xi } \right)\alpha _{k,m}^{\left( 2 \right)}\left( {a;i\xi } \right)}\\
{\quad  \times \left\{ {\left[ {{m^2} + {{\left( {kd} \right)}^2}} \right]{{\left[ {{K_m}\left( {kd} \right)} \right]}^2} + {{\left( {kd} \right)}^2}{{\left[ {{K_m}^\prime \left( {kd} \right)} \right]}^2}} \right\},}
\end{array}
\label{VIB6.3}
\end{equation}
where
\begin{equation}
\begin{array}{l}
\alpha _{k,m}^{\left( 2 \right)}\left( {a;i\xi } \right)\\
\quad \quad  = \frac{{\left( {1 - {{\tilde \varepsilon }_1}\left( {i\xi } \right)} \right){K_m}\left( {ka} \right){K_m}^\prime \left( {ka} \right)}}{{{I_m}^\prime \left( {ka} \right){K_m}\left( {ka} \right) - {{\tilde \varepsilon }_1}\left( {i\xi } \right){I_m}\left( {ka} \right){K_m}^\prime \left( {ka} \right)}}
\end{array}
\label{VIB6.4}
\end{equation}
is the multipole polarizability for inside a cylindrical vacuum cavity, Eq.\,(\ref{VIA9}). Note that in this section we defined the density of the gas shell so that the shell contained a single atom. Then we did not have to divide the energy with the number of atoms. The force on the atom is ${\bf{F}}\left( d \right) =  - {\bf{\hat r}}{\rm d}E(d)/{\rm d}d$.
\subsubsection{\label{atomcylindricalgapn}Force on an atom in a cylindrical gap (three layers)}
Here we study an atom in a cylindrical vacuum gap with the outer and inner radii $b$ and $a$, respectively. The medium outside the gap has dielectric function ${{\tilde \varepsilon }_1}\left( \omega  \right)$ and the medium inside the dielectric function  ${{\tilde \varepsilon }_2}\left( \omega  \right)$.  The atom is at the distance $r$ from the center. The matrix for this geometry is ${\bf{\tilde M}} = {{{\bf{\tilde M}}}_{{\rm{cavity}}}} \cdot {{{\bf{\tilde M}}}_{{\rm{shell}}}} \cdot {{{\bf{\tilde M}}}_{{\rm{core}}}}$, and the matrix element of interest is
\begin{equation}
\begin{array}{l}
{M_{11}} = \left( {M_{11}^{cav},M_{12}^{cav}} \right)\\
 \cdot \left( {\begin{array}{*{20}{c}}
{M_{11}^{shell}}&{M_{12}^{shell}}\\
{M_{21}^{shell}}&{M_{22}^{shell}}
\end{array}} \right) \cdot \left( {\begin{array}{*{20}{c}}
{M_{11}^{core}}\\
{M_{21}^{core}}
\end{array}} \right)\\
 = M_{11}^{cav}M_{11}^{shell}M_{11}^{core}\left( {1,\alpha _{k,m}^{{\rm{cav}}\left( {\rm{2}} \right)}} \right)\\
 \cdot \left( {\begin{array}{*{20}{c}}
1&{\alpha _{k,m}^{{\rm{shell}}\left( {\rm{2}} \right)}}\\
{ - \alpha _{k,m}^{shell}}&{\frac{{M_{22}^{{\rm{shell}}}}}{{M_{11}^{{\rm{shell}}}}}}
\end{array}} \right) \cdot \left( {\begin{array}{*{20}{c}}
1\\
{ - \alpha _{k,m}^{{\rm{core}}}}
\end{array}} \right)\\
 = M_{11}^{cav}M_{11}^{shell}M_{11}^{core}\left[ {1 - \alpha _{k,m}^{core}\alpha _{k,m}^{cav\left( 2 \right)} - \alpha _{k,m}^{shell}\alpha _{k,m}^{cav\left( 2 \right)}} \right.\\
\left. { - \alpha _{k,m}^{core}\left( {\alpha _{k,m}^{shell\left( 2 \right)} + \alpha _{k,m}^{cav\left( 2 \right)}\left( {\frac{{M_{22}^{shell}}}{{M_{11}^{shell}}} - 1} \right)} \right)} \right].
\end{array}
\label{VIB7.1}
\end{equation}
This leads to the following proper mode condition function
\begin{equation}
\begin{array}{*{20}{l}}
\begin{array}{l}
{{\tilde f}_{k,m}}\\
 = 1 - \frac{{\alpha _{k,m}^{shell}\alpha _{k,m}^{cav\left( 2 \right)} + \alpha _{k,m}^{core}\left( {\alpha _{k,m}^{shell\left( 2 \right)} + \alpha _{k,m}^{cav\left( 2 \right)}\left( {\frac{{M_{22}^{shell}}}{{M_{11}^{shell}}} - 1} \right)} \right)}}{{1 - \alpha _{k,m}^{core}\alpha _{k,m}^{cav\left( 2 \right)}}}
\end{array}\\
{ \approx 1 - \frac{{\alpha _{k,m}^{shell}\alpha _{k,m}^{cav\left( 2 \right)} + \alpha _{k,m}^{core}\left( {\alpha _{k,m}^{shell\left( 2 \right)} + 2\alpha _{k,m}^{cav\left( 2 \right)}\left( {M_{22}^{shell} - 1} \right)} \right)}}{{1 - \alpha _{k,m}^{core}\alpha _{k,m}^{cav\left( 2 \right)}}},}
\end{array}
\label{VIB7.2}
\end{equation}
where the reference system is the cylindrical gap in absence of the atom. The functions appearing in the expression are
\begin{equation}
\begin{array}{*{20}{l}}
{\alpha _{k,m}^{shell} \approx \left( {\delta n} \right)\frac{{4\pi {\alpha ^{at}}\left( \omega  \right)}}{r}}\\
{\quad \quad  \times \left\{ {\left[ {{m^2} + {{\left( {kr} \right)}^2}} \right]{{\left[ {{I_m}\left( {kr} \right)} \right]}^2} + {{\left( {kr} \right)}^2}{{\left[ {{I_m}^\prime \left( {kr} \right)} \right]}^2}} \right\};}\\
{\alpha _{k,m}^{shell\left( 2 \right)} \approx \left( {\delta n} \right)\frac{{4\pi {\alpha ^{at}}\left( \omega  \right)}}{r}}\\
{\quad \quad  \times \left\{ {\left[ {{m^2} + {{\left( {kr} \right)}^2}} \right]{{\left[ {{K_m}\left( {kr} \right)} \right]}^2} + {{\left( {kr} \right)}^2}{{\left[ {{K_m}^\prime \left( {kr} \right)} \right]}^2}} \right\};}\\
{M_{22}^{shell} - 1 \approx  - \left( {\delta n} \right)\frac{{4\pi {\alpha ^{at}}\left( \omega  \right)}}{r}}\\
{\quad \quad  \times \left\{ {\left[ {{m^2} + {{\left( {kr} \right)}^2}} \right]{I_m}\left( {kr} \right){K_m}\left( {kr} \right)} \right.}\\
{\left. {\quad \quad \quad \quad  + {{\left( {kr} \right)}^2}{I_m}^\prime \left( {kr} \right){K_m}^\prime \left( {kr} \right)} \right\};}\\
\begin{array}{l}
\alpha _{k,m}^{cav\left( 2 \right)} = \alpha _{k,m}^{\left( 2 \right)}\left( {b;\omega } \right)\\
\quad \quad  = \frac{{\left[ {1 - {{\tilde \varepsilon }_1}\left( \omega  \right)} \right]{K_m}\left( {kb} \right){K_m}^\prime \left( {kb} \right)}}{{{I_m}^\prime \left( {kb} \right){K_m}\left( {kb} \right) - {{\tilde \varepsilon }_1}\left( \omega  \right){I_m}\left( {kb} \right){K_m}^\prime \left( {kb} \right)}};
\end{array}\\
\begin{array}{l}
\alpha _{k,m}^{core} = {\alpha _{k,m}}\left( {a;\omega } \right)\\
\quad \quad  = \frac{{\left[ {{{\tilde \varepsilon }_2}\left( \omega  \right) - 1} \right]\left( {kb} \right){I_m}\left( {kb} \right){I_m}^\prime \left( {kb} \right)}}{{1 + \left[ {{{\tilde \varepsilon }_2}\left( \omega  \right) - 1} \right]\left( {kb} \right){K_m}\left( {kb} \right){I_m}^\prime \left( {kb} \right)}}.
\end{array}
\end{array}
\label{VIB7.3}
\end{equation}
Before we write down the expression for the energy per atom  we make the factor $\left( {\delta n} \right)4\pi {\alpha ^{at}}\left( \omega  \right)$ explicit, a factor that is common for all terms after $1-$ in the expression for ${{\tilde f}_{k,m}}$. We have
\begin{equation}
\begin{array}{l}
{{\tilde f}_{k,m}} \approx 1 - \left( {\delta n} \right)\frac{{4\pi {\alpha ^{at}}\left( \omega  \right)}}{{r\left[ {1 - \alpha _{k,m}^{core}\alpha _{k,m}^{cav\left( 2 \right)}} \right]}}\left\{ {\left[ {{m^2} + {{\left( {kr} \right)}^2}} \right]} \right.\\
\quad \quad \quad  \times \left[ {\alpha _{k,m}^{cav\left( 2 \right)}{{\left[ {{I_m}\left( {kr} \right)} \right]}^2} + \alpha _{k,m}^{core}{{\left[ {{K_m}\left( {kr} \right)} \right]}^2}} \right.\\
\left. {\quad \quad \quad \quad \quad \quad  - 2\alpha _{k,m}^{core}\alpha _{k,m}^{cav\left( 2 \right)}{I_m}\left( {kr} \right){K_m}\left( {kr} \right)} \right]\\
\quad  + {\left( {kr} \right)^2}\left[ {\alpha _{k,m}^{cav\left( 2 \right)}{{\left[ {{I_m}'\left( {kr} \right)} \right]}^2} + \alpha _{k,m}^{core}{{\left[ {{K_m}'\left( {kr} \right)} \right]}^2}} \right.\\
\left. {\left. {\quad \quad \quad \quad \quad \quad  - 2\alpha _{k,m}^{core}\alpha _{k,m}^{cav\left( 2 \right)}{I_m}'\left( {kr} \right){K_m}'\left( {kr} \right)} \right]} \right\}.
\end{array}
\label{VIB7.4}
\end{equation}
Now, the energy per atom is
\begin{equation}
\begin{array}{*{20}{l}}
\begin{array}{l}
\frac{E}{{2\pi \left( {\delta n} \right)L}}\\
 = \frac{\hbar }{{2\pi \left( {\delta n} \right)L}}\int\limits_0^\infty  {\frac{{d\xi }}{{2\pi }}} \sum\limits_{m =  - \infty }^\infty  {L\int\limits_{ - \infty }^\infty  {\frac{{dk}}{{2\pi }}} \ln \left[ {{{\tilde f}_{k,m}}\left( {k,i\xi } \right)} \right]} 
\end{array}\\
{ \approx \frac{\hbar }{{2\pi \left( {\delta n} \right)}}\int\limits_0^\infty  {\frac{{d\xi }}{{2\pi }}} \sum\limits_{m =  - \infty }^\infty  {\int\limits_{ - \infty }^\infty  {\frac{{dk}}{{2\pi }}} \left[ {{{\tilde f}_{k,m}}\left( {k,i\xi } \right) - 1} \right]} }\\
{ =  - \frac{{2\hbar }}{r}\int\limits_0^\infty  {\frac{{d\xi }}{{2\pi }}} \sum\limits_{m =  - \infty }^\infty  {\int\limits_{ - \infty }^\infty  {\frac{{dk}}{{2\pi }}} \frac{{{\alpha ^{at}}\left( {i\xi } \right)}}{{\left[ {1 - \alpha _{k,m}^{core}\alpha _{k,m}^{cav\left( 2 \right)}} \right]}}} }\\
{\quad \quad \quad  \times \left\{ {\left[ {{m^2} + {{\left( {kr} \right)}^2}} \right]} \right.\left[ {\alpha _{k,m}^{cav\left( 2 \right)}{{\left[ {{I_m}\left( {kr} \right)} \right]}^2}} \right.}\\
{\left. { + \alpha _{k,m}^{core}{{\left[ {{K_m}\left( {kr} \right)} \right]}^2} - 2\alpha _{k,m}^{core}\alpha _{k,m}^{cav\left( 2 \right)}{I_m}\left( {kr} \right){K_m}\left( {kr} \right)} \right]}\\
{ + {{\left( {kr} \right)}^2}\left[ {\alpha _{k,m}^{cav\left( 2 \right)}{{\left[ {{I_m}^\prime \left( {kr} \right)} \right]}^2} + \alpha _{k,m}^{core}{{\left[ {{K_m}^\prime \left( {kr} \right)} \right]}^2}} \right.}\\
{\left. {\left. {\quad \quad \quad \quad \quad \quad  - 2\alpha _{k,m}^{core}\alpha _{k,m}^{cav\left( 2 \right)}{I_m}^\prime \left( {kr} \right){K_m}\left( {kr} \right)} \right]} \right\}.}
\end{array}
\label{VIB7.5}
\end{equation}
The force on the atom is ${\bf{F}}\left( r \right) =  - {\bf{\hat r}}{\rm d}E(r)/{\rm d}r$.
\subsubsection{\label{atom-outsidecylindershellnonret}Force on an atom outside a 2D cylindrical shell. (three layers)}

In this section we derive the interaction between an atom and a very thin cylindrical shell, Fig.\,\ref{figu20}. It could approximate the interaction between an atom and a nano tube. We let the shell have the thickness $\delta$ and let $\delta$ be very small so that one keeps only terms linear in $\delta$. The 3D dielectric function of the material will then be inversely proportional to  $\delta$.\,\cite{grap,arx} The derivation proceeds along the lines in Sec.\,\ref{atom-cylindern} and the matrix ${{{\bf{\tilde M}}}^{core}}$ is replaced by ${{{\bf{\tilde M}}}^{{\rm{2D}}}}$. The matrix of the problem is just the matrix of the thin gas shell, Eq.\,(\ref{VIB3.4}), multiplying that for the 2D shell, given in Eq.\,(\ref{VIB4.7}), ${\bf{\tilde M}} = {{\bf{\tilde M}}_{{\rm{shell}}}} \cdot {{\bf{\tilde M}}_{{\rm{2D}}}}$. Note that the radius of the 2D cylindrical film is $a$ and the atom is at the distance $d$ from the film and distance $b$ from the cylinder axis. The element of interest is
\begin{equation}
{M_{11}} = M_{11}^{{\rm{shell}}}M_{11}^{{\rm{2D}}} + M_{12}^{{\rm{shell}}}M_{21}^{{\rm{2D}}}.
\label{VIB8.1}
\end{equation}
The mode condition function becomes
\begin{equation}
{{\tilde f}_{k,m}} = 1 + \frac{{M_{12}^{{\rm{shell}}}}}{{M_{11}^{{\rm{shell}}}}}\frac{{M_{21}^{{\rm{2D}}}}}{{M_{11}^{{\rm{2D}}}}} \approx 1 - M_{12}^{{\rm{shell}}}\alpha _{k,m}^{{\rm{2D}}},
\label{VIB8.2}
\end{equation}
where
\begin{equation}
\begin{array}{l}
M_{12}^{{\rm{shell}}} = \left( {\delta n} \right)\frac{{4\pi {\alpha ^{at}}\left( \omega  \right)}}{b}\left\{ {\left[ {{m^2} + {{\left( {kb} \right)}^2}} \right]} \right.{\left[ {{K_m}\left( {kb} \right)} \right]^2}\\
\left. {\quad \quad \quad \quad  + {{\left( {kb} \right)}^2}{{\left[ {{K_m}'\left( {kb} \right)} \right]}^2}} \right\};\\
\alpha _{k,m}^{2D} = \frac{{\delta {{\tilde \varepsilon }^{{\rm{3D}}}}\left( \omega  \right)\left[ {{m^2} + {{\left( {ka} \right)}^2}} \right]{I_m}\left( {ka} \right){I_m}\left( {ka} \right)}}{{a + \delta {{\tilde \varepsilon }^{{\rm{3D}}}}\left( \omega  \right)\left[ {{m^2} + {{\left( {ka} \right)}^2}} \right]{I_m}\left( {ka} \right){K_m}\left( {ka} \right)}}.
\end{array}
\label{VIB8.3}
\end{equation}
Now, the non-retarded (van der Waals) interaction energy between an atom and a 2D cylinder shell is given by
\begin{equation}
\begin{array}{l}
E \approx  - \frac{{2\hbar }}{{2\pi b\delta Ln}}\int\limits_0^\infty  {\frac{{d\xi }}{{2\pi }}} \sum\limits_{m =  - \infty }^\infty  {L\int\limits_0^\infty  {\frac{{dk}}{{2\pi }}} M_{12}^{{\rm{shell}}}\alpha _{k,m}^{{\rm{2D}}}} \\
\quad  =  - \frac{{4\hbar }}{{{b^2}}}\int\limits_0^\infty  {\frac{{d\xi }}{{2\pi }}} \sum\limits_{m =  - \infty }^\infty  {\int\limits_0^\infty  {\frac{{dk}}{{2\pi }}} {\alpha ^{at}}\left( {i\xi } \right)} \\
\quad \quad  \times \left\{ {\left[ {{m^2} + {{\left( {kb} \right)}^2}} \right]{{\left[ {{K_m}\left( {kb} \right)} \right]}^2} + {{\left( {kb} \right)}^2}{{\left[ {{K_m}'\left( {kb} \right)} \right]}^2}} \right\}\\
\quad \quad \quad \quad  \times \frac{{\delta {{\tilde \varepsilon }^{{\rm{3D}}}}\left( {i\xi } \right)\left[ {{m^2} + {{\left( {ka} \right)}^2}} \right]{I_m}\left( {ka} \right){I_m}\left( {ka} \right)}}{{a + \delta {{\tilde \varepsilon }^{{\rm{3D}}}}\left( {i\xi } \right)\left[ {{m^2} + {{\left( {ka} \right)}^2}} \right]{I_m}\left( {ka} \right){K_m}\left( {ka} \right)}}.
\end{array}
\label{VIB8.4}
\end{equation}
Two examples where the results apply are a cylinder made of a graphene like film and a thin metal film, respectively. Then the expressions for $\delta {\tilde \varepsilon }\left( {i\xi } \right)$ as given in Eq.\,(\ref{VB11.4}) can be used.\,\cite{grap,arx} The force on the atom is ${\bf{F}}\left( b \right) =  - {\bf{\hat r}}{\rm d}E(b)/{\rm d}b$. 
\subsubsection{\label{atom-insidecylindershellnonret}Force on an atom inside a 2D cylindrical shell. (three layers)}
In this section we derive the interaction of an atom inside a very thin cylindrical shell, Fig.\,\ref{figu21}. It could approximate the interaction of an atom inside a nano tube. We let the shell have the thickness $\delta$ and let $\delta$ be very small so that one keeps only terms linear in $\delta$. The 3D dielectric function of the material will then be inversely proportional to  $\delta$.\,\cite{grap,arx} The derivation proceeds along the lines in the previous section and the matrix of the problem is just the matrix of the 2D shell, Eq.\,(\ref{VIB4.7}), multiplying that for the  thin gas shell, given in Eq.\,(\ref{VIB3.4}), ${\bf{\tilde M}} = {{\bf{\tilde M}}_{{\rm{2D}}}}\cdot  {{\bf{\tilde M}}_{{\rm{shell}}}}$. Note that the radius of the 2D cylindrical film is $a$ and the atom is at the distance $d$ from the cylinder axis. The element of interest is
\begin{equation}
{M_{11}} = M_{11}^{{\rm{2D}}}M_{11}^{{\rm{shell}}} + M_{12}^{{\rm{2D}}}M_{21}^{{\rm{shell}}}.
\label{VIB9.1}
\end{equation}
The mode condition function becomes
\begin{equation}
{\tilde f_{k,m}} = 1 + \frac{{M_{12}^{{\rm{2D}}}}}{{M_{11}^{{\rm{2D}}}}}\frac{{M_{21}^{{\rm{shell}}}}}{{M_{11}^{{\rm{shell}}}}} \approx 1 + \alpha _{k,m}^{{\rm{2D}}\left( 2 \right)}M_{21}^{{\rm{shell}}},
\label{VIB9.2}
\end{equation}
where
\begin{equation}
\begin{array}{l}
M_{21}^{{\rm{shell}}} =  - \left( {\delta n} \right)\frac{{4\pi {\alpha ^{at}}\left( \omega  \right)}}{d}\\
\quad \quad  \times \left\{ {\left[ {{m^2} + {{\left( {kd} \right)}^2}} \right]{{\left[ {{I_m}\left( {kd} \right)} \right]}^2} + {{\left( {kd} \right)}^2}{{\left[ {{I_m}^\prime \left( {kd} \right)} \right]}^2}} \right\};\\
\alpha _{k,m}^{2D\left( 2 \right)} = \frac{{\delta {{\tilde \varepsilon }^{{\rm{3D}}}}\left( \omega  \right)\left[ {{m^2} + {{\left( {ka} \right)}^2}} \right]{K_m}\left( {ka} \right){K_m}\left( {ka} \right)}}{{a + \delta {{\tilde \varepsilon }^{{\rm{3D}}}}\left( \omega  \right)\left[ {{m^2} + {{\left( {ka} \right)}^2}} \right]{I_m}\left( {ka} \right){K_m}\left( {ka} \right)}}.
\end{array}
\label{VIB9.3}
\end{equation}
Now, the non-retarded (van der Waals) interaction energy of an atom inside a thin cylindrical shell is given by
\begin{equation}
\begin{array}{*{20}{l}}
{E \approx \frac{{2\hbar }}{{2\pi d\delta Ln}}\int\limits_0^\infty  {\frac{{d\xi }}{{2\pi }}} \sum\limits_{m =  - \infty }^\infty  {L\int\limits_0^\infty  {\frac{{dk}}{{2\pi }}} \alpha _{k,m}^{{\rm{2D}}\left( 2 \right)}M_{21}^{{\rm{shell}}}} }\\
{\quad  =  - \frac{{4\hbar }}{{{d^2}}}\int\limits_0^\infty  {\frac{{d\xi }}{{2\pi }}} \sum\limits_{m =  - \infty }^\infty  {\int\limits_0^\infty  {\frac{{dk}}{{2\pi }}} {\alpha ^{at}}\left( {i\xi } \right)} }\\
{\quad \quad  \times \left\{ {\left[ {{m^2} + {{\left( {kd} \right)}^2}} \right]{{\left[ {{I_m}\left( {kd} \right)} \right]}^2} + {{\left( {kd} \right)}^2}{{\left[ {{I_m}^\prime \left( {kd} \right)} \right]}^2}} \right\}}\\
{\quad \quad \quad \quad  \times \frac{{\delta {{\tilde \varepsilon }^{{\rm{3D}}}}\left( {i\xi } \right)\left[ {{m^2} + {{\left( {ka} \right)}^2}} \right]{I_m}\left( {ka} \right){I_m}\left( {ka} \right)}}{{a + \delta {{\tilde \varepsilon }^{{\rm{3D}}}}\left( {i\xi } \right)\left[ {{m^2} + {{\left( {ka} \right)}^2}} \right]{I_m}\left( {ka} \right){K_m}\left( {ka} \right)}}.}
\end{array}
\label{VIB9.4}
\end{equation}
Two examples where the results apply are a cylinder made of a graphene like film and a thin metal film, respectively. Then the expressions for $\delta {\tilde \varepsilon }\left( {i\xi } \right)$ as given in Eq.\,(\ref{VB11.4}) can be used.\,\cite{grap,arx} The force on the atom is ${\bf{F}}\left( d \right) =  - {\bf{\hat r}}{\rm d}E(d)/{\rm d}d$.
\subsubsection{\label{twoscylindricalfilmsn}Interaction between two 2D cylindrical shells (three layers)}
We consider two coaxial thin cylindrical shells, one outer of radius $b$ and one inner of radius $a$. The matrix for the system is ${\bf{\tilde M}} = {{\bf{\tilde M}}_{{\rm{2Do}}}} \cdot {{\bf{\tilde M}}_{{\rm{2Di}}}}$ and the matrix of interest is
\begin{equation}
\begin{array}{l}
{M_{11}} = M_{11}^{{\rm{2Do}}}M_{11}^{{\rm{2Di}}} + M_{12}^{{\rm{2Do}}}M_{21}^{{\rm{2Di}}}\\
\quad \quad  = M_{11}^{{\rm{2Do}}}M_{11}^{{\rm{2Di}}}\left( {1 - \alpha _{k,m}^{{\rm{2Do}}\left( 2 \right)}\alpha _{k,m}^{{\rm{2Di}}}} \right),
\end{array}
\label{VIB10.1}
\end{equation}
where all appearing functions are
\begin{equation}
\begin{array}{*{20}{l}}
{M_{11}^{{\rm{2Do}}} = 1 + \delta {{\tilde \varepsilon }^{{\rm{3D}}}}\left( \omega  \right)\frac{{\left[ {{m^2} + {{\left( {kb} \right)}^2}} \right]{I_m}\left( {kb} \right){K_m}\left( {kb} \right)}}{b};}\\
{M_{11}^{{\rm{2Di}}} = 1 + \delta {{\tilde \varepsilon }^{{\rm{3D}}}}\left( \omega  \right)\frac{{\left[ {{m^2} + {{\left( {ka} \right)}^2}} \right]{I_m}\left( {ka} \right){K_m}\left( {ka} \right)}}{a};}\\
{M_{12}^{{\rm{2Do}}} = \delta {{\tilde \varepsilon }^{{\rm{3D}}}}\left( \omega  \right)\frac{{\left[ {{m^2} + {{\left( {kb} \right)}^2}} \right]{{\left[ {{K_m}\left( {kb} \right)} \right]}^2}}}{b};}\\
{M_{21}^{{\rm{2Di}}} =  - \delta {{\tilde \varepsilon }^{{\rm{3D}}}}\left( \omega  \right)\frac{{\left[ {{m^2} + {{\left( {ka} \right)}^2}} \right]{{\left[ {{I_m}\left( {ka} \right)} \right]}^2}}}{a};}\\
\begin{array}{l}
\alpha _{k,m}^{{\rm{2Do}}\left( 2 \right)} = \alpha _{k,m}^{{\rm{2D}}\left( 2 \right)}\left( {b;\omega } \right)\\
\quad \quad  = \frac{{\delta {{\tilde \varepsilon }^{{\rm{3D}}}}\left( \omega  \right)\left[ {{m^2} + {{\left( {kb} \right)}^2}} \right]{{\left[ {{K_m}\left( {kb} \right)} \right]}^2}}}{{b + \delta {{\tilde \varepsilon }^{{\rm{3D}}}}\left( \omega  \right)\left[ {{m^2} + {{\left( {kb} \right)}^2}} \right]{I_m}\left( {kb} \right){K_m}\left( {kb} \right)}};
\end{array}\\
\begin{array}{l}
\alpha _{k,m}^{{\rm{2Di}}} = \alpha _{k,m}^{{\rm{2D}}}\left( {a;\omega } \right)\\
\quad \quad  = \frac{{\delta {{\tilde \varepsilon }^{{\rm{3D}}}}\left( \omega  \right)\left[ {{m^2} + {{\left( {ka} \right)}^2}} \right]{{\left[ {{I_m}\left( {ka} \right)} \right]}^2}}}{{a + \delta {{\tilde \varepsilon }^{{\rm{3D}}}}\left( \omega  \right)\left[ {{m^2} + {{\left( {ka} \right)}^2}} \right]{I_m}\left( {ka} \right){K_m}\left( {ka} \right)}}.
\end{array}
\end{array}
\label{VIB10.2}
\end{equation}
If one wants to find the electromagnetic normal modes of the system one finds the solutions to ${M_{11}} = 0$.  If one wants to find the energy it takes to bring the two thin cylindrical shells from infinite separation together and place the inner inside the outer one uses the proper mode condition function,
\begin{equation}	
{{\tilde f}_{k,m}}\left( {i\xi } \right) = 1 - \alpha _{k,m}^{{\rm{2D}}\left( 2 \right)}\left( {b;i\xi } \right)\alpha _{k,m}^{{\rm{2D}}}\left( {a;i\xi } \right).
\label{VIB10.3}
\end{equation}
The energy per per unit length is
\begin{equation}
\begin{array}{l}
E = 2\hbar \int\limits_0^\infty  {\frac{{d\xi }}{{2\pi }}} \sum\limits_{m =  - \infty }^\infty  {\int\limits_0^\infty  {\frac{{dk}}{{2\pi }}} \ln \left[ {{{\tilde f}_{k,m}}\left( {k,i\xi } \right)} \right]} \\
\quad \; = 2\hbar \int\limits_0^\infty  {\frac{{d\xi }}{{2\pi }}} \sum\limits_{m =  - \infty }^\infty  {\int\limits_0^\infty  {\frac{{dk}}{{2\pi }}} \ln \left[ {1 - \alpha _{k,m}^{{\rm{2D}}\left( 2 \right)}\left( {b;i\xi } \right)} \right.} \\
\quad \quad \quad \quad \quad \quad \quad \quad \quad \quad \quad \quad \quad \left. { \times \alpha _{k,m}^{{\rm{2D}}}\left( {a;i\xi } \right)} \right]\\
\quad \; = 2\hbar \int\limits_0^\infty  {\frac{{d\xi }}{{2\pi }}} \sum\limits_{m =  - \infty }^\infty  {\int\limits_0^\infty  {\frac{{dk}}{{2\pi }}} \ln \left[ {1 - } \right.} \\
\quad \quad \quad \quad \quad \quad \quad \frac{{\delta {{\tilde \varepsilon }^{{\rm{3D}}}}\left( \omega  \right)\left[ {{m^2} + {{\left( {kb} \right)}^2}} \right]{{\left[ {{K_m}\left( {kb} \right)} \right]}^2}}}{{b + \delta {{\tilde \varepsilon }^{{\rm{3D}}}}\left( \omega  \right)\left[ {{m^2} + {{\left( {kb} \right)}^2}} \right]{I_m}\left( {kb} \right){K_m}\left( {kb} \right)}}\\
\left. {\quad \quad \quad \quad \quad \quad \quad  \times \frac{{\delta {{\tilde \varepsilon }^{{\rm{3D}}}}\left( \omega  \right)\left[ {{m^2} + {{\left( {ka} \right)}^2}} \right]{{\left[ {{I_m}\left( {ka} \right)} \right]}^2}}}{{a + \delta {{\tilde \varepsilon }^{{\rm{3D}}}}\left( \omega  \right)\left[ {{m^2} + {{\left( {ka} \right)}^2}} \right]{I_m}\left( {ka} \right){K_m}\left( {ka} \right)}}} \right].
\end{array}
\label{VIB10.4}
\end{equation}
\subsubsection{\label{atom2cylindricalfilmsn}Force on an atom in between two 2D cylindrical films (five layers)}
Here we study an atom in a cylindrical vacuum gap between two 2D cylindrical films with the outer and inner radii $b$ and $a$, respectively. The ambient medium in which the films and the atom reside is vacuum.  The atom is at the distance $r$ from the center. Here we may make use of the results  in Sec.\,\ref{atomcylindricalgapn}. The matrix for this geometry is the product of those for the outer 2D film, the gas shell and the inner 2D film,  ${\bf{\tilde M}} = {{{\bf{\tilde M}}}_{{\rm{2Do}}}} \cdot {{{\bf{\tilde M}}}_{{\rm{shell}}}} \cdot {{{\bf{\tilde M}}}_{{\rm{2Di}}}}$, and the matrix element of interest is
\begin{equation}
\begin{array}{*{20}{l}}
{{M_{11}} = \left( {M_{11}^{{\rm{2Do}}},M_{12}^{{\rm{2Do}}}} \right)}\\
{ \cdot \left( {\begin{array}{*{20}{c}}
{M_{11}^{{\rm{shell}}}}&{M_{12}^{{\rm{shell}}}}\\
{M_{21}^{{\rm{shell}}}}&{M_{22}^{{\rm{shell}}}}
\end{array}} \right) \cdot \left( {\begin{array}{*{20}{c}}
{M_{11}^{{\rm{2Di}}}}\\
{M_{21}^{{\rm{2Di}}}}
\end{array}} \right)}\\
{ = M_{11}^{{\rm{2Do}}}M_{11}^{{\rm{shell}}}M_{11}^{{\rm{2D}}i}\left( {1,\alpha _{k,m}^{{\rm{2Do}}\left( {\rm{2}} \right)}} \right)}\\
{ \cdot \left( {\begin{array}{*{20}{c}}
1&{\alpha _{k,m}^{{\rm{shell}}\left( 2 \right)}}\\
{ - \alpha _{k,m}^{{\rm{shell}}}}&{\frac{{M_{22}^{{\rm{shell}}}}}{{M_{11}^{{\rm{shell}}}}}}
\end{array}} \right) \cdot \left( {\begin{array}{*{20}{c}}
1\\
{ - \alpha _{k,m}^{{\rm{2Di}}}}
\end{array}} \right)}\\
{ = M_{11}^{{\rm{2Do}}}M_{11}^{{\rm{shell}}}M_{11}^{{\rm{2Di}}}\left[ {1 - \alpha _{k,m}^{{\rm{2Di}}}\alpha _{k,m}^{{\rm{2Do}}\left( 2 \right)} - \alpha _{k,m}^{{\rm{2Di}}}\alpha _{k,m}^{{\rm{2Do}}\left( 2 \right)}} \right.}\\
{\left. { - \alpha _{k,m}^{{\rm{2Di}}}\left( {\alpha _{k,m}^{{\rm{shell}}\left( 2 \right)} + \alpha _{k,m}^{{\rm{2Do}}\left( 2 \right)}\left( {\frac{{M_{22}^{{\rm{shell}}}}}{{M_{11}^{{\rm{shell}}}}} - 1} \right)} \right)} \right].}
\end{array}
\label{VIB11.1}
\end{equation}
This leads to the following proper mode condition function
\begin{equation}
\begin{array}{*{20}{l}}
{{{\tilde f}_{k,m}} = 1 - \frac{{\alpha _{k,m}^{{\rm{shell}}}\alpha _{k,m}^{{\rm{2Do}}\left( 2 \right)} + \alpha _{k,m}^{{\rm{2Di}}}\left[ {\alpha _{k,m}^{{\rm{shell}}\left( 2 \right)} + \alpha _{k,m}^{{\rm{2Do}}\left( 2 \right)}\left( {\frac{{M_{22}^{{\rm{shell}}}}}{{M_{11}^{{\rm{shell}}}}} - 1} \right)} \right]}}{{1 - \alpha _{k,m}^{{\rm{2Di}}}\alpha _{k,m}^{{\rm{2Do}}\left( 2 \right)}}}}\\
{\quad \quad  \approx 1 - \frac{{\alpha _{k,m}^{{\rm{shell}}}\alpha _{k,m}^{{\rm{2Do}}\left( 2 \right)} + \alpha _{k,m}^{{\rm{2Di}}}\left[ {\alpha _{k,m}^{{\rm{shell}}\left( 2 \right)} + 2\alpha _{k,m}^{{\rm{2Do}}\left( 2 \right)}\left( {M_{22}^{{\rm{shell}}} - 1} \right)} \right]}}{{1 - \alpha _{k,m}^{{\rm{2Di}}}\alpha _{k,m}^{{\rm{2Do}}\left( 2 \right)}}},}
\end{array}
\label{VIB11.2}
\end{equation}
where the reference system is a system where all three shells are well separated from each other. The functions appearing in the expression are
\begin{equation}
\begin{array}{*{20}{l}}
{\alpha _{k,m}^{{\rm{shell}}} \approx \left( {\delta n} \right)\frac{{4\pi {\alpha ^{at}}\left( \omega  \right)}}{r}}\\
{\quad \quad  \times \left\{ {\left[ {{m^2} + {{\left( {kr} \right)}^2}} \right]{{\left[ {{I_m}\left( {kr} \right)} \right]}^2} + {{\left( {kr} \right)}^2}{{\left[ {{I_m}^\prime \left( {kr} \right)} \right]}^2}} \right\};}\\
{\alpha _{k,m}^{{\rm{shell}}\left( 2 \right)} \approx \left( {\delta n} \right)\frac{{4\pi {\alpha ^{at}}\left( \omega  \right)}}{r}}\\
{\quad \quad  \times \left\{ {\left[ {{m^2} + {{\left( {kr} \right)}^2}} \right]{{\left[ {{K_m}\left( {kr} \right)} \right]}^2} + {{\left( {kr} \right)}^2}{{\left[ {{K_m}^\prime \left( {kr} \right)} \right]}^2}} \right\};}\\
{M_{22}^{{\rm{shell}}} - 1 \approx  - \left( {\delta n} \right)\frac{{4\pi {\alpha ^{at}}\left( \omega  \right)}}{r}}\\
{\quad \quad  \times \left\{ {\left[ {{m^2} + {{\left( {kr} \right)}^2}} \right]{I_m}\left( {kr} \right){K_m}\left( {kr} \right)} \right.}\\
{\left. {\quad \quad \quad \quad  + {{\left( {kr} \right)}^2}{I_m}^\prime \left( {kr} \right){K_m}^\prime \left( {kr} \right)} \right\};}\\
\begin{array}{l}
\alpha _{k,m}^{{\rm{2Do}}\left( 2 \right)} = \alpha _{k,m}^{{\rm{2D}}\left( 2 \right)}\left( {b;\omega } \right)\\
\quad \quad  = \frac{{\delta {{\tilde \varepsilon }^{{\rm{3D}}}}\left( \omega  \right)\left[ {{m^2} + {{\left( {kb} \right)}^2}} \right]{{\left[ {{K_m}\left( {kb} \right)} \right]}^2}}}{{b + \delta {{\tilde \varepsilon }^{{\rm{3D}}}}\left( \omega  \right)\left[ {{m^2} + {{\left( {kb} \right)}^2}} \right]{I_m}\left( {kb} \right){K_m}\left( {kb} \right)}};
\end{array}\\
\begin{array}{l}
\alpha _{k,m}^{{\rm{2Di}}} = \alpha _{k,m}^{{\rm{2D}}}\left( {a;\omega } \right)\\
\quad \quad  = \frac{{\delta {{\tilde \varepsilon }^{{\rm{3D}}}}\left( \omega  \right)\left[ {{m^2} + {{\left( {ka} \right)}^2}} \right]{{\left[ {{I_m}\left( {ka} \right)} \right]}^2}}}{{a + \delta {{\tilde \varepsilon }^{{\rm{3D}}}}\left( \omega  \right)\left[ {{m^2} + {{\left( {ka} \right)}^2}} \right]{I_m}\left( {ka} \right){K_m}\left( {ka} \right)}}.
\end{array}
\end{array}
\label{VIB11.3}
\end{equation}
Before we write down the expression for the energy per atom  we make the factor $\left( {\delta n} \right)4\pi {\alpha ^{at}}\left( \omega  \right)$ explicit, a factor that is common for all terms after $1-$ in the expression for ${{\tilde f}_{k,m}}$. We have
\begin{equation}
\begin{array}{*{20}{l}}
{{{\tilde f}_{k,m}} \approx 1 - \left( {\delta n} \right)\frac{{4\pi {\alpha ^{at}}\left( \omega  \right)}}{{r\left[ {1 - \alpha _{k,m}^{{\rm{2Di}}}\alpha _{k,m}^{{\rm{2Do}}\left( 2 \right)}} \right]}}\left\{ {\left[ {{m^2} + {{\left( {kr} \right)}^2}} \right]} \right.}\\
{\quad \quad \quad  \times \left[ {\alpha _{k,m}^{{\rm{2Do}}\left( 2 \right)}{{\left[ {{I_m}\left( {kr} \right)} \right]}^2} + \alpha _{k,m}^{{\rm{2Di}}}{{\left[ {{K_m}\left( {kr} \right)} \right]}^2}} \right.}\\
{\left. {\quad \quad \quad \quad \quad \quad  - 2\alpha _{k,m}^{{\rm{2Di}}}\alpha _{k,m}^{{\rm{2Do}}\left( 2 \right)}{I_m}\left( {kr} \right){K_m}\left( {kr} \right)} \right]}\\
{\quad  + {{\left( {kr} \right)}^2}\left[ {\alpha _{k,m}^{{\rm{2Do}}\left( 2 \right)}{{\left[ {{I_m}^\prime \left( {kr} \right)} \right]}^2} + \alpha _{k,m}^{{\rm{2Di}}}{{\left[ {{K_m}^\prime \left( {kr} \right)} \right]}^2}} \right.}\\
{\left. {\left. {\quad \quad \quad \quad \quad \quad  - 2\alpha _{k,m}^{{\rm{2Di}}}\alpha _{k,m}^{{\rm{2Do}}\left( 2 \right)}{I_m}^\prime \left( {kr} \right){K_m}^\prime \left( {kr} \right)} \right]} \right\}.}
\end{array}
\label{VIB11.4}
\end{equation}
Now, the energy per atom is
\begin{equation}
\begin{array}{*{20}{l}}
\begin{array}{l}
\frac{E}{{2\pi \left( {\delta n} \right)L}}\\
 = \frac{\hbar }{{2\pi \left( {\delta n} \right)L}}\int\limits_0^\infty  {\frac{{d\xi }}{{2\pi }}} \sum\limits_{m =  - \infty }^\infty  {L\int\limits_{ - \infty }^\infty  {\frac{{dk}}{{2\pi }}} \ln \left[ {{{\tilde f}_{k,m}}\left( {k,i\xi } \right)} \right]} 
\end{array}\\
{ \approx \frac{\hbar }{{2\pi \left( {\delta n} \right)}}\int\limits_0^\infty  {\frac{{d\xi }}{{2\pi }}} \sum\limits_{m =  - \infty }^\infty  {\int\limits_{ - \infty }^\infty  {\frac{{dk}}{{2\pi }}} \left[ {{{\tilde f}_{k,m}}\left( {k,i\xi } \right) - 1} \right]} }\\
{ =  - \frac{{2\hbar }}{r}\int\limits_0^\infty  {\frac{{d\xi }}{{2\pi }}} \sum\limits_{m =  - \infty }^\infty  {\int\limits_{ - \infty }^\infty  {\frac{{dk}}{{2\pi }}} \frac{{{\alpha ^{at}}\left( {i\xi } \right)}}{{\left[ {1 - \alpha _{k,m}^{{\rm{2Di}}}\alpha _{k,m}^{{\rm{2Do}}\left( 2 \right)}} \right]}}} }\\
{\quad \quad \quad  \times \left\{ {\left[ {{m^2} + {{\left( {kr} \right)}^2}} \right]} \right.\left[ {\alpha _{k,m}^{{\rm{2Do}}\left( 2 \right)}{{\left[ {{I_m}\left( {kr} \right)} \right]}^2}} \right.}\\
{\left. { + \alpha _{k,m}^{{\rm{2Di}}}{{\left[ {{K_m}\left( {kr} \right)} \right]}^2} - 2\alpha _{k,m}^{{\rm{2Di}}}\alpha _{k,m}^{{\rm{2Do}}\left( 2 \right)}{I_m}\left( {kr} \right){K_m}\left( {kr} \right)} \right]}\\
{\quad  + {{\left( {kr} \right)}^2}\left[ {\alpha _{k,m}^{{\rm{2Do}}\left( 2 \right)}{{\left[ {{I_m}^\prime \left( {kr} \right)} \right]}^2} + \alpha _{k,m}^{{\rm{2Di}}}{{\left[ {{K_m}^\prime \left( {kr} \right)} \right]}^2}} \right.}\\
{\left. {\left. {\quad \quad \quad \quad \quad \quad  - 2\alpha _{k,m}^{{\rm{2Di}}}\alpha _{k,m}^{{\rm{2Do}}\left( 2 \right)}{I_m}^\prime \left( {kr} \right){K_m}\left( {kr} \right)} \right]} \right\}.}
\end{array}
\label{VIB11.5}
\end{equation}

The force on the atom is ${\bf{F}}\left( r \right) =  - {\bf{\hat r}}{\rm d}E(r)/{\rm d}r$.

\subsection{\label{cylindricalretmain}Retarded main results}
To find the normal modes for a layered cylinder including retardation effects we need to solve the wave equation for the electric and magnetic fields in all layers and use the proper boundary conditions at the interfaces. To solve the vector wave equation the vector Helmholtz equation, Eq.\,(\ref{III22}), is not a trivial task.  One can instead solve the problem by introducing Hertz-Debye potentials $\pi _1 $ and $\pi _2 $. They are solutions to the scalar wave equation, Eq.\,(\ref{III23}).  We let $\pi _1 $ be the potential that generates TM modes and $\pi _2 $ be the potential that generates TE modes.
Separation of variables, $\pi  = R\left( r \right)\Theta \left( \theta  \right)Z\left( z \right)$, leads to one differential equation for each of the variables, 
\begin{equation}
\begin{array}{*{20}{l}}
{rd{\rm{ /}}dr\left[ {rdR\left( r \right)/dr} \right] + \left[ {\left( {{q^2} - {h^2}} \right){r^2} - {m^2}} \right]\left[ {R\left( r \right)} \right] = 0};\\
{{d^2}\Theta \left( \theta  \right)/d{\theta ^2} + {m^2}\Theta \left( \theta  \right) = 0};\\
{{d^2}Z\left( z \right)/d{z^2} + {h^2}Z\left( z \right) = 0.}
\end{array}
\label{VIC1}
\end{equation}
 The variable $h$ is the projection of the incoming momentum on the cylinder axis. The general solution for the potentials is expressed in terms of
\begin{equation}
\begin{array}{l}
{\pi _{i,m}} = \left[ {{R_m}\left( {\sqrt {{q^2} - {h^2}} r} \right)} \right]\left[ {{e^{im\theta }}} \right]\left[ {{e^{ihz}}} \right]\left[ {{e^{ - i\omega t}}} \right],\\
\;m = 0, \pm 1, \pm 2 \ldots 
\end{array}
\label{VIC2}
\end{equation}
 The radial part $R(r)$ is a solution to the Bessel equation,
\begin{equation}
{z^2}\frac{{{d^2}\omega }}{{d{z^2}}} + z\frac{{d\omega }}{{dz}} + \left( {{z^2} - {\nu ^2}} \right)\omega  = 0.
\label{VIC3}
\end{equation}
The Bessel equation has many different solutions:
\begin{itemize}
\item Bessel functions of the first kind: ${J_{ \pm \nu}}\left( z \right)$.

\item Bessel functions of the second kind: ${Y_\nu}\left( z \right)$ (Weber's function, Neumann's function).

\item Bessel functions of the third kind: $H_\nu^{\left( 1 \right)}\left( z \right)$,  $H_\nu^{\left( 2 \right)}\left( z \right)$ (Hankel functions).

\end{itemize}

Each is a regular function of $z$ throughout the complex $z$-plane cut along the negative real axis. They are related to each other according to
\begin{equation}
\begin{array}{*{20}{l}}
{{Y_\nu }\left( z \right) = \left[ {{J_\nu }\left( z \right)\cos \left( {\nu \pi } \right) - {J_{ - \nu }}\left( z \right)} \right]/\sin \left( {\nu \pi } \right);}\\
{{H_\nu }^{\left( 1 \right)}\left( z \right) = {J_\nu }\left( z \right) + i{Y_\nu }\left( z \right);}\\
{{H_\nu }^{\left( 2 \right)}\left( z \right) = {J_\nu }\left( z \right) - i{Y_\nu }\left( z \right).}
\end{array}
\label{VIC4}
\end{equation}

Let us study a layered cylinder of radius $R$ consisting of $N$ layers and an inner cylindrical core. We have $N+2$ media and $N+1$ interfaces. Let the numbering be as follows. Medium $0$ is the medium surrounding the cylinder, medium $1$ is the outermost layer and medium $N+1$ the innermost layer and $N+2$ the innermost cylindrical core. Let ${r_n}$ be the inner radius of layer $n$. This is completely in line with the system represented by Fig.\,\ref{figu3}.
 
We will use the two Hankel versions  since they represent waves that go in either the positive or negative $r$-directions. We assume a time dependence of the form ${e^{ - i\omega t}}$. With this choice the first Hankel function, ${H_\nu}^{\left( 1 \right)}\left( {qr} \right){e^{ - i\omega t}} \propto {e^{i\left( {qr - \omega t} \right)}}$ , represents a wave moving in the positive radial direction (towards the left in Fig.\,\ref{figu3}) while the second, ${H_\nu}^{\left( 2 \right)}\left( {qr} \right){e^{ - i\omega t}} \propto {e^{ - i\left( {qr + \omega t} \right)}}$ , represents a wave moving in the negative radial direction (towards the right in Fig.\,\ref{figu3}.) 
Thus the general solution for the potentials is
\begin{equation}
\begin{array}{*{20}l}
   {\pi  = \sum\limits_{m = 0}^\infty  {\left[ {a_m H_m ^{\left( 2 \right)} \left( {k r} \right) + b_m H_m ^{\left( 1 \right)} \left( {k r} \right)} \right]e^{im\theta } e^{ihz} e^{ - i\omega t} ;} }  \\
   {k  = \sqrt {q^2  - h^2 } .}  \\
\end{array}
\label{VIC5}
\end{equation}
Let us now use the boundary conditions that the tangential components of ${\bf{E}}$ and ${\bf{H}}$ are continuous at the interface between layer $n$ and $n+1$. We get\,\cite{Kerker2}
\begin{equation}
\begin{array}{l}
{\left[ {\left( {{\partial/ {\partial r}}} \right)\pi _{_1}^n + \left( {{{imh}/ {{q_n}r}}} \right)\pi _{_2}^n} \right]_{r = {r_n}}}\\
 = {\left[ {\left( {{\partial/ {\partial r}}} \right)\pi _{_1}^{n + 1} + \left( {{{imh}/ {{q_{n + 1}}r}}} \right)\pi _2^{n + 1}} \right]_{r = {r_n}}};\\
{\left[ {\left( {q_n^2 - {h^2}} \right)\pi _1^n} \right]_{r = {r_n}}} = {\left[ {\left( {q_{n + 1}^2 - {h^2}} \right)\pi _1^{n + 1}} \right]_{r = {r_n}}};\\
{\left[ {{q_n}\left( {{\partial/ {\partial r}}} \right)\pi _{_2}^n - \left( {{{imh}/ r}} \right)\pi _{_1}^n} \right]_{r = {r_n}}}\\
 = {\left[ {{q_{n + 1}}\left( {{\partial/{\partial r}}} \right)\pi _{_2}^{n + 1} - \left( {{{imh}/ r}} \right)\pi _{_1}^{n + 1}} \right]_{r = {r_n}}};\\
{\left[ {\left[ {{{\left( {q_n^2 - {h^2}} \right)}/ {{q_n}}}} \right]\pi _2^n} \right]_{r = {r_n}}}\\
 = {\left[ {\left[ {{{{{\left( {q_{n + 1}^2 - {h^2}} \right)}/ q}}_{n + 1}}} \right]\pi _2^{n + 1}} \right]_{r = {r_n}}},
\end{array}
\label{VIC6}
\end{equation}
where ${q_n} = \sqrt {{{\tilde {\tilde \varepsilon }}_n}} \left( {\omega /c} \right)$.

This gives
\begin{equation}
\begin{array}{l}
a_{1,m}^n{k _n} {H_m^{\left( 2 \right)}}{'}\left( {{k _n}{r_n}} \right) + b_{1,m}^n{k _n} {H_m^{\left( 1 \right)}}{'}\left( {{k _n}{r_n}} \right)\\
 + a_{2,m}^n\left( {\frac{{imh}}{{{q_n}{r_n}}}} \right)H_m^{\left( 2 \right)}\left( {{k _n}{r_n}} \right)\\
 + b_{2,m}^n\left( {\frac{{imh}}{{{q_n}{r_n}}}} \right)H_m^{\left( 1 \right)}\left( {{k _n}{r_n}} \right)\\
 = a_{1,m}^{n + 1}{k _{n + 1}} {H_m^{\left( 2 \right)}}{'}\left( {{k _{n + 1}}{r_n}} \right)\\
 + b_{1,m}^{n + 1}{k _{n + 1}} {H_m^{\left( 1 \right)}}{'}\left( {{k _{n + 1}}{r_n}} \right)\\
 + a_{2,m}^{n + 1}\left( {\frac{{imh}}{{{q_{n + 1}}{r_n}}}} \right)H_m^{\left( 2 \right)}\left( {{k _{n + 1}}{r_n}} \right)\\
 + b_{2,m}^{n + 1}\left( {\frac{{imh}}{{{q_{n + 1}}{r_n}}}} \right)H_m^{\left( 1 \right)}\left( {{k _{n + 1}}{r_n}} \right);\\
a_{1,m}^nk _n^2H_m^{\left( 2 \right)}\left( {{k _n}{r_n}} \right) + b_{1,m}^nk _n^2H_m^{\left( 1 \right)}\left( {{k _n}{r_n}} \right)\\
 = a_{1,m}^{n + 1}k _{n + 1}^2H_m^{\left( 2 \right)}\left( {{k _{n + 1}}{r_n}} \right)\\
 + b_{1,m}^{n + 1}k _{n + 1}^2H_m^{\left( 1 \right)}\left( {{k _{n + 1}}{r_n}} \right);\\
 - a_{1,m}^n\left( {\frac{{imh}}{{{r_n}}}} \right)H_m^{\left( 2 \right)}\left( {{k _n}{r_n}} \right) - b_{1,m}^n\left( {\frac{{imh}}{{{r_n}}}} \right)H_m^{\left( 1 \right)}\left( {{k _n}{r_n}} \right)\\
 + a_{2,m}^n{q_n}{k _n} {H_m^{\left( 2 \right)}}{'}\left( {{k _n}{r_n}} \right) + b_{2,m}^n{q_n}{k _n} {H_m^{\left( 1 \right)}}{'}\left( {{k _n}{r_n}} \right)\\
 =  - a_{1,m}^{n + 1}\left( {\frac{{imh}}{{{r_n}}}} \right)H_m^{\left( 2 \right)}\left( {{k _{n + 1}}{r_n}} \right)\\
 - b_{1,m}^{n + 1}\left( {\frac{{imh}}{{{r_n}}}} \right)H_m^{\left( 1 \right)}\left( {{k _{n + 1}}{r_n}} \right)\\
 + a_{2,m}^{n + 1}{q_{n + 1}}{k _{n + 1}} {H_m^{\left( 2 \right)}}{'}\left( {{k _{n + 1}}{r_n}} \right)\\
 + b_{2,m}^{n + 1}{q_{n + 1}}{k _{n + 1}} {H_m^{\left( 1 \right)}}{'}\left( {{k _{n + 1}}{r_n}} \right);\\
a_{2,m}^n\left( {\frac{{k _n^2}}{{{q_n}}}} \right)H_m^{\left( 2 \right)}\left( {{k _n}{r_n}} \right) + b_{2,m}^n\left( {\frac{{k _n^2}}{{{q_n}}}} \right)H_m^{\left( 1 \right)}\left( {{k _n}{r_n}} \right)\\
 = a_{2,m}^{n + 1}\left( {\frac{{k _{n + 1}^2}}{{{q_{n+1}}}}} \right)H_m^{\left( 2 \right)}\left( {{k _{n + 1}}{r_n}} \right)\\
 + b_{2,m}^{n + 1}\left( {\frac{{k _{n + 1}^2}}{{{q_{n+1}}}}} \right)H_m^{\left( 1 \right)}\left( {{k _{n + 1}}{r_n}} \right).
\end{array}
\label{VIC7}
\end{equation}
This may be arranged as
\begin{equation}
{{{\bf{\tilde A}}}_n}\left( {\begin{array}{*{20}{c}}
{a_{1,m}^n}\\
{b_{1,m}^n}\\
{a_{2,m}^n}\\
{b_{2,m}^n}
\end{array}} \right) = {{{\bf{\tilde A}}}_{n + 1}}\left( {\begin{array}{*{20}{c}}
{a_{1,m}^{n + 1}}\\
{b_{1,m}^{n + 1}}\\
{a_{2,m}^{n + 1}}\\
{b_{2,m}^{n + 1}}
\end{array}} \right),
\label{VIC8}
\end{equation}
and we may now identify the matrix
\begin{equation}
\begin{array}{l}
 {\bf{\tilde A}}_n  =  \\ 
 \left( {\begin{array}{*{20}c}
   {k _n H_m^{\left( 2 \right)} {'}} & {k _n H_m^{\left( 1 \right)} {'}} & {\frac{{imh}}{{q_n r_n }}H_m^{\left( 2 \right)} } & {\frac{{imh}}{{q_n r_n }}H_m^{\left( 1 \right)} }  \\
   {k _n^2 H_m^{\left( 2 \right)} } & {k _n^2 H_m^{\left( 1 \right)} } & 0 & 0  \\
   { - \frac{{imh}}{{r_n }}H_m^{\left( 2 \right)} } & { - \frac{{imh}}{{r_n }}H_m^{\left( 1 \right)} } & {q_n k _n H_m^{\left( 2 \right)} {'}} & {q_n k _n H_m^{\left( 1 \right)} {'}}  \\
   0 & 0 & {\frac{{k _n^2 }}{{q_n }}H_m^{\left( 2 \right)} } & {\frac{{k _n^2 }}{{q_n }}H_m^{\left( 1 \right)} }  \\
\end{array}} \right), \\ 
 \end{array}
\label{VIC9}
\end{equation}
where we have omitted the argument $\left( {{k_n}{r_n}} \right)$ in all Hankel functions and their derivatives. 

For the special case when $h = 0$ we see that the matrices are on block form:
\begin{equation}
\begin{array}{l}
 {\bf{\tilde A}}_n  =  \\ 
 \left( {\begin{array}{*{20}c}
   {k _n H_m^{\left( 2 \right)} {'}} & {k _n H_m^{\left( 1 \right)} {'}} & 0 & 0  \\
   {k _n^2 H_m^{\left( 2 \right)} } & {k _n^2 H_m^{\left( 1 \right)} } & 0 & 0  \\
   0 & 0 & {q_n k _n H_m^{\left( 2 \right)} {'}} & {q_n k _n H_m^{\left( 1 \right)} {'}}  \\
   0 & 0 & {\frac{{k _n^2 }}{{q_n }}H_m^{\left( 2 \right)} } & {\frac{{k _n^2 }}{{q_n }}H_m^{\left( 1 \right)} }  \\
\end{array}} \right), \\ 
 \end{array}
\label{VIC10}
\end{equation}
which means that the TM and TE modes decouple for $h = 0$. 

Now, the resulting matrix ${{{\bf{\tilde M}}}_n} = {\bf{\tilde A}}_n^{ - 1} \cdot {{{\bf{\tilde A}}}_{n + 1}}$ becomes too large to write down in matrix form. Instead, we list each element:
%
\begin{equation}
\begin{array}{l}
{M_{11}} = \frac{1}{{Wk _n^3}}\left[ {k _n^2H_m^{\left( 1 \right)}{k _{n + 1}}H_m^{\left( 2 \right)+}{'}} \right.\\
\left. {\quad \quad  - {k _n}H_m^{\left( 1 \right)}{'}k _{n + 1}^2H_m^{\left( 2 \right){+}}} \right];\\
{M_{12}} = \frac{1}{{Wk _n^3}}\left[ {k _n^2H_m^{\left( 1 \right)}{k _{n + 1}}H_m^{\left( 1 \right)+}{'}} \right.\\
\quad \quad  - \left. {{k _n}H_m^{\left( 1 \right)}{'}k _{n + 1}^2H_m^{\left( 1 \right){+}}} \right];\\
{M_{13}} = \frac{1}{{Wk _n^3}}H_m^{\left( 1 \right)}H_m^{\left( 2 \right) + }\left( {imh/{q_{n + 1}}{r_n}} \right)\left[ {k _n^2} \right. - \left. {k _{n + 1}^2} \right];\\
{M_{14}} = \frac{1}{{Wk _n^3}}H_m^{\left( 1 \right)}H_m^{\left( 1 \right) + }\left( {imh/{q_{n + 1}}{r_n}} \right)\left[ {k _n^2} \right. - \left. {k _{n + 1}^2} \right];\\
{M_{21}} = \frac{1}{{Wk _n^3}}\left[ { - k _n^2H_m^{\left( 2 \right)}{k _{n + 1}}H_m^{\left( 2 \right)+}{'}} \right.\\
\quad \quad  + \left. {{k _n}H_m^{\left( 2 \right)}{'}k _{n + 1}^2H_m^{\left( 2 \right){+}}} \right];\\
{M_{22}} = \frac{1}{{Wk _n^3}}\left[ { - k _n^2H_m^{\left( 2 \right)}{k _{n + 1}}H_m^{\left( 1 \right)+}{'}} \right.\\
\quad \quad  + \left. {{k _n}H_m^{\left( 2 \right)}{'}k _{n + 1}^2H_m^{\left( 1 \right){+}}} \right];\\
{M_{23}} = \frac{1}{{Wk _n^3}}H_m^{\left( 2 \right)}H_m^{\left( 2 \right) + }\left( {imh/{q_{n + 1}}{r_n}} \right)\left[ {k _{n + 1}^2 - k _n^2} \right];\\
{M_{24}} = \frac{1}{{Wk _n^3}}H_m^{\left( 1 \right) + }H_m^{\left( 2 \right)}\left( {imh/{q_{n + 1}}{r_n}} \right)\left[ {k _{n + 1}^2 - k _n^2} \right];\\
{M_{31}} = \frac{1}{{Wk_n^3}}H_m^{\left( 1 \right)}H_m^{\left( 2 \right) + }\left( {{{imh} \mathord{\left/
 {\vphantom {{imh} {{q_n}{r_n}}}} \right.
 \kern-\nulldelimiterspace} {{q_n}{r_n}}}} \right)\left[ {k_{n + 1}^2 - k_n^2} \right];\\
{M_{32}} = \frac{1}{{Wk_n^3}}H_m^{\left( 1 \right)}H_m^{\left( 1 \right) + }\left( {{{imh} \mathord{\left/
 {\vphantom {{imh} {{q_n}{r_n}}}} \right.
 \kern-\nulldelimiterspace} {{q_n}{r_n}}}} \right)\left[ {k_{n + 1}^2 - k_n^2} \right];\\
{M_{33}} = \frac{1}{{Wk _n^3}}\left[ {\left( {{{k _n^2}/ {{q_n}}}} \right)H_m^{\left( 1 \right)}{q_{n + 1}}{k _{n + 1}}H_m^{\left( 2 \right)+}{'}} \right.\\
\quad \quad  - \left. {{q_n}{k _n}H_m^{\left( 1 \right)}{'}\left( {{{k _{n + 1}^2}/ {{q_{n + 1}}}}} \right)H_m^{\left( 2 \right){+}}} \right];\\
{M_{34}} = \frac{1}{{Wk _n^3}}\left[ {\left( {{{k _n^2}/ {{q_n}}}} \right)H_m^{\left( 1 \right)}{q_{n + 1}}{k _{n + 1}}H_m^{\left( 1 \right)+}{'}} \right.\\
\quad \quad  - \left. {{q_n}{k _n}H_m^{\left( 1 \right)}{'}\left( {{{k _{n + 1}^2}/ {{q_{n + 1}}}}} \right)H_m^{\left( 1 \right){+}}} \right];\\
{M_{41}} = \frac{1}{{Wk _n^3}}H_m^{\left( 2 \right)}H_m^{\left( 2 \right) + }\left( {imh/{q_n}{r_n}} \right)\left[ {k _n^2 - k _{n + 1}^2} \right];\\
{M_{42}}= \frac{1}{{Wk _n^3}}H_m^{\left( 1 \right) + }H_m^{\left( 2 \right)}\left( {imh/{q_n}{r_n}} \right)\left[ {k _n^2 - k _{n + 1}^2} \right];\\
{M_{43}} = \frac{1}{{Wk _n^3}}\left[ { - \left( {{{k _n^2}/{{q_n}}}} \right)H_m^{\left( 2 \right)}{q_{n + 1}}{k _{n + 1}}H_m^{\left( 2 \right)}{{'}}^{+}} \right.\\
\quad \quad  + \left. {{q_n}{k _n}H_m^{\left( 2 \right)}{'}\left( {{{k _{n + 1}^2}/ {{q_{n + 1}}}}} \right)H_m^{\left( 2 \right){+}}} \right];\\
{M_{44}} = \frac{1}{{Wk_n^3}}\left[ { - \left( {k_n^2/{q_n}} \right)H_m^{\left( 2 \right)}{q_{n + 1}}{k_{n + 1}}H_m^{\left( 1 \right) + }{'}} \right.\\
\quad \quad  + \left. {{q_n}{k_n}H_m^{\left( 2 \right)}{'}\left( {k_{n + 1}^2/{q_{n + 1}}} \right)H_m^{\left( 1 \right) + }} \right].
\end{array}
\label{VIC11}
\end{equation}
We have suppressed all arguments of the Hankel functions and their derivatives. All functions with a + added as a superscript have the argument $\left( {{k_{n + 1}}{r_n}} \right)$ and the ones without the superscript have the argument  $\left( {{k_n}{r_n}} \right)$. $W$ is short for the Wronskian, $W\left[ {H_m^{\left( 1 \right)}\left( x \right),H_m^{\left( 2 \right)}\left( x \right)} \right] = H_m^{\left( 1 \right)}\left( x \right)H_m^{\left( 2 \right)}{\rm{{'}}}\left( x \right){\rm{ - H}}_{\rm{m}}^{\left( {\rm{1}} \right)}{\rm{'}}\left( x \right){\rm{H}}_{\rm{m}}^{\left( {\rm{2}} \right)}\left( x \right){\rm{ =  - 4i/}}\pi x$
Now we have all we need to determine the fully retarded normal modes in a layered cylindrical structure. We give some examples in the following sections.

Since it is inconvenient to work with $4\times 4$ matrices we divide a general matrix into 4 sub matrices,
\begin{equation}
{\bf{\tilde M}} = \left( {\begin{array}{*{20}{c}}
{{\bf{\tilde B}}}&{{\bf{\tilde C}}}\\
{{\bf{\tilde D}}}&{{\bf{\tilde F}}}
\end{array}} \right).
\label{VIC12}
\end{equation}
With this alternative notation the condition for modes becomes
\begin{equation}
\left( {{B_{11}} + {B_{12}}} \right)\left( {{F_{11}} + {F_{12}}} \right) - \left( {{C_{11}} + {C_{12}}} \right)\left( {{D_{11}} + {D_{12}}} \right) = 0,
\label{VIC13}
\end{equation}
and the product of two adjacent matrices becomes
\begin{equation}
\begin{array}{*{20}{l}}
{{{{\bf{\tilde M}}}_n} \cdot {{{\bf{\tilde M}}}_{n + 1}}}\\
{ = \left( {\begin{array}{*{20}{c}}
{{{{\bf{\tilde B}}}_n}}&{{{{\bf{\tilde C}}}_n}}\\
{{{{\bf{\tilde D}}}_n}}&{{{{\bf{\tilde F}}}_n}}
\end{array}} \right) \cdot \left( {\begin{array}{*{20}{c}}
{{{{\bf{\tilde B}}}_{n + 1}}}&{{{{\bf{\tilde C}}}_{n + 1}}}\\
{{{{\bf{\tilde D}}}_{n + 1}}}&{{{{\bf{\tilde F}}}_{n + 1}}}
\end{array}} \right)}\\
{ = \left( {\begin{array}{*{20}{c}}
{{{{\bf{\tilde B}}}_n} \cdot {{{\bf{\tilde B}}}_{n + 1}} + {{{\bf{\tilde C}}}_n} \cdot {{{\bf{\tilde D}}}_{n + 1}}}&{{{{\bf{\tilde C}}}_n} \cdot {{{\bf{\tilde F}}}_{n + 1}}}\\
{{{{\bf{\tilde D}}}_n} \cdot {{{\bf{\tilde B}}}_{n + 1}}}&{{{{\bf{\tilde F}}}_n} \cdot {{{\bf{\tilde F}}}_{n + 1}} + {{{\bf{\tilde D}}}_n} \cdot {{{\bf{\tilde C}}}_{n + 1}}}
\end{array}} \right)}\\
\begin{array}{l}
 = \left( {\begin{array}{*{20}{c}}
{{{{\bf{\tilde B}}}_n} \cdot {{{\bf{\tilde B}}}_{n + 1}}}&0\\
0&{{{{\bf{\tilde F}}}_n} \cdot {{{\bf{\tilde F}}}_{n + 1}}}
\end{array}} \right)\\
 \quad\quad +\left( {\begin{array}{*{20}{c}}
{{{{\bf{\tilde C}}}_n} \cdot {{{\bf{\tilde D}}}_{n + 1}}}&{{{{\bf{\tilde C}}}_n} \cdot {{{\bf{\tilde F}}}_{n + 1}}}\\
{{{{\bf{\tilde D}}}_n} \cdot {{{\bf{\tilde B}}}_{n + 1}}}&{{{{\bf{\tilde D}}}_n} \cdot {{{\bf{\tilde C}}}_{n + 1}}}
\end{array}} \right).
\end{array}
\end{array}
\label{VIC14}
\end{equation}
\subsection{\label{cylindricalretspecial}Retarded special results}

\subsubsection{\label{cylinderret}Solid cylinder (no layer)}
For a solid cylinder of radius  $a$ and dielectric function ${{\tilde \varepsilon }_1}\left( \omega  \right)$ in an ambient of dielectric function ${{\tilde \varepsilon }_0}\left( \omega  \right)$, as illustrated in Fig.\,\ref{figu16}, we have ${\bf{\tilde M}} = {{{\bf{\tilde M}}}_0}$, and the type of combinations of matrix elements that appear in the mode condition are:
%
%
\begin{equation}
\begin{array}{l}
{M_{11}} + {M_{12}} = {B_{11}} + {B_{12}}\\
 = \frac{2}{{Wk_0^3}}\left[ {k_0^2H_m^{\left( 1 \right)}{k_1}{J_m}^\prime- {k_0}{H_m^{\left( 1 \right)}}{\rm{^\prime }} k_1^2{J_m}} \right];\\
{M_{13}} + {M_{14}} = {C_{11}} + {C_{12}}\\
 = \frac{2}{{Wk_0^3}}H_m^{\left( 1 \right)}{J_m}\left( {imh/{q_1}a} \right)\left[ {k_0^2 - k_1^2} \right];\\
{M_{21}} + {M_{22}} = {B_{21}} + {B_{22}}\\
 = \frac{2}{{Wk_0^3}}\left[ { - k_0^2H_m^{\left( 2 \right)}{k_1}{J_m}\prime+ {k_0}{H_m^{\left( 2 \right)}}{\rm{^\prime }} k_1^2{J_m}} \right];\\
{M_{23}} + {M_{24}} = {C_{21}} + {C_{22}}\\
 = \frac{2}{{Wk_0^3}}H_m^{\left( 2 \right)}{J_m}\left( {imh/{q_1}a} \right)\left[ {k_1^2 - k_0^2} \right];\\
{M_{31}} + {M_{32}} = {D_{11}} + {D_{12}}\\
 = \frac{2}{{Wk_0^3}}H_m^{\left( 1 \right)}{J_m}\left( {imh/{q_0}a} \right)\left[ {k_1^2 - k_0^2} \right];\\
{M_{33}} + {M_{34}} = {F_{11}} + {F_{12}} = \frac{2}{{Wk_0^3}}\\
\times \left[ {\left( {k_0^2/{q_0}} \right)H_m^{\left( 1 \right)}{q_1}{k_1}{J_m}{\rm{^\prime }} - {q_0}{k_0}H_m^{\left( 1 \right)}{\rm{^\prime }}\left( {k_1^2/{q_1}} \right){J_m}} \right];\\
{M_{41}} + {M_{42}} = {D_{21}} + {D_{22}}\\
 = \frac{2}{{Wk_0^3}}H_m^{\left( 2 \right)}{J_m}\left( {imh/{q_0}a} \right)\left[ {k_0^2 - k_1^2} \right];\\
{M_{43}} + {M_{44}} = {F_{21}} + {F_{22}} = \frac{2}{{Wk_0^3}}\\
\times\left[ { - \left( {k_0^2/{q_0}} \right)H_m^{\left( 2 \right)}{q_1}{k_1}{J_m}{\rm{^\prime }} + {q_0}{k_0}H_m^{\left( 2 \right)}{\rm{^\prime }}\left( {k_1^2/{q_1}} \right){J_m}} \right],
\end{array}
\label{VID1.1}
\end{equation}
where we have used the relation $2{J_m}\left( z \right) = H_m^{\left( 1 \right)}\left( z \right) + H_m^{\left( 2 \right)}\left( z \right)$. We have suppressed all arguments of the functions. The suppressed arguments are $\left( {{k _0}{a}} \right)$ for the $H$-functions and their derivatives and $\left( {{k _1}{a}} \right)$ for the $J$-functions and their derivatives.

The condition for modes is according to Eq.\,(\ref{III15}) $\left( {{M_{11}} + {M_{12}}} \right)\left( {{M_{33}} + {M_{34}}} \right) = \left( {{M_{13}} + {M_{14}}} \right)\left( {{M_{31}} + {M_{32}}} \right)$, or according to Eq.\,(\ref{VIC13}), $\left( {{B_{11}} + {B_{12}}} \right)\left( {{F_{11}} + {F_{12}}} \right) = \left( {{C_{11}} + {C_{12}}} \right)\left( {{D_{11}} + {D_{12}}} \right)$, which leads to
\begin{equation}
\begin{array}{l}
\left( {\frac{1}{{{k _1}}}\frac{{{J_m}{'}\left( {{k _1}{a}} \right)}}{{{J_m}\left( {{k _1}{a}} \right)}} - \frac{1}{{{k _0}}}\frac{{H_m^{\left( 1 \right)}{'}\left( {{k _0}{a}} \right)}}{{H_m^{\left( 1 \right)}\left( {{k _0}{a}} \right)}}} \right)\\
 \quad \quad \quad \quad \quad \quad   \times \left( {\frac{{q_1^2}}{{{k _1}}}\frac{{{J_m}{'}\left( {{k _1}{a}} \right)}}{{{J_m}\left( {{k _1}{a}} \right)}} - \frac{{q_0^2}}{{{k _0}}}\frac{{H_m^{\left( 1 \right)}{'}\left( {{k _0}{a}} \right)}}{{H_m^{\left( 1 \right)}\left( {{k _0}{a}} \right)}}} \right)\\
 = {\left( {{{mh}/ {{a}}}} \right)^2}{\left( {\frac{1}{{k _1^2}} - \frac{1}{{k _0^2}}} \right)^2}.
\end{array}
\label{VID1.2}
\end{equation}
This is in complete agreement with Ruppin in Eq.\,(107) on page 389 in Ref.\,[\onlinecite{Board}]. When either $m$ or $h$ or both are zero the TM and TE modes decouple. If they are decoupled letting the first factor on the left hand side be equal to zero defines the TE  modes and letting  the second  factor be equal to zero defines the TM modes.
%
\subsubsection{\label{atom-cylinderret}Force on an atom outside a cylinder (two layers)}
Here we proceed in the same way as in Sec.\ref{atom-cylindern} but the matrices are now much more involved. Here both $\alpha$ and $\delta$ appear in the arguments of the functions. In the non-retarded case only $\delta$ did. The geometry of this problem is illustrated in Fig.\,\ref{figu18}. We first expand the matrix for the gas layer in $\alpha$ and keep terms up to linear in $\alpha$. We have
\begin{equation}
\begin{array}{l}
{\bf{\tilde M}} = {{{\bf{\tilde M}}}_0} \cdot {{{\bf{\tilde M}}}_1} \cdot {{{\bf{\tilde M}}}_2} = {{{\bf{\tilde M}}}^{layer}} \cdot {{{\bf{\tilde M}}}_2};\\
{{{\bf{\tilde M}}}_0} \approx \tilde 1 + \alpha {\bf{\tilde M}}_0^1;\,\,{{{\bf{\tilde M}}}_1} \approx \tilde 1 + \alpha {\bf{\tilde M}}_1^1;\\
{{{\bf{\tilde M}}}^{layer}} \approx \tilde 1 + \alpha \left( {{\bf{\tilde M}}_0^1 + {\bf{\tilde M}}_1^1} \right).
\end{array}
\label{VID2.1}
\end{equation}

The elements of ${{\bf{\tilde M}}_0^1}$ are in the first row
\begin{equation}
\begin{array}{l}
{M_{11}} = \frac{{i\pi {k_0}\left( {b + \delta } \right)}}{8}{\left( {\frac{q_0 }{{k_0}}} \right)^2}\left[ {H_m^{\left( 1 \right)}H_m^{\left( 2 \right)}{'} - 2H_m^{\left( 1 \right)}{'}H_m^{\left( 2 \right)}} \right.\\
\quad \quad  + \left. {{k_0}\left( {b + \delta } \right)\left( {H_m^{\left( 1 \right)}H_m^{\left( 2 \right)}{'}{'} - H_m^{\left( 1 \right)}{'}H_m^{\left( 2 \right)}{'}} \right)} \right];\\
{M_{12}} = \frac{{i\pi {k_0}\left( {b + \delta } \right)}}{8}{\left( {\frac{q_0 }{{k_0}}} \right)^2}\left[ { - H_m^{\left( 1 \right)}{'}H_m^{\left( 1 \right)}} \right.\\
\quad \quad  + \left. {{k_0}\left( {b + \delta } \right)\left( {H_m^{\left( 1 \right)}H_m^{\left( 1 \right)}{'}{'} - H_m^{\left( 1 \right)}{'}H_m^{\left( 1 \right)}{'}} \right)} \right];\\
{M_{13}} = \frac{{m\pi }}{4}H_m^{\left( 1 \right)}H_m^{\left( 2 \right)}\frac{h}{{{k_0}}}\left( {\frac{q_0 }{{k_0}}} \right);\\
{M_{14}} = \frac{{m\pi }}{4}H_m^{\left( 1 \right)}H_m^{\left( 1 \right)}\frac{h}{{{k_0}}}\left( {\frac{q_0 }{{k_0}}} \right),
\end{array}
\label{VID2.2}
\end{equation}
in the second row
\begin{equation}
\begin{array}{l}
{M_{21}} = \frac{{i\pi {k_0}\left( {b + \delta } \right)}}{8}{\left( {\frac{q_0 }{{k_0}}} \right)^2}\left[ {H_m^{\left( 2 \right)}{'}H_m^{\left( 2 \right)}} \right.\\
\quad \quad  - \left. {{k_0}\left( {b + \delta } \right)\left( {H_m^{\left( 2 \right)}H_m^{\left( 2 \right)}{'}{'} - H_m^{\left( 2 \right)}{'}H_m^{\left( 2 \right)}{'}} \right)} \right];\\
{M_{22}} = \frac{{i\pi {k_0}\left( {b + \delta } \right)}}{8}{\left( {\frac{q_0 }{{k_0}}} \right)^2}\left[ { - H_m^{\left( 2 \right)}H_m^{\left( 1 \right)}{'} + 2H_m^{\left( 2 \right)}{'}H_m^{\left( 1 \right)}} \right.\\
\quad \quad  - \left. {{k_0}\left( {b + \delta } \right)\left( {H_m^{\left( 2 \right)}H_m^{\left( 1 \right)}{'}{'} - H_m^{\left( 2 \right)}{'}H_m^{\left( 1 \right)}{'}} \right)} \right];\\
{M_{23}} =  - \frac{{m\pi }}{4}H_m^{\left( 2 \right)}H_m^{\left( 2 \right)}\frac{h}{{{k_0}}}\left( {\frac{q_0 }{{k_0}}} \right);\\
{M_{24}} =  - \frac{{m\pi }}{4}H_m^{\left( 1 \right)}H_m^{\left( 2 \right)}\frac{h}{{{k_0}}}\left( {\frac{q_0 }{{k_0}}} \right),
\end{array}
\label{VID2.3}
\end{equation}
in the third row
\begin{equation}
\begin{array}{l}
{M_{31}} =  - \frac{{m\pi }}{4}H_m^{\left( 1 \right)}H_m^{\left( 2 \right)}\frac{h}{{{k_0}}}\left( {\frac{q_0 }{{k_0}}} \right);\\
{M_{32}} =  - \frac{{m\pi }}{4}H_m^{\left( 1 \right)}H_m^{\left( 1 \right)}\frac{h}{{{k_0}}}\left( {\frac{q_0 }{{k_0}}} \right);\\
{M_{33}} = \frac{{i\pi {k_0}\left( {b + \delta } \right)}}{8}\left\{ {H_m^{\left( 1 \right)}H_m^{\left( 2 \right)}{\rm{{'} + }}H_m^{\left( 1 \right)}{'}H_m^{\left( 2 \right)}} \right.\\
\quad \quad  + {\left( {\frac{q_0 }{{k_0}}} \right)^2}\left[ {H_m^{\left( 1 \right)}H_m^{\left( 2 \right)}{\rm{{'} - }}2H_m^{\left( 1 \right)}{'}H_m^{\left( 2 \right)}} \right.\\
\quad \quad  + \left. {\left. {{k_0}\left( {b + \delta } \right)\left( {H_m^{\left( 1 \right)}H_m^{\left( 2 \right)}{\rm{{'}{'} - }}H_m^{\left( 1 \right)}{'}H_m^{\left( 2 \right)}{\rm{{'}}}} \right)} \right]} \right\};\\
{M_{34}} = \frac{{i\pi {k_0}\left( {b + \delta } \right)}}{8}\left[ {2H_m^{\left( 1 \right)}H_m^{\left( 1 \right)}{\rm{{'}}}} \right. + {\left( {\frac{q_0 }{{k_0}}} \right)^2} \times\\
\left. {\left( {- H_m^{\left( 1 \right)}H_m^{\left( 1 \right)}{\rm{{'} + }}{k_0}\left( {b + \delta } \right)\left( {H_m^{\left( 1 \right)}H_m^{\left( 1 \right)}{\rm{{'}{'} - }}H_m^{\left( 1 \right)}{'}H_m^{\left( 1 \right)}{\rm{{'}}}} \right)} \right)} \right],
\end{array}
\label{VID2.4}
\end{equation}
and in the fourth row
\begin{equation}
\begin{array}{l}
{ÄM_{41}} = \frac{{m\pi }}{4}H_m^{\left( 2 \right)}H_m^{\left( 2 \right)}\frac{h}{{{k_0}}}\left( {\frac{q_0 }{{k_0}}} \right);\\
{M_{42}} = \frac{{m\pi }}{4}H_m^{\left( 1 \right)}H_m^{\left( 2 \right)}\frac{h}{{{k_0}}}\left( {\frac{q_0 }{{k_0}}} \right);\\
{M_{43}} =  - \frac{{i\pi {k_0}\left( {b + \delta } \right)}}{8}\left\{ {2H_m^{\left( 2 \right)}H_m^{\left( 2 \right)}{\rm{' - }}{{\left( {\omega /c{k_0}} \right)}^{\rm{2}}} \times } \right.\\
\left. {\left[ {H_m^{\left( 2 \right)}H_m^{\left( 2 \right)}{\rm{'}} + {k_0}\left( {b + \delta } \right)\left( {H_m^{\left( 2 \right)}H_m^{\left( 2 \right)}{{\rm{'}}{\rm{'}}} - H_m^{\left( 2 \right)}{\rm{'}}H_m^{\left( 2 \right)}{\rm{'}}} \right)} \right]} \right\};\\
{M_{44}} =  - \frac{{i\pi {k_0}\left( {b + \delta } \right)}}{8}\left\{ {H_m^{\left( 2 \right)}H_m^{\left( 1 \right)}{\rm{'}} + H_m^{\left( 2 \right)}{\rm{'}}H_m^{\left( 1 \right)}} \right.\\
\quad \quad  + {\left( {\frac{q_0 }{{k_0}}} \right)^2}\left[ {H_m^{\left( 2 \right)}H_m^{\left( 1 \right)}{\rm{'}} - 2H_m^{\left( 2 \right)}{\rm{'}}H_m^{\left( 1 \right)}} \right.\\
\quad \quad \quad \quad  + \left. {\left. {{k_0}\left( {b + \delta } \right)H_m^{\left( 2 \right)}H_m^{\left( 1 \right)}{{\rm{'}}{\rm{'}}} - H_m^{\left( 2 \right)}{\rm{'}}H_m^{\left( 1 \right)}{\rm{'}}} \right]} \right\}.
\end{array}
\label{VID2.5}
\end{equation}
The suppressed arguments are in all elements ${k_0}\left( {b + \delta } \right)$.
The elements of ${{\bf{\tilde M}}_1^1}$ are in the first row
\begin{equation}
\begin{array}{l}
{M_{11}} = \frac{{i\pi {k_0}b}}{8}{\left( {\frac{q_0 }{{k_0}}} \right)^2}\left[ {H_m^{\left( 1 \right)}{\rm{'}}H_m^{\left( 2 \right)}} \right.\\
\quad \quad  + \left. {\left( {{k_0}b} \right)\left( {H_m^{\left( 1 \right)}{\rm{'}}H_m^{\left( 2 \right)}{\rm{'}} - H_m^{\left( 1 \right)}{{\rm{'}}{\rm{'}}}{{\rm{'}}{\rm{'}}}H_m^{\left( 2 \right)}} \right)} \right];\\
{M_{12}} = \frac{{i\pi {k_0}b}}{8}{\left( {\frac{q_0 }{{k_0}}} \right)^2}\left[ {H_m^{\left( 1 \right)}H_m^{\left( 1 \right)}{\rm{'}}} \right.\\
\quad \quad  + \left. {\left( {{k_0}b} \right)\left( {H_m^{\left( 1 \right)}{\rm{'}}H_m^{\left( 1 \right)}{\rm{' - }}H_m^{\left( 1 \right)}H_m^{\left( 1 \right)}{\rm{'}\rm{'}}} \right)} \right];\\
{M_{13}} =  - \frac{{m\pi }}{4}H_m^{\left( 1 \right)}H_m^{\left( 2 \right)}\frac{h}{{{k_0}}}\left( {\frac{q_0 }{{k_0}}} \right);\\
{M_{14}} =  - \frac{{m\pi }}{4}H_m^{\left( 1 \right)}H_m^{\left( 1 \right)}\frac{h}{{{k_0}}}\left( {\frac{q_0 }{{k_0}}} \right),
\end{array}
\label{VID2.6}
\end{equation}
in the second row
\begin{equation}
\begin{array}{l}
{M_{21}} = -\frac{{i\pi {k_0}b}}{8}{\left( {\frac{q_0 }{{k_0}}} \right)^2}\left[ {H_m^{\left( 2 \right)}H_m^{\left( 2 \right)}{\rm{'}}} \right.\\
\quad \quad  + \left. {\left( {{k_0}b} \right)\left( {H_m^{\left( 2 \right)}{\rm{'}}H_m^{\left( 2 \right)}{\rm{'}} - H_m^{\left( 2 \right)}H_m^2{{\rm{'}}{\rm{'}}}} \right)} \right];\\
{M_{22}} =  - \frac{{i\pi {k_0}b}}{8}{\left( {\frac{q_0 }{{k_0}}} \right)^2}\left[ {H_m^{\left( 1 \right)}H_m^{\left( 2 \right)}{\rm{'}}} \right.\\
\quad \quad  + \left. {\left( {{k_0}b} \right)\left( {H_m^{\left( 1 \right)}{\rm{'}}H_m^{\left( 2 \right)}{\rm{' - }}H_m^{\left( 1 \right)}H_m^{\left( 2 \right)}{{\rm{'}}{\rm{'}}}} \right)} \right];\\
{M_{23}} = \frac{{m\pi }}{4}H_m^{\left( 2 \right)}H_m^{\left( 2 \right)}\frac{h}{{{k_0}}}\left( {\frac{q_0 }{{k_0}}} \right);\\
{M_{24}} = \frac{{m\pi }}{4}H_m^{\left( 1 \right)}H_m^{\left( 2 \right)}\frac{h}{{{k_0}}}\left( {\frac{q_0 }{{k_0}}} \right),
\end{array}
\label{VID2.7}
\end{equation}
in the third row
\begin{equation}
\begin{array}{l}
{M_{31}} = \frac{{m\pi }}{4}H_m^{\left( 1 \right)}H_m^{\left( 2 \right)}\frac{h}{{{k_0}}}\left( {\frac{q_0 }{{k_0}}} \right);\\
{M_{32}} = \frac{{m\pi }}{4}H_m^{\left( 1 \right)}H_m^{\left( 1 \right)}\frac{h}{{{k_0}}}\left( {\frac{q_0 }{{k_0}}} \right);\\
{M_{33}} = \frac{{i\pi {k_0}b}}{8}\left\{ {{\rm{ - }}H_m^{\left( 1 \right)}H_m^{\left( 2 \right)}{\rm{' - }}H_m^{\left( 1 \right)}{\rm{'}}H_m^{\left( 2 \right)}} \right.\\
\quad \quad  + {\left( {\frac{q_0 }{{k_0}}} \right)^2}\left[ {H_m^{\left( 1 \right)}{\rm{'}}H_m^{\left( 2 \right)}} \right.\\
\quad \quad  + \left. {\left. {\left( {{k_0}b} \right)\left( {H_m^{\left( 1 \right)}{\rm{'}}H_m^{\left( 2 \right)}{\rm{' - }}H_m^{\left( 1 \right)}{{\rm{'}}{\rm{'}}}H_m^{\left( 2 \right)}} \right)} \right]} \right\};\\
{M_{34}} = \frac{{i\pi {k_0}b}}{8}\left\{ {{\rm{ - }}H_m^{\left( 1 \right)}H_m^{\left( 1 \right)}{\rm{' - }}H_m^{\left( 1 \right)}{\rm{'}}H_m^{\left( 1 \right)}} \right.\\
\quad \quad  + {\left( {\frac{q_0 }{{k_0}}} \right)^2}\left[ {H_m^{\left( 1 \right)}{\rm{'}}H_m^{\left( 1 \right)}} \right.\\
\quad \quad  + \left. {\left. {\left( {{k_0}b} \right)\left( {H_m^{\left( 1 \right)}{\rm{'}}H_m^{\left( 1 \right)}{\rm{' - }}H_m^{\left( 1 \right)}{{\rm{'}}{\rm{'}}}H_m^{\left( 1 \right)}} \right)} \right]} \right\},
\end{array}
\label{VID2.8}
\end{equation}
and in the fourth row
\begin{equation}
\begin{array}{l}
{M_{41}} =  - \frac{{m\pi }}{4}H_m^{\left( 2 \right)}H_m^{\left( 2 \right)}\frac{h}{{{k_0}}}\left( {\frac{q_0 }{{k_0}}} \right);\\
{M_{42}} =  - \frac{{m\pi }}{4}H_m^{\left( 1 \right)}H_m^{\left( 2 \right)}\frac{h}{{{k_0}}}\left( {\frac{q_0 }{{k_0}}} \right);\\
{M_{43}} = \frac{{i\pi {k_0}b}}{8}\left\{ {H_m^{\left( 1 \right)}H_m^{\left( 1 \right)}{\rm{' + }}H_m^{\left( 1 \right)}{\rm{'}}H_m^{\left( 1 \right)}} \right.\\
\quad \quad  - {\left( {\frac{q_0 }{{k_0}}} \right)^2}\left[ {H_m^{\left( 2 \right)}{\rm{'}}H_m^{\left( 2 \right)}} \right.\\
\quad \quad  + \left. {\left. {\left( {{k_0}b} \right)\left( {H_m^{\left( 2 \right)}{\rm{'}}H_m^{\left( 2 \right)}{\rm{' - }}H_m^{\left( 2 \right)}{{\rm{'}}{\rm{'}}}H_m^{\left( 2 \right)}} \right)} \right]} \right\};\\
{M_{44}} = \frac{{i\pi {k_0}b}}{8}\left\{ {H_m^{\left( 1 \right)}{\rm{'}}H_m^{\left( 2 \right)}{\rm{ + }}H_m^{\left( 1 \right)}H_m^{\left( 2 \right)}{\rm{'}}} \right.\\
\quad \quad  - {\left( {\frac{q_0 }{{k_0}}} \right)^2}\left[ {H_m^{\left( 1 \right)}H_m^{\left( 2 \right)}{\rm{'}}} \right.\\
\quad \quad  + \left. {\left. {\left( {{k_0}b} \right)\left( {H_m^{\left( 1 \right)}{\rm{'}}H_m^{\left( 2 \right)}{\rm{' - }}H_m^{\left( 1 \right)}H_m^{\left( 2 \right)}{{\rm{'}}{\rm{'}}}} \right)} \right]} \right\}.
\end{array}
\label{VID2.9}
\end{equation}
The suppressed arguments are in all elements $\left( {{k_0}b} \right)$.
Next we expand ${{{\bf{\tilde M}}}^{layer}}$ from Eq.\,(\ref{VID2.1}) in $\delta$ up to the linear term. We note that the zeroth order term of ${\bf{\tilde M}}_0^1$ exactly cancels ${\bf{\tilde M}}_1^1$. Thus
\begin{equation}
{{{\bf{\tilde M}}}^{layer}} \approx \tilde 1 + \alpha \delta {\left[ {\frac{{\partial {\bf{\tilde M}}_0^1}}{{\partial \delta }}} \right]_{\delta  = 0}}.
\label{VID2.10}
\end{equation}

The elements of the resulting matrix linear in $\delta$ are in the first row

\begin{equation}
\begin{array}{l}
{M_{11}} = \frac{{i\pi \alpha   \left( {{k_0}\delta } \right)}}{8}{\left( {\frac{q_0 }{{k_0}}} \right)^2}\left\{ {H_m^{\left( 1 \right)}H_m^{\left( 2 \right)}{\rm{'}} - 2H_m^{\left( 1 \right)}{\rm{'}}H_m^{\left( 2 \right)}} \right.\\
\quad  + \left( {{k_0}b} \right)\left( {3H_m^{\left( 1 \right)}H_m^{\left( 2 \right)}{{\rm{'}}{\rm{'}}} - 3H_m^{\left( 1 \right)}{\rm{'}}H_m^{\left( 2 \right)}{\rm{'}} - 2H_m^{\left( 1 \right)}{{\rm{'}}{\rm{'}}}H_m^{\left( 2 \right)}} \right)\\
\quad \quad \quad \quad \quad  + \left. {{{\left( {{k_0}b} \right)}^2}\left( { - H_m^{\left( 1 \right)}{{\rm{'}}{\rm{'}}}H_m^{\left( 2 \right)}{\rm{' + }}H_m^{\left( 1 \right)}H_m^{\left( 2 \right)}{{\rm{'}}{\rm{'}}{\rm{'}}}} \right)} \right\};\\
{M_{12}} = \frac{{i\pi \alpha   \left( {{k_0}\delta } \right)}}{8}{\left( {\frac{q_0 }{{k_0}}} \right)^2}\left[ { - H_m^{\left( 1 \right)}{\rm{'}}H_m^{\left( 1 \right)}} \right.\\
\quad \quad  + \left( {{k_0}b} \right)\left( {H_m^{\left( 1 \right)}H_m^{\left( 1 \right)}{{\rm{'}}{\rm{'}}} - 3H_m^{\left( 1 \right)}{\rm{'}}H_m^{\left( 1 \right)}{\rm{'}}} \right)\\
\quad \quad  + \left. {{{\left( {{k_0}b} \right)}^2}\left( { - H_m^{\left( 1 \right)}{\rm{'}}H_m^{\left( 1 \right)}{\rm{'' + }}H_m^{\left( 1 \right)}H_m^{\left( 1 \right)}{{\rm{'}}{\rm{'}}{\rm{'}}}} \right)} \right];\\
{M_{13}} = \frac{{m\pi \alpha   }}{4}\left( {{k_0}\delta } \right)\left[ {H_m^{\left( 1 \right)}{\rm{'}}H_m^{\left( 2 \right)} + H_m^{\left( 1 \right)}H_m^{\left( 2 \right)}{\rm{'}}} \right]\frac{h}{{{k_0}}}\left( {\frac{q_0 }{{k_0}}} \right);\\
{M_{14}} = \frac{{m\pi \alpha   }}{4}\left( {{k_0}\delta } \right)\left[ {2H_m^{\left( 1 \right)}H_m^{\left( 1 \right)}{\rm{'}}} \right]\frac{h}{{{k_0}}}\left( {\frac{q_0 }{{k_0}}} \right),
\label{VID2.11}
\end{array}
\end{equation}
in the second row
\begin{equation}
\begin{array}{l}
{M_{21}} = \frac{{i\pi \alpha \left( {{k_0}\delta } \right)}}{8}{\left( {\frac{q_0 }{{k_0}}} \right)^2}\left[ {H_m^{\left( 2 \right)}H_m^{\left( 2 \right)}{\rm{'}}} \right.\\
\quad \quad  - \left( {{k_0}b} \right)\left( {H_m^{\left( 2 \right)}H_m^{\left( 2 \right)}{{\rm{'}}{\rm{'}}} - 3H_m^{\left( 2 \right)}{\rm{'}}H_m^{\left( 2 \right)}{\rm{'}}} \right)\\
\quad \quad  - \left. {{{\left( {{k_0}b} \right)}^2}\left( { - H_m^{\left( 2 \right)}{\rm{'}}H_m^{\left( 2 \right)}{\rm{'' + }}H_m^{\left( 2 \right)}H_m^{\left( 2 \right)}{{\rm{'}}{\rm{'}}{\rm{'}}}} \right)} \right];\\
{M_{22}} =  - \frac{{i\pi \alpha   \left( {{k_0}\delta } \right)}}{8}{\left( {\frac{q_0 }{{k_0}}} \right)^2}\left\{ {H_m^{\left( 2 \right)}H_m^{\left( 1 \right)}{\rm{'}} - 2H_m^{\left( 2 \right)}{\rm{'}}H_m^{\left( 1 \right)}} \right.\\
\quad  + \left( {{k_0}b} \right)\left( {3H_m^{\left( 2 \right)}H_m^{\left( 1 \right)}{{\rm{'}}{\rm{'}}} - 3H_m^{\left( 2 \right)}{\rm{'}}H_m^{\left( 1 \right)}{\rm{'}} - 2H_m^{\left( 2 \right)}{{\rm{'}}{\rm{'}}}H_m^{\left( 1 \right)}} \right)\\
\quad \quad \quad \quad \quad  + \left. {{{\left( {{k_0}b} \right)}^2}\left( { - H_m^{\left( 2 \right)}{{\rm{'}}{\rm{'}}}H_m^{\left( 1 \right)}{\rm{' + }}H_m^{\left( 2 \right)}H_m^{\left( 1 \right)}{{\rm{'}}{\rm{'}}{\rm{'}}}} \right)} \right\};\\
{M_{23}} =  - \frac{{m\pi \alpha   }}{4}\left( {{k_0}\delta } \right)\left[ {2H_m^{\left( 2 \right)}H_m^{\left( 2 \right)}{\rm{'}}} \right]\frac{h}{{{k_0}}}\left( {\frac{q_0 }{{k_0}}} \right);\\
{M_{24}} =  - \frac{{m\pi \alpha   }}{4}\left( {{k_0}\delta } \right)\left[ {H_m^{\left( 1 \right)}{\rm{'}}H_m^{\left( 2 \right)} + H_m^{\left( 1 \right)}H_m^{\left( 2 \right)}{\rm{'}}} \right]\frac{h}{{{k_0}}}\left( {\frac{q_0 }{{k_0}}} \right),
\end{array}
\label{VID2.12}
\end{equation}
in the third row
\begin{equation}
\begin{array}{l}
{M_{31}} =  - \frac{{m\pi \alpha }}{4}\left( {{k_0}\delta } \right)\left[ {H_m^{\left( 1 \right)}{\rm{'}}H_m^{\left( 2 \right)} + H_m^{\left( 1 \right)}H_m^{\left( 2 \right)}{\rm{'}}} \right]\frac{h}{{{k_0}}}\left( {\frac{q_0 }{{k_0}}} \right);\\
{M_{32}} =  - \frac{{m\pi  \alpha}}{4}\left( {{k_0}\delta } \right)\left[ {2H_m^{\left( 1 \right)}H_m^{\left( 1 \right)}{\rm{'}}} \right]\frac{h}{{{k_0}}}\left( {\frac{q_0 }{{k_0}}} \right);\\
{M_{33}} = \frac{{i\pi  \alpha\left( {{k_0}\delta } \right)}}{8}\left[ {H_m^{\left( 1 \right)}H_m^{\left( 2 \right)}{\rm{' + }}H_m^{\left( 1 \right)}{\rm{'}}H_m^{\left( 2 \right)}} \right.\\
\quad \quad  + \left. {\left( {{k_0}b} \right)\left( {H_m^{\left( 1 \right)}H_m^{\left( 2 \right)}{\rm{'' + }}2H_m^{\left( 1 \right)}{\rm{'}}H_m^{\left( 2 \right)}{\rm{' + }}H_m^{\left( 1 \right)}{{\rm{'}}{\rm{'}}}H_m^{\left( 2 \right)}} \right)} \right]\\
\quad \quad  + \frac{{i\pi \alpha \left( {{k_0}\delta } \right)}}{8}{\left( {\frac{q_0 }{{k_0}}} \right)^2}\left[ {H_m^{\left( 1 \right)}H_m^{\left( 2 \right)}{\rm{'}} - 2H_m^{\left( 1 \right)}{\rm{'}}H_m^{\left( 2 \right)}} \right.\\
\quad \quad  + \left( {{k_0}b} \right)\left( {3H_m^{\left( 1 \right)}H_m^{\left( 2 \right)}{{\rm{'}}{\rm{'}}} - 2H_m^{\left( 1 \right)}{{\rm{'}}{\rm{'}}}H_m^{\left( 2 \right)} - 3H_m^{\left( 1 \right)}{\rm{'}}H_m^{\left( 2 \right)}{\rm{'}}} \right)\\
\quad \quad \quad \quad \quad \quad  + \left. {{{\left( {{k_0}b} \right)}^2}\left( { - H_m^{\left( 1 \right)}{{\rm{'}}{\rm{'}}}H_m^{\left( 2 \right)}{\rm{' + }}H_m^{\left( 1 \right)}H_m^{\left( 2 \right)}{{\rm{'}}{\rm{'}}{\rm{'}}}} \right)} \right];\\
{M_{34}} = \frac{{i\pi \alpha \left( {{k_0}\delta } \right)}}{8}\left[ {2H_m^{\left( 1 \right)}H_m^{\left( 1 \right)}{\rm{'}}} \right.\\
\quad \quad  + \left. {\left( {{k_0}b} \right)\left( {2H_m^{\left( 1 \right)}{\rm{'}}H_m^{\left( 1 \right)}{\rm{'}} + 2H_m^{\left( 1 \right)}H_m^{\left( 1 \right)}{{\rm{'}}{\rm{'}}}} \right)} \right]\\
\quad \quad  + \frac{{i\pi  \alpha\left( {{k_0}\delta } \right)}}{8}{\left( {\frac{q_0 }{{k_0}}} \right)^2}\left[ {-H_m^{\left( 1 \right)}H_m^{\left( 1 \right)}{\rm{'}}} \right.\\
\quad \quad \quad \quad  + \left( {{k_0}b} \right)\left( {H_m^{\left( 1 \right)}H_m^{\left( 1 \right)}{\rm{'' - 3}}H_m^{\left( 1 \right)}{\rm{'}}H_m^{\left( 1 \right)}{\rm{'}}} \right)\\
\quad \quad  + \left. {{{\left( {{k_0}b} \right)}^2}\left( { - H_m^{\left( 1 \right)}{\rm{'}}H_m^{\left( 1 \right)}{{\rm{'}}{\rm{'}}} + H_m^{\left( 1 \right)}H_m^{\left( 1 \right)}{{\rm{'}}{\rm{'}}{\rm{'}}}} \right)} \right],
\end{array}
\label{VID2.13}
\end{equation}
and in the fourth row
\begin{equation}
\begin{array}{l}
{M_{41}} = \frac{{m\pi  \alpha}}{4}\left( {{k_0}\delta } \right)\left[ {2H_m^{\left( 2 \right)}H_m^{\left( 2 \right)}{\rm{'}}} \right]\frac{h}{{{k_0}}}\left( {\frac{q_0 }{{k_0}}} \right);\\
{M_{42}} = \frac{{m\pi \alpha }}{4}\left( {{k_0}\delta } \right)\left[ {H_m^{\left( 1 \right)}{\rm{'}}H_m^{\left( 2 \right)} + H_m^{\left( 1 \right)}H_m^{\left( 2 \right)}{\rm{'}}} \right]\frac{h}{{{k_0}}}\left( {\frac{q_0 }{{k_0}}} \right);\\
{M_{43}} = \frac{-{i\pi  \alpha\left( {{k_0}\delta } \right)}}{8}\left[ {2H_m^{\left( 2 \right)}H_m^{\left( 2 \right)}{\rm{'}}} \right.\\
\quad \quad  + \left. {\left( {{k_0}b} \right)\left( {2H_m^{\left( 2 \right)}{\rm{'}}H_m^{\left( 2 \right)}{\rm{'}} + 2H_m^{\left( 2 \right)}H_m^{\left( 2 \right)}{{\rm{'}}{\rm{'}}}} \right)} \right]\\
\quad \quad  + \frac{-{i\pi \alpha \left( {{k_0}\delta } \right)}}{8}{\left( {\frac{q_0 }{{k_0}}} \right)^2}\left[ {-H_m^{\left( 2 \right)}H_m^{\left( 2 \right)}{\rm{'}}} \right.\\
\quad \quad \quad \quad  + \left( {{k_0}b} \right)\left( {H_m^{\left( 2 \right)}H_m^{\left( 2 \right)}{\rm{'' - 3}}H_m^{\left( 2 \right)}{\rm{'}}H_m^{\left( 2 \right)}{\rm{'}}} \right)\\
\quad \quad  + \left. {{{\left( {{k_0}b} \right)}^2}\left( { - H_m^{\left( 2 \right)}{\rm{'}}H_m^{\left( 2 \right)}{{\rm{'}}{\rm{'}}} + H_m^{\left( 2 \right)}H_m^{\left( 2 \right)}{{\rm{'}}{\rm{'}}{\rm{'}}}} \right)} \right];\\
{M_{44}} =  - \frac{{i\pi  \alpha\left( {{k_0}\delta } \right)}}{8}\left[ {H_m^{\left( 2 \right)}H_m^{\left( 1 \right)}{\rm{' + }}H_m^{\left( 2 \right)}{\rm{'}}H_m^{\left( 1 \right)}} \right.\\
\quad \quad  + \left. {\left( {{k_0}b} \right)\left( {H_m^{\left( 2 \right)}H_m^{\left( 1 \right)}{\rm{'' + }}2H_m^{\left( 2 \right)}{\rm{'}}H_m^{\left( 1 \right)}{\rm{' + }}H_m^{\left( 2 \right)}{{\rm{'}}{\rm{'}}}H_m^{\left( 1 \right)}} \right)} \right]\\
\quad \quad  - \frac{{i\pi  \alpha\left( {{k_0}\delta } \right)}}{8}{\left( {\frac{q_0 }{{k_0}}} \right)^2}\left[ {H_m^{\left( 2 \right)}H_m^{\left( 1 \right)}{\rm{'}} - 2H_m^{\left( 2 \right)}{\rm{'}}H_m^{\left( 1 \right)}} \right.\\
\quad \quad  + \left( {{k_0}b} \right)\left( {3H_m^{\left( 2 \right)}H_m^{\left( 1 \right)}{{\rm{'}}{\rm{'}}} - 2H_m^{\left( 2 \right)}{{\rm{'}}{\rm{'}}}H_m^{\left( 1 \right)} - 3H_m^{\left( 2 \right)}{\rm{'}}H_m^{\left( 1 \right)}{\rm{'}}} \right)\\
\quad \quad \quad \quad \quad \quad  + \left. {{{\left( {{k_0}b} \right)}^2}\left( { - H_m^{\left( 2 \right)}{{\rm{'}}{\rm{'}}}H_m^{\left( 1 \right)}{\rm{' + }}H_m^{\left( 2 \right)}H_m^{\left( 1 \right)}{{\rm{'}}{\rm{'}}{\rm{'}}}} \right)} \right].
\end{array}
\label{VID2.14}
\end{equation}

The suppressed arguments are in all elements ${k_0}b = {k_0}\left( {a + d} \right)$.
To go further we introduce some short hand notation. Let
\begin{equation}
\begin{array}{*{20}{l}}
{{\bf{\tilde A}} = {{{\bf{\tilde M}}}_2};}\\
{{\bf{\tilde C}} = {\bf{\tilde B}} \cdot {\bf{\tilde A}} = {{\left[ {\frac{{\partial {\bf{\tilde M}}_0^1}}{{\partial \delta }}} \right]}_{\delta  = 0}} \cdot {\bf{\tilde A}};}\\
{{\bf{\tilde M}} \approx {\bf{\tilde A}} + \delta \alpha {\bf{\tilde C}}.}
\end{array}
\label{VID2.15}
\end{equation}
Expressed in terms of the elements of these matrices the condition for modes becomes
\begin{equation}
\begin{array}{*{20}{l}}
{\left[ {\left( {{A_{11}} + {A_{12}}} \right) + \delta \alpha \left( {{C_{11}} + {C_{12}}} \right)} \right]}\\
{ \times \left[ {\left( {{A_{33}} + {A_{34}}} \right) + \delta \alpha \left( {{C_{33}} + {C_{34}}} \right)} \right]}\\
{ - \left[ {\left( {{A_{13}} + {A_{14}}} \right) + \delta \alpha \left( {{C_{13}} + {C_{14}}} \right)} \right]}\\
{ \times \left[ {\left( {{A_{31}} + {A_{32}}} \right) + \delta \alpha \left( {{C_{31}} + {C_{32}}} \right)} \right] = 0},
\end{array}
\label{VID2.16}
\end{equation}
and to linear order in $\delta$
\begin{equation}
\begin{array}{*{20}{l}}
{\left( {{A_{11}} + {A_{12}}} \right)\left( {{A_{33}} + {A_{34}}} \right) - \left( {{A_{13}} + {A_{14}}} \right)\left( {{A_{31}} + {A_{32}}} \right)}\\
\begin{array}{l}
 + \delta \alpha \left[ {\left( {{A_{11}} + {A_{12}}} \right)\left( {{C_{33}} + {C_{34}}} \right)} \right.\\
 - \left( {{A_{13}} + {A_{14}}} \right)\left( {{C_{31}} + {C_{32}}} \right)
\end{array}\\
{ + \left. {\left( {{C_{11}} + {C_{12}}} \right)\left( {{A_{33}} + {A_{34}}} \right) - \left( {{C_{13}} + {C_{14}}} \right)\left( {{A_{31}} + {A_{32}}} \right)} \right]}\\
{ = 0.}
\end{array}
\label{VID2.17}
\end{equation}
From this we find the mode condition function is
\begin{equation}
\begin{array}{*{20}{l}}
{f = 1}\\
\begin{array}{l}
 + \delta \alpha \left[ {\left( {{A_{11}} + {A_{12}}} \right)\left( {{C_{33}} + {C_{34}}} \right)} \right.\\
 - \left( {{A_{13}} + {A_{14}}} \right)\left( {{C_{31}} + {C_{32}}} \right)\\
 + \left( {{C_{11}} + {C_{12}}} \right)\left( {{A_{33}} + {A_{34}}} \right)
\end{array}\\
{\left. { - \left( {{C_{13}} + {C_{14}}} \right)\left( {{A_{31}} + {A_{32}}} \right)} \right]/}\\
{\begin{array}{*{20}{l}}
{\left[ {\left( {{A_{11}} + {A_{12}}} \right)\left( {{A_{33}} + {A_{34}}} \right) - \left( {{A_{13}} + {A_{14}}} \right)\left( {{A_{31}} + {A_{32}}} \right)} \right]}\\
{ - \delta \alpha \left( {{B_{11}} + {B_{12}} + {B_{33}} + {B_{34}}} \right),}
\end{array}}
\end{array}
\label{VID2.18}
\end{equation}
or expressed in the ${{\bf{\tilde A}}}$ and ${{\bf{\tilde B}}}$ elements
\begin{equation}
\begin{array}{*{20}{l}}
{f = 1}\\
{ + \delta \alpha \left\{ {{B_{32}}\left[ {\left( {{A_{11}} + {A_{12}}} \right)\left( {{A_{23}} + {A_{24}}} \right)} \right.} \right.}\\
{\left. {\quad \quad  - \left( {{A_{13}} + {A_{14}}} \right)\left( {{A_{21}} + {A_{22}}} \right)} \right]}\\
 + {B_{34}}\left[ {\left( {{A_{11}} + {A_{12}}} \right)\left( {{A_{43}} + {A_{44}}} \right)} \right.\\
\quad \quad  - \left( {{A_{13}} + {A_{14}}} \right)\left( {{A_{41}} + {A_{42}}} \right)\\
\quad \quad  - \left( {{A_{11}} + {A_{12}}} \right)\left( {{A_{33}} + {A_{34}}} \right)\\
{\left. {\quad \quad  + \left( {{A_{13}} + {A_{14}}} \right)\left( {{A_{31}} + {A_{32}}} \right)} \right]}\\
 + {B_{12}}\left[ {\left( {{A_{33}} + {A_{34}}} \right)\left( {{A_{21}} + {A_{22}}} \right)} \right.\\
\quad \quad  - \left( {{A_{31}} + {A_{32}}} \right)\left( {{A_{23}} + {A_{24}}} \right)\\
\quad \quad  - \left( {{A_{11}} + {A_{12}}} \right)\left( {{A_{33}} + {A_{34}}} \right)\\
{\left. {\quad \quad  + \left( {{A_{13}} + {A_{14}}} \right)\left( {{A_{31}} + {A_{32}}} \right)} \right]}\\
{ + {B_{13}}\left[ {\left( {{A_{33}} + {A_{34}}} \right)\left( {{A_{41}} + {A_{42}}} \right)} \right.}\\
{\left. {\left. {\quad \quad  - \left( {{A_{31}} + {A_{32}}} \right)\left( {{A_{43}} + {A_{44}}} \right)} \right]} \right\}/}\\
{\left[ {\left( {{A_{11}} + {A_{12}}} \right)\left( {{A_{33}} + {A_{34}}} \right) - \left( {{A_{13}} + {A_{14}}} \right)\left( {{A_{31}} + {A_{32}}} \right)} \right].}
\end{array}
\label{VID2.19}
\end{equation}
Note that we have chosen as reference system a system of two independent ones, one with the solid cylinder alone and the other with the thin cylinder alone. The calculated energy is then the interaction energy between the two objects.

Now, the retarded (Casimir) interaction energy between an atom and a cylinder is given by
\begin{equation}
\begin{array}{*{20}{l}}
{E = \hbar \int\limits_0^\infty  {\frac{{d\xi }}{{2\pi }}} \sum\limits_{m =  - \infty }^\infty  {L\int\limits_{ - \infty }^\infty  {\frac{{dh}}{{2\pi }}} \ln \left[ {{f_m}\left( {h,i\xi } \right)} \right]} }\\
{ \approx \hbar \int\limits_0^\infty  {\frac{{d\xi }}{{2\pi }}} \sum\limits_{m =  - \infty }^\infty  {L\int\limits_{ - \infty }^\infty  {\frac{{dh}}{{2\pi }}} \left[ {{f_m}\left( {h,i\xi } \right) - 1} \right]} }\\
{ \approx \hbar \int\limits_0^\infty  {\frac{{d\xi }}{{2\pi }}} \sum\limits_{m =  - \infty }^\infty  {\int\limits_{ - \infty }^\infty  {\frac{{dh}}{{2\pi }}\frac{{2{\alpha ^{at}}}}{b}} } }\\
{ \times \left\{ {{B_{32}}\left[ {\left( {{A_{11}} + {A_{12}}} \right)\left( {{A_{23}} + {A_{24}}} \right)} \right.} \right.}\\
{\left. {\quad \quad  - \left( {{A_{13}} + {A_{14}}} \right)\left( {{A_{21}} + {A_{22}}} \right)} \right]}\\
{ + {B_{34}}\left[ {\left( {{A_{11}} + {A_{12}}} \right)\left( {{A_{43}} + {A_{44}}} \right)} \right.}\\
{\quad \quad  - \left( {{A_{13}} + {A_{14}}} \right)\left( {{A_{41}} + {A_{42}}} \right)}\\
{\quad \quad  - \left( {{A_{11}} + {A_{12}}} \right)\left( {{A_{33}} + {A_{34}}} \right)}\\
{\left. {\quad \quad  + \left( {{A_{13}} + {A_{14}}} \right)\left( {{A_{31}} + {A_{32}}} \right)} \right]}\\
{ + {B_{12}}\left[ {\left( {{A_{33}} + {A_{34}}} \right)\left( {{A_{21}} + {A_{22}}} \right)} \right.}\\
{\quad \quad  - \left( {{A_{31}} + {A_{32}}} \right)\left( {{A_{23}} + {A_{24}}} \right)}\\
{\quad \quad  - \left( {{A_{11}} + {A_{12}}} \right)\left( {{A_{33}} + {A_{34}}} \right)}\\
{\left. {\quad \quad  + \left( {{A_{13}} + {A_{14}}} \right)\left( {{A_{31}} + {A_{32}}} \right)} \right]}\\
{ + {B_{13}}\left[ {\left( {{A_{33}} + {A_{34}}} \right)\left( {{A_{41}} + {A_{42}}} \right)} \right.}\\
{\left. {\left. {\quad \quad  - \left( {{A_{31}} + {A_{32}}} \right)\left( {{A_{43}} + {A_{44}}} \right)} \right]} \right\}/}\\
{\left[ {\left( {{A_{11}} + {A_{12}}} \right)\left( {{A_{33}} + {A_{34}}} \right) - \left( {{A_{13}} + {A_{14}}} \right)\left( {{A_{31}} + {A_{32}}} \right)} \right],}
\end{array}
\label{VID2.20}
\end{equation}
where we have used that $\alpha  = 2{\alpha ^{at}}/\left( {bL\delta } \right)$. Remember that $b=a+d$ is the distance to the atom from the cylinder axis and $d$ is the closest distance from the atom to the cylinder. The radial coordinate in the $A$ elements is $a$ and in the $B$ elements is $b$. The ${{\bf{\tilde A}}}$ elements are equal to the ${{\bf{\tilde M}}}$ elements in Eq.\,(\ref{VID1.1}) and the ${{\bf{\tilde B}}}$ elements are equal to the ${{\bf{\tilde M}}}$ elements in Eqs.\,(\ref{VID2.11}) and (\ref{VID2.13}) after removal of a factor ${\delta \alpha }$.
\begin{equation}
\begin{array}{l}
\left( {{A_{11}} + {A_{12}}} \right)\left( {{A_{23}} + {A_{24}}} \right) - \left( {{A_{13}} + {A_{14}}} \right)\left( {{A_{21}} + {A_{22}}} \right)\\
 = {\left( {\frac{2}{{Wk_0^3}}} \right)^2}\frac{{k_1^4k_0^4}}{{{q_1}{q_0}}}\left[ {\frac{1}{{k_0^2}} - \frac{1}{{k_1^2}}} \right]\frac{{{q_0}}}{{{k_0}}}iW\left( {mh/a} \right)J_m^2\\
 = {\left( {\frac{2}{{Wk_0^3}}} \right)^2}\frac{{k_1^4k_0^4}}{{{q_1}{q_0}}}\left[ {\frac{1}{{k_0^2}} - \frac{1}{{k_1^2}}} \right]\frac{{4{q_0}}}{{\pi k_0^2a}}\left( {mh/a} \right)J_m^2;\\
\left( {{A_{11}} + {A_{12}}} \right)\left( {{A_{33}} + {A_{34}}} \right) - \left( {{A_{13}} + {A_{14}}} \right)\left( {{A_{31}} + {A_{32}}} \right)\\
 = {\left( {\frac{2}{{Wk_0^3}}} \right)^2}\frac{{k_0^4k_1^4}}{{{q_1}{q_0}}}{\left( {H_m^{\left( 1 \right)}{J_m}} \right)^2}\\
\quad \quad  \times \left[ {\left( {\frac{1}{{{k_1}}}\frac{{{J_m}'}}{{{J_m}}} - \frac{1}{{{k_0}}}\frac{{H_m^{\left( 1 \right)}{\rm{'}}}}{{H_m^{\left( 1 \right)}}}} \right)\left( {\frac{{q_1^2}}{{{k_1}}}\frac{{{J_m}'}}{{{J_m}}} - \frac{{q_0^2}}{{{k_0}}}\frac{{H_m^{\left( 1 \right)}{\rm{'}}}}{{H_m^{\left( 1 \right)}}}} \right)} \right.\\
\left. {\quad \quad \quad \quad \quad \quad \quad \quad \quad \quad \quad \quad  + {{\left( {mh/a} \right)}^2}{{\left[ {\frac{1}{{k_0^2}} - \frac{1}{{k_1^2}}} \right]}^2}} \right];\\
\left( {{A_{11}} + {A_{12}}} \right)\left( {{A_{43}} + {A_{44}}} \right) - \left( {{A_{13}} + {A_{14}}} \right)\left( {{A_{41}} + {A_{42}}} \right)\\
 = {\left( {\frac{2}{{Wk_0^3}}} \right)^2}\frac{{k_0^4k_1^4}}{{{q_1}{q_0}}}H_m^{\left( 1 \right)}H_m^{\left( 2 \right)}J_m^2\\
\quad \quad  \times \left[ {\left( {\frac{1}{{{k_1}}}\frac{{{J_m}'}}{{{J_m}}} - \frac{1}{{{k_0}}}\frac{{H_m^{\left( 1 \right)}{\rm{'}}}}{{H_m^{\left( 1 \right)}}}} \right)\left( { - \frac{{q_1^2}}{{{k_1}}}\frac{{{J_m}{\rm{'}}}}{{{J_m}}} + \frac{{q_0^2}}{{{k_0}}}\frac{{H_m^{\left( 2 \right)}{\rm{'}}}}{{H_m^{\left( 2 \right)}}}} \right)} \right.\\
\left. {\quad \quad \quad \quad \quad \quad \quad \quad \quad \quad \quad \quad  + {{\left( {mh/a} \right)}^2}{{\left[ {\frac{1}{{k_0^2}} - \frac{1}{{k_1^2}}} \right]}^2}} \right];\\
\left( {{A_{33}} + {A_{34}}} \right)\left( {{A_{21}} + {A_{22}}} \right) - \left( {{A_{31}} + {A_{32}}} \right)\left( {{A_{23}} + {A_{24}}} \right)\\
 = {\left( {\frac{2}{{Wk_0^3}}} \right)^2}\frac{{k_0^4k_1^4}}{{{q_1}{q_0}}}H_m^{\left( 1 \right)}H_m^{\left( 2 \right)}J_m^2\\
\quad \quad  \times \left[ {\left( {\frac{{q_1^2}}{{{k_1}}}\frac{{{J_m}{\rm{'}}}}{{{J_m}}} - \frac{{q_0^2}}{{{k_0}}}\frac{{H_m^{\left( 1 \right)}{\rm{'}}}}{{H_m^{\left( 1 \right)}}}} \right)\left( { - \frac{1}{{{k_1}}}\frac{{{J_m}{\rm{'}}}}{{{J_m}}} + \frac{1}{{{k_0}}}\frac{{H_m^{\left( 2 \right)}{\rm{'}}}}{{H_m^{\left( 2 \right)}}}} \right)} \right.\\
\left. {\quad \quad \quad \quad \quad \quad \quad \quad \quad \quad \quad \quad  + {{\left( {mh/a} \right)}^2}{{\left[ {\frac{1}{{k_0^2}} - \frac{1}{{k_1^2}}} \right]}^2}} \right];\\
\nonumber
\end{array}
\end{equation}
\begin{equation}
\begin{array}{l}
\left( {{A_{33}} + {A_{34}}} \right)\left( {{A_{41}} + {A_{42}}} \right) - \left( {{A_{31}} + {A_{32}}} \right)\left( {{A_{43}} + {A_{44}}} \right)\\
 = {\left( {\frac{2}{{Wk_0^3}}} \right)^2}\frac{{k_0^4k_1^4}}{{{q_1}{q_0}}}H_m^{\left( 1 \right)}H_m^{\left( 2 \right)}J_m^2\frac{{imh}}{a}\left[ {\frac{1}{{k_0^2}} - \frac{1}{{k_1^2}}} \right]\\
\quad \quad  \times \left[ {\frac{1}{{{q_1}}}\left( { - \frac{{q_1^2}}{{{k_1}}}\frac{{{J_m}{\rm{'}}}}{{{J_m}}} + \frac{{q_0^2}}{{{k_0}}}\frac{{H_m^{\left( 1 \right)}{\rm{'}}}}{{H_m^{\left( 1 \right)}}}} \right)} \right.\\
\left. {\quad \quad \quad \quad \quad \quad \quad \quad \quad \quad  + \frac{1}{{{q_0}}}\left( {\frac{{q_1^2}}{{{k_1}}}\frac{{{J_m}{\rm{'}}}}{{{J_m}}} - \frac{{q_0^2}}{{{k_0}}}\frac{{H_m^{\left( 2 \right)}{\rm{'}}}}{{H_m^{\left( 2 \right)}}}} \right)} \right].
\end{array}
\label{VID2.21}
\end{equation}

We have suppressed all arguments of the functions. The suppressed arguments are $\left( {{k _0}{a}} \right)$ for the $H$-functions and their derivatives and $\left( {{k _1}{a}} \right)$ for the $J$-functions and their derivatives. Note that the factor ${\left( {2/Wk_0^3} \right)^2}$ in common of all three $A$ expressions cancels out in the integrand of Eq.\,(\ref{VID2.20}).

Now, the arguments of the functions are all imaginary on the imaginary axis. It may be favorable to have real-valued arguments, $\left( {{\gamma _0}ha} \right)$, and $\left( {{\gamma _1}ha} \right)$ instead of  $\left( {{k _0}{a}} \right)$  and $\left( {{k _1}{a}} \right)$, respectively, where ${\gamma _0}\left( \omega  \right) = \sqrt {1 - {{\left( {\omega /ch} \right)}^2}} $ and ${\gamma _1}\left( \omega  \right) = \sqrt {1 - {\tilde \varepsilon }\left( \omega  \right){{\left( {\omega /ch} \right)}^2}} $, respectively. On the imaginary frequency axis these become real valued, ${\gamma _0}\left( {i\xi } \right) = \sqrt {1 + {{\left( {\xi /ch} \right)}^2}} $ and ${\gamma _1}\left( {i\xi } \right) = \sqrt {1 + {\tilde \varepsilon }\left( {i\xi } \right){{\left( {\xi /ch} \right)}^2}} $, respectively. To achieve real valued arguments we transform the functions to the modified Bessel functions ${I_m}\left( z \right)$ and ${K_m}\left( z \right)$.  The transformation rules are:\,\cite{Steg}
\begin{equation}
\begin{array}{*{20}{l}}
{H_m^{\left( 1 \right)}\left( {ix} \right) = \frac{2}{\pi }\frac{1}{{{i^{m + 1}}}}{K_m}\left( x \right);}\\
{H_m^{\left( 2 \right)}\left( {ix} \right) = 2{i^m}{S_m}\left( x \right);}\\
{{J_m}\left( {ix} \right) = {i^m}{I_m}\left( x \right);}\\
{H_m^{\left( 1 \right)}\left( {ix} \right){J_m}\left( {ix} \right) =  - \frac{2}{\pi }i{K_m}\left( x \right){I_m}\left( x \right);}\\
{H_m^{\left( 2 \right)}\left( {ix} \right){J_m}\left( {ix} \right) = 2{{\left( { - 1} \right)}^m}{S_m}\left( x \right){I_m}\left( x \right);}\\
{H_m^{\left( 1 \right)}\left( {ix} \right)H_m^{\left( 2 \right)}\left( {ix} \right) =  - \frac{4}{\pi }i{K_m}\left( x \right){S_m}\left( x \right),}
\end{array}
\label{VID2.22}
\end{equation}
where we have introduced the complex valued function, ${S_m}\left( x \right)$,  
\begin{equation}
{S_m}\left( x \right) ={\frac{1}{\pi }i{{\left( { - 1} \right)}^m}{K_m}\left( x \right) + {I_m}\left( x \right)}.
\label{VID2.23}
\end{equation}
The modified Bessel functions of real valued arguments are real valued. With these transformations and after a removal of a common factor the factors containing $A$ elements in Eq.(\ref{VID2.20}) become
\begin{equation}
\allowdisplaybreaks[4]
\begin{array}{*{20}{l}}
{\left( {{A_{11}} + {A_{12}}} \right)\left( {{A_{23}} + {A_{24}}} \right) - \left( {{A_{13}} + {A_{14}}} \right)\left( {{A_{21}} + {A_{22}}} \right)}\\
{ = \left[ {\frac{1}{{\gamma _0^2}} - \frac{1}{{\gamma _1^2}}} \right]\frac{{4m\left( {{q_0}/h} \right)}}{\pi }\frac{1}{{{{\left( {{\gamma _0}ha} \right)}^2}}};}\\
{\left( {{A_{11}} + {A_{12}}} \right)\left( {{A_{33}} + {A_{34}}} \right) - \left( {{A_{13}} + {A_{14}}} \right)\left( {{A_{31}} + {A_{32}}} \right)}\\
{ =  - {{\left( {\frac{2}{\pi }} \right)}^2}{{\left( { - 1} \right)}^m}{{\left( {{K_m}\left( {{\gamma _0}ha} \right)} \right)}^2}}\\
{\quad \quad  \times \left\{ {{{\left( {\frac{{{q_0}}}{h}} \right)}^2}\left[ {\frac{1}{{{\gamma _1}}}\frac{{{I_m}^\prime \left( {{\gamma _1}ha} \right)}}{{{I_m}\left( {{\gamma _1}ha} \right)}} - \frac{1}{{{\gamma _0}}}\frac{{{K_m}^\prime \left( {{\gamma _0}ha} \right)}}{{{K_m}\left( {{\gamma _0}ha} \right)}}} \right]} \right.}\\
{\quad \quad \quad  \times \left[ {\frac{{\tilde \varepsilon }}{{{\gamma _1}}}\frac{{{I_m}^\prime \left( {{\gamma _1}ha} \right)}}{{{I_m}\left( {{\gamma _1}ha} \right)}} - \frac{1}{{{\gamma _0}}}\frac{{{K_m}^\prime \left( {{\gamma _0}ha} \right)}}{{{K_m}\left( {{\gamma _0}ha} \right)}}} \right]}\\
{\left. {\quad \quad \quad \quad \quad \quad \quad \quad \quad  + {{\left( {\frac{m}{{ha}}} \right)}^2}{{\left[ {\frac{1}{{\gamma _0^2}} - \frac{1}{{\gamma _1^2}}} \right]}^2}} \right\};}\\
\nonumber
\end{array}
\end{equation} 
\begin{equation}
\begin{array}{l}
\left( {{A_{11}} + {A_{12}}} \right)\left( {{A_{43}} + {A_{44}}} \right) - \left( {{A_{13}} + {A_{14}}} \right)\left( {{A_{41}} + {A_{42}}} \right)\\
 =  - \frac{{4i}}{\pi }{K_m}\left( {{\gamma _0}ha} \right){S_m}\left( {{\gamma _0}ha} \right)\\
\quad \quad \quad  \times \left\{ {{{\left( {\frac{{{q_0}}}{h}} \right)}^2}\left[ {\frac{1}{{{\gamma _1}}}\frac{{{I_m}^\prime \left( {{\gamma _1}ha} \right)}}{{{I_m}\left( {{\gamma _1}a} \right)}} - \frac{1}{{{\gamma _0}}}\frac{{{K_m}^\prime \left( {{\gamma _0}ha} \right)}}{{{K_m}\left( {{\gamma _0}ha} \right)}}} \right]} \right.\\
\quad \quad \quad \quad \quad \quad  \times \left[ { - \frac{{\tilde \varepsilon }}{{{\gamma _1}}}\frac{{{I_m}^\prime \left( {{\gamma _1}ha} \right)}}{{{I_m}\left( {{\gamma _1}ha} \right)}} + \frac{1}{{{\gamma _0}}}\frac{{{S_m}^\prime \left( {{\gamma _0}ha} \right)}}{{{S_m}\left( {{\gamma _0}ha} \right)}}} \right]\\
\quad \quad \quad \quad \quad \quad \quad \quad \quad \quad \quad \quad \quad \left. { + {{\left( {\frac{{mh}}{a}} \right)}^2}{{\left[ {\frac{1}{{\gamma _0^2}} - \frac{1}{{\gamma _1^2}}} \right]}^2}} \right\};\\
\left( {{A_{33}} + {A_{34}}} \right)\left( {{A_{21}} + {A_{22}}} \right) - \left( {{A_{31}} + {A_{32}}} \right)\left( {{A_{23}} + {A_{24}}} \right)\\
 =  - \frac{{4i}}{\pi }{K_m}\left( {{\gamma _0}ha} \right){S_m}\left( {{\gamma _0}ha} \right)\\
\quad \quad \quad  \times \left\{ {{{\left( {\frac{{{q_0}}}{h}} \right)}^2}\left[ {\frac{{\tilde \varepsilon }}{{{\gamma _1}}}\frac{{{I_m}^\prime \left( {{\gamma _1}ha} \right)}}{{{I_m}\left( {{\gamma _1}ha} \right)}} - \frac{1}{{{\gamma _0}}}\frac{{{K_m}^\prime \left( {{\gamma _0}ha} \right)}}{{{K_m}\left( {{\gamma _0}ha} \right)}}} \right]} \right.\\
\quad \quad \quad \quad \quad \quad  \times \left[ { - \frac{1}{{{\gamma _1}}}\frac{{{I_m}^\prime \left( {{\gamma _1}ha} \right)}}{{{I_m}\left( {{\gamma _1}ha} \right)}} + \frac{1}{{{\gamma _0}}}\frac{{{S_m}^\prime \left( {{\gamma _0}ha} \right)}}{{{S_m}\left( {{\gamma _0}ha} \right)}}} \right];\\
\quad \quad \quad \quad \quad \quad \quad \quad \quad \quad \quad \left. {\quad \quad  + {{\left( {\frac{m}{{ha}}} \right)}^2}{{\left[ {\frac{1}{{\gamma _0^2}} - \frac{1}{{\gamma _1^2}}} \right]}^2}} \right\};\\
\left( {{A_{33}} + {A_{34}}} \right)\left( {{A_{41}} + {A_{42}}} \right) - \left( {{A_{31}} + {A_{32}}} \right)\left( {{A_{43}} + {A_{44}}} \right)\\
 = \frac{4}{\pi }\frac{m}{{ha}}\left( {\frac{{{q_0}}}{h}} \right)\left[ {\frac{1}{{\gamma _0^2}} - \frac{1}{{\gamma _1^2}}} \right]{K_m}\left( {{\gamma _0}ha} \right){S_m}\left( {{\gamma _0}ha} \right)\\
\quad \quad \quad  \times \left\{ {\frac{1}{{\sqrt {\tilde \varepsilon } }}\left[ { - \frac{{\tilde \varepsilon }}{{{\gamma _1}}}\frac{{{I_m}^\prime \left( {{\gamma _1}ha} \right)}}{{{I_m}\left( {{\gamma _1}ha} \right)}} + \frac{1}{{{\gamma _0}}}\frac{{{K_m}^\prime \left( {{\gamma _0}ha} \right)}}{{{K_m}\left( {{\gamma _0}ha} \right)}}} \right]} \right.\\
\left. {\quad \quad \quad \quad \quad \quad \quad \quad \quad  + \left[ {\frac{{\tilde \varepsilon }}{{{\gamma _1}}}\frac{{{I_m}^\prime \left( {{\gamma _1}ha} \right)}}{{{I_m}\left( {{\gamma _1}ha} \right)}} - \frac{1}{{{\gamma _0}}}\frac{{{S_m}^\prime \left( {{\gamma _0}ha} \right)}}{{{S_m}\left( {{\gamma _0}ha} \right)}}} \right]} \right\}.
\end{array}
\label{VID2.24}
\end{equation}

The $B$ elements in Eq.\,(\ref{VID2.20}) become
\begin{equation}
\begin{array}{*{20}{l}}
{{B_{11}} = \frac{1}{b}{{\left( {\frac{{{q_0}}}{{{\gamma _0}h}}} \right)}^2}}\\
{ \times \left[ {{m^2}{K_m}{S_m} - \left( {{\gamma _0}hb} \right)\frac{1}{2}{K_m}{S_m}^\prime  + {{\left( {{\gamma _0}hb} \right)}^2}{K_m}^\prime {S_m}^\prime } \right]}\\
{{B_{12}} = \frac{{ - i{{\left( { - 1} \right)}^m}}}{{b\pi }}{{\left( {\frac{{{q_0}}}{{{\gamma _0}h}}} \right)}^2}}\\
{ \times \left[ {{m^2}{{\left( {{K_m}} \right)}^2} - \left( {{\gamma _0}hb} \right)\frac{1}{2}{K_m}{K_m}^\prime  + {{\left( {{\gamma _0}hb} \right)}^2}{{\left( {{K_m}^\prime } \right)}^2}} \right]}\\
{{B_{13}} = \frac{1}{b}\frac{{mhi}}{{{q_0}}}{{\left( {\frac{{{q_0}}}{{{\gamma _0}h}}} \right)}^2}\left[ {1 + 2\left( {{\gamma _0}hb} \right){K_m}^\prime {S_m}} \right];}\\
{{B_{32}} = \frac{{ - {{\left( { - 1} \right)}^m}}}{{b\pi }}\frac{{mh}}{{{q_0}}}{{\left( {\frac{{{q_0}}}{{{\gamma _0}h}}} \right)}^2}2\left( {{\gamma _0}hb} \right){K_m}{K_m}^\prime ;}\\
{{B_{34}} = \frac{{ - i{{\left( { - 1} \right)}^m}}}{{b\pi }}}\\
{ \times \left\{ {\left[ {{m^2}{{\left( {{K_m}} \right)}^2} + {{\left( {{\gamma _0}hb} \right)}^2}\left[ {{{\left( {{K_m}^\prime } \right)}^2} + {{\left( {{K_m}} \right)}^2}} \right]} \right] + {{\left( {\frac{{{q_0}}}{{{\gamma _0}h}}} \right)}^2}} \right.}\\
{\left. { \times \left[ {{m^2}{{\left( {{K_m}} \right)}^2} - \left( {{\gamma _0}hb} \right)\frac{1}{2}{K_m}{K_m}^\prime  + {{\left( {{\gamma _0}hb} \right)}^2}{{\left( {{K_m}^\prime } \right)}^2}} \right]} \right\},}
\end{array}
\label{VID2.25}
\end{equation}
where all functions have the same suppressed argument, $\left( {{\gamma _0}b} \right)$. To arrive at these expressions we have made use of the modified Bessel equation and its derivative to rid us of second and third oder derivatives of the modified Bessel functions.

Since the derivations are rather involved one should make as many checks as possible. We have checked our results by taking the non-retarded limit of the resulting integrand in Eq.\,(\ref{VID2.20}) and have reproduced the integrand of Eq.\,(\ref{VIB5.3}). The force on the atom is ${\bf{F}}\left( b \right) =  - {\bf{\hat r}}{\rm d}E(b)/{\rm d}b$.

\subsubsection{\label{atom-outsidecylindershellret}Force on an atom outside a 2D cylindrical shell. (three layers)}
In this section we derive the interaction between an atom and a very thin cylindrical shell. It could approximate the interaction between an atom and a nano tube. The geometry is illustrated in Fig.\,\ref{figu20}. We let the shell have the thickness $\delta$ and let $\delta$ be very small so that one keeps only terms linear in $\delta$. The 3D dielectric function of the material will then be inversely proportional to  $\delta$.\,\cite{grap,arx} The derivation proceeds along the lines in the previous section and the matrix ${{\bf{\tilde A}}}$ is replaced according to  ${\bf{\tilde A}} \to \tilde 1 + {\bf{\tilde F}} \cdot {\bf{\tilde G}}$. This result is obtained in the following way. The matrix for the thin layer is obtained as the product of two matrices of the form of Eq.\,(\ref{VIC11}). The first is the matrix of the left interface at   $a+\delta$, the second of the next interface at  $a$. We keep terms up to linear order in $\delta$ only. The unit vector is the result of the zeroth order term and ${\bf{\tilde F}} \cdot {\bf{\tilde G}}$ of the first order term,  respectively.
\begin{equation}
\begin{array}{l}
\left( {1 + F{G_{11}} + F{G_{12}}} \right)\left( {F{G_{23}} + F{G_{24}}} \right)\\
 - \left( {F{G_{13}} + F{G_{14}}} \right)\left( {F{G_{21}} + 1 + F{G_{22}}} \right)\\
 = \delta \tilde \varepsilon {\left( { - 1} \right)^m}\frac{{\pi m{q_0}}}{{{\gamma _0}}}{I_m}\left( {{\gamma _0}ha} \right){I_m}^\prime \left( {{\gamma _0}ha} \right)\\
 - {\left( {\delta \tilde \varepsilon } \right)^2}{\left( { - 1} \right)^m}\left[ {i\pi {a^2}{\gamma _0}^2{h^2}{q_0}^2} \right]\\
 \times {K_m}^\prime \left( {{\gamma _0}ha} \right){I_m}^\prime \left( {{\gamma _0}ha} \right){S_m}\left( {{\gamma _0}ha} \right){I_m}\left( {{\gamma _0}ha} \right);\\
\left( {1 + F{G_{11}} + F{G_{12}}} \right)\left( {1 + F{G_{33}} + F{G_{34}}} \right)\\
 - \left( {F{G_{13}} + F{G_{14}}} \right)\left( {F{G_{31}} + F{G_{32}}} \right)\\
 = 1 + \delta \tilde \varepsilon \left\{ {{K_m}^\prime \left( {{\gamma _0}ha} \right){I_m}^\prime \left( {{\gamma _0}ha} \right)\left[ {{q_0}^2a} \right]} \right.\\
\left. { + {K_m}\left( {{\gamma _0}ha} \right){I_m}\left( {{\gamma _0}ha} \right)\left[ {\frac{{{m^2} + {{\left( {ah{\gamma _0}^2} \right)}^2}}}{{a{\gamma _0}^2}}} \right]} \right\};\\
\left( {1 + F{G_{11}} + F{G_{12}}} \right)\left( {F{G_{43}} + 1 + F{G_{44}}} \right)\\
 - \left( {F{G_{13}} + F{G_{14}}} \right)\left( {F{G_{41}} + F{G_{42}}} \right)\\
 = 1 + \delta \tilde \varepsilon \left\{ {{K_m}^\prime \left( {{\gamma _0}ha} \right){I_m}^\prime \left( {{\gamma _0}ha} \right)\left[ {{q_0}^2a} \right]} \right.\\
\left. { - {S_m}\left( {{\gamma _0}ha} \right){I_m}\left( {{\gamma _0}ha} \right){{\left( { - 1} \right)}^m}i\pi \left[ {\frac{{{m^2} + {{\left( {ah{\gamma _0}^2} \right)}^2}}}{{a{\gamma _0}^2}}} \right]} \right\}\\
 - {\left( {\delta \tilde \varepsilon } \right)^2}{\left( { - 1} \right)^m}i\pi {h^2}{a^2}{\gamma _0}^2{q_0}^2\\
 \times {K_m}^\prime \left( {{\gamma _0}ha} \right){I_m}^\prime \left( {{\gamma _0}ha} \right){I_m}\left( {{\gamma _0}ha} \right){S_m}\left( {{\gamma _0}ha} \right);\\
\left( {1 + F{G_{33}} + F{G_{34}}} \right)\left( {F{G_{21}} + 1 + F{G_{22}}} \right)\\
 - \left( {F{G_{31}} + F{G_{32}}} \right)\left( {F{G_{23}} + F{G_{24}}} \right)\\
 = 1 + \delta \tilde \varepsilon \left\{ {{S_m}^\prime \left( {{\gamma _0}ha} \right){I_m}^\prime \left( {{\gamma _0}ha} \right)\left[ { - {{\left( { - 1} \right)}^m}i\pi {q_0}^2a} \right]} \right.\\
\left. { + {K_m}\left( {{\gamma _0}ha} \right){I_m}\left( {{\gamma _0}ha} \right)\left[ {\frac{{{m^2} + {{\left( {ha{\gamma _0}^2} \right)}^2}}}{{a{\gamma _0}^2}}} \right]} \right\};\\
\left( {1 + F{G_{33}} + F{G_{34}}} \right)\left( {F{G_{41}} + F{G_{42}}} \right)\\
 - \left( {F{G_{31}} + F{G_{32}}} \right)\left( {F{G_{43}} + 1 + F{G_{44}}} \right)\\
 =  - \delta \tilde \varepsilon {\left( { - 1} \right)^m}\frac{{\pi m{q_0}}}{{{\gamma _0}}}{I_m}\left( {{\gamma _0}ha} \right){I_m}^\prime \left( {{\gamma _0}ha} \right).
\end{array}
\label{VID3.1}
\end{equation}
To find the interaction energy between an atom and the thin cylindrical shell we use Eq.\,(\ref{VID2.20}) and replace the factors containing $A$ elements, given in Eq.\,(\ref{VID2.24}) with the factors containing $FG$ elements, given in Eq.\,(\ref{VID3.1}). 

Two examples where the results apply are a cylinder made of a graphene like film and a thin metal film, respectively. Then the expressions for $\delta {\tilde \varepsilon }\left( {i\xi } \right)$ as given in Eq.\,(\ref{VB11.4}) can be used.\,\cite{grap,arx} 

We have checked our results by taking the non-retarded limit of the resulting integrand in Eq.\,(\ref{VID2.20}) and have reproduced the integrand of Eq.\,(\ref{VIB5.3}).




\section{\label{summary}Summary and discussion}

We have presented a general formalism for determining the electromagnetic normal modes in layered structures. We have furthermore shown how to calculate the dispersion energy and forces for these structures, both at zero and finite temperature. For the convenience of the reader we have derived in detail what is needed to address the three most common geometrical classes viz. the planar, spherical and cylindrical.
We have presented both non-retarded and fully retarded treatments. We have also given the resulting relations for a large number of illustrating examples.

Systems with a general number of layers can be handled and the thickness of each layer can have any value; even 2D layers are allowed which means that graphene, graphene-like, and 2D electron gases can be treated. 

Within the formalism it is possible to obtain the force on an atom inside or outside the layered structures. We have given many examples of this in the text. We have even derived the van der Waals and Casimir-Polder interactions between two atoms using the formalism for spherical structures in Secs. \ref{vdW} and \ref{Casimir}, respectively.

\end{document}